# Economic-based Distributed Resource Management and Scheduling for Grid Computing

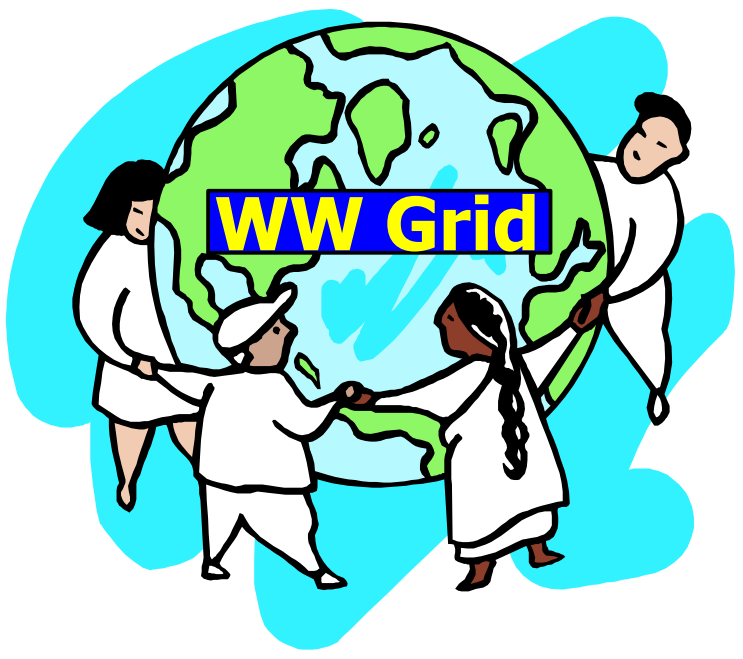

*by*

**Rajkumar Buyya**
B.E. (Mysore Univ.) and M.E. (Bangalore Univ.)
http://www.buyya.com

A thesis submitted in fulfillment of
the requirements for the Degree of

**Doctor of Philosophy**

MONASH UNIVERSITY

School of Computer Science and Software Engineering
Monash University, Melbourne, Australia

April 2002

# Table of Contents

# Table of Figures

# Economic-based Distributed Resource Management and Scheduling for Grid Computing

Rajkumar Buyya

Monash University
Melbourne, Australia

## Abstract

Computational Grids, emerging as an infrastructure for next generation computing, enable the sharing, selection, and aggregation of geographically distributed resources for solving large-scale problems in science, engineering, and commerce. As the resources in the Grid are heterogeneous and geographically distributed with varying availability and a variety of usage and cost policies for diverse users at different times and, priorities as well as goals that vary with time. The management of resources and application scheduling in such a large and distributed environment is a complex task. This thesis proposes a distributed computational economy as an effective metaphor for the management of resources and application scheduling. It proposes an architectural framework that supports resource trading and quality of services based scheduling. It enables the regulation of supply and demand for resources and provides an incentive for resource owners for participating in the Grid and motives the users to trade-off between the deadline, budget, and the required level of quality of service. The thesis demonstrates the capability of economic-based systems for peer-to-peer distributed computing by developing users' quality-of-service requirements driven scheduling strategies and algorithms. It demonstrates their effectiveness by performing scheduling experiments on the World-Wide Grid for solving parameter sweep applications.

# Declaration

I declare that this thesis is my own work and has not been submitted in any form for another degree or diploma at any university or other institute of tertiary education. Information derived from the published and unpublished work of others has been acknowledged in the text, and a list of references is given.

Rajkumar Buyya
April 10, 2002

# Acknowledgements

I am grateful to numerous local and global "peers" who have contributed towards shaping this thesis.

At the outset, I would like to express my appreciation to Professor David Abramson for his advice during my doctoral research endeavor for the past three years. As my supervisor, he has constantly enforced me to be in focus towards achieving the goal. His observations and comments helped me to establish the overall direction of the research and move forward with investigation in depth. I thank him for providing me with the opportunity to work with a talented team of researchers.

I would like to thank my former supervisor Professor Clemens Szyperski who helped me to start my doctoral research in Australia (Queensland University of Technology) with a smile. Surprisingly, within six months of my candidature, he announced his plan for joining Microsoft Research in the United States, which triggered my migration to Monash University in search of a new advisor. This migration resulted going beyond clusters: building technologies for world-wide computing and demonstrating their suitability for real applications. This happens to coincide with my original theme for PhD research dreamt in early 1990s!

I am grateful to Jonathan Giddy (Distributed Systems Technology Centre, Melbourne) for contributing significantly towards the implementation of a prototype Nimrod-G system based on the architecture and scheduling policies that we proposed and investigated in our research. He has been the greatest helping hand and a source of knowledge with real world programming skills specifically in Python!

I would like to express my sincere thanks to Manzur Murshed (Monash University, Gippsland Campus) for generously sharing his time and knowledge in our cooperative work on building a foundation infrastructure for grid simulation. He has played a major role in making me understand the concept of discrete-event simulation.

I thank Kim Branson (Walter and Eliza Hall Institute for Medical Research) for sharing information on the molecular docking application and data model that enabled me to develop a virtual laboratory environment. It helped me in demonstrating the suitability of our Grid technologies for enabling drug design application processing on the Grid.

I would like to thank Ian Foster (Argonne National Laboratory, USA) for his critical comments and suggestions during the initial phases of thesis work.

I am grateful to Mark Baker (University of Portsmouth, UK), Domenico Laforenza (CNR, Italy), Heinz Stockinger (CERN, Switzerland), Klaus Krauter (Redfern Networks, Australia), Muthucumaru Maheswaran (University of Manitoba, Canada), Steve Chapin (Syracuse University, USA), and David DiNucci (Elepar, USA) for sharing their knowledge that influenced the work and resulted in cooperative joint publications.

I greatly appreciate Rob Gray (Distributed Systems Technology Centre) for his generous time in proof reading. I would like to express my special thanks to him for being a great soul. He is always ready to help with a smile.

I thank Omer Rana (Cardiff University), Franck Cappello (University of Paris-Sud), Marcelo Pasin (Federal University of Santa Maria), Shoaib Ali Burq (University of Melbourne), Ajith Abraham (Monash University), Jagan Kommineni (Monash University), Rajanikanth Batchu (Mississippi State University), Slavisa Garic (DSTC), Andrew Hunter (Monash University), Rohit Patil, Sohel Khan (Icode Software), and Abhishek Bhagat (Digital India) for their comments and proof reading some of the chapters.

I thank John Crossley (Monash University), Jack Dongarra (University of Tennessee, Knoxville), Wolfgang Gentzsch (Sun Microsystems), Francine Berman (University of California, San Diego), Geoffrey Fox (Indiana University), Toni Cortes (Universitat Politecnica de Catalunya, Barcelona), Hai Jin (University of Southern California, Los Angeles), Savithri S (Motorola India Electronics Ltd., Bangalore), Sudharshan Vazhkudai, Spyros Lalis (FORTH, Greece), Jim Basney (University of Wisconsin), Colin Gan (University of Minnesota), Nirav Kapadia (Purdue University), John Karpovich (Virginia University), Andre Merzky (ZIB, Berlin), Jarek Nabrzyski (Poznan Supercomputing Centre), Hidemoto Nakada (Tokyo Institute of Technology), Michael Neary (UCSB), Joern Gehring (Paderborn Center for Parallel Computing, Germany), Achim Streit (Paderborn), Henri Casanova (UCSD), Lance Norskog (Enron), Scott Jackson (Pacific Northwest National Laboratory), and Craig Lee (Aerospace Corporation) for their comments on the early work.

I would like to thank colleagues and organisations for sharing their resources that have been loosely aggregated to create a World-Wide Grid (WWG) testbed. The organizations whose resources I have used in scheduling experiments reported in this thesis are: Monash University (Melbourne, Australia), Victorian Partnership for Advanced Computing (Melbourne, Australia), Argonne National Laboratories (Chicago, USA), University of Southern California's Information Sciences Institute (Los Angeles, USA), Tokyo Institute of Technology (Tokyo, Japan), National Institute of Advanced Industrial Science and Technology (Tsukuba, Japan), University of Lecce (Italy), and CNUCE-Institute of the Italian National Research Council (Pisa, Italy), Zuse Institute Berlin (Berlin, Germany), Charles University, (Prague, Czech Republic), University of Portsmouth (UK), and University of Manchester (UK).

I acknowledge Australian Government, Queensland University of Technology, Monash University, Cooperative Research Centre for Enterprise Distributed Systems Technology (DSTC), Distributed Systems and Software Engineering Centre, and the IEEE Computer Society for providing scholarships to pursue doctoral studies.

I would like to thank administrative and technical staff members of the school who have been kind enough to advice and help in their respective roles. I thank Thomas Peachey for making life fun while working.

Last, but not least, I would like to dedicate this thesis to my family (my wife Smrithi and daughter Soumya) for their love, patience, and understanding—they allowed me to spend most of the time on thesis. For the past six months, my daughter kept on reminding me: "Appaja, go to office and write your thesis"!

Rajkumar Buyya

April 2002

# Chapter 1

# Introduction

This chapter provides a high-level overview of the application of an economic-based model to Grid resource management and scheduling. It briefly presents the inspiration for computational Grids and our work. Then, it summarises the key components of economic-based distributed resource management and application scheduling and presents primary contributions of our research. The chapter ends with a discussion on the organization of the rest of this dissertation.

## 1.1 Inspiration for Computational Grids

Following Alessandro Volta's invention of the electrical battery in 1800, Thomas Edison and Nikola Tesla paved the way for electricity's widespread use by inventing the electric bulb and the alternating current (AC) respectively. Figure 1.1 shows Volta demonstrating the battery for Napoleon I in 1801 at the French National Institute, Paris. Whether or not Volta envisioned it, his invention evolved into a worldwide electrical power Grid that provides dependable, consistent, and pervasive access to utility power and has become an integral part of modern society.

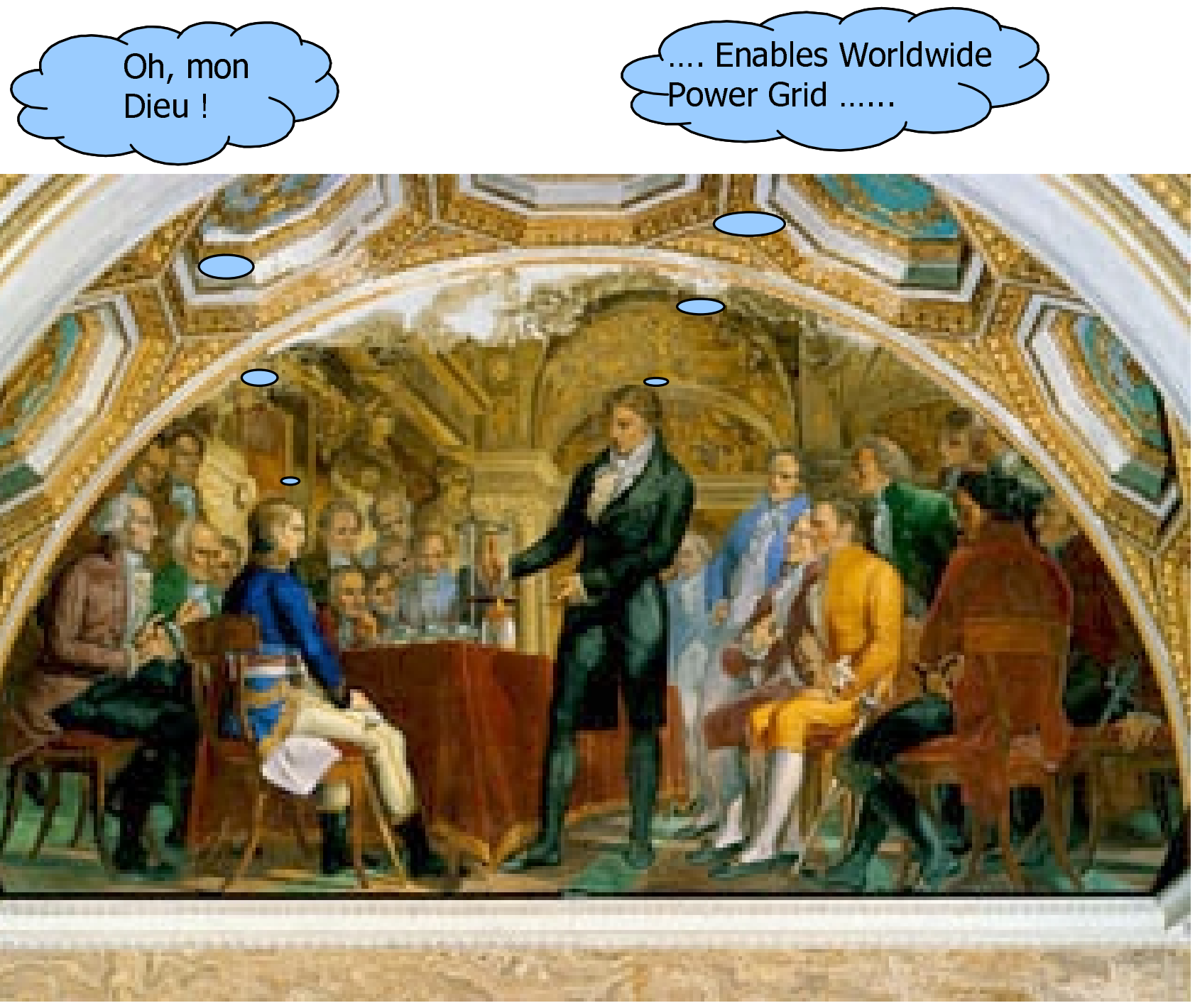

**Figure 1.1: Volta demonstrates the battery for Napoleon I at the French National Institute, Paris, in 1801.** The painting (by N. Cianfanelli, 1841) is from the Zoological Section of "La Specula" at the National History Museum, Florence University, Italy.

Inspired by the electrical power Grid's pervasiveness, ease of use, and reliability, computer scientists in the mid-1990s began exploring the design and development of an analogous infrastructure called the *computational power Grid* [48] for wide-area parallel and distributed computing. The motivation for computational Grids was initially driven by large-scale, resource (computational and data) intensive scientific applications that require more resource than a single computer (PC, workstation, supercomputer, or cluster) could provide in a single administrative domain. A Grid enables the sharing, selection, and aggregation of a wide variety of geographically distributed resources including supercomputers, storage systems, data sources, and specialized devices owned by different organizations for solving large-scale resource intensive problems in science, engineering, and commerce.

To build a Grid, the development and deployment of a number of services is required. They include *low-level services* such as security, information, directory, resource management (resource trading, resource allocation, quality of services) and *high-level services/tools* for application development, resource management and scheduling (resource discovery, access cost negotiation, resource selection, scheduling strategies, quality of services, and execution management) [48][73][100][99][106]. Among them, the two most challenging aspects of Grid computing are resource management and scheduling. This thesis presents a distributed computational economy framework and algorithms for the management of resources and scheduling of applications that are driven by the Quality-of-Service (QoS) requirements of the users.

## 1.2 Economic-based Grid Resource Management and Scheduling

Grid [48] and Peer-to-Peer (P2P) [4] computing platforms enable the sharing, selection, and aggregation of geographically distributed heterogeneous resources—such as computers and data sources—for solving large-scale problems in science, engineering, and commerce. However, resource management and scheduling in these environments is a complex undertaking. The geographic distribution of resources owned by different organizations with different usage policies, cost models and varying loads and availability patterns is problematic. The *producers (*resource owners) and *consumers* (resource users) have different goals, objectives, strategies, and requirements. To address these resource management challenges, we have proposed and developed a distributed computational economy-based[1] framework, called the **Gr**id **A**rchitecture for **C**omputational **E**conomy (GRACE) [98][99], for resource allocation and to regulate supply and demand of the available resources. This economic-based framework offers an incentive to resource owners for contributing and sharing resources; and motivates resource users to think about trade-offs between the processing time (e.g., deadline) and computational cost (e.g., budget), depending on their QoS requirements. We believe that this approach is essential for promoting the Grids as a mainstream computing paradigm, which can lead to the emergence of a new service-oriented computing industry.

The resource management and scheduling systems for Grid computing need to manage resources and application execution depending on resource consumers and owners requirements, and they need to continuously adapt to changes in the availability of resources. This requirement introduces a number of challenging issues that need to be addressed namely, site autonomy, heterogeneous substrate, policy extensibility, resource allocation or co-allocation, online control, resource trading, and quality of service-based scheduling. A number of Grid systems (such as Globus [63]) have addressed many of these issues with the exception of *resource trading* and *quality of service-based scheduling*. The GRACE framework has been proposed to address, particularly these two issues. The GRACE architecture leverages existing technologies such as Globus, and it provides new services that are essential for resource trading and aggregation, depending on their availability, capability, cost, and users QoS requirements. Therefore, we mainly focus on three aspects of resource trading and quality of service-based scheduling. First, to develop a generic distributed computational economy architectural framework and strategies for resource trading using different economic models. Second, to use supported resource trading services along with other middleware services in developing advanced user-centric Grid resource brokers with QoS driven scheduling algorithms. Finally, to develop a comprehensive Grid simulation toolkit to support *repeatable* performance evaluation of scheduling strategies for a range of Grid scenarios.

The idea of applying economics to resource management in distributed systems has been explored in previous research to help understand the potential benefits of market-based systems. For example, Spawn [14], Popcorn [87], Java Market [146], Enhanced MOSIX [147], JaWS [127], Xenoservers [23], D'Agents

[1] The terms "economic/economy-based" and "market-based" are synonymous and interchangeable.

[55], Rexec/Anemone [8], Mojo Nation [84], Mariposa [83], and Mungi [33]. Unfortunately, many of them were limited to experimental simulations. The systems that were implemented followed a monolithic architecture, which means they are hard to extend. They expect the users to develop resource-aware applications explicitly for their platform using their new programming interface (e.g., Spawn and Popcorn). Consequently, developing applications for such platforms is difficult because programmers have to address both the application development and resource allocation issues concurrently. This problem is overcome in Nimrod-G by separating application development and resource management issues. To enable the creation of parameter parallel/sweep applications, Nimrod-G provides a simple parameter specification language. The execution of such applications is managed by Nimrod-G, where the users define their quality of service requirements such the deadline, budget, and optimisation preference; and the Nimrod-G broker automatically handles the allocation of budget for each job and execution.

### 1.2.1 Assessing Wants and Needs

In an economic-based Grid computing environment, resource management systems need to provide mechanisms and tools that allow resource consumers (end users) and providers (resource owners) to express their requirements and facilitate the realization of their goals. Resource consumer's need:

- a utility model—how consumers demand resources and their preference parameters, and
- brokers that support strategies for resource discovery and application scheduling, depending on user requirements and that manage all issues associated with application execution.

The resource/service providers need tools and mechanisms for price generation schemes to increase system utilization and protocols that help them offer competitive services. For the market to be competitive and healthy, coordination mechanisms are required to help reach equilibrium price—the market price at which the supply of a service equals the quantity demanded.

Numerous economic theories and models including micro and macroeconomic principles have been proposed. They include,

- commodity market models,
- posted price models,
- bargaining models,
- tendering or contract-net models,
- auction models,
- bid-based proportional resource sharing models,
- cooperative bartering models, and
- monopoly and oligopoly models.

A detailed discussion on the use of these economic models within the GRACE framework can be found in Chapter 3.

### 1.2.2 The Nimrod-G Grid Resource Broker

Nimrod-G is a global resource management and scheduling system that supports deadline and budget-constrained algorithms for scheduling parameter sweep (task and data parallel) applications on global Grids [100]. It provides a simple *parameter specification language* for creating parameter-sweep applications. The domain experts (application-specific experts) can create a *plan* for parameter studies and use the Nimrod-G broker to handle all the issues related to the seamless management and execution, including resource discovery, mapping jobs to appropriate resources, data and code staging and gathering results from multiple Grid nodes back to the home node[2]. Depending on the user's requirements, it dynamically leases Grid services at runtime based on their availability, capability, and cost.

A diagram of high-level architecture and components of Nimrod-G is shown in Figure 1.2. The components of Nimrod-G are:

- A persistent task farming engine,

[2] a node from which a job request originates.

- A Grid explorer for resource discovery,
- A resource trading manager for establishing access price,
- A schedule advisor that maps jobs to resources using deadline and budget constrained scheduling algorithms.
- A dispatcher and actuators for deploying agents on Grid resources; and
- Agents for managing execution of Nimrod-G jobs on Grid resources.

When Nimrod-G deploys its agents on the Grid node at runtime, it is submitted to the local resource manager, which then allocates a compute node[3] to it for executing the job.

Nimrod-G provides a persistent Task-Farming Engine (TFE), which supports job management protocols and APIs. It can be used to create and plug-in user-defined scheduling policies and customized problem solving environments. For example, ActiveSheets [20] uses the Nimrod-G broker services to execute Microsoft Excel computations and cells on the Grid. The TFE coordinates resource trading, scheduling, data staging, execution, and gathering results from remote Grid nodes to the user's home transparently.

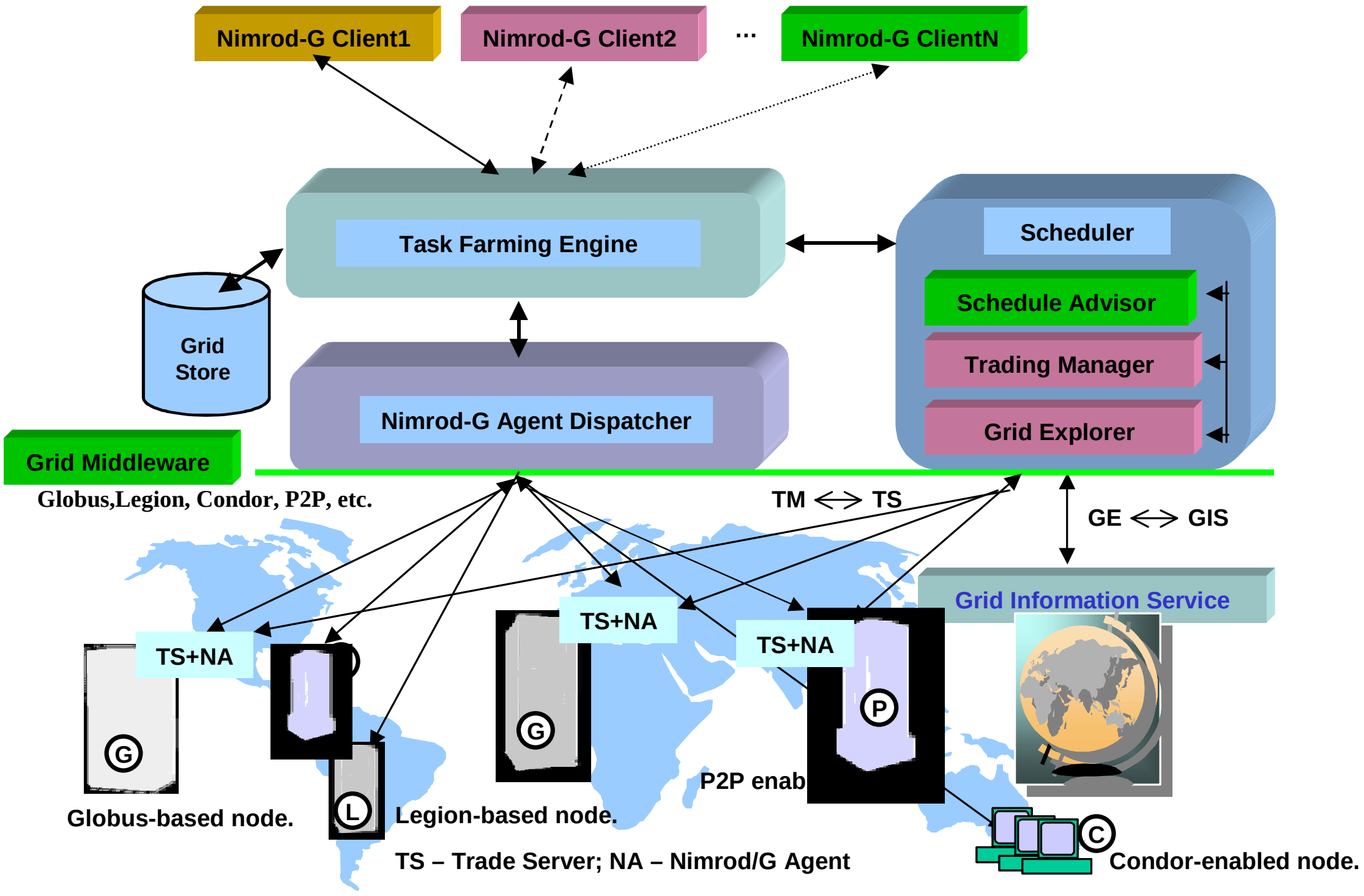

**Figure 1.2: Nimrod-G Grid resource broker.**

An associated dispatcher is capable of deploying computations (jobs) on Grid resources enabled by Globus [49], Legion [2], and Condor [79]. We have developed a number of deadline-based, market-driven scheduling algorithms: cost, time, conservative-time, and cost-time optimizations with deadline and budget constraints [107]. Once a price is established using a suitable economic model, depending on users requirements, our scheduling algorithms can lease resource services depending on their availability, capability, cost, and user-level performance.

We have used the Nimrod-G broker to schedule data-intensive computational applications (such as molecular modelling for drug design) on the World-Wide Grid (WWG) testbed that has resources located in five continents. The results of our scheduling experiments can be found in Chapter 4. A complete discussion of the use of Nimrod-G tools for formulating drug design application as a parameter sweep

[3] It is a node/CPU that the local resource manager allocates to the Nimrod-G agent for job processing.

application and processing them on the Grid can be found in Chapter 7.

### 1.2.3 GridSim Toolkit and Economic Grid Broker Simulator

In order to prove the effectiveness of resource brokers and associated scheduling algorithms, their performance needs to be evaluated under different scenarios such as varying the number of resources and users with different requirements. In a real Grid environment, it is hard and perhaps even impossible to perform scheduler performance evaluation in a repeatable and controllable manner for different scenarios—the availability of resources and their load continuously varies from time to time and it is impossible for an individual user/domain to control activities of other users in different administrative domains. To overcome this limitation, we have developed a Java-based discrete-event Grid simulation toolkit called GridSim. This toolkit supports modeling and simulation of heterogeneous Grid resources (both time- and space-shared), users, brokers, and application models. It provides primitives for the creation of application tasks, the mapping of tasks to resources, as well as their management. A detailed discussion of GridSim can be found in Chapter 5.

Using the GridSim toolkit, we developed an economic Grid resource broker that simulates the Nimrod-G broker. The simulator implements various "Deadline and Budget Constrained" (DBC) cost, time, cost-time, and conservative-time optimisation scheduling algorithms. We have simulated the WWG testbed resources and hypothetical task farming applications, to evaluate the performance of the scheduling algorithms through a series of simulations, by varying the number of users, deadlines, budgets, and optimisation strategies. The results of scheduling simulations can be found in Chapter 6.

## 1.3 Contributions

To support the thesis that an economic-based Grid resource management and scheduling systems can deliver significant value to users, resource providers and consumers, compared to traditional approaches, we have made several novel research contributions. They are as follows:

1. The thesis identifies and proposes distributed computational economy as a metaphor for effective management of resources in Grid computing environments, due to a large-scale heterogeneity, distribution, and decentralization present in them. It enables the regulation of supply and demand for resources. It offers an incentive to resource owners to contribute resources and motivates resource users to consider the trade-off between the time of results delivery and computational cost, depending on their quality of service requirements. To realize this, we have developed a distributed computational economy framework called the **Gr**id **A**rchitecture for **C**omputational **E**conomy (GRACE). The architecture is generic enough to accommodate different economic models and maps well onto the architecture of modern wide-area distributed systems. Its implementation leverages many existing technologies and provides additional services for resource trading and their aggregation. Thereby, we are able to abstract away implementation details and focus on how the system is delivering value to resource owners and consumers.
2. The thesis articulates the three key functionalities that economic-based Grid resource management and scheduling systems must support in order to increase the value of the utility and quality of services delivered to users (i.e., both resource providers/owners and consumers/end-users). To deliver enhanced value to users than traditional approaches, our framework provides (i) a means to express their requirements, valuations, and objectives, (ii) scheduling policies to translate them to resource allocations, and (iii) mechanisms to enforce the selection and allocation of differential services, and dynamic adaptation to changes in their availability at runtime. Grids need to use *competitive economic models* as different resource providers and resource consumers have different goals, objectives, strategies, and requirements that vary with time. Essentially, in market-based Grid systems, resource consumers adopt the strategy of solving their problems at low cost within a required timeframe and resource providers adopt the strategy of obtaining the best possible return on their investment. The users valuation of a utility of the same work and resources are time dependent. For example, end users with an immediate production schedule, value their computations much higher than others; the resource owners charge a higher price when there is a high demand for resources, and lower price when the demand is low.
3. The thesis presents the design and development of the Nimrod-G prototype system that realises the

system architecture by exploiting lower-level services provided by existing Grid technologies. It uses component-based layered architecture that enables the deployment of Nimrod-G on any emerging low-level Grid and P2P technologies with a minimal development effort. To realise the ultimate goal of delivering quality of services to users, the Nimrod-G Grid resource broker supports the deadline and budget constrained scheduling with user selected optimisation strategies, for processing large-scale task-and-data parallel (parameter-sweep) applications on globally distributed resources. The broker is capable of dynamically leasing Grid services/resources at runtime depending on their cost, capabilities, availability, and users' requirements. The broker has been deployed and used for scheduling large-scale applications (such as molecular modelling for drug design) on the WWG testbed.

4. The thesis presents **D**eadline and **B**udget **C**onstrained (DBC) scheduling algorithms with four different strategies: cost, time, conservative-time, and cost-time optimisations. The DBC *cost-optimisation* scheduling algorithm completes application processing by the deadline and minimizes the computational cost; the *time-optimisation* scheduling algorithm strives to complete application processing earlier than the deadline and within the budget limit; *conservative-time optimisation* is similar to time-optimisation, but ensures that each job has a minimum budget-per-job allocated and surplus is moved to other jobs only after their completion; and the *cost-time optimisation scheduling* algorithm is similar to cost-optimisation and tries to optimise for time without incurring extra expenses. The first three algorithms have been implemented within the Nimrod-G broker and a series of scheduling experiments have been conducted on the WWG testbed for different deadline, budget, optimisation strategies, application workloads, and resource groups.
5. The thesis presents the design and development of a toolkit, called GridSim, that supports discrete-event based simulation of Grid environments that allows *repeatable* performance evaluation under different scenarios, which is not possible in a real Grid environment as the availability of resources and their load continuously varies with time. The toolkit supports modeling and simulation of heterogeneous Grid resources (both time- and space-shared), users, brokers, and application models. It provides primitives for creation of application tasks, mapping of tasks to resources, and their management. A Nimrod-G like economic Grid resource broker is being simulated using the GridSim toolkit to evaluate the performance of deadline and scheduling algorithms through a series of simulations. It is achieved by varying the number of users, deadline, budget, and optimisation strategies and simulating geographically distributed Grid resources that resemble the WWG testbed. The results at microscopic level reveal their impact on the application processing cost and time, depending on user's requirements and valuations. They demonstrate the usefulness of allowing users to trade-off between the timeframe and processing cost depending on their QoS requirements.
6. The thesis demonstrates the effectiveness and application of Grid technologies for solving real-world problems by creating a Virtual Laboratory (Vlab) environment, which leverages existing tools and technologies including the Nimrod-G broker. We use molecular modelling for drug design as an example application to demonstrate the creation of a parameter parallel application and processing of its molecular docking jobs, that screen compounds in the Chemical DataBases (CDBs) to identify their potential as drug candidates. The Virtual Laboratory enables remote access to domain-specific databases (e.g., CDB) as a network service. The two scheduling experiments with cost and time optimisations using Nimrod-G for processing molecular docking jobs on the WWG testbed, demonstrate that users can express their valuations naturally by defining deadline, budget limits, and optimisation preference. The fact that users have the option of expressing their requirements, allows them to trade-off between the time for results delivery and the cost of computations, depending on the perceived value of utility at that time. They also prove that Grids indeed enable *sharing* and *aggregation* of geographically distributed resources for solving large-scale, resource intensive problems faster and cheaper.

## 1.4 Organization

The rest of this thesis is organized as follows. Chapter 2 presents state-of-the-art Grid technologies from areas concerned with traditional and computational economy based resource management systems. Chapter 3 proposes computational economy as a metaphor for effective management of distributed resources and application scheduling. A distributed computational economy framework, called the Grid Architecture for Computational Economy (GRACE), leverages existing technologies and provides additional services for resource trading and aggregation. It discusses the use of real-world economic models and strategies: commodity market, posted prices, bargaining, tendering, auction, proportional resource sharing, and cooperative bartering for resource management and scheduling within the GRACE framework.

Chapter 4 presents architecture and implementation of the Nimrod-G resource broker that uses a computational economy driven framework for managing resources and scheduling applications. It discusses the deadline and budget constrained scheduling algorithms and the results of scheduling parameter sweep applications on the World-Wide Grid resources using the Nimrod-G resource broker.

Chapter 5 discusses the design and implementation of GridSim, a toolkit for modelling and simulation of resources and application scheduling in large-scale parallel and distributed computing environments. Given this simulation toolkit, Chapter 6 briefly presents the development of an economic Grid broker simulator. Then, we discuss the DBC scheduling algorithms and their performance evaluation through a series of simulations by varying the number of users, deadline, budget, and optimisation strategies and simulating geographically distributed Grid resources.

Chapter 7 presents the design and development of a virtual laboratory environment that enables molecular modelling for drug design on the Grid using Nimrod-G tools. Finally, Chapter 8 presents conclusions and provides directions for future work.

## 1.5 Acknowledgements

During the three years of my Ph.D. candidature at Monash University, I have collaborated with a number of colleagues within the university and outside. The work resulted in joint publications from which some of the chapters are *partially* derived. Note that *"I have only used the material that I contributed"*, which is directly relevant to my research.

Chapter 2 is *partially* derived from the following joint publications.

- M. Baker, R. Buyya, and D. Laforenza, *The Grid: International Efforts in Global Computing*, International Conference on Advances in Infrastructure for Electronic Business, Science, and Education on the Internet (SSGRR 2000), l'Aquila, Rome, Italy, July 2000.
- K. Krauter, R. Buyya, and M. Maheswaran, *A Taxonomy and Survey of Grid Resource Management Systems for Distributed Computing*, International Journal of Software: Practice and Experience (SPE), Wiley Press, New York, USA, May 2002. (to appear).

Chapter 3 is *partially* derived from the following joint publications.

- R. Buyya, D. Abramson, and J. Giddy, *Economy Driven Resource Management Architecture for Computational Power Grids*, International Conference on Parallel and Distributed Processing Techniques and Applications (PDPTA 2000), Las Vegas, USA, 2000.
- R. Buyya, J. Giddy, and D. Abramson, *A Case for Economy Grid Architecture for Service-Oriented Grid Computing*, 10th IEEE International Heterogeneous Computing Workshop (HCW 2001), In conjunction with IPDPS 2001, San Francisco, California, USA, April 2001.
- R. Buyya, H. Stockinger, J. Giddy, and D. Abramson, *Economic Models for Management of Resources in Peer-to-Peer and Grid Computing*, SPIE International Conference on Commercial Applications for High-Performance Computing, August 20-24, 2001, Denver, USA. An extended version to appear in *The Journal of Concurrency and Computation: Practice and Experience* (CCPE), Wiley Press, 2002.

  **Comments:** H. Stockinger (CERN) has helped me to enhance the paper and contributed a section on Data Grid economy, which I have *not* included in the chapter.

Chapter 4 is *partially* derived from the following joint publications.

- R. Buyya, D. Abramson, and J. Giddy, *Nimrod-G: An Architecture for a Resource Management and Scheduling System in a Global Computational Grid*, The 4$^{th}$ International Conference on High Performance Computing in Asia-Pacific Region (HPC Asia 2000), Beijing, China.

- R. Buyya, J. Giddy, and D. Abramson, *An Evaluation of Economy-based Resource Trading and Scheduling on Computational Power Grids for Parameter Sweep Applications*, The Second Workshop on Active Middleware Services (AMS 2000), In conjunction with HPDC 2001, August 1, 2000, Pittsburgh, USA (Kluwer Academic Press).
- R. Buyya, J. Giddy, and D. Abramson, *A Case for Economy Grid Architecture for Service-Oriented Grid Computing*, 10th IEEE International Heterogeneous Computing Workshop (HCW 2001), In conjunction with IPDPS 2001, San Francisco, California, USA, April 2001.

  **Comments:** J. Giddy and I worked as members of the Nimrod-G project led by my PhD supervisor (Prof. D. Abramson). I have contributed towards the Nimrod-G architecture, distributed computational economy methodologies, scheduling strategies, and performed scheduling experiments on the World-Wide Grid testbed. J. Giddy has played a major role in implementing some of those concepts within Nimrod-G.

Chapter 5 and 6 are *partially* derived from the following joint publications.

- R. Buyya and M. Murshed, *GridSim: A Toolkit for the Modeling and Simulation of Distributed Resource Management and Scheduling for Grid Computing*, The Journal of Concurrency and Computation: Practice and Experience (CCPE), Wiley Press, USA, May 2002 (to appear).
- R. Buyya, M. Murshed, and D. Abramson, *A Deadline and Budget Constrained Cost-Time Optimization Algorithm for Scheduling Task Farming Applications on Global Grids*, The Eleventh IEEE International Symposium on High Performance Distributed Computing (HPDC-11), Edinburgh, Scotland, UK, July 24-26, 2002 (submitted).

  **Comments:** I have collaborated with M. Murshed in developing the GridSim Toolkit discussed in Chapter 5. While developing GridSim base module, I worked with M. Murshed and explored the use of discrete-event simulation techniques with SimJava package. I have developed and simulated a Nimrod-G like economic Grid resource broker using the GridSim toolkit, designed, developed, and evaluated performance of scheduling algorithms.

Chapter 7 is *partially* derived from the following joint publication.

- R. Buyya, K. Branson, J. Giddy, and D. Abramson, *The Virtual Laboratory: Enabling Molecular Modeling for Drug Design on the World Wide Grid*, Technical Report, Monash-CSSE-2001-103, Monash University, December 2001. An extended version to appear in *The Journal of Concurrency and Computation: Practice and Experience* (CCPE), Wiley Press, May 2002.

  **Comments:** I have collaborated with K. Branson and developed the *Virtual Laboratory* environment for enabling the processing of Drug Design application on the Grid using the Nimrod-G broker. He has provided me necessary application data and Nimrod scripts, which he explored on a Linux cluster.

# Chapter 2

# Grid Technologies and Resource Management Systems

This chapter presents an overview of Grid technologies with major emphasis on resource management and scheduling systems. It discusses some of the important technological advances that have led to the emergence of Grid computing. It presents the taxonomy of Grid resource management systems briefly followed by a survey some representative example systems.

## 2.1 Introduction

The last decade has seen a substantial increase in commodity computer and network performance, mainly as a result of faster hardware and more sophisticated software. Nevertheless, there are still problems, in the fields of science, engineering, and business, which cannot be effectively dealt with using the current generation of supercomputers. In fact, due to their size and complexity, these problems are often resource (computational and data) intensive and consequently require a variety of heterogeneous resources that are not available in a single organisation.

The ubiquity of the Internet as well as the availability of powerful computers and high-speed network technologies as low-cost commodity components is rapidly changing the computing landscape and society. These technology opportunities have led to the possibility of using wide-area distributed computers for solving large-scale problems, leading to what is popularly known as Grid computing [48]. The term Grid is chosen as an analogy to the electric power Grid that provides consistent, pervasive, dependable, transparent access to electricity irrespective of its source. Such an approach to network computing is known by several names: metacomputing, scalable computing, global computing, Internet computing, and more recently Peer-to-Peer (P2P) computing [4].

Grids enable the sharing, selection, and aggregation of a wide variety of resources including supercomputers, storage systems, data sources, and specialized devices (see Figure 2.1) that are geographically distributed and owned by different organizations for solving large-scale computational and data intensive problems in science, engineering, and commerce.

The concept of Grid computing started as a project to link geographically dispersed supercomputers, but now it has grown far beyond its original intent. The Grid infrastructure can benefit many applications, including collaborative engineering, data exploration, high throughput computing, distributed supercomputing, and service-oriented computing. Moreover, due to the rapid growth of the Internet and Web, there has been a growing interest in Web-based distributed computing, and many projects have been started and aim to exploit the Web as an infrastructure for running coarse-grained distributed and parallel applications. In this context, the Web has the capability to act as a platform for parallel and collaborative work as well as a key technology to create a pervasive and ubiquitous Grid-based infrastructure.

Grid applications (typically multi-disciplinary and large-scale processing applications) often couple resources that cannot be replicated at a single site, or may be globally located for other practical reasons (see Figure 2.1). These are some of the driving forces behind the foundation of global Grids. In this light, the Grid allows users to solve larger-scale problems by pooling together resources that could not be coupled easily before. Hence, the Grid is not only a computing infrastructure, for large applications, it is a technology that can bond and unify remote and diverse distributed resources ranging from meteorological sensors to data vaults, and from parallel supercomputers to personal digital organizers. As such, it will provide pervasive services to all users that need them.

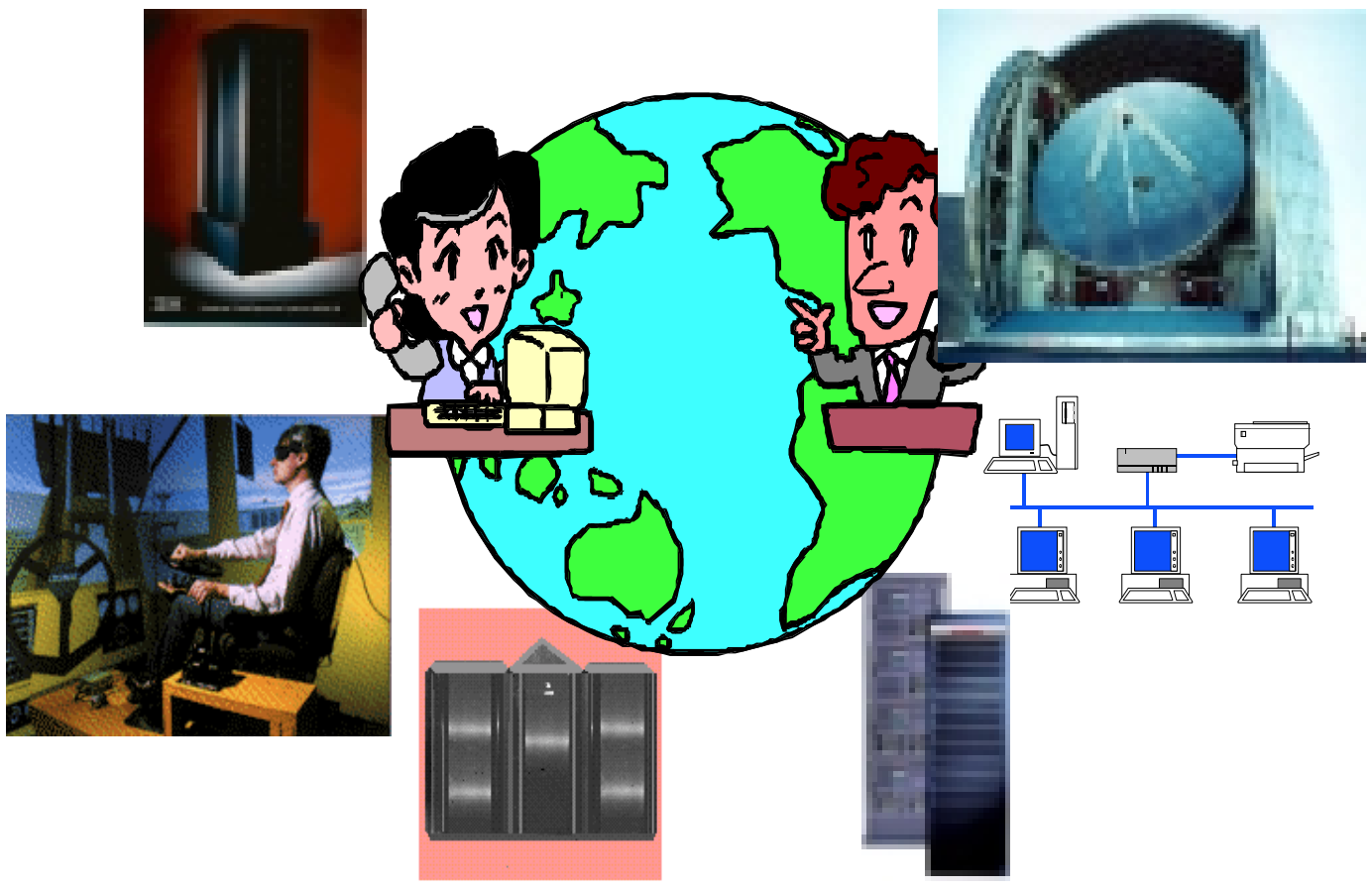

**Figure 2.1: Towards Grid computing: A conceptual view.**

A Grid can be viewed as a seamless, integrated computational and collaborative environment and a high level view of activities within the Grid is shown in Figure 2.3. The users interact with the Grid resource broker for solving problems, which in turn performs resource discovery, scheduling, and processing application jobs on the distributed Grid resources.

## 2.2 Major Technological Milestones: Enabling Grid and P2P Computing

The major technological advancements that happened from 1960 to date in computing and networking technologies that led to the emergence of P2P and Grid computing is shown in Figure 2.2. There has been the rise and fall of different systems. In 1960, mainframes mainly from IBM were serving the needs of computing users, but a decade later DEC introduced less expensive minicomputers that took over mainframes market share. During 1980s, vector computers (e.g., Crays) and later parallel computers (e.g., MPP systems) were serving the needs of grand challenging applications. We briefly discuss technological milestones in networking followed by computing.

The communication infrastructure for computational Grids is the Internet that began as a modest research network, supported by the Advanced Research Projects Agency (ARPA) of the US Defense Department. The ARPA's effort started as a response to the USSR's launch of Sputnik, the first artificial earth satellite in 1957 [121]. The ARPANET with four nodes was first established in 1969 at the University of California at Los Angeles, Stanford Research Institute, University of California Santa Barbara (UCSB), and University of Utah during September, October, November, and December months respectively. By the mid-1970s, the ARPANET *Internet* work embraced more than 30 universities, military sites and government contractors and its user base expanded to include the larger computer science research community. Bob Metcalfe's Harvard PhD Thesis outlines the idea for the Ethernet in 1973 that came into existence in 1976 [116]. Vint Cerf and Bob Kahn proposed the Transmission Control Program (TCP) in 1974, which was split into TCP/IP in 1978. By 1983, the network still consisted of a network of several hundred computers on only a few local area networks. In 1985, the National Science Foundation (NSF) arranged with ARPA to support a collaboration of supercomputing centers and computer science researchers across the ARPANET. In 1989, responsibility and management for the ARPANET, was officially passed from military interests to the academically oriented NSF. Much of the Internet's etiquette and rules for behavior were established during this time. The Internet Engineering Task Force (IETF) was formed during 1986 as a loosely self-organized group of people who contribute to the engineering and evolution of Internet technologies [125].

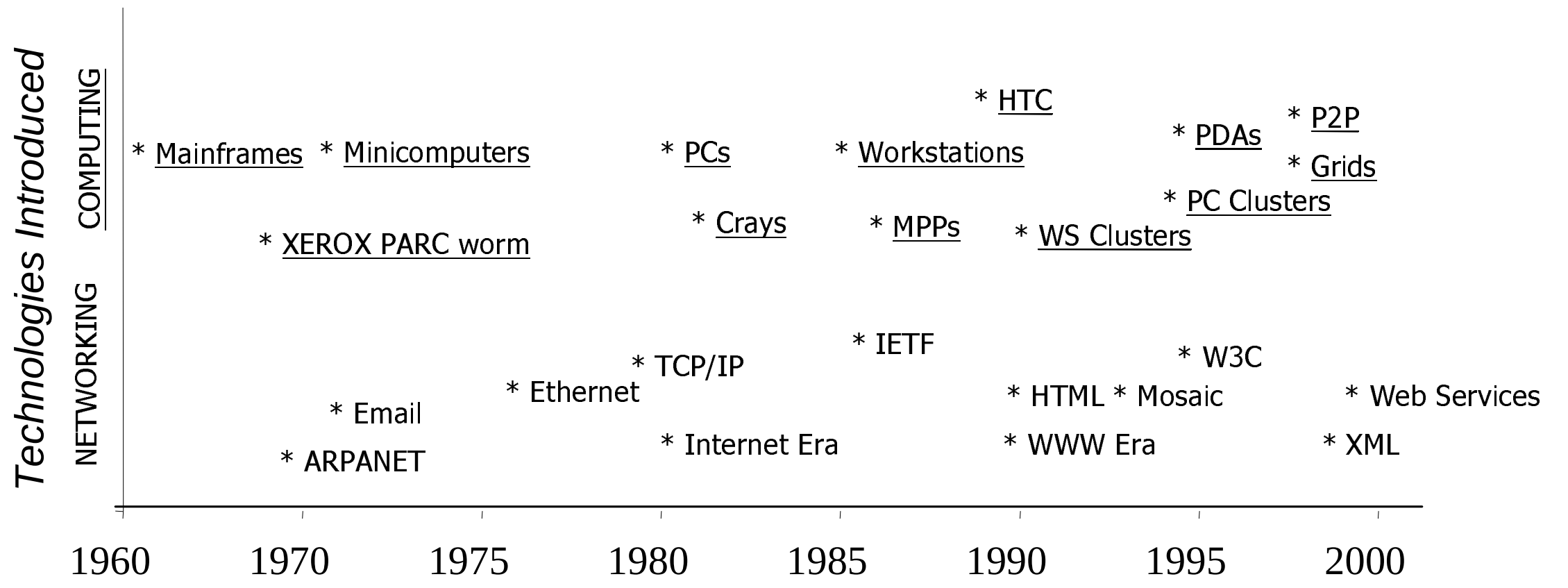

**Figure 2.2: Major milestones in networking and computing technologies from the year 1960 onwards.**

The invention of the Web [131] in 1989 by Tim Berners-Lee of CERN, Switzerland, for sharing information with ease has fueled a major revolution in computing. It provided the means for creating and organizing documents (using HTML language) with links and accessing them online transparently, irrespective of their location (using http protocols, browsers, and servers). The World-Wide Web consortium (W3C) [145] formed in 1994 is engaged in developing new standards for information interchange such as XML (eXtended Markup Language) Web services for providing remote access to software and applications as a service.

In the early 1970s when computers were first linked by networks, the idea of harnessing unused CPU cycles was born [138]. A few early experiments with distributed computing—including a pair of programs called *Creeper and Reaper*—ran on the Internet's predecessor, the ARPAnet. In 1973, the Xerox Palo Alto Research Center (PARC) installed the first Ethernet network and the first fully-fledged distributed computing effort was underway. Scientists at PARC developed a program called “worm” that routinely cruised about 100 Ethernet-connected computers. They envisioned their worm migrating from one machine to another to harness idle resources for beneficial purposes. The worm would roam throughout the PARC network, replicating itself in each machine's memory. Each worm used idle resources to perform a computation and had the ability to reproduce and transmit clones to other nodes of the network. With the worms, developers distributed graphic images and shared computations for rendering realistic computer graphics.

Since 1990, with the maturation and ubiquity of the Internet and Web technologies along with the availability of powerful computers and system area networks as commodity components, distributed computing scaled to a new global level. The availability of powerful PCs and workstations; and high-speed networks (e.g., Gigabit Ethernet) as commodity components has lead to the emergence of clusters [92] serving the needs of high performance computing (HPC) users. The ubiquity of the Internet and Web technologies along with the availability of many low-cost and high-performance commodity clusters within many organizations has prompted the exploration of aggregating distributed resources for solving large scale problems of multi-institutional interest. This has led to the emergence of computational Grids and P2P networks for sharing distributed resources. The Grid community is generally focused on aggregation of distributed high-end machines such as clusters, whereas the P2P community (e.g., SETI@Home [143]) is looking into sharing low-end systems such as PCs connected to the Internet and contents (e.g., exchange music files via Napster and Gnuetella networks). Given the number of projects and forums [74][91] started all over the world in early 2000, it is clear that interest in the research, development, and deployment of Grid and P2P computing technologies, tools, and applications is rapidly growing.

Already application domains like Monte Carlo simulations and parameter sweep applications (e.g., ionization chamber calibration [19], drug design [108], operations research, electronic CAD, and ecological modeling), where large processing problems can easily be divided into sub-problems and solved independently, are taking great advantage of Grid computing.

## 2.3 Grid Computing Environments

### 2.3.1 Resource Management Challenges

The Grid environment contains heterogeneous resources, fabric management systems (single system image OS, queuing systems, etc.) and policies, and applications (scientific, engineering, and commercial) with varied requirements (CPU, I/O, memory, and/or network intensive). The *producers* (also called resource owners) and *consumers* (who are the users) have different goals, objectives, strategies, and demand patterns [99]. More importantly both resources and end-users are geographically distributed with different time zones. A number of approaches for resource management architectures have been proposed and the prominent ones are: centralized, decentralized, and hierarchical.

In managing complexities present in large-scale Grid-like systems, traditional approaches are not suitable as they attempt to optimize system-wide measure of performance. Traditional approaches use centralized policies that need complete state information and a common fabric management policy, or decentralized consensus based policy. Due to the complexity in constructing successful Grid environments, it is impossible to define an acceptable system-wide performance matrix and common fabric management policy. Therefore, hierarchical and decentralized approaches are suitable for Grid resource and operational management [99]. Within these approaches, there exist different economic models for management and regulation of supply-and-demand for resources [104]. The Grid resource broker mediates between producers and consumers (see Figure 2.3). The resources are Grid enabled by deploying low-level middleware systems on them. The core middleware deployed on producer's Grid resources support the ability to handle resource access authorization and permits only authorized users to access them. The user-level and core middleware on consumer's resources support the ability to create Grid enabled applications or necessary tools to support the execution of legacy applications on the Grid. Upon authenticating to the Grid, consumers interact with resource brokers for executing their applications on remote resources. The resource broker takes care of resource discovery, selection, aggregation, data and program transportation, initiating execution on remote resources and gathering results.

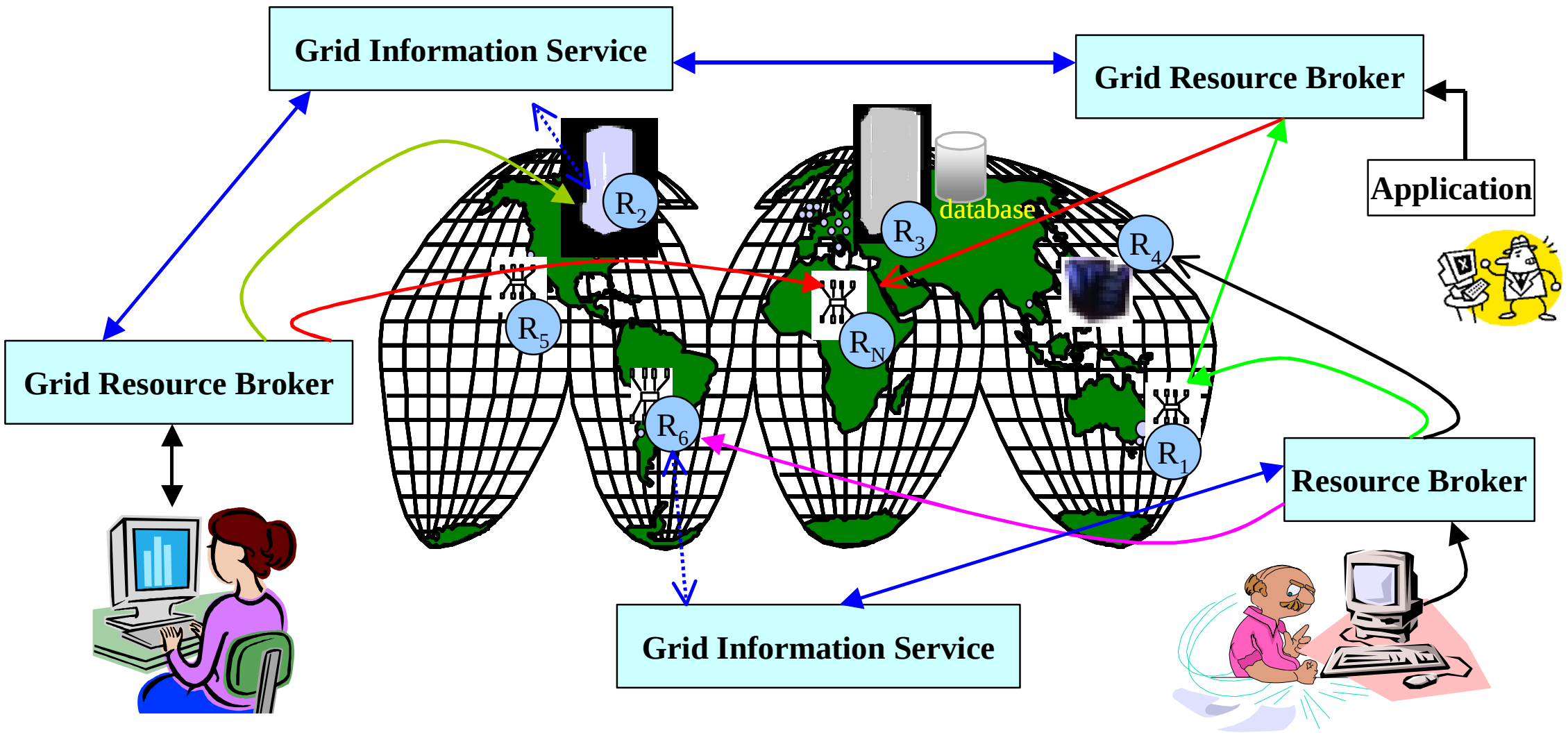

**Figure 2.3: A high-level view of the Grid and interaction between its entities.**

For the operation of a computational Grid, the broker discovers properties of resources that the user can access through the Grid information server(s), negotiates with (Grid-enabled) resources or their agents using middleware services, maps tasks to resources (scheduling), stages the application and data for processing (deployment) and finally gathers results [100]. It is also responsible for monitoring application execution progress along with managing changes in the Grid infrastructure and resource failures. There are a number of projects, worldwide, actively exploring the development of various Grid computing system components, services, and applications.

### 2.3.2 Grid Components

This section briefly highlights some of the general principles that underlie the construction of the Grid. In particular, the idealized design features that are required by a Grid to provide users with a seamless computing environment are discussed. Four main aspects characterise a Grid:

- *Multiple Administrative Domains and Autonomy:* Grid resources are geographically distributed across multiple administrative domains and owned by different organizations. The autonomy of resource owners needs to be honored along with their local resource management and usage policies.
- *Heterogeneity*: A Grid involves a multiplicity of resources that are heterogeneous in nature and will encompass a vast range of technologies.
- *Scalability*: A Grid might grow from a few integrated resources to millions. This raises the problem of potential performance degradation. Consequently, applications that require a large number of geographically located resources must be designed to be latency and bandwidth tolerant.
- *Dynamicity or Adaptability*: In a Grid, resource failure is the rule rather than the exception. In fact, with so many resources in a Grid, the probability of some resource failing is high. Resource managers or applications must tailor their behavior dynamically and use the available resources and services efficiently and effectively.

The steps necessary to realize a Grid include:

- The integration of individual software and hardware components into a combined networked resource (e.g., a single system image cluster).
- The deployment of:
  - Low-level middleware to provide a secure and transparent access to resources.
  - User-level middleware and tools for application development and the aggregation of distributed resources.
- The development and optimization of distributed applications to take advantage of the available resources and infrastructure.

The Grid is made up of a number of components from enabling resources to end user applications. A layered architecture of the Grid is shown in Figure 2.4. The key components of a Grid are:

- **Grid Fabric:** This consists of all the globally distributed resources that are accessible from anywhere on the Internet. These resources could be computers (such as PCs, SMPs, clusters) running a variety of operating systems (such as UNIX or Windows) as well as resource management systems such as LSF (Load Sharing Facility), Condor, PBS (Portable Batch System) or SGE (Sun Grid Engine), storage devices, databases, and special scientific instruments such as a radio telescope or particular heat sensor.
- **Core Grid Middleware:** This offers core services such as remote process management, co-allocation of resources, storage access, information registration and discovery, security, and aspects of Quality of Service (QoS) such as resource reservation and trading.
- **User-Level Grid Middleware:** This includes application development environments, programming tools, and resource brokers for managing resources and scheduling application tasks for execution on global resources.
- **Grid Applications and Portals:** Grid applications are typically developed using Grid-enabled languages and utilities such as MPI (message-passing interface) or Nimrod parameter specification language. An example application, such as parameter simulation or grand-challenge problem would require computational power, access to remote data sets, and may need to interact with scientific instruments. Grid portals offer Web-enabled application services, where the users can submit and collect results for their jobs on remote resources through the Web.

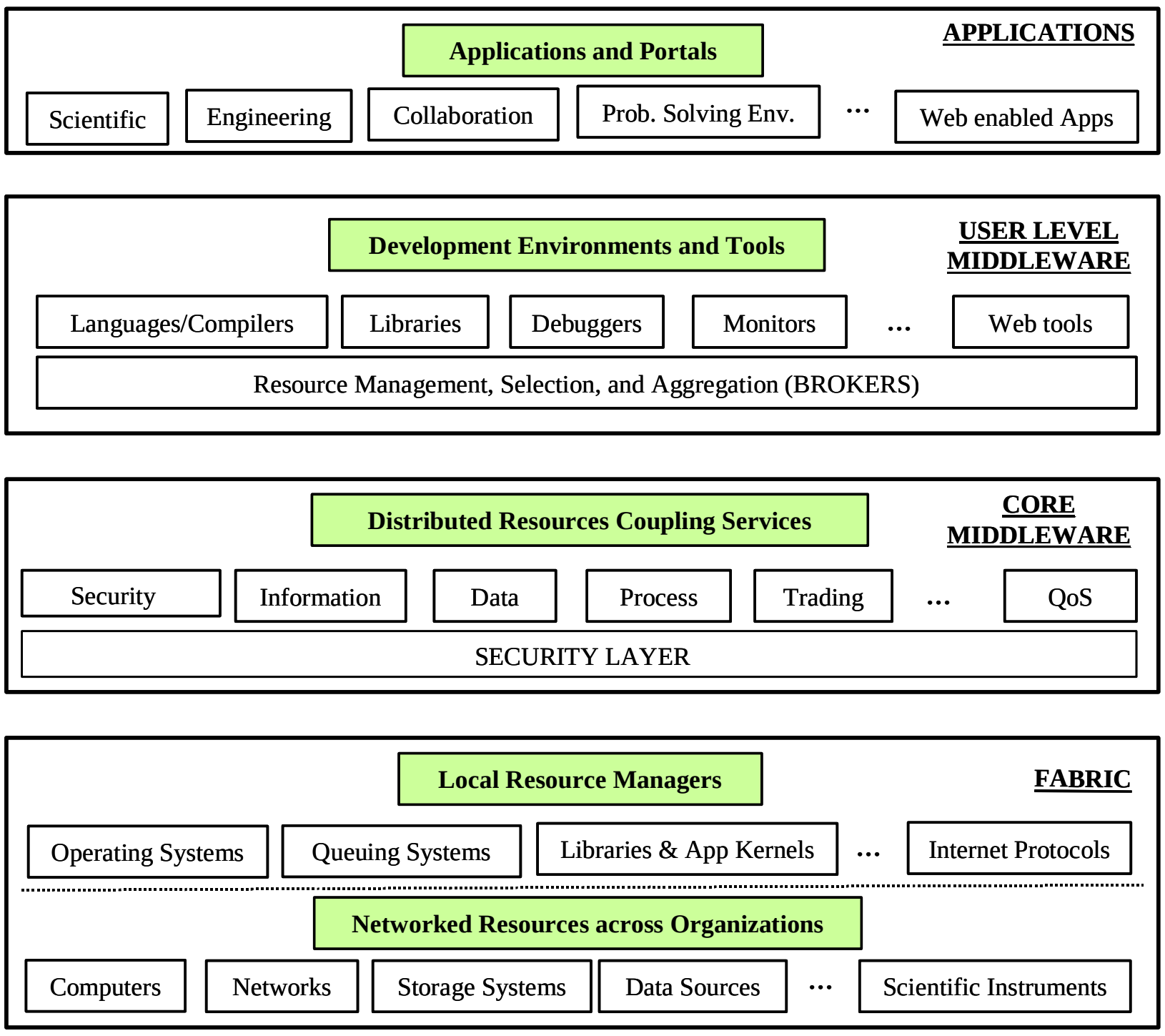

**Figure 2.4: A layered Grid architecture and components.**

### 2.3.3 Grid Computing Projects

There are many international Grid projects worldwide, which are hierarchically categorized as integrated Grid systems, core middleware, user-level middleware, and applications/application driven efforts (see Table 1). A listing of the majority of projects in Grid computing worldwide along with pointers to their websites can be found in [91][72]. Selected projects further grouped into country/continents and discussed in [74].

**Table 1: Hierarchical organization of major Grid efforts.**

| Category | Project | Organisation | Remarks |
|---|---|---|---|
| Integrated Grid Systems | NetSolve | U. Tennessee | A programming and runtime system for accessing high-performance libraries and resources transparently. |
| | Ninf | U. Tokyo | Functionality is similar to NetSolve. |
| | ST-ORM | UPC, Barcelona | A scheduler for distributed batch systems. |
| | MOL | | A scheduler for distributed batch systems. |
| | Albatross | Vrije U. | Object oriented programming system. |
| | PUNCH | Purdue U. | A portal computing environment and service for applications. |
| | Javelin | UCSB | Java-based programming and runtime system. |
| | XtremWeb | Paris-Sud U. | A global computing environment |
| | MILAN | Arizona and NY | Aims to provide end-to-end services for transparent utilization and management of networked resources |
| | DISCWorld | U. of Adelaide | A distributed information-processing environment. |
| | Unicore | Germany | Java-based environment for accessing remote supercomputers. |
| Core Middleware | Cosm | Mithral | A toolkit building P2P applications. |
| | Globus | ANL and ISI | Globus provides uniform and secure environment for accessing remote computational and storage resources. |
| | GRACE | Monash U. | A distributed computational economy framework for service oriented Grid computing. |
| | GridSim | Monash U. | A toolkit for Grid simulation. |
| | JXTA | Sun Microsystems | A Java-based framework and infrastructure for P2P computing. |
| | Legion | U. of Virginia | A Grid operating system providing transparent access to distributed resources. |
| | P2P Accelerator | Intel | A basic infrastructure for creating P2P applications for .NET platform. |
| User-level Middleware: *Schedulers* | AppLeS | UCSD | Application specific scheduler. |
| | Condor-G | U. of Wisconsin | A wide area job processing system. |
| | Nimrod-G | Monash U. | Economic-based Grid resource broker for parameter sweep/task farming applications. |
| User-level Middleware: *Programming Environments* | MPICH-G | Northern Illinois U. | MPI implementation on Globus. |
| | Nimrod parameter programming tools | Monash U. | A declarative language parametric programming. |
| | MetaMPICH | RWTH, Aachen | MPI programming and runtime environment. |
| | Cactus | Max Planck Institute for Gravitational Physics | A framework for writing parallel applications. It is developed using the MPICH-G and Globus. |
| | GrADS | Rice U. | Grid application development tools. |
| | GridPort | SDSC | Tools for creating computing portals. |
| Applications and application driven Grid efforts | European Data Grid | CERN | High Energy Physics, Earth Observation, Biology |
| | GriPhyN | UCF and ANL | High Energy Physics |
| | PPDG | Caltech and ANL | High Energy Physics |
| | Virtual Laboratory | Monash U and WEHI | Molecular modeling for drug design |
| | HEPGrid | Melbourne U | High Energy Physics applications |
| | NEESGrid | NCSA | Earthquake Engineering |
| | Geodise | Southampton U. | Aerospace Design Optimisation |
| | Fusion Grid | Princeton/ANL/ | Magnetic fusion |
| | IPG | NASA | Aerospace |
| | Active Sheets | Monash, QUT, & DSTC | Spread sheet processing |
| | Earth System Grid | LLNL, ANL, &NCAR | Climate Modeling |
| | Virtual Instruments | UCSD | Neuroscience |
| | National Virtual Observatory | Johns Hopkins U. & Caltech | Access to distributed astronomical databases and processing. |

Grids can be used to solve grand challenge problems in areas such as biophysics, chemistry, biology, scientific instrumentation [19], drug design [112], tomography [129], high energy physics [64], data mining, financial analysis, nuclear simulations, material science, chemical engineering, environmental studies, climate modeling [6], weather prediction, molecular biology, structural analysis, mechanical CAD/CAM and astrophysics. Although wide-area distributed supercomputing has been a popular application of the Grid, there are a large number of other applications that can benefit from the Grid [142] [75].

## 2.4 Resource Management Systems Taxonomy

Depending on the focus and application target, the Grid Resource Management Systems (RMSs) are broadly classified into Computational Grids, Data Grids, and Service Grids. In [65], taxonomy for Grid resource management systems is developed, which classifies resource management systems by characterizing different attributes as summarized in Table 2.2. The taxonomy focuses on the type of Grid system, machine organization, resource model characterization, and scheduling characterization.

**Table 2.2: Taxonomy of Grid resource management systems**

| Attributes of Resource Management Systems | Taxonomy |
|---|---|
| Grid Type (Service focus) | Computational Grids, Data Grids, Service Grids |
| Machine organization | Flat, cell (flat cells and hierarchical cells), hierarchical |
| Resource model | Schema, Object model (fixed or extensible) |
| Namespace organization | Relational, Hierarchical, Graph |
| QoS | Soft, Hard, None |
| Resource Information Store | Network Directory and Distributed Objects |
| Resource discovery | Query and Agents |
| Resource Info Dissemination | Batch/Period (push or pull), Online/On-demand |
| Scheduler organization | Centralised, Hierarchical, Decentralised |
| Scheduling policy | System-Centric, User Centric |
| State estimation | Predictive (Heuristics, Pricing models, machine learning) and Non-Predictive |
| Rescheduling | Periodic, Event Driven |

The organization of the machines in the Grid affects the communication patterns of the RMS and thus determines the scalability of the resultant architecture. In a flat organization all machines can directly communicate with each other directly. In a hierarchal organization machines at the same level can directly communicate with the machines directly above them or below them, or peer to them in the hierarchy. The fan out below a machine in the hierarchy is not relevant to the classification. Most current Grid systems use this organization since it has proven scalability. In a cell structure, the machines within the cell communicate between themselves using a flat organization. Designated machines within the cell function as boundary elements that are responsible for all communication outside the cell (e.g., a cluster with a master node directly accessible from outside and manages internal nodes).

The resource model determines how applications and the RMS describe and manage Grid resources. In a schema based approach the data that comprises a resource is described in a description language along with some integrity constraints (e.g., Condor ClassAd). In the object model extensible approach the resource model provides a mechanism to extend the definition of the object model managed by the RMS (e.g., Legion object model [2]). The resource namespace influences the design of the resource management protocols and affects the discovery methods. The quality of service (QoS) allows users to specify the level of service they are expecting from a resource and an RMS. The resource information store, which is updated based on dissemination model, provides Grid information services. The brokers can discover resources by querying the information store.

The Grid scheduling systems can be classified into centralized, hierarchical, and decentralized. As the resource availability in the Grid changes with time, the scheduling systems need to be adaptive. This is achieved by evaluating the current schedule (state estimation) based on predictive techniques; and then developing a new schedule (rescheduling) to meet the users requirements. The re-scheduling can be initiated periodically or whenever some event occurs (e.g., a notification of job completion).

## 2.5 Grid Resource Management Systems Survey

There are many existing Grid computing projects currently underway. They include Globus, Legion, Nimrod-G, NetSolve, AppLeS, and Condor. This section provides a brief description of each system and then classifies the resource management system attributes according to our taxonomy. A summary of architectural design choices made by a few popular Grid resource management systems is shown in Table 2.3.

**Table 2.3: Grid resource management systems and their architecture choices.**

| System | Grid Type | Organization | Resource: model, namespace, QoS, information store, discovery, dissemination | Scheduling: organisation, state-estimation, rescheduling, and policy. |
|---|---|---|---|---|
| **AppLeS** | Computational Grid (scheduling) | Hierarchical | Uses resource model provided by the underlying Globus, Legion, or NetSolve middleware services | Decentralized scheduler, predictive heuristic state estimation, online rescheduling, fixed application oriented policy (system-centric) |
| **DataGrid** | Data Grid<br>Computational Grid | Hierarchical | Extensible schema model, hierarchical namespace, no QoS, LDAP network directory store, distributed query-based discovery, periodic push dissemination. | Hierarchical schedulers, predictive heuristic state estimation, online rescheduling, extensible scheduling policy (user-centric) |
| **Condor** | Computational Grid | Flat | Extensible schema model, hybrid namespace, no QoS, other network directory store, centralized query based discovery, periodic push dissemination | Cooperative/Centralized scheduler |
| **Globus** | Grid Toolkit | Hierarchical Cells | Extensible schema model, hierarchical namespace, soft QoS, LDAP network directory store, distributed query based discovery, periodic push dissemination | Hierarchical scheduler, ad-hoc extensible policy |
| **Javelin** | Computational Grid | Hierarchical | Fixed object model, graph namespace, soft QoS, other network directory store, distributed query based discovery, periodic push dissemination | Decentralized scheduler, fixed application oriented policy |
| **Legion** | Computational Grid | Flat<br>Hierarchical | Extensible object model, graph namespace, soft QoS, object model store, distributed query-based discovery, periodic pull dissemination. | Hierarchical scheduler, ad-hoc extensible scheduling policies |
| **MOL** | Computational Grid | Hierarchical Cells | Extensible schema model, hierarchical namespace, no QoS, object model store, distributed query based discovery, | Decentralized scheduler, extensible ad-hoc scheduling policies |
| | | | periodic push dissemination | policies |
| **NetSolve** | Computational & Service Grid | Hierarchical | Extensible schema model, hierarchical namespace, soft QoS, centralized query-based discovery, periodic push dissemination. | Decentralized scheduler, fixed application oriented policy |
| **Nimrod-G** | Computational & Service Grid | Hierarchical Cells | Uses resource model provided by the underlying Globus or Legion middleware services and extends with computational economy approach | Decentralized scheduler, predictive pricing models, event driven rescheduling, fixed application oriented scheduling policy |
| **Ninf** | Computational & Service Grid | Hierarchical | Extensible schema model, relational namespace, no QoS, centralized query based resource discovery, periodic push for dissemination. | Decentralized scheduler |
| **PUNCH** | Computational & Service Grid | Hierarchical | Extensible schema model, hybrid namespace, soft QoS, distributed query-based discovery, periodic push dissemination. | Decentralized scheduler, machine learning, fixed system oriented policy |

### 2.5.1 AppLeS: A Network Enabled Scheduler

The AppLeS [28] (Application Level Scheduling) project at the University of California, San Diego primarily focuses on developing scheduling agents for individual applications on production computational Grids. It uses the services of Network Weather Service (NWS) to monitor changes in performance of resources dynamically. AppLeS agents use static and dynamic application and system information while selecting a viable set of resources and resource configurations. It interacts with other resource management systems such as Globus, Legion, and NetSolve to implement application tasks. The applications have embedded AppLeS agents and thus become self-schedulable on the Grid. The concept of AppLeS has been applied to many application areas including Magnetohydrodynamics [44], Gene Sequence Comparison, and Tomography [129].

Another effort within AppLeS project framework is the development of AppLeS templates. It is similar to Nimrod-G framework and resource broker, but it does not support quality of services-driven scheduling since it does not take Grid economy into consideration.

As the focus of AppLeS project is on scheduling, it follows the resource management model supported by the underlying Grid middleware systems. An AppLeS scheduler is central to the application that performs mapping of jobs to resources, but the local resource schedulers perform the actual execution of application units similar to Nimrod-G. AppLeS schedulers do not offer QoS support and build on resource model supported by an underlying system. AppLeS can be considered to have predictive heuristic state estimation model with online rescheduling and application oriented scheduling policies.

### 2.5.2 Condor: Cycle Stealing Technology for High Throughput Computing

Condor [79][54] is a high-throughput computing environment developed at the University of Wisconsin at Madison, USA. It can manage a large collection of computers such as PCs, workstations, and clusters that are owned by different individuals. Although it is popularly known for harnessing idle computers CPU cycles (cycle stealing), it can be configured to share resources. The Condor environment follows a layered architecture and offers powerful and flexible resource management services for sequential and parallel applications. The Condor system pays special attention to the computer owner's rights and allocates their resources to the Condor pool as per the usage conditions defined by resource owners. Through its unique remote system call capabilities, Condor preserves the job's originating machine environment on the execution machine, even if the originating and execution machines do not share a common file system and/or user ID scheme. Condor jobs with a single process are automatically checkpointed and migrated between workstations as needed to ensure eventual completion. The Condor has been extended to support submission of jobs to resources Grid-enabled using Globus servcies [57].

Condor can have multiple Condor pools and each pool follows a flat machine organization. The Condor *collector*, which provides the resource information store, listens for advertisements of resource availability. A Condor resource agent runs on each machine periodically advertising its services to the collector.

Customer agents advertise their requests for resources to the collector. The Condor matchmaker queries the collector for resource discovery that it uses to determine compatible resource requests and offers. The agents are then notified of their compatibility. The compatible agents then contact each other directly and if they are satisfied, then the customer agent initiates computation on the resource.

Resource requests and offers are described in the Condor classified advertisement (ClassAd) language [117]. ClassAds use a semi-structured data model for resource description. Thus no specific schema is required by the matchmaker allowing it to work naturally in a heterogeneous environment. The ClassAd language includes a query language as part of the data model, allowing advertising agents to specify their compatibility by including constraints in their resource offers and requests.

The matchmaker performs scheduling in a Condor pool. The matchmaker is responsible for initiating contact between compatible agents. Customer agents may advertise resource requests to multiple pools with a mechanism called flocking, allowing a computation to utilize resources distributed across different Condor pools.

The Condor system has recently been enhanced to support creation of personal condor pools. It allows the user to include their Globus-enabled nodes into the Condor pool to create a “personal condor” pool along with public condor pool nodes. The Grid nodes that are included in a personal condor pool are only accessible to the user who created the pool.

Condor can be considered as a computational Grid with a flat organization. It uses an extensible schema with a hybrid namespace. It has no QoS support and the information store is a network directory that does not use X.500/LDAP technology. Resource discovery is achieved through centralized queries with periodic push dissemination. The scheduler is centralized.

### 2.5.3 Data Grid

CERN, the European Organization for Nuclear Research, and the High-Energy Physics (HEP) community have established an International Data Grid project [141] with intent to apply the work to other scientific communities such as Earth Observation and Bioinfomatics. The project objectives are to establish a research network for data Grid technology development, demonstrate data Grid effectiveness through the large-scale real world deployment of end-to-end application experiments, and to demonstrate the ability to use low-cost commodity components to build, connect, and manage large general-purpose, data intensive computer clusters.

The Data Grid project focuses on the development of middleware services in order to enable a distributed analysis of physics data. The core middleware system is the Globus toolkit with extensions for data Grids. Data in the order of several Petabytes will be distributed in a hierarchical fashion to multiple sites worldwide. Global namespaces are required to handle the creation, access, distribution, and replication of data items. Special workload distribution facilities will balance analysis of jobs in the Grid to maximize the throughput from several hundred physicists. Application and user access monitoring will be used to optimize data distribution.

The Data Grid project has a hierarchical machine organization with less data stored at lower levels of the hierarchy. CERN, which is Tier 0, stores almost all relevant data with several Tier 1 regional centers in Italy, France, UK, USA, and Japan supporting smaller amounts of data. It has an extensible schema based resource model with a hierarchical namespace organization. It does not offer any QoS and the resource information store is expected to be based on an LDAP network directory. Resource dissemination is batched and periodically pushed to other parts of the Grid. Resource discovery in the Data Grid is decentralized and query based. The scheduler uses a hierarchical organization with an extensible scheduling policy.

### 2.5.4 Globus: A Toolkit for Grid Computing

Globus [49] provides a software infrastructure that enables applications to view distributed heterogeneous computing resources as a single virtual machine. The Globus project is an American multi-institutional research effort that seeks to enable the construction of computational Grids. Currently the Globus researchers are working together with the High-Energy Physics and the Climate Modeling community to build a data Grid [1]. A central element of the Globus system is the Globus Toolkit, which defines the basic services and capabilities required for constructing computational Grids. The toolkit consists of a set of components that implement basic services, such as security, resource location, resource management, data management, resource reservation, and communications. The toolkit provides a bag of services from which developers of specific tools or applications can select from, to meet their own particular needs. Globus is constructed as a layered architecture in which higher-level services can be developed using the lower level core services [63]. Its emphasis is on the hierarchical integration of Grid components and their services. This feature encourages the usage of one or more lower level services in developing higher-level services.

Resource and status information is provided via an LDAP-based network directory called Metacomputing Directory Services (MDS) [124]. MDS consists of two components, Grid Index Information Service (GIIS) and Grid Resource Information Service (GRIS). GRIS implements a uniform interface for querying resource providers on a Grid for their current configuration, capabilities, and status. GIIS pulls the information from multiple GRIS services and integrate it into a single coherent resource information database. The resource information providers use a push protocol to update GRIS.

Thus MDS follows both push and pull protocols for resource dissemination. Higher-level tools such as resource brokers can perform resource discovery by querying MDS using LDAP protocols. The MDS namespace is organized hierarchically in the form of a tree structure. Globus offers QoS in the form of resource reservation. Globus provides scheduling components as part of its toolkit approach but does not supply scheduling policies relying instead on higher-level schedulers. Globus services have been used in developing many global schedulers including, Nimrod-G, AppLeS, and Condor/G.

### 2.5.5 Javelin

Javelin [82] is a Java based infrastructure for internet-wide parallel computing. The three key components of Javelin system are the clients or applications, hosts, and brokers. A client is a process seeking computing resources, a host is a process offering computing resources, and a broker is a process that coordinates the allocation of computing resources. Javelin supports piecework and branch and bound models of computation. In the piecework model, adaptively parallel computations are decomposed into a set of sub-computations. The sub-computations are each autonomous in terms of communication, apart from scheduling work and communicating results. This model is suitable for parameter sweep or master-worker applications such as ray tracing and Monte Carlo simulations. The latest Javelin system, Javelin 2.0, supports branch-and-bound computations. It achieves scalability and fault-tolerance by integrating distributed deterministic work stealing with a distributed deterministic eager scheduler. An additional fault-tolerance mechanism is implemented for replacing hosts that have failed or retreated.

The Javelin system can be considered a computational Grid for high-throughput computing. It has a hierarchical machine organization where each broker manages a tree of hosts. Resources are simple fixed objects with a tree namespace organization. The resources are simply the hosts that are attached to a broker.

Any host that wants to be part of Javelin contacts JavelinBNS system, a Javelin information backbone that maintains list of available brokers. The host then communicates with brokers, chooses a suitable broker, and then becomes part of the broker-managed resources. Thus the information store is a network directory implemented by JavelinBNS. Hosts and brokers update each other as a result of scheduling work thus Javelin uses demand resource dissemination. The broker manages the host-tree or resource information through a heap-like data structure. Resource discovery uses the decentralized query based approach since queries are handled by the distributed set of brokers.

Javelin follows a decentralized approach in scheduling using work stealing and fixed application oriented scheduling policy. Whenever a host completes an assigned job, it requests work from peers and thus load balancing is achieved.

### 2.5.6 Legion: A Grid Operating System

Legion [123] is an object-based metasystem or Grid operating system developed at the University of Virginia. Legion provides the software infrastructure so that a system of heterogeneous, geographically distributed, high performance machines can seamlessly interact. Legion provides application users with a single, coherent, virtual machine. The Legion system is organized into classes and metaclasses.

Legion objects represent all components of the Grid. Legion objects are defined and managed by their class object or metaclass. Class objects create new instances, schedule them for execution, activate or deactivate the object, and provide state information to client objects. Each object is an active process that responds to method invocations from other objects within the system. Objects can be deactivated and saved to persistent storage. Objects are reactivated automatically when another object wants to communicate with it. Legion defines an API for object interaction, but does not specify the programming language or communication protocol.

Although Legion appears as a complete vertically integrated system, its architecture follows the hierarchical model. It uses an object based information store organization through the Collection objects. Collections periodically pull resource state information from host objects. Host objects track load and users can call the individual host directly to get the resource information. Information about multiple objects is aggregated into Collection objects. Users or system administrators can organize collections into suitable arrangements. Currently, there is a global collection named “/etc/Collection” for the system that tracks HostObjects and VaultObjects which embody the notion of persistent storage. The users or their agents can obtain information about resources by issuing queries to a Collection.

All Classes in Legion are organized hierarchically with LegionClass at the top and the host and vault classes at the bottom. It supports a mechanism to control the load on hosts. It provides resource reservation capability and the ability for application level schedulers to perform periodic or batch scheduling. Legion resource management architecture is hierarchical with decentralized scheduling policies. Legion supplies default system oriented scheduling policies, but it allows policy extensibility through resource brokers. That is, application level schedulers such as Nimrod-G [107] and AppLeS [44] can change Legion default scheduling policies with user-centric policies.

### 2.5.7 MOL: Metacomputing Online Kernel

The MOL initiative is developing technologies that aim at utilizing multiple WAN-connected high performance systems as a computational resource for solving large-scale problems that are intractable on a single supercomputer. One of the key components of MOL toolbox is the MOL-Kernel [58]. It offers basic generic infrastructure and core services for robust resource management that can be used to construct higher-level services, tools and applications. The MOL-Kernel manages the resources of an institution’s computing centers, provides a dynamic infrastructure for interconnecting these institutions, manages network faults, and provides access points for users.

The MOL-Kernel follows a three-tier architecture consisting of resource abstraction, management, and access layers containing resource modules, center management modules (CMM), and access modules respectively. The resource modules encapsulate metacomputing resources such as computing devices, scientific devices, applications and databases. All resource modules in a center are coordinated by CMM. This module is responsible for keeping its resources in a consistent state and makes them accessible outside of an institution. It acts as a gatekeeper and controls the flow of data between the center resources and external networks.

There is usually one CMM per institution, but it is possible to have multiple CMM in the case of large organizations. Failure of MOL-Kernel components results in only one institute becoming inaccessible. As long as a single CMM is available, the MOL-kernel remains operational. That means, organizations can leave or enter the metacomputing environment as they wish. The MOL-Kernel dynamically reconfigures itself to include or exclude the corresponding resources. In the MOL-kernel, CMM consistency is achieved by using a transaction-oriented protocol on top of virtual shared memory objects associated with each CMM. In order to make the global state available at all entry points, mirrored instances of shared active objects are maintained at each CMM. Whenever the state of a shared object changes, the new information is automatically distributed to the corresponding mirror instances. Extension of the MOL-Kernel is provided via typed messages and event handlers. Events are generated by user interaction with an access module or resource state changes. Messages are routed to either predefined or dynamically loaded custom event handlers.

The MOL follows a service Grid model with hierarchical cell-based machine organization. It adopts the schema based resource model and hierarchical name space organization. The global state is maintained in shared objects of each CMM (i.e., object based resource information storage). The resources and services themselves announce their initial presence to MOL (push protocol in information dissemination). The access modules/schedulers perform resource discovery and scheduling by querying shared objects. Although the resource model is schema based, its primary mode is service based. For example, if users request an application (e.g. CFD-simulation) with a certain quality of service. MOL then finds those computers, which have this application, installed and asks them "which of you are powerful enough to provide the requested quality of service?" (i.e., decentralized scheduler). It then selects one or more to execute the request.

### 2.5.8 NetSolve: A Network Enabled Computational Kernel

Netsolve [41] is a client-agent-server paradigm based network enabled application server. It is designed to solve computational science problems in a distributed environment. The Netsolve system integrates network resources and provides a desktop application interface. The intent of Netsolve is to hide parallel processing complexity from user applications and deliver parallel processing power to desktop users. Netsolve clients can be written in C, Fortran, Matlab, or use Web pages to interact with the server. A Netsolve server can use any scientific package to provide its computational software. All component communications use TCP/IP. Netsolve provides resource discovery, fault tolerance, and load balancing.

The Netsolve system follows the service Grid model with hierarchical cell-based machine organization. The Netsolve-agents act as an information repository and maintain the record of resources available in the network. As new nodes come up, information such as its location and its services are sent to the Netsolve agent. Thus Netsolve Agent uses push resource dissemination. The Netsolve agent also acts as a resource broker and performs resource discovery and scheduling. The user requests are passed to an agent that identifies the best resource, initiates computations on that resource, and returns the results. Agents may request the assistance of other Agents in identifying the best resources and scheduling. Thus Netsolve has decentralized scheduler organization.

### 2.5.9 Nimrod-G Grid Resource Broker

Nimrod-G [103] [107] is a Grid resource broker that allows managing and steering task farming applications on computational Grids. It uses an economic model for resource management and scheduling. Users formulate parameter studies using a declarative parametric modeling language or GUI with the experiment being run on the Grid. Nimrod-G provides resource discovery, resource trading, scheduling, resource staging on Grid nodes, result gathering, and final presentation to the user. Nimrod-G uses GRACE services to dynamically trade with resource owner agents to select appropriate resources. GRACE enabled Nimrod-G has been used for scheduling parameter sweep application jobs on the WWG testbed resources [106].

Nimrod-G follows hierarchical and computational market model in resource management [107]. It uses the services of Grid middleware systems such as Globus and Legion for resource discovery and uses either a network directory or object model based data organization. It supports resource reservation and QoS through the computational economy services of the GRACE infrastructure. The users specify QoS requirements such as the deadline, budget, and preferred optimisation strategy. The Grid resource estimation is performed through heuristics and historical load profiling. Scheduling policy is application oriented and is driven by user defined requirements such as deadline and budget limitations. The load balancing is performed through periodic rescheduling.

### 2.5.10 Ninf: A Network Enabled Server

Ninf is a client server based network infrastructure for global computing [45] similar to NetSolve. It allows access to multiple remote compute and database servers. Ninf clients access remote computational resources from languages such as C and Fortran using the Ninf client library. The Ninf client library calls can be synchronous or asynchronous in nature. The key components of the Ninf system are the Ninf client library, the Ninf metaserver, and the Ninf remote libraries. Ninf applications invoke Ninf library functions that generate requests. Requests are sent to the Ninf metaserver that maintains the information of Ninf servers in the network using an LDAP directory. The Ninf metaserver allocates resources on the appropriate servers for load balancing or scheduling. Ninf computational resources register details of available library services with the Ninf metaserver thus using a push protocol for resource dissemination. Ninf follows a flat model in machine organization, schema for resource model, and relational name space organization. Ninf-metaservers performs resource brokering, but actual scheduling is done using extensible policies.

#### 2.5.11 PUNCH: The Purdue University Network Computing Hubs

PUNCH [85][86] is a middleware testbed that provides operating system services in a network-based computing environment. The PUNCH infrastructure allows seamless management of applications, data, and machines distributed across wide-area networks. Users can run applications via standard Web browsers without requiring application changes.

PUNCH employs a hierarchically distributed architecture with several layers. A computing portal services layer provides Web-based access to a distributed, network-computing environment. This layer primarily deals with content management and user-interface issues. A network OS layer provides distributed process management and data browsing services. An application middleware layer allows the infrastructure to interoperate with other application-level support systems such as PVM [139] and MPI [140]. A virtual file system layer consists of services that provide local access to distributed data in an application-transparent manner. Finally, an OS middleware layer interfaces with local OS services available on individual machines or clusters of machines. The layers interoperate with a distributed resource management system and a predictive performance modeling sub-system in order to make intelligent resource allocation decisions.

## 2.6 Summary and Comments

There are currently a large number of projects and diverse range of new and emerging Grid developmental approaches being pursued. These systems range from Grid frameworks to application testbeds, and from collaborative environments to batch submission mechanisms.

There are many approaches and models [110] for developing Grid resource management systems. The systems we surveyed have for the most part focused on either a computational Grid or a service Grid. The only data Grid project that we have surveyed is the CERN Data Grid, which is in the initial stages of development. The other category of system is the Grid scheduler such as Nimrod-G and AppLeS that is integrated with another Grid RMS such as Globus or Legion. These combinations are then used to create application oriented computational Grids with a certain degree of QoS. Among the various *Grid scheduling* systems, it can be observed that the Nimrod-G broker is the only system that uses user-centric scheduling policies to meet the users' quality of service requirements.

# Chapter 3

# Service-Oriented Grid Architecture for Distributed Computational Economies

This chapter identifies challenges in managing resources in a Grid computing environment and proposes computational economy as a metaphor for effective management of resources and application scheduling. To realize this, we propose a framework called Grid Architecture for Computational Economy (GRACE) that leverages the existing technologies and provides additional services for resource trading and aggregation. We present related systems, both historical and emerging, for cooperative and competitive trading of resources such as CPU cycles, storage, and network bandwidth. The chapter discusses the use of real-world economic models and strategies such as commodity market, posted prices, bargaining, tendering, auction, proportional resource sharing, and cooperative bartering for resource management and scheduling within the GRACE framework. For each model, high-level protocols and strategies that can be used by both resource owners and consumers to meet their respective objectives and goals will be highlighted.

## 3.1 Introduction

Computational Grids [48] and Peer-to-Peer (P2P) [4] computing systems are emerging as a new paradigm for solving large-scale problems in science, engineering, and commerce. They enable the *sharing* and *aggregation* of millions of resources (e.g., SETI@Home [143]) geographically distributed across organizations and administrative domains. They comprise heterogeneous resources (PCs, work-stations, clusters, and supercomputers), fabric management systems (single system image OS, queuing systems, etc.) and policies, and applications (scientific, engineering, and commercial) with varied requirements (CPU, I/O, memory, and/or network intensive). The resources are owned by different organizations with their own management policies, usage and cost models for different users at different times. Also, the availability of resources and the load on them dynamically varies with time.

In such Grid environments, the *producers* (resource owners) and *consumers* (resource users) have different goals, objectives, strategies, and supply-and-demand patterns. More importantly both resources and end-users are geographically distributed with different time zones. In managing such complex environments, traditional approaches to resource management, that attempt to optimize system-wide measure of performance, cannot be employed. Traditional approaches use centralized policies that need complete state information and a common fabric management policy, or a decentralized consensus based policy. Due to the complexity in constructing a Grid environment, it is impossible to define an acceptable system-wide performance matrix and common fabric management policy [22]. (The concepts discussed in this chapter apply to both P2P and Grid systems although we can argue about some of their technical, social, and political differences. However, the term Grid will be used for simplicity and brevity).

We propose and explore the use of an economic framework, called Grid Architecture for Computational Economy (GRACE), for managing resources and scheduling applications in Grid computing environments. The economic approach provides a fair basis in successfully managing decentralization and heterogeneity that is present in human economies. Competitive economic models provide algorithms/policies and tools for resource sharing or allocation in Grid systems. These models can be based on bartering or prices. In the *bartering-based model*, all participants need to own resources and trade resources by exchanges (e.g., storage space for CPU time). In the *price-based model*, the resources have a price, based on the demand, supply, value, and the wealth in the economic system.

Most of the related systems for Grid resource management and scheduling (such as Legion [123], Condor [79], AppLeS PST [28][42], NetSolve [41], PUNCH [85], and XtremWeb [32]) adopt a *conventional strategy* where a scheduling component decides which jobs are to be executed at which resource based on cost functions driven by system-centric parameters. They aim to enhance the system throughput, utilization, and complete execution at the earliest possible time rather than improving the utility of application processing. They do not take resource access cost (price) into consideration, which means that the value of processing applications at any time is treated the same, which is not the case in reality—the value should be higher when there is a production schedule deadline. The end user does not want to pay the highest price but wants to negotiate a particular price based on the demand, value, priority, and available budget. In an economic-based approach, the scheduling decisions are made dynamically at runtime and they are driven and directed by the end-users requirements. Whereas a conventional cost model often deals with software and hardware costs for running applications, the economic model primarily charges the end user for services that they consume based on the value they derive from it. Pricing based on the demand of users and the supply of resources is the main driver in the competitive, economic market model. Therefore, in the Grid environments, a user is in competition with other users and a resource owner with other resource owners.

The main contribution of this chapter is to provide a generic distributed computational economy for Grids along with system architecture and policies for resource management for different economic models, Currently, the user community and the technology are still rather new and not well accepted and established in commercial settings. However, we believe the Grid can become established in such settings by providing incentive to both consumers and resource owners for being part of the Grid. Since the Grid uses the Internet as a carrier for providing remote services, it is well positioned to create a cooperative problem solving environment, and means for sharing computational and data resources in a seamless manner. Up to now, the idea of using Grids for solving large computationally intensive applications has been more or less restricted to the scientific community. However, even if, in the scientific community, the pricing aspect seems to be of minor importance, funding agencies need to support the hardware and software infrastructure for Grids. Economic models can help them manage and evaluate resource allocations to user communities. For example, the system managers may impose quota limitations and value different resources with a different number of tokens [56]. In such environments, resource consumers certainly prefer to use economic driven schedulers to effectively utilize their tokens by using lightly loaded cheaper resources.

An economic approach to Grid computing introduces a number of new issues to be addressed in addition to those already addressed by existing Grid systems. The Grid economy framework needs to provide an infrastructure that offers the following:

- An Information and Market directory for publicizing Grid entities
- Models for establishing the value of resources
- Resource pricing schemes and publishing mechanisms
- Economic models and negotiation protocols
- Mediators to act as a regulatory agency for establishing resource value, currency standards, and crisis handling.
- Accounting, Billing, and Payment Mechanisms
- Users Quality of Service (QoS) requirements driven brokering/scheduling systems.

In this chapter we identify requirements of users (resource providers and consumers) in the Grid economy and various resource management issues that need to be addressed in realizing such a Grid system. We briefly discuss popular economic models for resource trading and present related work that employs computational economy in resource management. We propose a scalable architecture and new services for the Grid that provide mechanisms for addressing user requirements. The implementation of computational economy with the Nimrod-G resource broker will be discussed in the next chapter.

## 3.2 Computational Economy and its Benefits

The current research and investment into computational Grids is motivated by an assumption that coordinated access to diverse and geographically distributed resources is valuable. In this paradigm, we need mechanisms that allow such coordinated access, but also sustainable, scalable models and policies that promote precious Grid resource sharing. Based on the success of economic institutions in the real world as a sustainable model for exchanging and regulating resources, goods and services, we propose a computational economy framework. Among other things, this framework provides a mechanism to indicate which users should receive priority.

Like all systems involving goals, resources, and actions, computations can be viewed in economic terms. With the proliferation of networks, high-end computing systems architecture has moved from centralized toward decentralized models of control and action; the use of economic driven market mechanisms would be a natural extension of this development. The ability of trade and price mechanisms to combine local decisions by diverse entities into globally effective characteristics, imply their value for organizing computations in large systems such as Internet-scale computational Grids.

The need for an economy driven resource management and scheduling system comes from the answers to the following questions:

- What comprises the Grid and who owns its resources?
- What motivates resource owners to contribute their resource to the Grid?
- Is it possible to have access to all resources in the Grid by contributing our resource?
- If not, how do we have access to all Grid resources?
- If we have access to resources through collaboration, are we allowed to solve commercial problems?
- Do resource owners charge the same or different price for different users?
- Is access cost the same for peak and off-peak hours?
- How can resource owners maximize their profit?
- How can users solve their problems within a minimum cost?
- How can a user get high priority over others?
- If the user relaxes the deadline by which results are required, can solution cost be reduced?

To date, individuals or organizations that have contributed resources to the Grid have been largely motivated by the public good, prizes, fun, fame, or collaborative advantage. This is clearly evident from the construction of private Grids (but on volunteer resources) or research test-beds such as Distributed.net [24], SETI@Home [143], Condor pool [54], GUSTO [38], DAS (Distributed ASCI Supercomputer) [40], eGrid [31], and World-Wide Grid [113]. The computational resource contributors to these test-beds are mostly motivated by the aforementioned reasons. The chances of gaining access to such computational test-beds for solving commercial problems are low. Furthermore, contributing resources to a testbed does not guarantee access to all of the other resources in the testbed. For example, although we were part of the GUSTO testbed, gaining automatic access to all of its resources was not possible. Unless we have some kind of collaboration with contributors, it is difficult to get access to their resources. In this situation, we believe that a model that encourages resource owners to let their resources for others use can lead to computational economy – wherein users are charged for access at a rate that varies with time. This necessitates the need for a mechanism where one can buy compute power on-demand from computational Grids. As both resource owners and users want to maximize their profit (i.e., the owners wish to earn more money and the users wish to solve their problems within a minimum possible cost), the Grid computing environment needs to support this economy of computations.

Even commercial companies such as Entropia, ProcessTree, Popular Power, United Devices, and Parabon are exploiting idle CPU cycles from desktop machines to build a commercial computational Grid infrastructure based on P2P networks [97]. These companies are able to develop large-scale infrastructure for Internet computing and use it for their own financial gain by charging for access to CPU cycles for their customers without offering fiscal incentive to all resource contributors. In the long run, this model is less likely to succeed in creating a maintainable and sustainable infrastructure. Therefore, a Grid economy seems a better model for managing and handling requirements of both Grid providers and consumers. It is interesting to note that, even in electricity Grids, bid-based electricity trading over the Internet has been adopted to develop competitive forces in the electricity marketplace [53].

The Grid resource management systems must dynamically trade for the best resources based on a metric of the price and performance available and schedule computations on these resources such that they meet user requirements. The Grid middleware needs to offer services that help resource brokers and resource owners to trade for resource access.

The benefits of economic-based resource management include the following:

- It helps in building a large-scale Grid as it motivates resource owners to contribute their (idle) resources for others to use and profit from it.
- It provides a fair basis for access to Grid resources for everyone.
- It helps in regulating the supply and demand for resources.
- It offers an economic incentive for users to back off when solving low priority problems and thus encourages the solution of time critical problems first.
- It removes the need for a central coordinator (during negotiation).
- It offers uniform treatment of all resources. That is, it allows trading of everything including computational power, memory, storage, network bandwidth/latency [114], data, and devices or instruments.
- It allows users to express their requirements and objectives.
- It helps in developing scheduling policies that are user-centric rather than system-centric.
- It offers an efficient mechanism for allocation and management of resources.
- It helps in building a highly scalable system as the decision-making process is distributed across all users and resource owners.
- Finally, it places the power in the hands of both resource owners and users—they can make their own decisions to maximize the utility and profit.

## 3.3 Requirements for Economic-based Grid Systems

We envision a future in which economically intelligent and economically motivated peer-to-peer and Grid-like software systems will play an important role in establishing a distributed service-oriented computing paradigm. To deliver greater value to users than traditional systems, an economic-based resource management systems need to provide mechanisms and tools that allow resource consumers (end users) and providers (resource owners) to express their requirements and facilitate the realization of their goals. That is, they need (i) the means to express their requirements, valuations, and objectives [*value expression*], (ii) scheduling policies to translate them to resource allocations [*value translation*], and (iii) mechanisms to enforce selection and allocation of differential services, and dynamic adaptation to changes in their availability at runtime [*value enforcement*]. Similar requirements are raised [9] for market-based systems in a single administrative domain environment such as clusters and they are limited to co-operative economic models since they aim for social welfare. Grids need to use *competitive economic models* as different resource providers and resource consumers have different goals, objectives, strategies, and requirements that vary with time.

Essentially, resource consumers need a *utility model*—to allow them to specify resource requirements and constraints. For example, the Nimrod-G broker allows the users to specify the deadline and budget constraints along with optimisation parameters such as optimise for time [*value expression*]. They need *brokers* that provide strategies for choosing appropriate resources [*value translation*] and dynamically adapt to changes in resource availability at runtime to meet user requirements [*value enforcement*]. The Nimrod-G broker, discussed in the next chapter, supports all these requirements. The resource owners need mechanisms for *price generation schemes* to increase system utilization and *protocols* that help them offer competitive services [*value expression*]. For the market to be competitive and healthy, coordination mechanisms are required that help the market reach an equilibrium price—the price at which the supply of a service equals the quantity demanded. Grid resources have their schedulers (e.g., OS or queuing system) that allocate resources [*value translation*]. Some research systems support resource reservation in advance (e.g., reserving a slot from time t1 to t2 using the Globus GARA [50] and bid a job to it) and allocate resources during reserved time [*value enforcement*]. A number of research systems have explored QoS based resource (e.g., CPU time and network bandwidth [114][3]) allocation in operating systems and queuing systems, but the inclusion of QoS into mainstream systems has been slow paced (e.g., Internet mostly uses the best effort allocation policy [71], but this is changing with IPv6 [7]).

## 3.4 Grid Architecture for Computational Economy (GRACE)

A distributed Grid Architecture for Computational Economy (GRACE) is shown in Figure 3.1. This architecture is generic enough to accommodate different economic models used for resource trading for determining the service access cost. The key components of the Grid include,

- Grid Use with Applications (sequential, parametric, parallel, or collaborative applications)
- User-Level Middleware—Higher Level Services and Tools
    - Programming Environments
    - Grid Resource Brokers
- Core Grid Middleware (services resource trading and coupling distributed wide area resources)
- Grid Service Providers

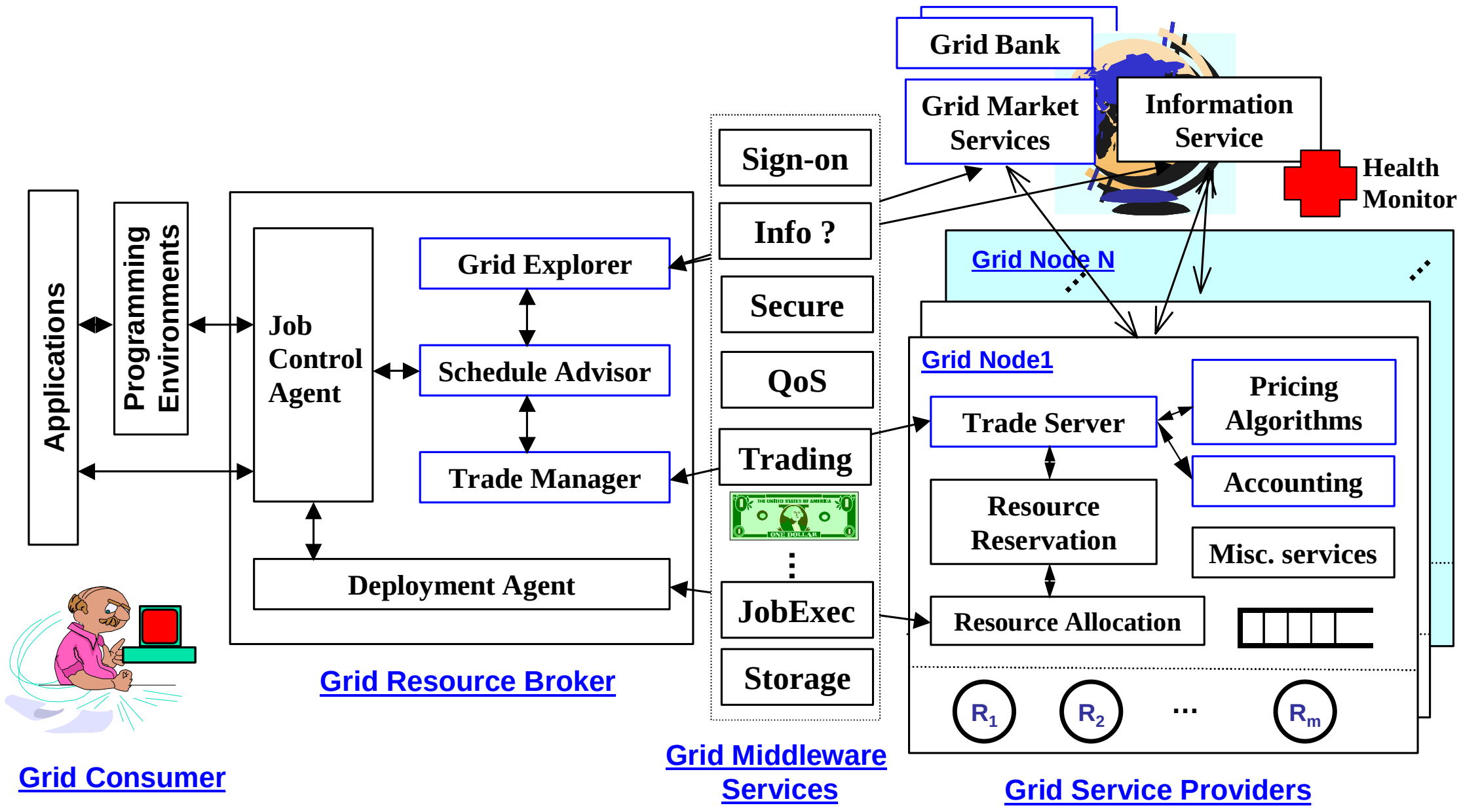

**Figure 3.1: A generic Grid architecture for computational economy.**

The two key players in market oriented computational Grids are *resource providers* (hereafter referred as GSPs—Grid Service Providers) and *resource consumers* (hereafter referred as GRBs—Grid Resource Broker that acts as a consumer's representative or software agent). Both have their own expectations and strategies for being part of the Grid. In the Grid economy, resource consumers adopt the strategy of solving their problems as cheaply as possible within a required timeframe. Resource providers adopt the strategy of obtaining best possible return on their investment. The resource owners try to maximize their resource utilization by offering a competitive service access cost in order to attract consumers. The users (resource consumers) have an option of choosing the providers that best meet their requirements. If resource providers have local users, they will try to recoup the best possible return on "idle/leftover" resources. In order to achieve this, the Grid systems need to offer tools and mechanisms that allow both resource providers and consumers to express their requirements. The Grid resource consumers interact with brokers to express their requirements such as the *budget* that they are willing to invest for solving a given problem and a *deadline*, a timeframe by which they need results. They also need capability to trade between these two requirements and steer the computations accordingly. The GSPs need tools for expressing their pricing policies and mechanisms that help them to maximize the profit and resource utilization. Various economic models, ranging from commodity market to auction-based, can be adopted for resource trading in Grid computing environments. The Grid resource management framework and strategies for these economic models is discussed in Section 3.6.

GRACE provides services that help both resource owners and user-agents maximize their objective functions. The resource providers can contribute their resource to the Grid and charge for services. They can use GRACE mechanisms to define their charging and access policies and the GRACE resource trader works according to those policies. The users interact with the Grid by defining their requirements through high-level tools such as resource brokers (also known as Grid schedulers). The resource brokers work for the consumers and attempt to maximize user utility. They can use GRACE services for resource trading and identifying GSPs that meets its requirements.

As mentioned earlier, our goal is to realize this Grid economy architecture by leveraging existing technologies such as Globus and Legion and develop new services that are particularly missing in them. Therefore, we mainly focus on two things: first, to develop middleware services for resource trading using different economic models; second to use these services along with other middleware services in developing advanced user-centric Grid resource brokers. The remainder of this section presents how we are realizing the Grid economy vision and show co-existence of our modules with other systems.

### 3.4.1 Grid Resource Broker (GRB)

The resource broker acts as a mediator between the user and Grid resources using middleware services. It is responsible for resource discovery, resource selection, binding of software, data, and hardware resources, initiating computations, adapting to the changes in Grid resources and presenting the Grid to the user as a single, unified resource. The components of resource broker are the following:

- **Job Control Agent (JCA):** This is a persistent control engine responsible for shepherding a job through the system. It coordinates with schedule adviser for schedule generation, handles actual creation of jobs, maintenance of job status, interacting with clients/users, schedule advisor, and dispatcher.
- **Schedule Advisor (Scheduler):** This is responsible for resource discovery (using the Grid explorer), resource selection and job assignment (schedule generation) to ensure that the user requirements are met.
- **Grid Explorer (GE):** This is responsible for resource discovery by interacting with the Grid-information server and identifying the list of authorized machines, and keeping track of resource status information.
- **Trade Manager (TM):** This works under the direction of resource selection algorithm (the schedule advisor) to identify resource access costs. It uses market directory services and GRACE negotiation services for trading with Grid service provides (i.e., their representative trade servers).
- **Deployment Agent (DA):** It is responsible for activating task execution on the selected resource as per the scheduler's instruction and periodically updates the status of task execution to JCA.

The Nimrod-G resource broker has components that support similar functions. It allows users to submit their application created using its parameter specification language; and express their requirements and objectives in the form of: deadline, budget with time or cost as the optimisation parameter [value expression]. The broker uses scheduling algorithms to select resources dynamically at runtime depending on their availability, capability, and cost to meet user requirements [value translation]. It continuously adapts to changes in resource availability conditions by performance profiling (establishing job completion rate) and reschedules jobs appropriately to ensure that users requirements are met [value enforcement].

### 3.4.2 GRACE Framework—Leveraging Globus Tools

The Grid middleware offers services that help in coupling a Grid user and remote resources through a resource broker or Grid enabled application. It offers core services such as remote process management, co-allocation of resources, storage access, directory information, security, authentication, and Quality of Service (QoS) such as resource reservation for guaranteed availability and trading for minimizing computational cost. Many of these services are already offered by Globus [49] components and they include,

- Resource allocation and process management (GRAM).
- Resource Co-allocation services (DUROC)
- Unicast and multicast communications services (Nexus)
- Authentication and related security services (GSI)
- Distributed access to structure and state information (MDS)
- Status and Health Monitoring components (HBM)
- Remote access to data via sequential and parallel interfaces (GASS)

- Construction, caching, and location of executables (GEM)
- Advanced resource reservation (GARA)

We provide components (see Figure 3.2) that offer services required for constructing the Grid economy and that can co-exist with systems like Globus:

- Applications (e.g., Molecular modelling for drug design as parameter sweep application)
- Problem solving environments built on schedulers (e.g., ActiveSheets [20] on Nimrod-G)
- Programming frameworks and development tools (e.g., Nimrod parameter specification language[21])
- A resource broker (e.g., Nimrod-G)
- Various resource trading protocols
- A mediator for negotiating between users and Grid service providers (Grid Market Directory)
- A deal template for specifying resource requirements and services offers
- A trade server
- A pricing policy specification
- Accounting (e.g., QBank [126]) and payment management (GBank)

The new middleware services being proposed are designed to offer low-level services that co-exist with Globus services and infrastructure. Higher-level services and tools such as the Nimrod-G Resource Broker, which uses economic models suitable for meeting the user requirements, can use these core services.

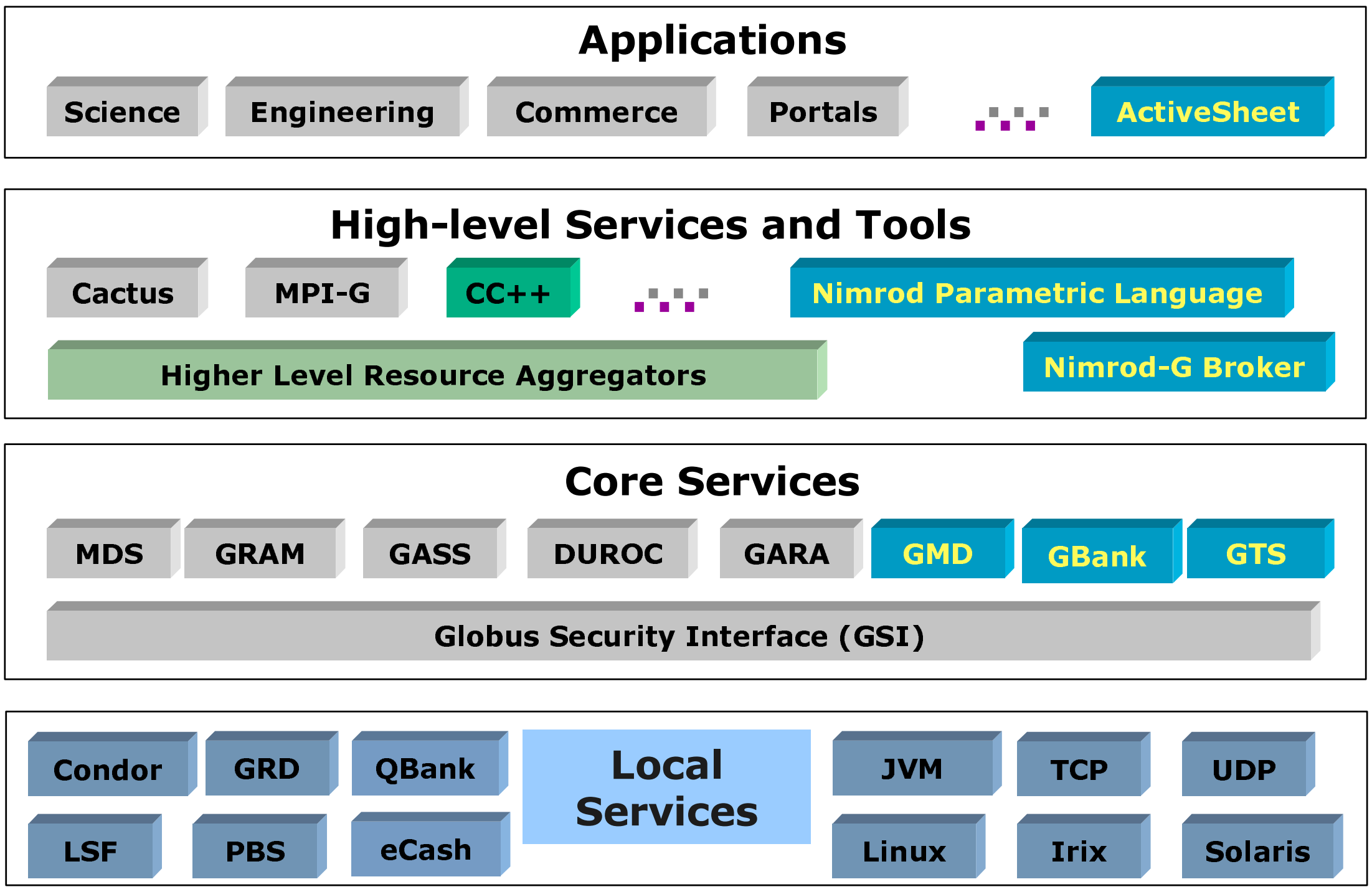

**Figure 3.2: GRACE framework realization within Globus context.**

The Grid service providers specifically deal with the following components along with Globus components:

- **Grid Market Directory: (GMD)**: It allows resource owners to publish their services in order to attract consumers.
- **Grid Trade Server (GTS)**: This is a resource owner agent that negotiates with resource users and sells access to resources. It aims to maximize the resource utility and profit for its owner i.e., earn as much money as possible. It consults pricing policies during negotiation and directs the accounting system for recording resource consumption and billing the user according to the agreed pricing policy.

- **Pricing Policies**: These define the prices that resource owners would like to charge users. The resource owners may follow various policies to maximise their profit and resource utilisation and the price they charge may vary with time and one user to another user. The pricing can also be driven by demand and supply like in the real market environment. That is, in this commodity market model, pricing is essentially determined by objective functions of service providers and users. The pricing policy can also be based on auction. In this auction based economic model, pricing is driven by how much users value the service and the highest bidder wins the access to Grid services.
- **Resource Accounting and Charging** components (such as GBank along with QBank) are responsible for recording resource usage and bill the user as per the usage agreement between resource broker (TM, a user agent) and trade server (resource owner agent).

The service providers publish their services through the Grid market directory (GMD). They use Grid trading services' declarative language for defining cost specification and their objectives such as access price for various users for different times and durations, along with possibilities of offering discounts to attract users during off-peak hours. The Grid trading server (GTS) can employ different economic models in providing services. The simplest would be a commodity model wherein the resource owners define pricing strategies including those driven by the demand and resource availability. The GTS can act as auctioneer if the Auction-based model is used in deciding the service access price or an external auctioneer service can be used.

### 3.4.3 Grid Open Trading Protocols and Deal Template

The resource trading protocols define the rules and format for exchanging commands between a GRACE client (Trade Manager), which is part of the Grid broker and a Trade Server, which is part of the Grid service providers. Figure 3.3 shows a sample multilevel negotiation protocol that both client and server need to follow while trading for the cost of resource access. The wire-level (low-level) details of these protocols are skipped, as they are obvious.

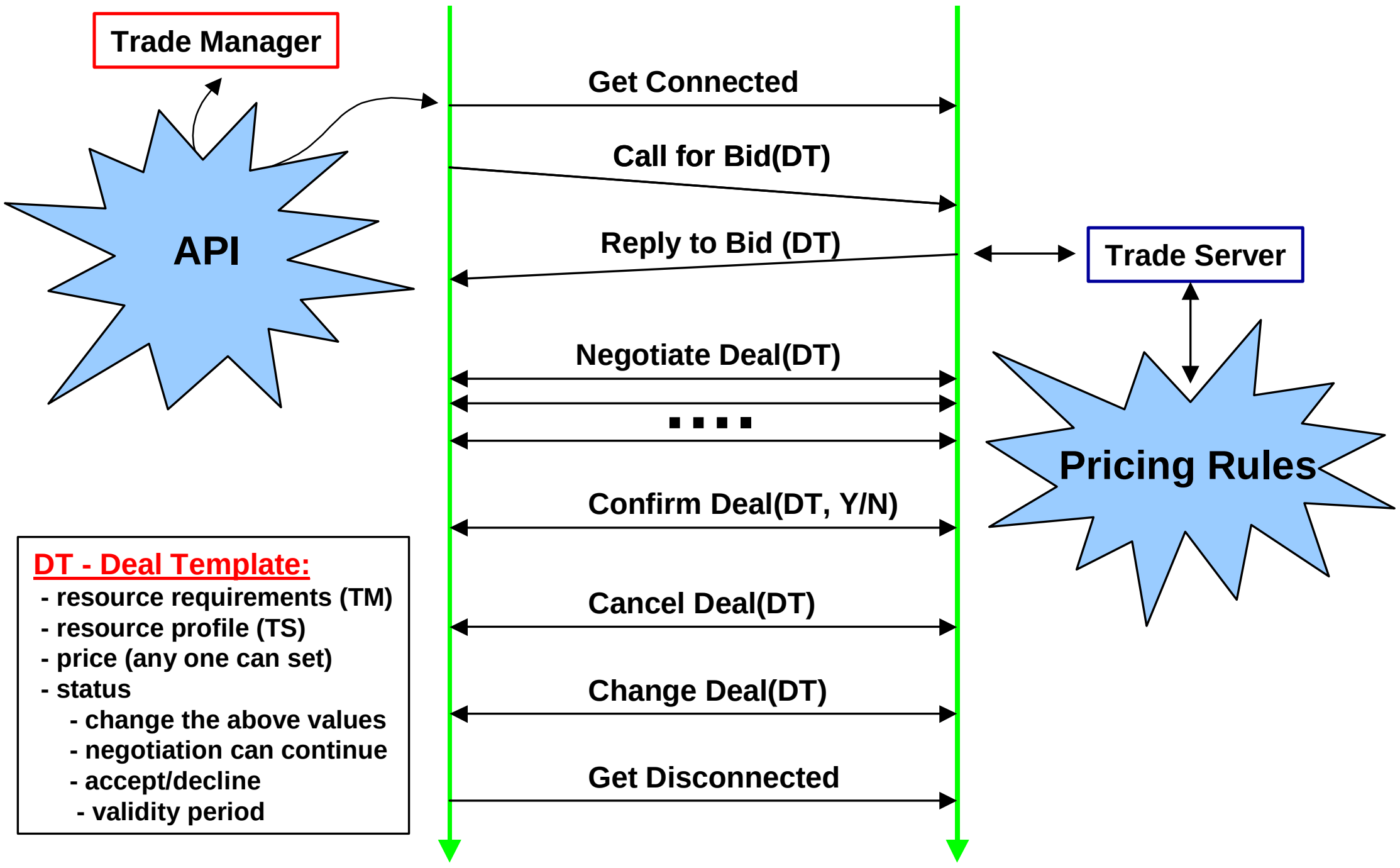

**Figure 3.3: GRACE Open Trading Protocols.**

A finite state machine representation of multilevel negotiation protocols that both client and server need to follow for the bargaining/tender model is shown in Figure 3.4. In this model, the broker's Trade Manager (TMs) contacts the resource owner's Trade Server (TS) with a request for a quote. The TM specifies resource requirements in a Deal Template (DT), which can be represented by a simple structure with its fields corresponding to deal items or by a “Deal Template Specification Language”, similar to the *ClassAds* mechanism employed by the Condor [54] system. The contents of DT include expected start time, usage duration, memory, and storage requirements along with its initial offer. The TM looks into DT and updates its contents and sends back to TS. This negotiation between TM and TS continues until one of them indicates that its offer is final. Following this, the other party decides whether to accept or reject the deal. If accepted, then both work as per the agreement mentioned in the deal. The overhead introduced by the multilevel point-to-point protocol can be reduced when resource access prices are announced through Grid information services (e.g., MDS) or market directory.

A number of interaction protocols for a business negotiation on the Internet have been presented in [78]. It highlights some commonalities in the structure of different price negotiation mechanisms such as fixed price sales, auctions, and brokerages. These business negotiation models and protocols are also applicable for our resource trading and we have explored such models and protocols in our resource management and scheduling system.

### *Grid Open Trading APIs*

The GRACE infrastructure supports generic Application Programming Interfaces (APIs) that can be used by the Grid tools and application programmers to develop software supporting the computational economy. The following trading APIs are C-like functions (high level view of trading protocols) that GRACE clients/brokers can use to communicate with trading servers:

- `grid_trade_connect(resource_id, tid)`
- `grid_request_quote(tid, DT)`
- `grid_trade_negotiate (tid, DT)`
- `grid_trade_confirm(tid, DT)`
- `grid_trade_cancel(tid, DT)`
- `grid_trade_change( tid, DT)`
- `grid_trade_reconnect(tid, resource_id)`
- `grid_trade_disconnect(tid)`

where,

```
tid = Trade Identification code
DT = Deal Template
```

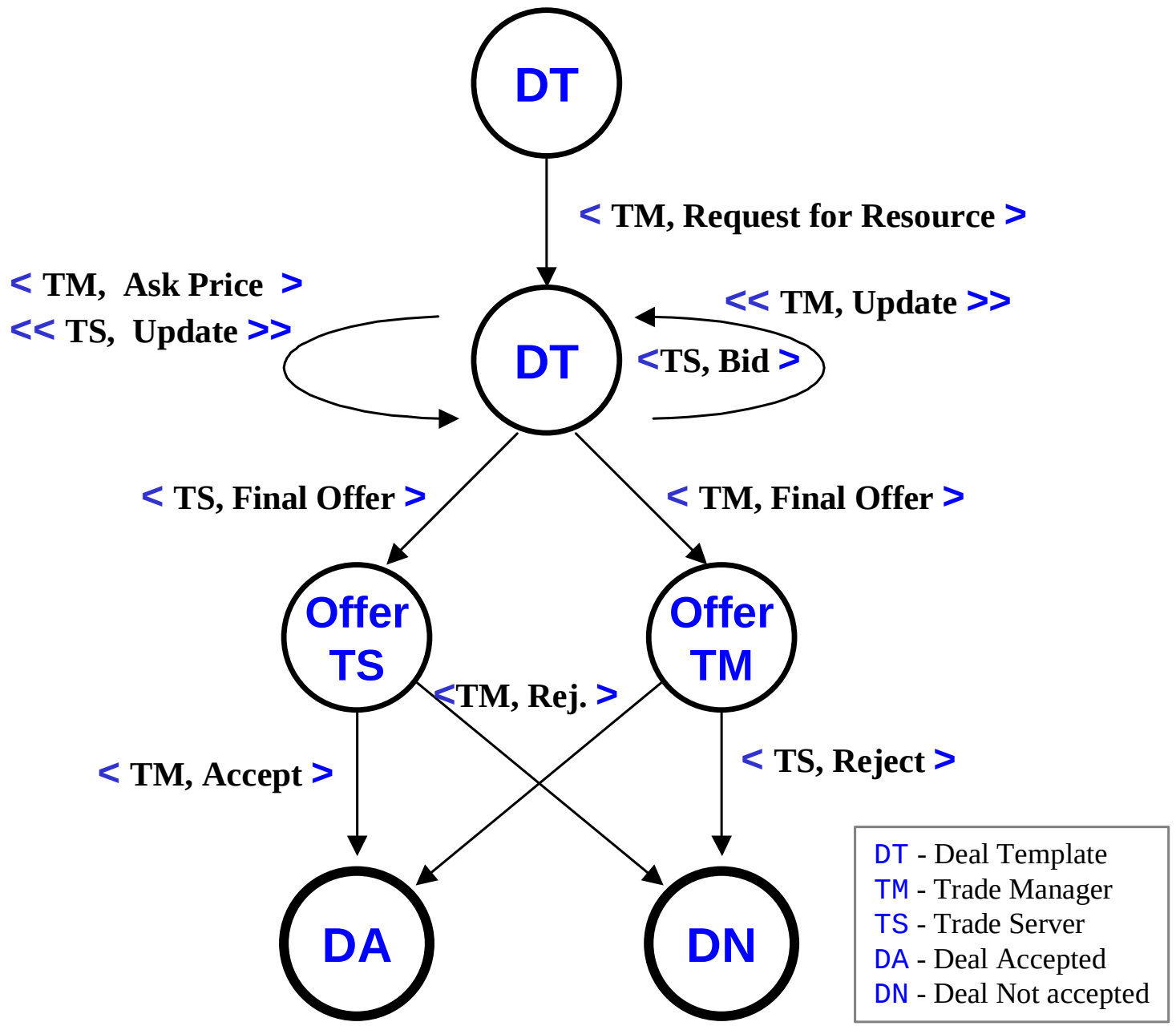

**Figure 3.4: A finite state representation of resource trading (for bargain model).**

### 3.4.4 Pricing, Accounting, and Payment Mechanisms

In a computational Grid economy environment, both resource owners and users want to maximize their benefits. As there will be many GSPs offering similar services, they need to have a competitive pricing structure in order to attract users, efficiently utilize resources, and maximize profit. The resources consumed by the user applications need to be accounted for and charged. Various payment mechanisms need to be supported. The users can purchase resource access credits in advance or pay-after-usage. Each GSP can maintain this by using systems like QBank or there can be global Grid-wide bank called GridBank (GBank) that mediates payment for services accessed by the user. Figure 3.5 shows various components at GSP node and their interactions during resource trading, consumption, metering (measuring), billing, and payment handling.

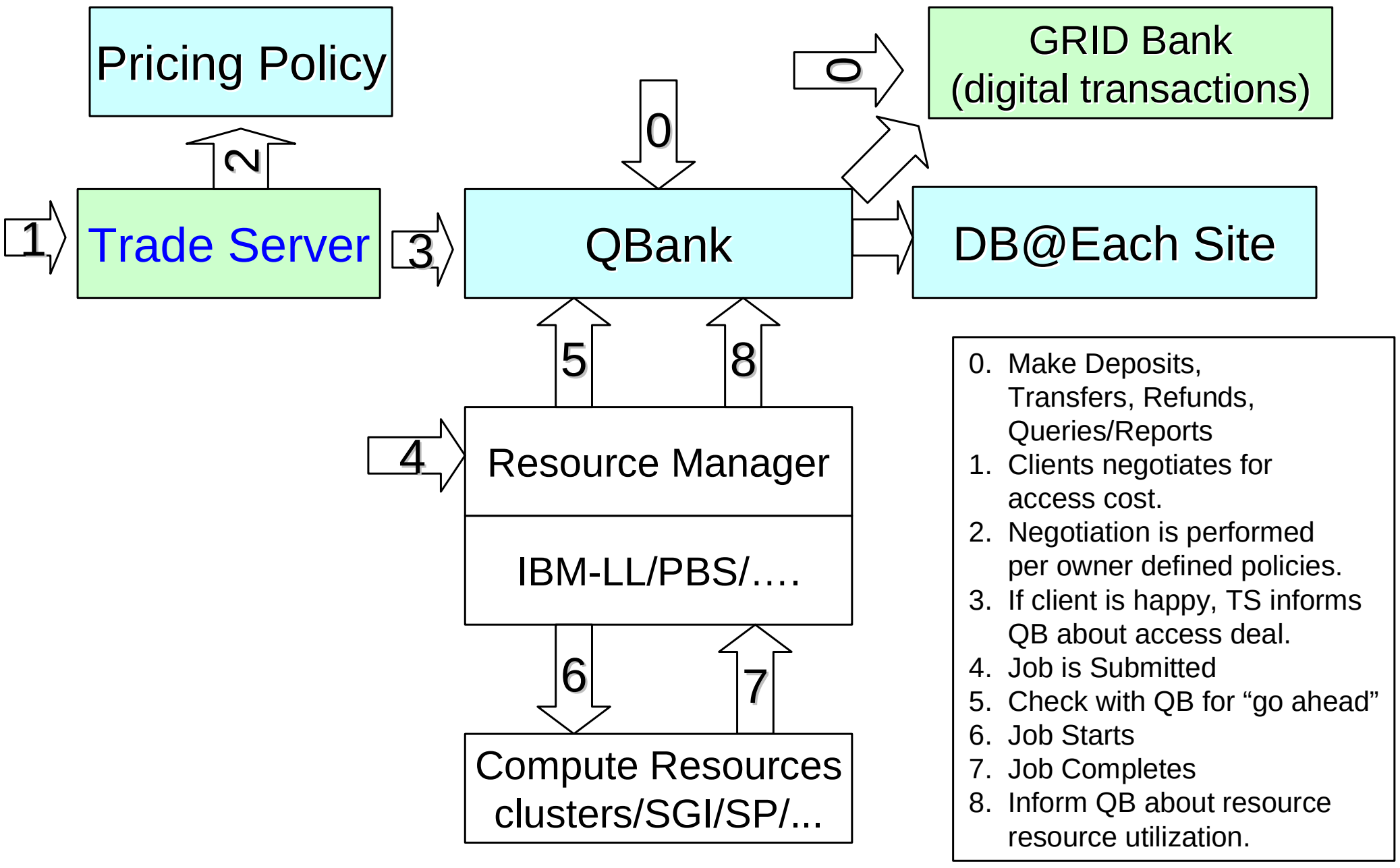

**Figure 3.5: An Interaction between GSP resource management components.**

#### *How to Determine the Price?*

A simple pricing scheme is a fixed price model, but this does not work when the users demand QoS. This requirement changes between applications and across time. In the context of software Agents [36][60], many researchers have investigated pricing schemes based on the supply and demand for resources and the QoS requirements. The pricing schemes based on different parameters include,

- A flat price model (the same cost for applications and no QoS like in today's Internet [71])
- Competitive economic models (e.g., auctions and contract-net)
- Usage timing (peak, off-peak, lunch time like pricing telephone services)
- Usage period and duration (short/long)
- Demand and supply (e.g., Smale model [128])
- Foresight-based [36] (i.e., an ability to model and predict responses by competitors)
- Loyalty of Customers (like Airlines favoring frequent flyers!)
- Historical data
- Advance agreement/contract with service provides
- Calendar based
- Bulk Purchase
- Voting in which trade unions decide pricing structure
- Resource capability as benchmarked in the capital market
- Application areas in which academic R&D or public good applications can be offered at cheaper rate compared to commercial applications.

In [60], five different pricing strategies, ranging widely from ones that require perfect knowledge and unlimited computational power to ones that require very little knowledge or computational capability, are employed in two different buyer populations, namely quality-sensitive and price sensitive buyers. The resulting collective dynamics have been investigated using a combination of analysis and simulation. In a population of *quality-sensitive buyers*, all pricing strategies lead to a price equilibrium predicted by a game-theoretic analysis. However, in a population of *price-sensitive buyers*, most pricing strategies lead to large-amplitude cyclical price wars. These pricing strategies and issues are also applicable to the Grid and strategies need to be designed such that the resource providers benefit through efficient resource utilization and consumers will have the ability to trade-off between cost and timeframe in the Grid marketplace.

### Service Items to be Charged and Accounted

User applications have different resource requirements depending on computations performed and algorithms used in solving problems. Some applications can be CPU intensive while others can be I/O intensive or a combination. For example, in CPU intensive applications it may be sufficient to charge only for CPU time whilst offering free I/O operations. This scheme cannot be applied for I/O intensive applications. Therefore, the consumption of the following resources needs to be accounted and charged:

- CPU - User time (consumed by user App.) and System time (consumed while serving user App.)
- Memory
- Maximum resident set size - page size
- Amount of memory used
- Page faults
- Storage used
- Network activity
- Signals received, context switches
- Software and Libraries accessed (particularly required for the emerging ASP world).

Access to each of these entities can be charged individually or in combination. Combined pricing schemes need to have a costing matrix that takes a request for multiple resources in pricing. An economic model proposed by Smale [128] allows formulation of such pricing schemes for resource allocation. G-commerce [120] is one such framework investigating the use and enhancing the Smale model for devising pricing strategies in the context of allocating resources for Grid users. By simulating hypothetical resource consumers and resource producers, they measured the efficiency of resource allocation under two different market conditions: commodities markets and auctions. By comparing the results of both market strategies in terms of price stability, market equilibrium, consumer efficiency, and producer efficiency, the G-commerce concludes that commodities markets are a better choice for controlling Grid resources than the existing auction strategies.

### Payment Mechanisms

A computational economy Grid framework needs to support various payment mechanisms. They include:

- Prepaid – Pay and use in which users need to buy credits in advance from GSPs or Grid Bank
- Use and pay later
- Pay as you go
- Grants based

Each GSP can bill their users directly and handle all payment processing issues themselves. This method introduces a great burden for both providers and users in a large-scale Grid environment. This can be simplified by having mediators like a scalable Grid Bank. The user brokers can (automatically) inform the GSPs about the user Grid Bank account details for which they can charge directly or users can pay by other electronic cash systems. This can be achieved by using digital currency mechanisms such as:

- NetCheque: [13] – Users registered with NetCheque accounting servers can write electronic cheques and send them to service providers. When deposited, the balance is transferred from sender to receiver account automatically.
- NetCash [34] – This supports anonymity and it uses the NetCheque system to clear payments between currency servers.
- Paypal [89] – This is an example of a credit-card based automated mediator for payments processing.

- Tokens – This mechanism can be used when (a) resources are allocated based on grants; (b) resources are bartered and used by earning or exchanging credits like in the MojoNation peer networks.

Such electronic payment mechanisms satisfy the diverse requirements of service providers and their users. We believe that these payment mechanisms can be easily integrated into our Grid economy infrastructure.

## 3.5 Related Work

As in the conventional marketplace, the users' community (GRBs) represents the demand, whereas the resource owners' community (GSPs) represents the supply. Consumers interact with their own brokers (such as Nimrod-G) for managing and scheduling their computations on the Grid. The GSPs make their resources Grid enabled by running software systems (such as Globus [49] and Legion [123]) along with *Grid Trading Services* (GTS) to enable resource trading and execution of consumer requests directed through GRBs. The interaction between GRBs and GSPs during resource trading is mediated through a *Grid Market Directory* (GMD) (see Figure 3.6 to Figure 3.10). They use various economic models or interaction protocols for deciding service access price.

Numerous economic models including microeconomic and macroeconomic principles for resource management have been proposed in the literature [99][80][53][3][77][119]. Some of the commonly used economic models that can be employed for managing resources in Grid environment, include:

- Commodity Market Model
- Posted Price Model
- Bargaining Model
- Tendering/Contract-Net Model
- Auction Model
- Bid-based Proportional Resource Sharing Model
- Cooperative Bartering Model
- Monopoly and Oligopoly

Various criteria used for judging effectiveness of a market model are [133]: social welfare (global good of all), Pareto efficiency (global perspective), individual rationality (better off by participating in negotiation), stability (mechanisms that cannot be manipulated, i.e., behave in the desired manner), computational efficiency (protocols should not consume too much computation time), and distribution and communication efficiency (communication overhead to capture a desirable global solution).

Several research systems (see Table 3.1) have explored the use of different economic models for trading resources to manage resources in different application domains: CPU cycles, storage space, database query processing, and distributed computing, They include Spawn [14], Popcorn [87], Java Market [146], Enhanced MOSIX [147], JaWS [127], Xenoservers [23], D'Agents [55], Rexec/Anemone [8], Mojo Nation [84], Mariposa [83], Mungi[33], Stanford Peers [10], G-Commerce [120], OCEAN [76], Nimrod-G [100], and GridSim [95]. These systems have been targeted to manage single or multiple resources for application domains as follows:

- *Single domain computing systems:* Enhanced MOSIX and Rexec/Anemone.
- *Agent-based systems:* Xenoservers and D'Agents.
- *Distributed database management system:* Mariposa
- *Shared storage management system:* Mungi.
- *Storage Space Trading system:* Stanford Peers
- *Web-based distributed systems:* Popcorn, Java Market, and JaWS.
- *Multi-domain distributed Grid system:* Nimrod-G and GridSim Resource Broker

Among the above, the three most recent systems are Stanford Peers, G-Commerce, and GridSim Resource Broker and all of them happen to use simulation techniques to demonstrate distributed resource trading. However, each have a different focus: the Stanford Peers project is exploring storage trading for data replication; the G-Commerce project is exploring pricing strategies for selling access to resources; and the GridSim project provides a general purpose Grid simulation toolkit and the economic resource broker that allows exploration of quality of services driven algorithms for scheduling task and data parallel applications in large-scale distributed environments.

**Table 3.1: Computational economy based distributed resource management systems.**

| SYSTEM NAME | ECONOMIC MODEL | PLATFORM | REMARKS |
|---|---|---|---|
| Mariposa [83] (UC Berkeley) | Bidding (Tendering/ ContractNet). Pricing based on load and historical info. | Distributed database. | It supports budget-based query processing and storage management. |
| Mungi [33] (University of New South Wales) | Commodity market (renting storage space that increases as available storage runs low, forcing users to release unneeded storage.) | Storage servers. | It supports storage objects based on bank accounts from which rent is collected for the storage occupied by objects. . |
| Popcorn [87] (Hebrew University) | Auction. (Highest bidder gets access to resource and it transfers credits from buyer to the seller account.) | Web browsers. (P*opcorn parallel code* runs within a browser of CPU cycles seller.) | Popcorn API-based parallel applications need to specify a budget for processing each of its modules. |
| Java Market [146] (John Hopkins University) | QoS based computational market. (The resource owner receives f(j, t) award for completing f in time t.) | Web browsers. (JavaMarket runs *standard Java Applets* within a browser). | One can sell CPU cycles by pointing Java-enabled browser to Portal & allow execution of Applets. |
| Enhanced MOSIX [147] (Hebrew U., Israel) | Commodity market (resource cost of each node is known) | Clusters of computers (Linux PCs) | It supports process migration such that overall cost of job execution is kept low. |
| JaWS [127] (University of Crete) | Bidding (Tendering) | Web browsers | It is similar to Popcorn. |
| Xenoservers [23] (University of Cambridge) | Bidding (Proportional resource sharing) | Single computer | Accounted execution of un-trusted code. |
| D'Agents [55] (Dartmouth College) | Bidding (Proportional resource sharing) | Single computer or Mobile Agents | Agents bid function is proportional to benefit. |
| Rexec/Anemone [8] (UC Berkeley) | Bidding/Auction (for proportional resource sharing) | Clusters (A market-based Cluster Batch Queue System) | Users assign utility value to their application and system allocates resources proportionally. |
| Mojo Nation [84]<br><br>(Autonomous Zone Industries, CA) | A Credit-based partnership and/or bartering model. (Contributors earn credits by sharing storage and spend them when required) | Network storage. | It is a content-sharing community network. It combines marketplace and bartering approach for file/resource sharing. |
| Spawn [14] (Xerox PARC) | Second-price/Vickery auction (uses sponsorship model for funding money to each task depending on some requirements) | Network on workstations. Each WS executes a single task per time slice | It supports execution of concurrent program expressed in the form of hierarchy of processes that expand and shrink size depending on the resource cost. |
| CSAR Supercomputing center [56] | Commodity market and priority-based model (they charge for CPU, memory, | MPPs, Crays, and Clusters, and Storage servers. | Any application can use this service and QoS is proportional to user priority |
| (University of Manchester) | storage, and human support services) | | and scheduling mechanisms. |
| Nimrod-G [100] (Monash University) | It supports economy models such as commodity market, spot market, and contract-net for price establishment. | World Wide Grid (having resource Grid enabled using middleware systems like Globus) | It is a *real* system that supports deadline and budget constrained algorithms for scheduling task-farming and data parallel applications on world-wide distributed resources depending on their cost, power, availability and users quality of service requirements. |
| GridSim [95] (Monash University) | Currently, it supports economic models similar to those used in Nimrod-G, but limited to them. | A Java-based discrete event toolkit for simulating Grid resources, users, applications, and brokers. | The economic Grid resource broker supports deadline and budget based time, cost, cost-time, and conservative time optimisation scheduling algorithms. |
| G-Commerce [120] (U. of California Santa Barbara) | Commodity and auctions | Simulates hypothetical consumers and produces. | It is exploring strategies for pricing Grid resources to enable resource trading. |
| OCEAN [76] (U. of Florida) | Continuous double auction | A Java based platform with distributed PCs. | It is exploring the use of continuous double auction for trading computational resources – in development. |
| Stanford Peers [10] (Stanford University) | Auctions with cooperative bartering in a cooperative sharing environment. | Simulates storage trading for content replication and archiving. | It demonstrates distributed resource trading policies based on auctions by *simulation* – in development. |

Each of the resource management systems presented in Table 3.1 follows a single model for resource trading. They have been designed with a specific goal in mind either for CPU or storage management. In order to use some of these systems, applications have to be designed using their proprietary programming models, which is generally discouraging, as applications need to be specifically developed for executing on those systems. Also, resource trading and job management modules have been developed as integrated monolithic systems, which limits their extensibility.

In the GRACE framework, we have separated these two concerns through a layered design approach to support different middleware technologies that co-exist with trading strategies and user-level resource brokers. The resource trading services are offered as core services and they can be used by different higher-level services/tools such as resource brokers and resource-aware applications. Another key advantage of Nimrod-G system is that it allows the execution of legacy applications on large wide-area distributed systems.

Typically, in a Grid marketplace, the resource owners, and users can use any one or more of these models or even combinations of them in meeting their objectives [98]. Both have their own expectations and strategies for being part of the Grid. The resource consumers adopt the strategy of solving their problems at low cost within a required timeframe. The resource providers adopt the strategy of obtaining best possible return on their investment while trying to maximize their resource utilization by offering a competitive service access cost in order to attract consumers. The resource consumers can choose providers that best meet their requirements. The design and architecture for the development of Grid systems using these economic models is discussed in Section 3.6.

Both GRBs and GSPs can initiate resource trading and participate in the interaction depending on their requirements and objectives. GRBs may invite bids from a number of GSPs and select those that offer the lowest service costs and meet their deadline and budget requirements. Alternatively, GSPs may invite bids in an auction and offer services to the highest bidder as long as its objectives are met. Both GSPs and GRBs have their own utility functions that must be satisfied and maximized. The GRBs perform a cost-benefit analysis depending on the deadline (by which the results are required) and budget available (the amount of money the user is willing to invest for solving the problem). The resource owners decide their pricing based on various factors. They may charge different prices for different users for the same service or it can vary depending on the specific user demands. Resources may have different prices based on environmental influences such as the availability of larger core memory and better communication bandwidth with an outside world.

Grid brokers (note that in a Grid environment each user has his own broker as his agent) may have different goals (e.g., different deadlines and budgets), and each broker tries to maximize its own good without concern for the global good. This needs to be taken into consideration in building automated negotiation infrastructure. In a *cooperative distributed computing or problem-solving environment* (like cluster computers), the system designers impose an *interaction protocol* (possible actions to take at different points) and a *strategy* (a mapping from one state to another and a way to use the protocol). This model aims for global efficiency as nodes cooperate towards a common goal. On the other hand, in Grid systems, brokers and GSPs are provided with an interaction protocol, but they choose their own private strategy (like in multi-agent systems), which cannot be imposed from outside. Therefore, the negotiation protocols need to be designed assuming a *non-cooperative, strategic* perspective. In this case, the main concern is what social outcomes follow given a protocol, which *guarantees that each broker/GSP's desired local strategy is best for that broker/GSP and hence the broker/GSP will use it.*

## 3.6 Economic Models in the Context of GRACE Framework

In the previous section we identified a few popular models that are used in human economies. In this section we discuss the use of different economic models and propose architecture for realizing them. The discussion on realizing negotiation protocols based on different economic models is kept as generic as possible. This ensures that our proposed architecture is free from any specific implementation and provides a general framework for any other Grid middleware and tools developers. Particular emphasis will be placed on framework and heuristics that Grid resource brokers (*G-Brokers*) can employ for establishing a service price depending on their customers' requirements.

For each of the economic models, the economic model theory, its parameters and strategies are discussed and then a possible solution is given for a current Grid environment and how they can be mapped to existing Grid tools and architectures or what needs to be extended. In the classical economic theory there are different models for specific environmental situations and computing applications. Since the end-user interaction is the main interest of this chapter, we point out possible interactions with the broker.

### 3.6.1 Commodity Market (Flat or Supply-and-Demand Driven Pricing) Model

In the commodity market model, resource owners specify their service price and charge users according to the amount of resource they consume. The pricing policy can be derived from various parameters and can be *flat* or *variable* depending on the resource *supply and demand*. In general, services are priced in such a way that supply and demand equilibrium is maintained. In the *flat price model*, once pricing is fixed for a certain period, it remains the same irrespective of service quality. It is not significantly influenced by the demand, whereas in a *supply and demand* model, prices change very often based on supply and demand changes. In principle, when the demand increases or supply decreases, prices are increased until there exists equilibrium between supply and demand. Pricing schemes in a Commodity Market Model can be based on:

- Flat fee
- Usage Duration (Time)
- Subscription
- Demand and Supply-based [71]

The resource owners publish their prices and rules in the Grid market directory (GMD) service (see Figure 3.6) similar to publishing through yellow pages. This is accomplished by defining the price specification that GTS can use for publishing service access price in the market directory. A simple price specification may contain the following parameters.

- consumer_id, which is the as same Grid-ID
- peak_time_price (say, between 9am-6pm: office hours on working days)
- lunch_time_price
- offpeak_time_price
- discount_when_lightly_loaded (i.e., if the load is less than 50% at any time)
- raise_price_high_demand (i.e., raise in price if the average load is above 50%)
- price_holiday_time (i.e., during holidays and week ends)

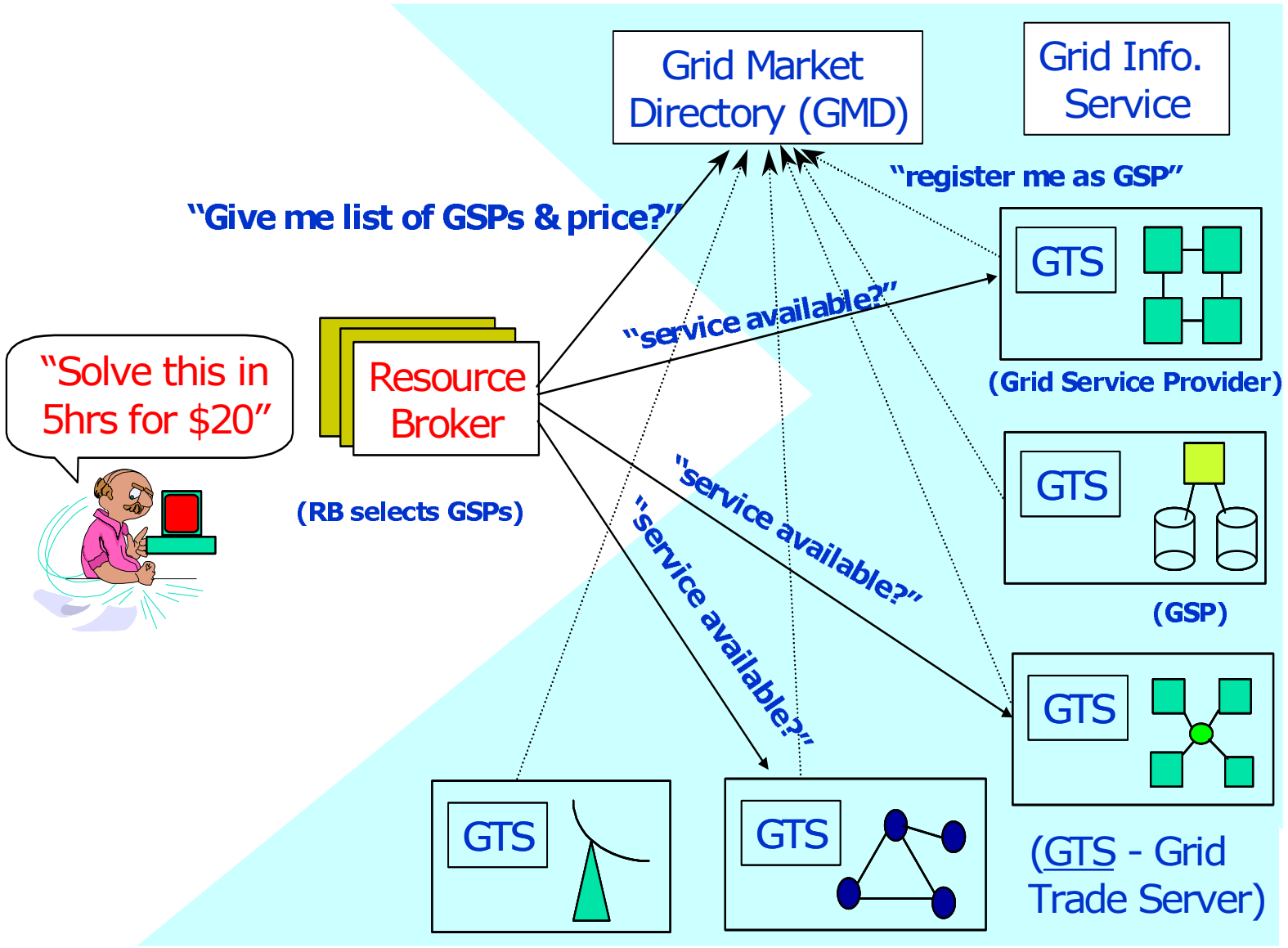

**Figure 3.6: Interaction between GSPs and users in a commodity market Grid for resource trading.**

Traditionally, computational services are priced based on their production cost and desired profit margin. However, the consumers' perception of value is based on parameters such as supply and demand for resources, priority and service quality requirements. Therefore, the resource value in Grid economy needs to be defined as a function of many parameters as follows:

- Resource Value = Function (Resource strength, Cost of physical resources, Service overhead, Demand, Value perceived by the user, Preferences);

The last three parameters are difficult to capture from consumers unless they see any benefit in disclosing them and vary from time to time, from one application to another. However, there are consumers who prefer regular access to resources during a particular period of the day. For example, those involved in making regular decisions on supply chain management of goods shipping from inventory to the departmental stores prefer calendar-based guaranteed access, and stable but competitive pricing to resources unlike spot-market based access to services [69]. In this case demand and preferences are clear, pricing policy can be easily negotiated in advance in a competitive and reasonable manner and resource quality of services can be guaranteed through reservation during the required period as agreed in advance.

Consumers can be charged for access to various resources including CPU cycles, storage, software, and network. The users compose their application using higher-level Grid programming languages. For example, in our Nimrod problem-solving environment we provide a declarative programming language for composing parameter sweep applications and defining application and user requirements such as deadline and budget. The resource broker (working for the user) can carry out the following steps for executing applications:

a. The broker identifies service providers.
b. It identifies suitable resources and establishes their prices (by interacting with GMD and GTS).

c. It selects resources that meet its utility function and objectives. It uses heuristics and/or historical knowledge base while choosing resources and mapping jobs to them.
d. It uses resource services for job processing and issues payments as agreed.

As we are focusing on a generic framework, implementation specific details for releasing the above steps are not presented. For example, implementation specific details of our Nimrod-G resource broker vary from other related systems.

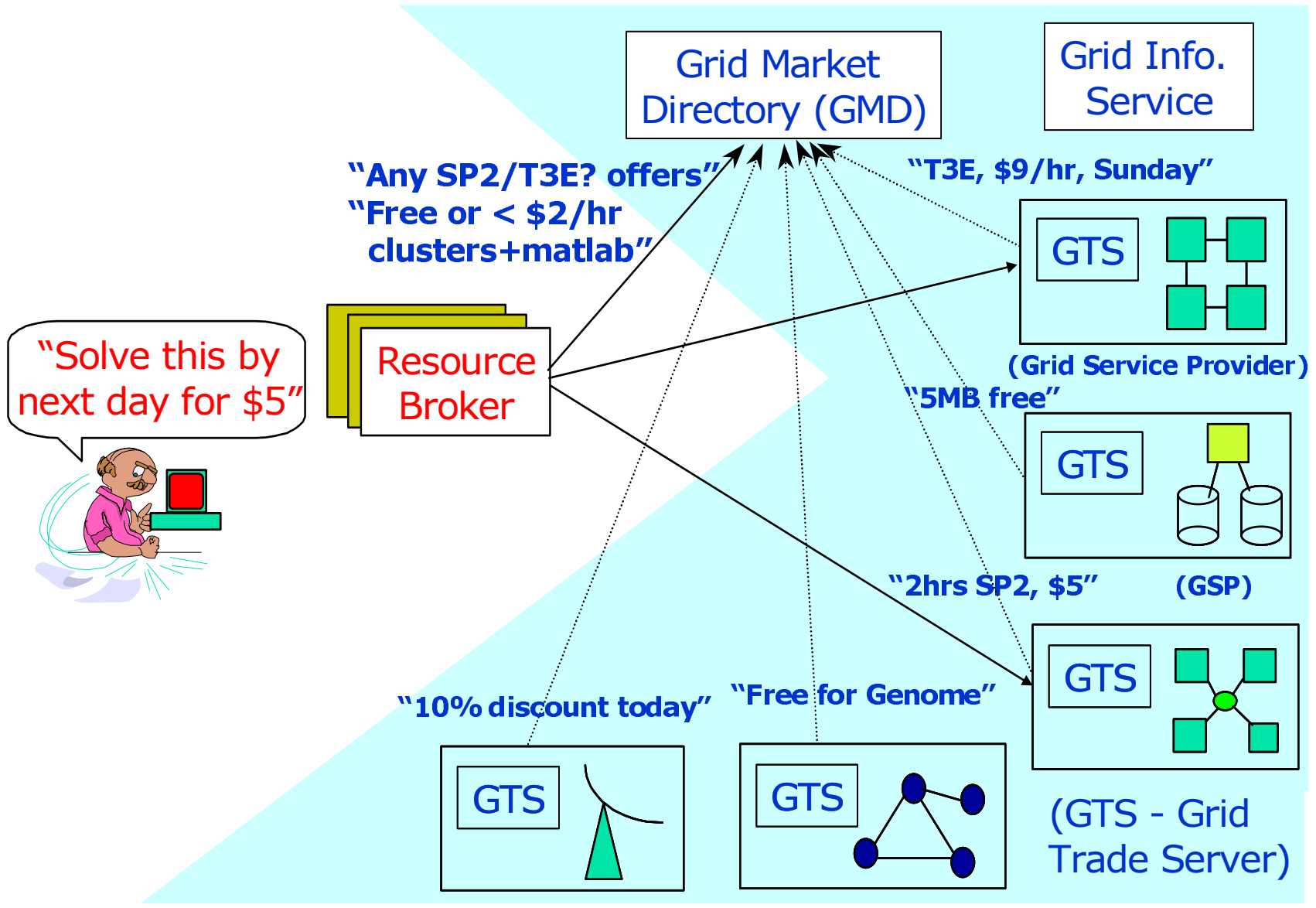

**Figure 3.7: Posted price model and resource trading in a computational market environment.**

### 3.6.2 Posted Price Model

The posted price model is similar to the commodity market model, except that it advertises special offers (see Figure 3.7) in order to attract (new) consumers to establish market share or motivate users to consider using cheaper slots. In this case, brokers need not negotiate directly with GSPs for price, but use posted prices as they are generally cheaper compared to regular prices. The posted-price offers will have usage conditions, but they might be attractive for some users. For example, during holiday periods, demand for resources is likely to be limited and GSPs can post tempting offers or prices aiming to attract users to increase resource utilization. The activities that are specifically related to the posted-price model in addition to those related to the commodity market model are:

a. Grid Service Providers (GSPs) post their special offers and associated conditions etc. in Grid Market Directory.
b. Broker looks at GMD to identify if any of these posted services are available and fits its requirements.
c. Broker enquires (GSP) for availability of posted services.
d. Other steps are similar to those pointed out in commodity market model.

### 3.6.3 Bargaining Model

In the previous models, the brokers pay access prices, which are fixed by GSPs. In the bargaining model, resource brokers bargain with GSPs for lower access price and higher usage duration. Both brokers and GSPs have their own objective functions and they negotiate with each other as long as their objectives are met. The brokers might start with a very low price and GSPs with a higher price. They both negotiate until they reach a mutually agreeable price (see Figure 3.8) or one of them is not willing to negotiate any further. This negotiation is guided by user requirements (e.g., deadline is too relaxed) and brokers can take risk and negotiate for cheaper prices as much as possible and can discard expensive machines. This might lead to lower utilization of resources, so GSPs might be willing to reduce the price instead of wasting resource cycles. Brokers and GSPs generally employ this model when market *supply-and-demand* and service prices are not clearly established. The users can negotiate for a lower price with promise of some kind favour or use of GSPs services in the future.

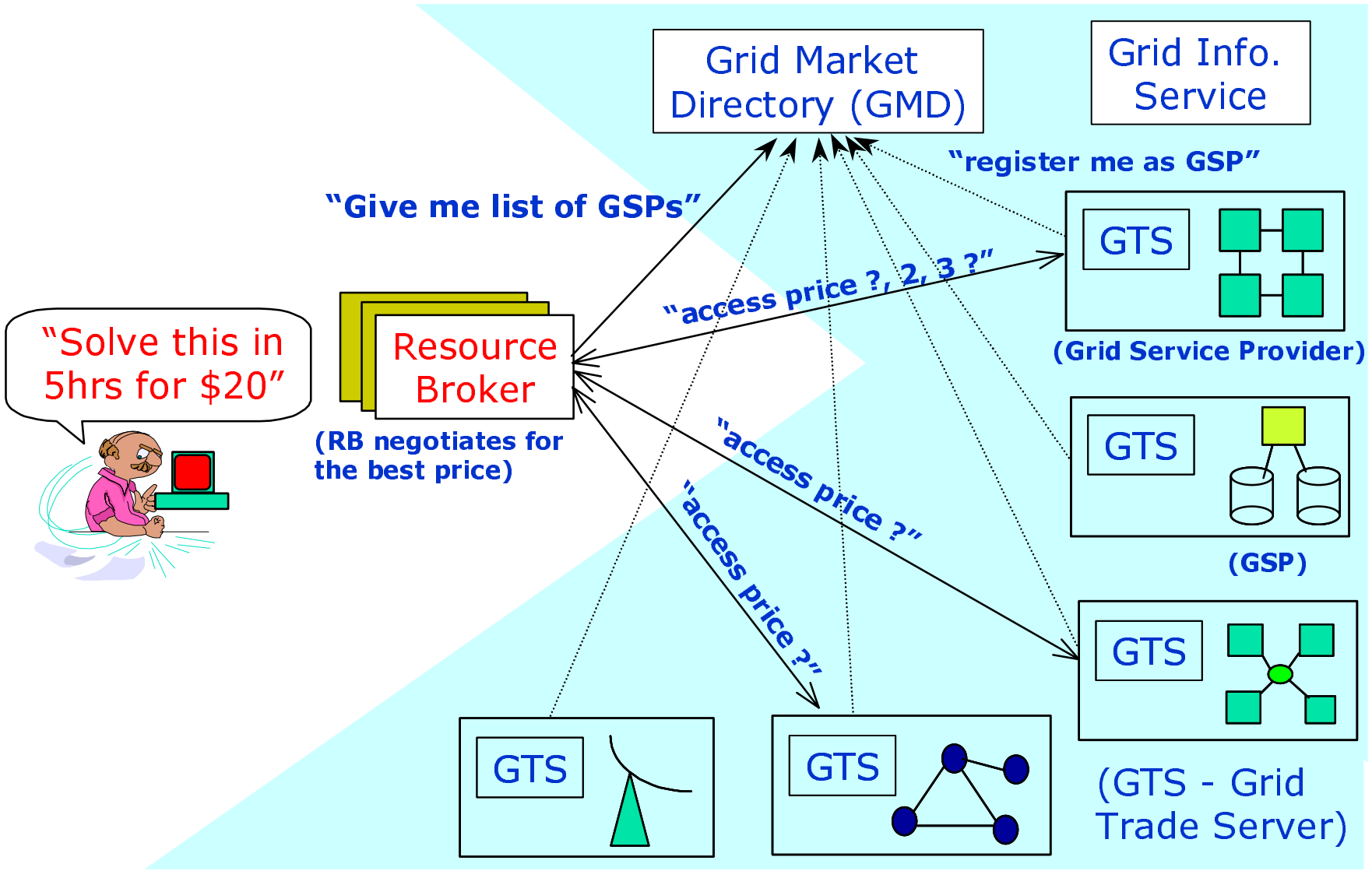

**Figure 3.8: Brokers bargaining for lower access price for minimizing computational cost.**

### 3.6.4 Tender/Contract-Net Model

Tender/Contract-Net model is one of the most widely used models for service negotiation in a distributed problem-solving environment [119]. It is modeled on the contracting mechanism used by businesses to govern the exchange of goods and services. It helps in finding an appropriate service provider to work on a given task. Figure 3.9 illustrates the interaction between brokers and GSPs in their bid to meet their objectives. A user/resource broker asking for a task to be solved is called the *manager* and a resource that might be able to solve the task is called potential *contractor*.

From a manager's perspective, the process is:

1. Consumer (Broker) announces its requirements (using deal template) and invites bids from GSPs.
2. Interested GSPs evaluate the announcement and respond by submitting their bids.
3. Broker evaluates and awards the contract to the most appropriate GSP(s).
4. The broker and GSP communicate privately and use the resource (R).

The contents of the deal template used for work announcement include, addressee (user), eligibility requirements specifications (for instance, Linux, x86arch, and 128MB memory), task/service abstraction, optional price that the user is willing to invest, bid specification (what should offer contain), expiration time (deadline for receiving bids).

From a contractor's/GSP perspective, the process is:

1. Receive tender announcements/advertisements (say in GMD).
2. Evaluate service capability.
3. Respond with bid.
4. Deliver service if bid is accepted.
5. Report results and bill the broker/user as per the usage and agreed bid.

The advantage of this model is that if the selected GSP is unable to deliver a satisfactory service, the brokers can seek services of other GSPs. This protocol has certain disadvantages. A task might be awarded to a less capable GSP if a more capable GSP is busy at award time. Another limitation is that the GRB manager has no obligation to inform potential contractors that an award has already been made. Sometimes, a manager may not receive bids for several reasons: (a) all potential GSPs are busy with other tasks, (b) a potential GSP is idle but ranks the proposed tender/task below the other tasks under consideration, (c) no GSPs, even if idle, are capable of offering service (e.g., resource is Windows NT-based, but user wants Linux). To handle such cases, a GRB can request quick response bids to which GSPs respond with messages such as *eligible, busy, ineligible* or *not interested*. This helps the GRB in making changes to its work plan. For example, the user can change deadline or budget to wait for new GSPs or attract existing GSPs to submit bids.

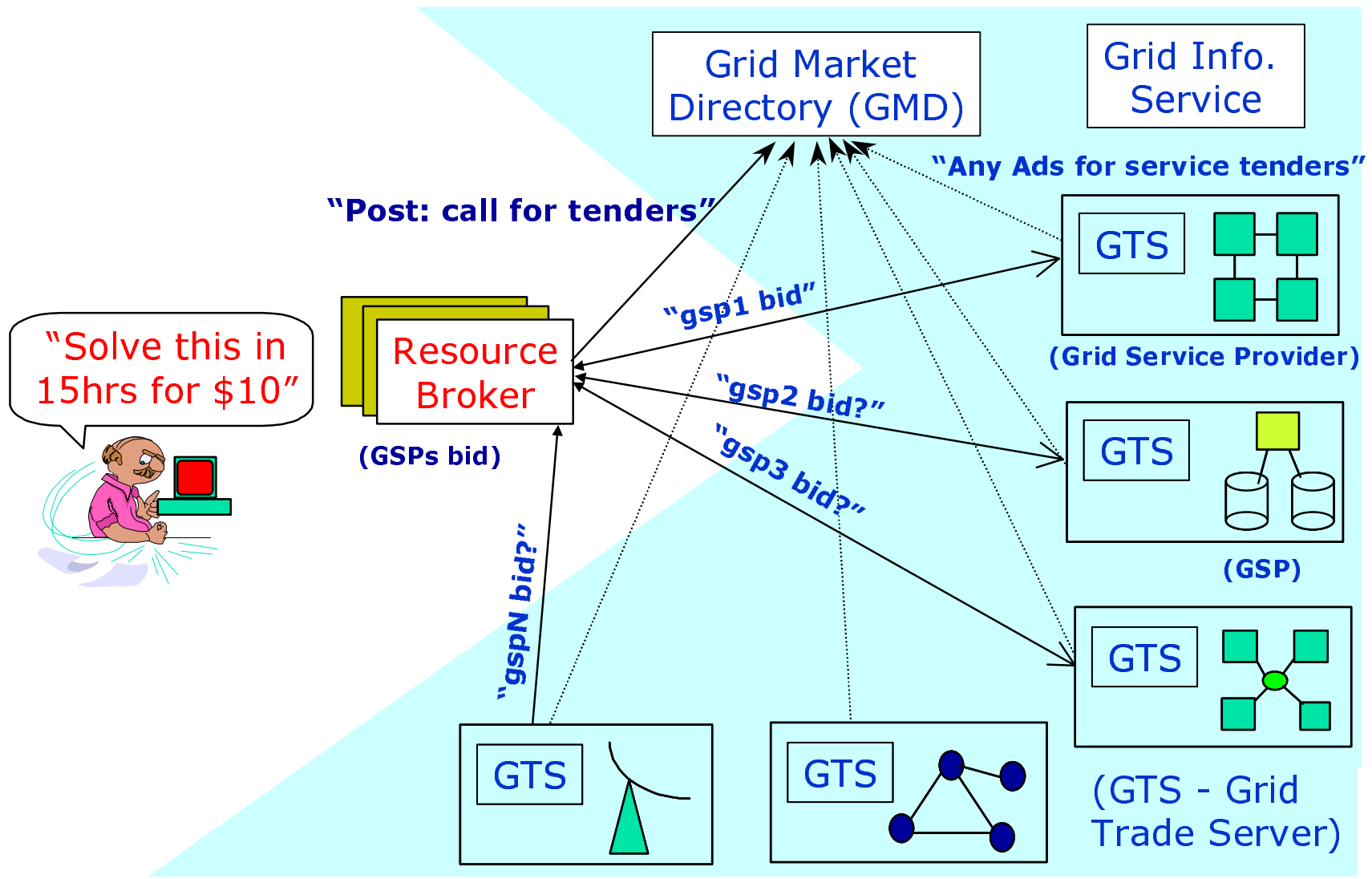

**Figure 3.9: Tender/ContractNet model for resource trading.**

The tender model allows directed contracts to be issued without negotiation. The selected GSP responds with an *acceptance* or *refusal* of award. This capability can simplify the protocol and improve the efficiency of certain services.

### 3.6.5 Auction Model

The auction model supports one-to-many negotiation, between a service provider (seller) and many consumers (buyers), and reduces negotiation to a single value (i.e., price). The auctioneer sets the rules of auction, acceptable for the consumers and the providers. Auctions basically use market forces to negotiate a clearing price for the service.

In the real world, auctions are used extensively, particularly for selling goods/items within a set duration. The three key players involved in auctions are: resource owners, auctioneers (mediators), and buyers (see Figure 3.10). Many e-commerce portals such as Amazon.com and eBay.com are serving as mediators (auctioneers). Both buyers' and sellers' roles can also be automated. In a Grid environment, providers can use an auction protocol for deciding service value/price (see Figure 3.11). The steps involved in the auction process are:

a. GSPs announce their services and invite bids.
b. Brokers offer their bids (and they can see what other consumers offer if they like - depending on open/closed).
c. Step (b) goes on until no one is willing to bid higher price or auctioneer stops if the minimum price line is not met.
d. GSP offers service to the one who wins.
e. Consumer uses the resource.

Auctions can be conducted as open or closed depending on whether they allow back-and-forth offers and counter offers. The consumer may update the bid and the provider may update the offered sale price. Depending on these parameters, auctions can be classified into four types:

- English Auction (first-price open cry)
- First-price sealed-bid auction
- Vickrey (Second-price sealed-bid) auction [144]

- Dutch Auction
- Double Auction (Continuous)

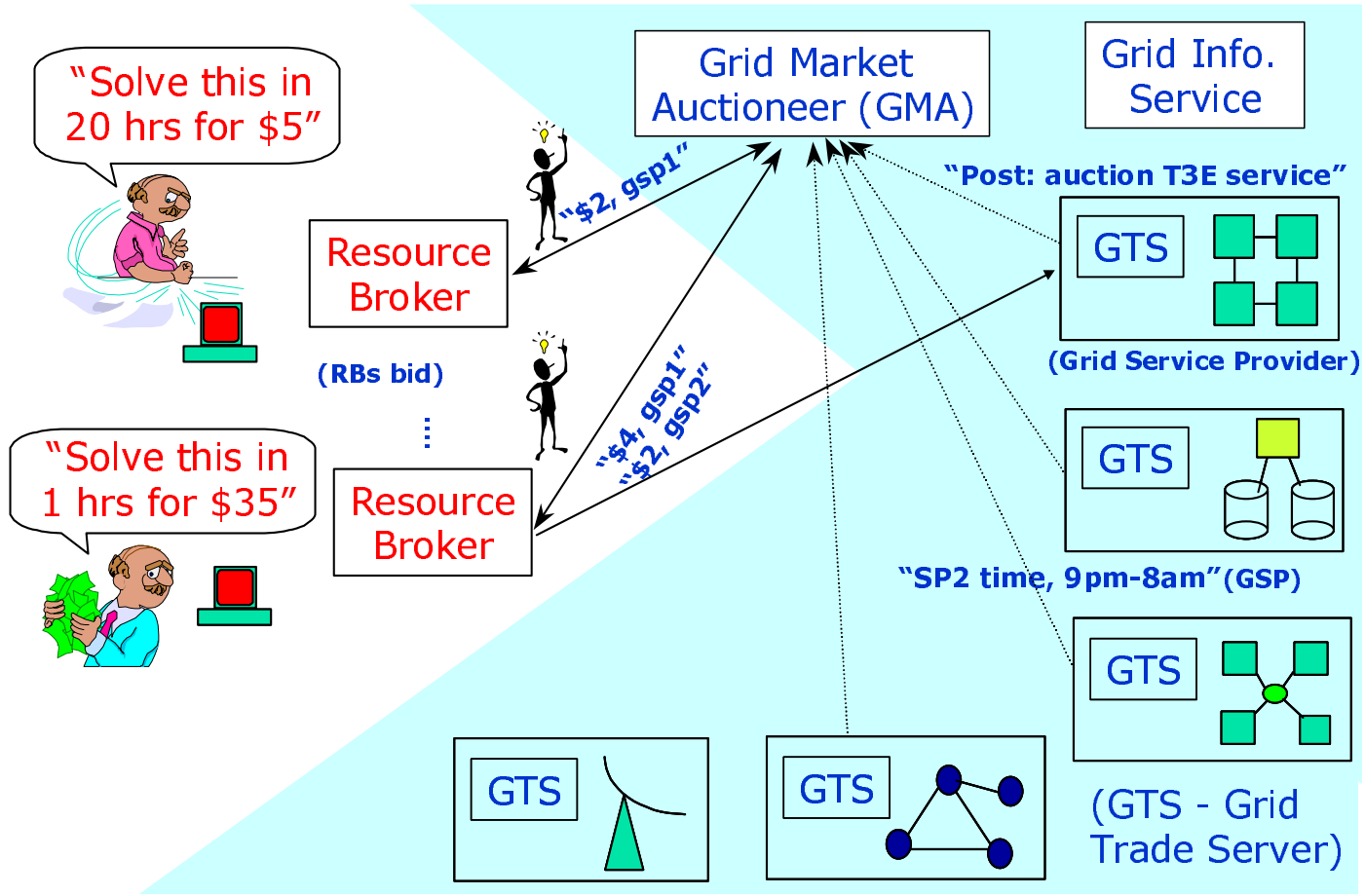

**Figure 3.10: Auctions using external auctioneer.**

*English Auction (first-price open cry)* — all bidders are free to increase their bids exceeding other offers. When none of the bidders are willing to raise the price anymore, the auction ends, and the highest bidder wins the item at the price of his bid. In this model, the key issue is how GRBs decide how much to bid. A GRB has a private value (as defined by the user) and can have a strategy for a series of bids as a function of its private value and prior estimation of other bidder's valuations, and the past bids of others. The GRB decides the private value depending on the user-defined requirements (mainly deadline and budget that he is willing to invest for solving the problem). In the case of private value English auctions, a GRB's dominant strategy is to always bid a small amount "higher" than the current highest bid, and stop when its private value price is reached. In correlated value auctions, the policies are different and allow the auctioneer to increase the price at a constant rate or at the rate he wishes. Those not interested in bidding anymore can openly declare so (open-exit) without re-entry possibility. This information helps other bidders and gives a chance to adjust their valuation.

*First-price sealed-bid auction* — each bidder submits one bid without knowing the others' bids. The highest bidder wins the item at the price of his bid. In this case a broker bid strategy is a function of the private value and the prior beliefs of other bidders' valuations. The best strategy is bid less than its true valuation and it might still win the bid, but it all depends on what the others bid.

*Vickrey (Second-price sealed-bid) auction* — each bidder submits one bid without knowing the others' bids. The highest bidder wins the item at the price of the second highest bidder [144]. The implementation architecture and strategies are similar to the ContractNet/Tender model discussed earlier.

*Dutch Auction* — the auctioneer starts with a high bid/price and continuously lowers the price until one of the bidders takes the item at the current price. It is similar to first-price sealed-bid auction because in both cases the bid matters only if it is the highest, and no relevant information is revealed during the auction process. From the broker's bidding strategic point of view, Dutch auction is similar to English (first-price sealed-bid auction). The key difference between them is that in an English auction bids start with low opening and increase progressively until demand falls whereas, in a Dutch auction bids start with high opening and decrease progressively until demand rises to match supply.

The interaction protocols for Dutch auction are as follows: the auction attempts to find market price for a good/service by starting at a price much higher than the expected market value, then progressively reducing the price until one of the buyers accepts the price. The rate of reduction in price is up to the auctioneer and they have a reserve price below which not to go. If the auction reduces the price to reserve price with no buyers, the auction terminates. In terms of real time, Dutch auction is much more efficient as the auctioneer can decrease the price at a strategic rate and first higher bidder wins. In an Internet wide auction, it is appealing in terms of automating the process wherein all parties can define their strategies for agents that can participate in multiple auctions to optimize their objective functions.

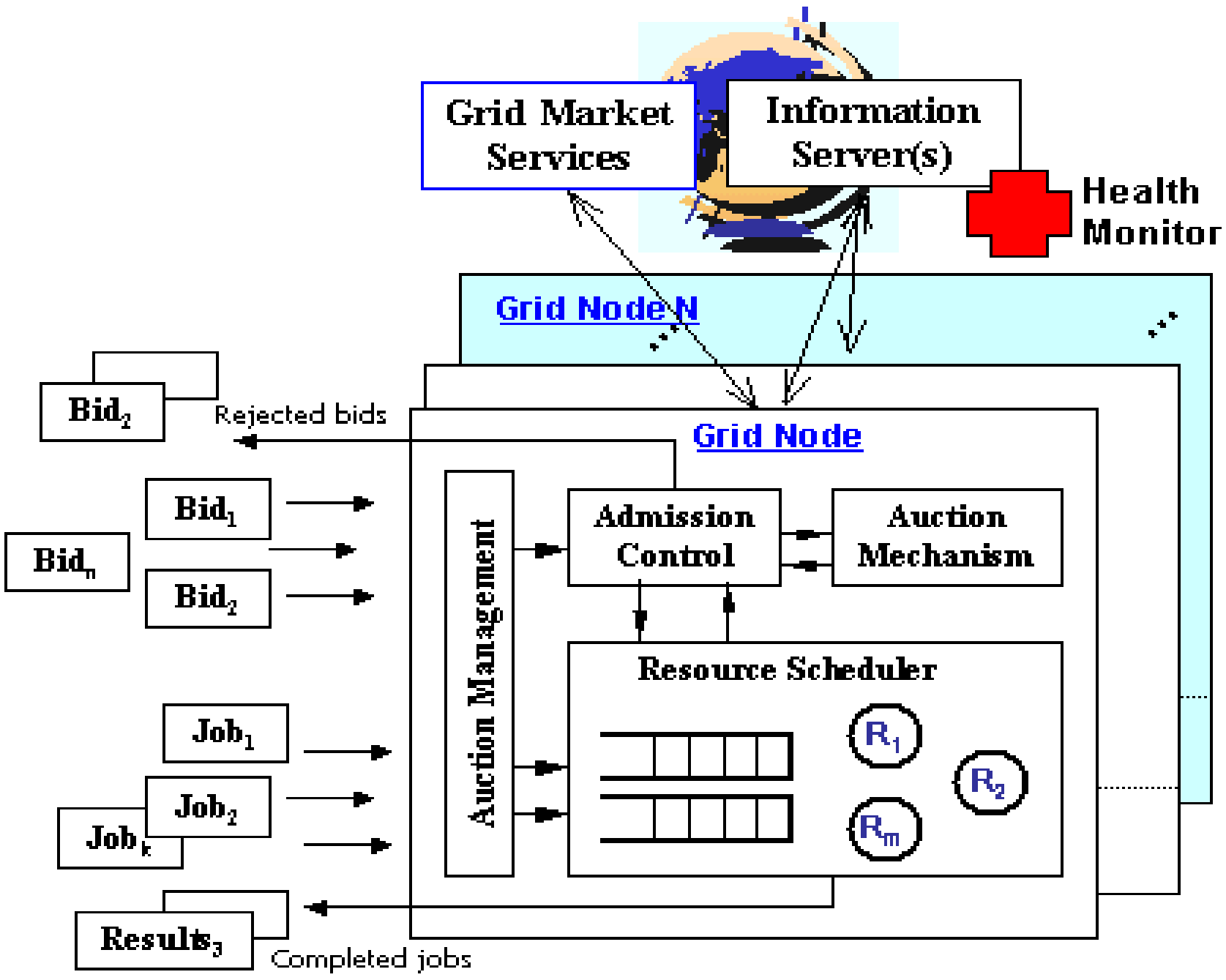

**Figure 3.11: Auction using their own Auctioneer.**

*Double Auction* — This is one of the most common exchange institutions in the marketplace whose roots go back to ancient Egypt and Mesopotamia [66]. In fact, it is the primary economic model for trading of equities, commodities, and derivatives in stock markets (e.g., NASDAQ). In the double auction model, buy orders (*bids*) and sell orders (*asks*) may be submitted at anytime during the trading period. If at any time there are open bids and asks that match or are compatible in terms of price and requirements (e.g., quantity of goods or shares), a trade is executed immediately. In this auction orders are ranked highest to lowest to generate demand and supply profiles. From the profiles, the maximum quantity exchanged can be determined by matching *asks* (starting with lowest price and moving up) with demand *bids* (starting with highest price and moving down). Researchers have developed software-based agents mechanisms to automate a double auction for stock trading with or without human interaction [115].

The double auction model has high potential for Grid computing. The brokers can easily be enabled to issue *bids* depending on budget, deadline, job complexity, scheduling strategy, and resource characteristics requirements and GSPs can issue *asks* depending on current load and perceived demand, and price constraints. Both orders can be submitted to GMD agents that provide continuous clearance or matching services. Since bids are cleared continuously, both GRBs and GSPs can make instant decisions with less computational overhead and complexity.

All the above auctions differ in terms of whether they are performed as open or closed auctions and the offer price for the highest bidder. In open auctions, bidding agents can know the bid value of other agents and will have an opportunity to offer competitive bids. In closed auctions, the participants' bids are not disclosed to others. Auctions can suffer from collusion (if bidders coordinate their bid prices so that the bids stay artificially low), deceptive auctioneers in the case of a Vickrey auction (auctioneer may overstate the second highest bid to the highest bidder unless that bidder can vary it), deceptive bidders, counter speculation, etc.

### 3.6.6 Bid-based Proportional Resource Sharing Model

Market-based proportional resource sharing systems are quite popular in cooperative problem-solving environments like clusters (in single administrative domain). In this model, the percentage of resource share allocated to the user application is proportional to the bid value in comparison to other users' bids. The users are allocated credits or tokens, which they can use for having access to resources. The value of each credit depends on the resource demand and the value that other users place on the resource at the time of usage. For example, consider two users wishing to access a resource with similar requirements, but the first user is willing to spend 2 tokens and the second user is willing to spend 4 tokens. In this case, the first user gets 1/3 of resource share whereas the second user gets 2/3 of resource share, which is proportional to the value that both users place on the resource for executing their applications.

This can be a good way of managing a large shared resource in an organization or resource owned by multiple individuals (like multiple departments in a university) who can have credit allocation depending on the investment they made. They can specify how much of a credit they are willing to offer for running their applications on the resource. For example, a user might specify low credits for non-interactive batch jobs and high credits for interactive jobs with high response times. GSPs can employ this model for offering a QoS for higher price paying customers in a shared resource environment (as shown in Figure 3.12). Systems such as Rexec/Anemone and Xenoservers, D'Agents CPU market employ proportional resource sharing model in managing resource allocations [98].

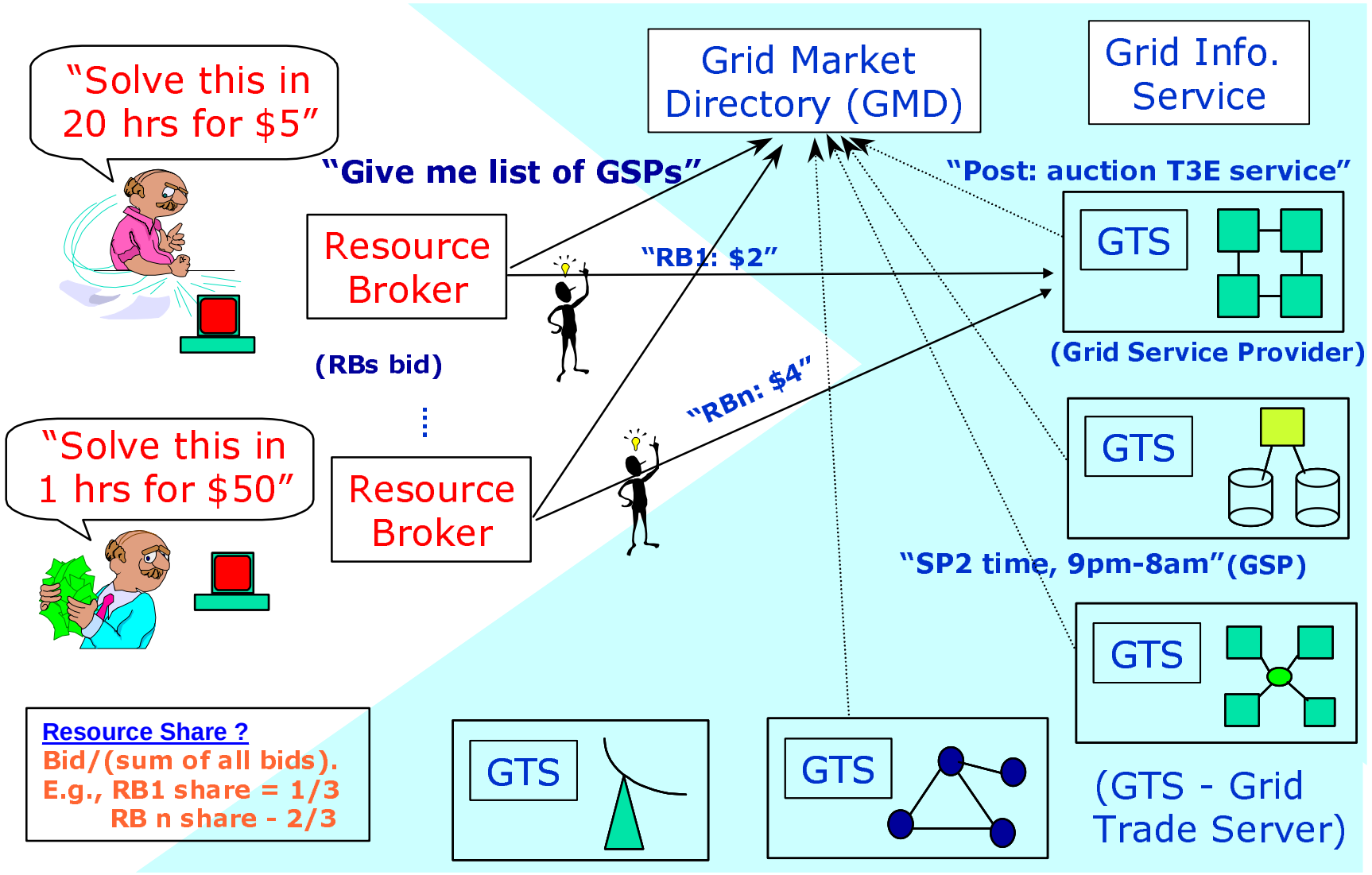

**Figure 3.12: Market-based proportional resource sharing.**

### 3.6.7 Cooperative Bartering Model

A community of individuals shares each other's resources to create a cooperative computing environment. Those who are contributing their resources to a common pool can get access to that pool. A sophisticated model can also be employed here for deciding how much resources share contributors can get. It can involve credits that one can earn by sharing a resource, which can then be used when needed. A system like Mojonation.net employs this model for storage sharing. This model works when those participating in the Grid act as both service providers and consumers.

### 3.6.8 Monopoly/Oligopoly

In the previously mentioned models we have assumed a competitive market where several GSPs and brokers/consumers determine the market price. However, there exist cases where a single GSP dominates the market and is the single provider of a particular service. In economic theory this model is known as a monopoly. Users cannot influence the prices of services and have to choose the service at the price given by the single GSP who monopolized the Grid marketplace. As regards the technical realization of this model, the single site puts the prices into the GMD or information services and brokers consult it without any possibility to negotiate prices.

The competitive markets are one extreme and monopolies are the other extreme. In most of the cases, the market situation is *oligopoly*, which is in between these two extreme cases: a small number of GSPs dominate the market and set the prices.

### 3.6.9 Other Influences on Market Prices

We now state more influences on price setting strategies in competitive, international markets. Supply and demand is the most common one but one also has to take into account national borders and different pricing policies within different countries such as taxation, consumer price index, inflation, etc. These factors are not dealt with in this chapter, however implementations may need to consider them. There are micro and macro-economic factors that play an important role. One can also neglect them and build a price model on which all the Grid consumers have to agree. So this would correspond to an international market with special rules. Then, a model has to be formed for price changes. What is the factor for that change? Is there a monopoly that can decide what to do? Is the market transparent with optimally adapted prices? These are some of the main questions that need to be answered by GSPs when they decide their prices in an international market. A broker may consult the Grid Information Service to find out where the price for a particular service is minimal. For instance, one country might impose special taxes on a service whereas another country does not.

There are occasions where resources are not valued as per the actual cost of resources and overhead involved in offering services. When new players enter the market, in order to attract customers from the existing GSPs they are likely to offer access at minimal price by under valuing resources. This leads to price wars as GSPs are caught in a price cutting round to compete with each other. Measures such as intervention of *price regulation authorities* can be in place to prevent the market from collapsing or leaving it to the to the market to consolidate naturally.

## 3.7 Summary and Conclusion

We have discussed motivations for the use of computational economy as a metaphor for the management of resources and application scheduling in Grid computing environments. We proposed the Grid Architecture for Computational Economy (GRACE) framework that leverages existing middleware services and tools and offers new services that are essential for realizing a Grid market place. We have presented economic models such as commodity market, posted prices, bargaining, tendering, auction, proportional resource sharing or shareholder, and cooperative bartering models along with architecture and strategies for releasing them within the GRACE framework.

The computational economies driven brokering system can be applied to peer-to-peer computing [12] applications that enable content sharing. Systems like Napster [88] or Gnutella [39] could use infrastructure that is similar to GRACE for encouraging people to share files, contents, or music in larger scale by providing them economic incentive. The brokering systems like Nimrod-G can discover the best content provider that meets consumers QoS requirements. We believe this approach, of providing an economic incentive for resource owners to share their resources and resource users to trade-off between the deadline and budget, promotes the Grid as a platform for mainstream computing, which can lead to the emergence of a new service oriented computing industry.

In the next chapter, we present an implementation of computational economy within the Nimrod-G resource broker along with economic-based scheduling algorithms and application scheduling experiments on the World Wide Grid.

## Chapter 4

# The Nimrod-G Grid Resource Broker and Economic Scheduling Algorithms

This chapter presents the Nimrod-G Grid resource broker as an example of a Grid system that uses a computational economy driven architecture for managing resources and scheduling task farming applications on large-scale distributed resources. It discusses a layered and component-oriented modular architecture for the Nimrod-G broker design and development. The architecture is generic enough to leverage services provided by various Grid middleware systems such as Globus, Legion, and Condor for uniform access to diverse resources; and the GRACE trading mechanisms. It briefly discusses the deadline and budget constrained scheduling algorithms that we developed and incorporated into the Nimrod-G broker. We evaluate the performance of scheduling strategies and their ability in meeting the user quality of service requirements. Finally, we discuss the results of scheduling parameter sweep applications on the World-Wide Grid resources using the Nimrod-G resource broker to demonstrate its ability in allocating resources depending on their availability, capability, user-level performance, and cost.

## 4.1 Introduction

Computational Grids enable the coordinated and aggregated use of geographically distributed resources, often owned by autonomous organizations, for solving large-scale problems in science, engineering, and commerce. However, application composition, resource management and scheduling in these environments is a complex undertaking [98]. This is due to the geographic distribution of resources that are often owned by different organizations having different usage policies and cost models, and varying loads and availability patterns. To address these resource management challenges, we have developed a distributed computational economy framework for quality of service-driven resource allocation and regulation of supply and demand for resources. The new framework offers incentive to resource owners for being part of the Grid and motivates resource users to trade off between time for results delivery and economic cost, i.e., deadline and budget [99].

Resource management systems need to provide mechanisms and tools that realize the goals of both service providers and consumers. The resource consumers need a *utility model*, representing their resource demand and preferences, and *brokers* that automatically generate strategies for choosing providers based on this model. Further, the brokers need to manage all issues associated with the execution of the underlying application. The service providers need *price generation schemes* so as to increase system utilization, as well as economic *protocols* that help them to offer competitive services. For the market to be competitive and efficient, coordination mechanisms are required that help the market reach an equilibrium price, that is, the market price at which the supply of a service equals the quantity demanded [22]. Numerous economic theories have been proposed in the literature and many commonly used economic models for selling goods and services can be employed as negotiation protocols in Grid computing. Some of these market or social driven economic models are shown in Table 1 along with the identity of the distributed system that adopted the approach [104].

These economic models regulate the supply and demand for resources in Grid-based virtual enterprises. We demonstrate the power of these models in scheduling computations using the Nimrod-G resource broker on a Grid testbed, called the World Wide Grid (WWG) spanning across five continents. Whilst it is not the goal of the system to earn revenue for the resource providers, this approach does provide an economic incentive for resource owners to share their resources on the Grid. Further, it encourages the emergence of a new service oriented computing industry. Importantly, it provides mechanisms to trade-off QoS parameters, deadline and computational cost, and offers incentive for users to relax their requirements. For example, a user may be prepared to accept a later deadline if the computation can be achieved more cheaply.

**Table 4.1: Economics models and their use in some distributed computing scheduling systems.**

| Economic Model | Adopted by |
|---|---|
| Commodity Market | Mungi [33], MOSIX [147],& Nimrod-G [100] |
| Posted Price | Nimrod-G |
| Bargaining | Mariposa [83] & Nimrod-G |
| Tendering or Contract-Net Model | Mariposa |
| Auction Model | Spawn [14] & Popcorn [87] |
| Bid-based Proportional Resource Sharing | Rexec & Anemone [9] |
| Community and Coalition | Condor and SETI@Home [143] |
| Cooperative Bartering | MojoNation [84] |
| Monopoly and Oligopoly | Nimrod-G broker can be used to choose between resources offered at different quality and prices. |

The rest of this chapter explores the use of an economic paradigm for Grid computing with particular emphasis on providing the tools and mechanisms that support economic-based scheduling. The emphasis will be placed on the Nimrod-G resource broker that supports soft-deadline and budget based scheduling of parameter-sweep applications on the Grid [98]. Depending on the users' Quality of Service (QoS) requirements, the resource broker dynamically leases Grid services at runtime depending on their cost, quality, and availability. The broker supports the optimisation of time or cost within specified deadline and budget constraints. The results of a series of scheduling experiments that we conducted on the WWG testbed using the Nimrod broker will be reported.

## 4.2 The Nimrod-G Resource Broker: An Economic based Grid Scheduler

### 4.2.1 Objectives and goals

Nimrod-G is a tool for automated modeling and execution of parameter sweep applications (parameter studies) over global computational Grids [100][98][103]. It provides a simple declarative parametric modeling language for expressing parametric experiments. A domain expert can easily create a plan for a parametric experiment and use the Nimrod system to submit jobs for execution. It uses novel resource management and scheduling algorithms based on economic principles. Specifically, it supports user-defined deadline and budget constraints for schedule optimisations and manages supply and demand of resources in the Grid using a set of resource trading services [99].

Nimrod-G provides a persistent and programmable *task-farming engine* (TFE) that enables "plugging" of user-defined schedulers and customised applications or problem solving environments (e.g., ActiveSheets) in place of default components. The task-farming engine is a coordination point for processes performing resource trading, scheduling, data and executable staging, remote execution, and result collation. In the past, the major focus of our project was on creating *tools* that help domain experts to compose their legacy serial applications for parameter studies and run them on computational clusters and manually managed Grids [21][18]. Our current focus is on the use of economic principles in resource management and scheduling on the Grid in order to provide some measurable quality of service to the end user. Real-world economic methods provide incentives for owners to contribute their resources to markets, and it also provides consumers with a basis for trading the quality of service they receive against cost. That is, our focus revolves within an intersection area of Grid architectures, economic principles, and scheduling optimizations  (see Figure 4.1), which is essential for pushing the Grid into the mainstream computing.

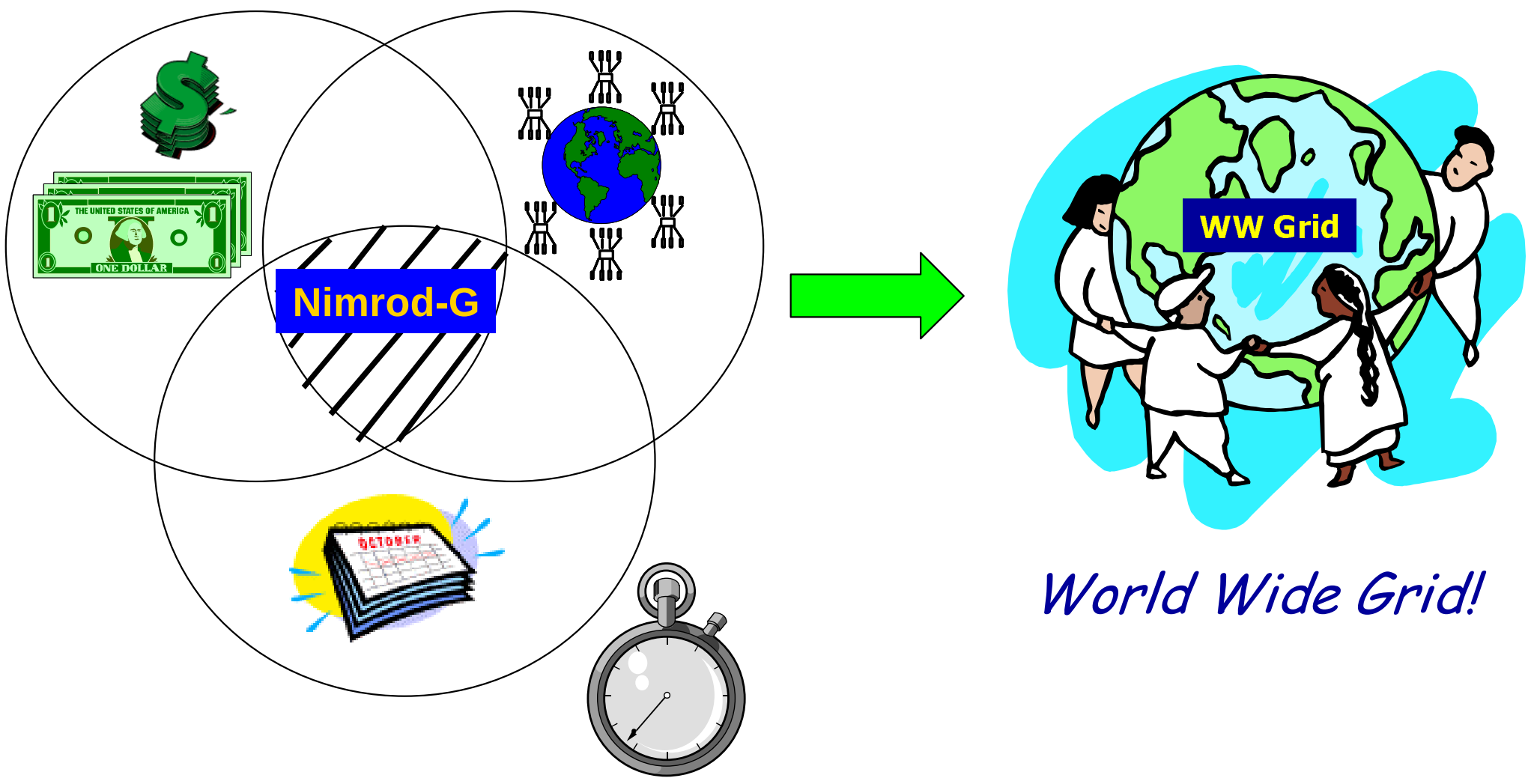

**Figure 4.1: QoS based resource management: intersection of economic, scheduling, and Grid worlds.**

### 4.2.2 Services and End Users

Nimrod-G provides a suite of tools and services for creating parameter sweep applications, performing resource management, and scheduling applications. They are, a simple declarative programming language and associated GUI tools for creating scripts and parameterization of application input data files, and a Grid resource broker with programmable entities for processing jobs on Grid resources. The resource broker is made of a number of components, namely a persistent and programmable task farming engine, a schedule advisor, and a dispatcher, whose functionalities are discussed later. It also provides job management services that can be used for creating user-defined schedulers, steering and monitoring tools, and customized applications. Therefore, the end users that benefit from Nimrod-G tools, protocols, and services are:

- **Domain Experts:** This group includes scientific, engineering, and commercial users with large-scale data-set processing requirements. Parameter applications can use Nimrod-G tools to compose them as coarse-grained data-parallel, parameter sweep applications for executing on distributed resources. They can also take advantage of the Nimrod-G broker features to trade off between a deadline and the cost of computation while scheduling application execution on the Grid. This quality of service aspect is important to end users, because the results are only useful if they are returned in a timely manner. Previous Grid work has largely ignored this aspect of running real applications.

- **Problem Solving Environments Developers:** Application developers can Grid-enable their applications with their own mechanisms to submit jobs to the Nimrod-G resource broker at runtime depending on user requirements for processing on the Grid. This gives them the ability to create applications capable of directly using Nimrod-G tools and job management services, which in turn enables their applications for Grid execution.

- **Task Farming or Master-Worker Programming Environments Designers:** These users can focus on designing and developing easy to use and powerful application creation primitives for task farming and master-work style programming model; developing translators and application execution environments by taking advantage of Nimrod-G runtime machinery for executing jobs on distributed Grid resources.

- **Scheduling Researchers:** The scheduling policy developers generally use simulation techniques and tools such as GridSim [81] and Simgrid [43] for evaluating performance of their algorithms. In simulation, it is very difficult to capture the complete property and behavior of a real world system and hence, evaluation results may be inaccurate. Accordingly, to prove the usefulness of scheduling algorithms on actual systems, researchers need to develop runtime machinery, which is a resource intensive and time-consuming task. This can be overcome by using Nimrod-G broker programmable capability. Researchers can use Nimrod-G job management protocols and services to develop their own scheduler and associated scheduling algorithms. The new scheduler can be used to run actual applications on distributed resources and then evaluate ability of scheduling algorithms in optimally mapping jobs to resources.

### 4.2.3 Architecture

The Nimrod-G toolkit and resource broker is developed by leveraging services provided by Grid middleware systems such as Globus, Legion, Condor/G, and the GRACE trading mechanisms. These middleware systems provide a set of low-level protocols for secure and uniform access to remote resources; and services for accessing resources information and storage management,. The modular and layered architecture of Nimrod-G is shown in Figure 4.2.

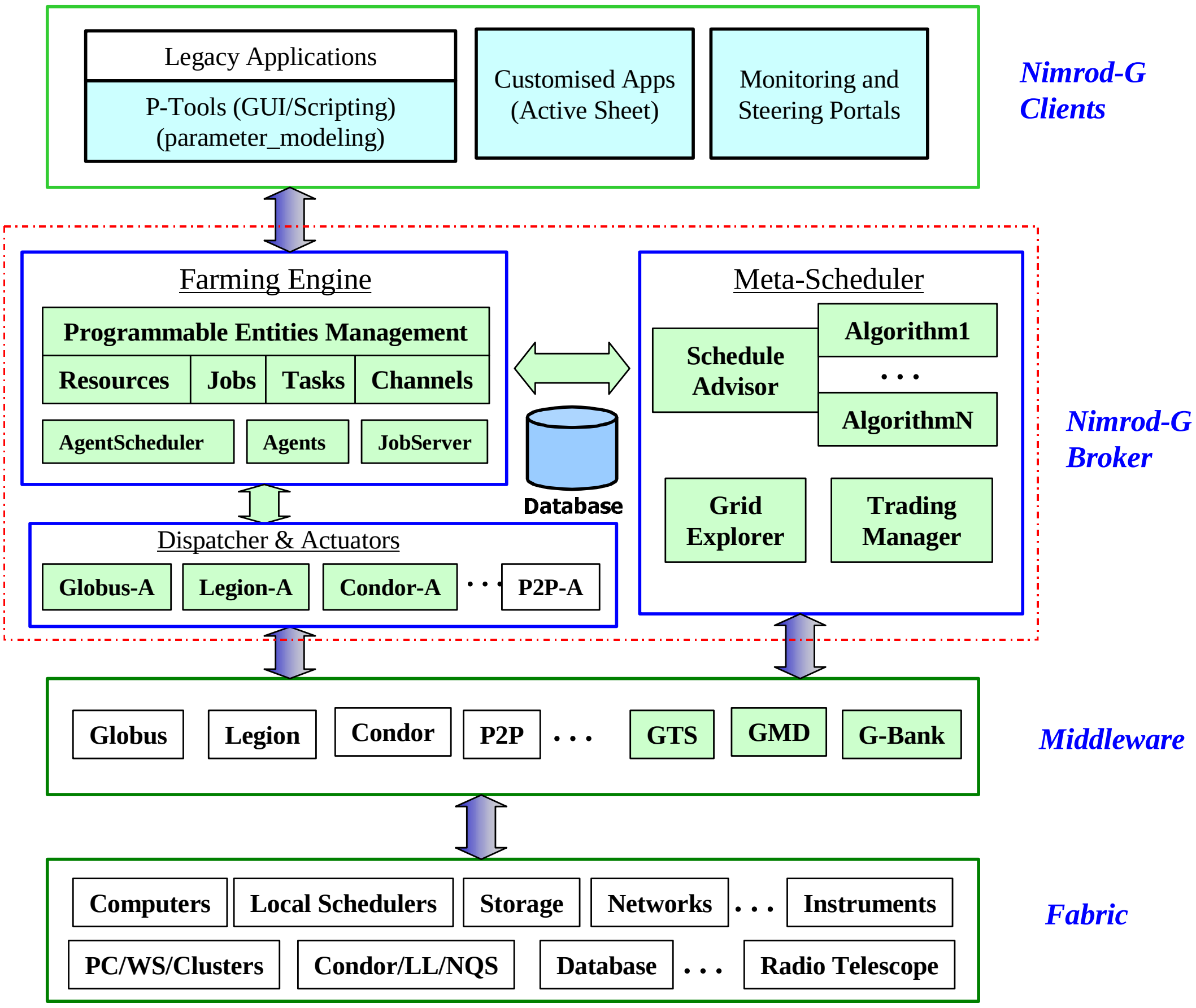

**Figure 4.2: A layered with hourglass shape architecture of Nimrod-G system.**

The key components of Nimrod-G resource broker consist of:

- Nimrod-G Clients, which can be:
  - Tools for creating parameter sweep applications.
  - Steering and control monitors, and
  - Customised end user applications (e.g., ActiveSheets [20]).
- The Nimrod-G Resource Broker, that consists of:
  - A Task Farming Engine (TFE),
  - A Scheduler that performs resource discovery, trading, and scheduling,
  - A Dispatcher and Actuator, and
  - Agents for managing the execution of jobs on resources.

The Nimrod-G broker architecture leverages services provided by lower-level different Grid middleware solutions to perform resource discovery, trading, and deployment of jobs on Grid resources.

### 4.2.4 Nimrod-G Clients

#### *Tools for Creating Parameter Sweep Applications*

Nimrod supports GUI tools and declarative programming language that assist in creation of parameter sweep applications [21]. They allow the user to: a) parameterise input files, b) prepare a plan file containing the commands that define parameters and their values, c) generate a run file, which converts the generic plan file to a detailed list of jobs, and d) control and monitor execution of the jobs. The application execution environment handles online creation of input files and command line arguments through parameter substitution.

#### *Steering and Control Monitors*

These components act as a user-interface for controlling and monitoring a Nimrod-G experiment. The user can vary constraints related to time and cost that influence the direction the scheduler takes while selecting resources. It serves as a monitoring console and lists the status of all jobs, which a user can view and control. A Nimrod-G monitoring and steering client snapshot is shown in Figure 4.3. Another feature of the Nimrod-G client is that it is possible to run multiple instances of the same client at different locations. That means the experiment can be started on one machine, monitored on another machine by the same or different user, and the experiment can be controlled from yet another location. We have used this feature to monitor and control an experiment from Monash University and Pittsburgh Supercomputing Centre at Carnegie Melon University simultaneously during HPDC-2000 research demonstrations.

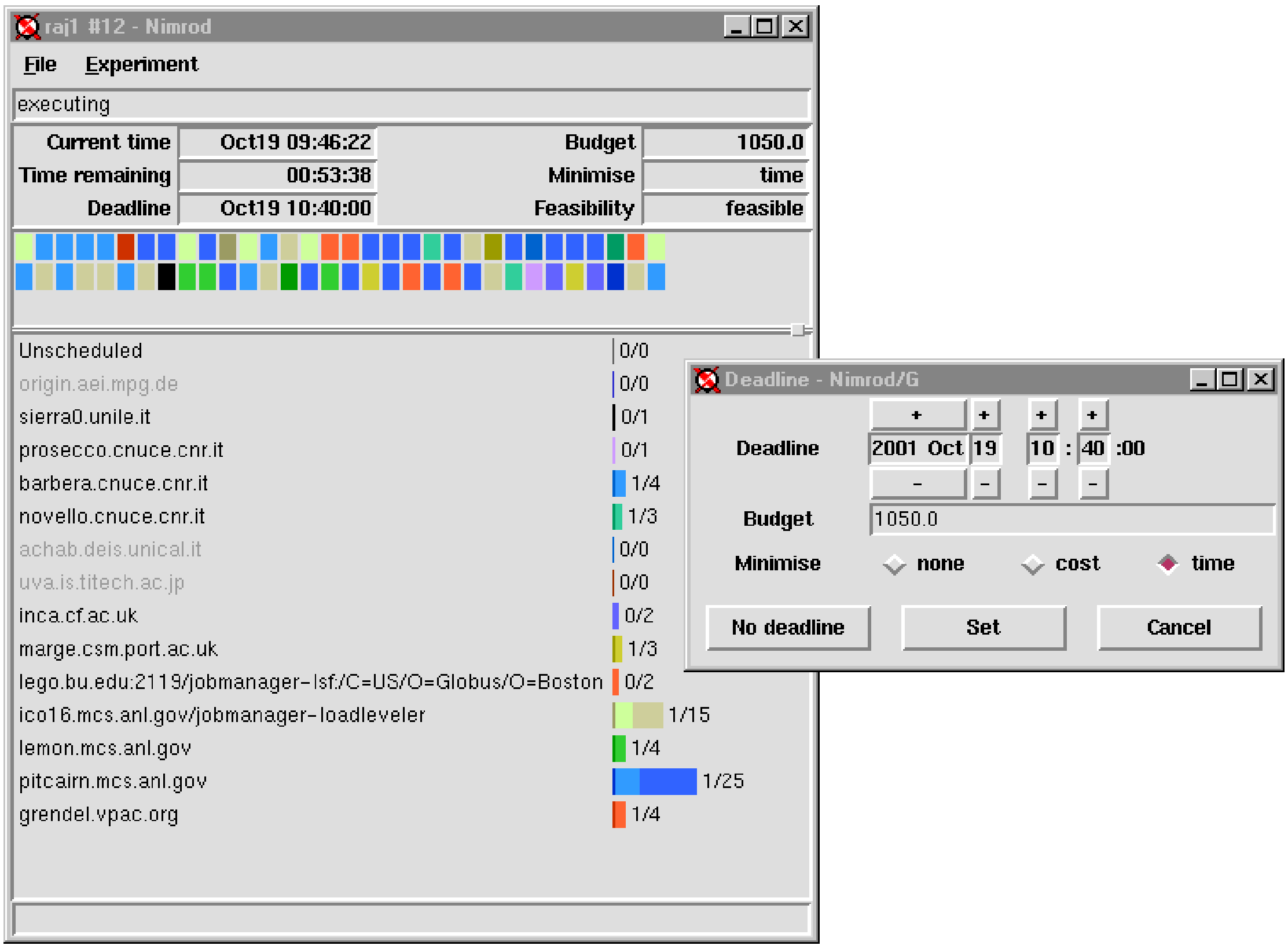

**Figure 4.3: A Snapshot of Nimrod-G Execution Monitoring and Steering Client.**

#### *Customised End User Applications*

Specialized applications can be developed to create jobs at runtime and add jobs to the Nimrod-G Engine for processing on the Grid. These applications can use the Nimrod-G job management services (APIs and protocols) for adding and managing jobs. One such application is ActiveSheets [20], an extended Microsoft Excel spreadsheet that submits cell functions as jobs to the Nimrod-G broker for parallel execution on the Grid (see Figure 4.4). Another example is the Nimrod/O system, a tool that uses non-linear optimization algorithms to facilitate automatic optimal design [16]. This tool has been used on a variety of case studies, including antenna design, smog modeling, durability optimization, aerofoil design, and computational fluid dynamics [17].

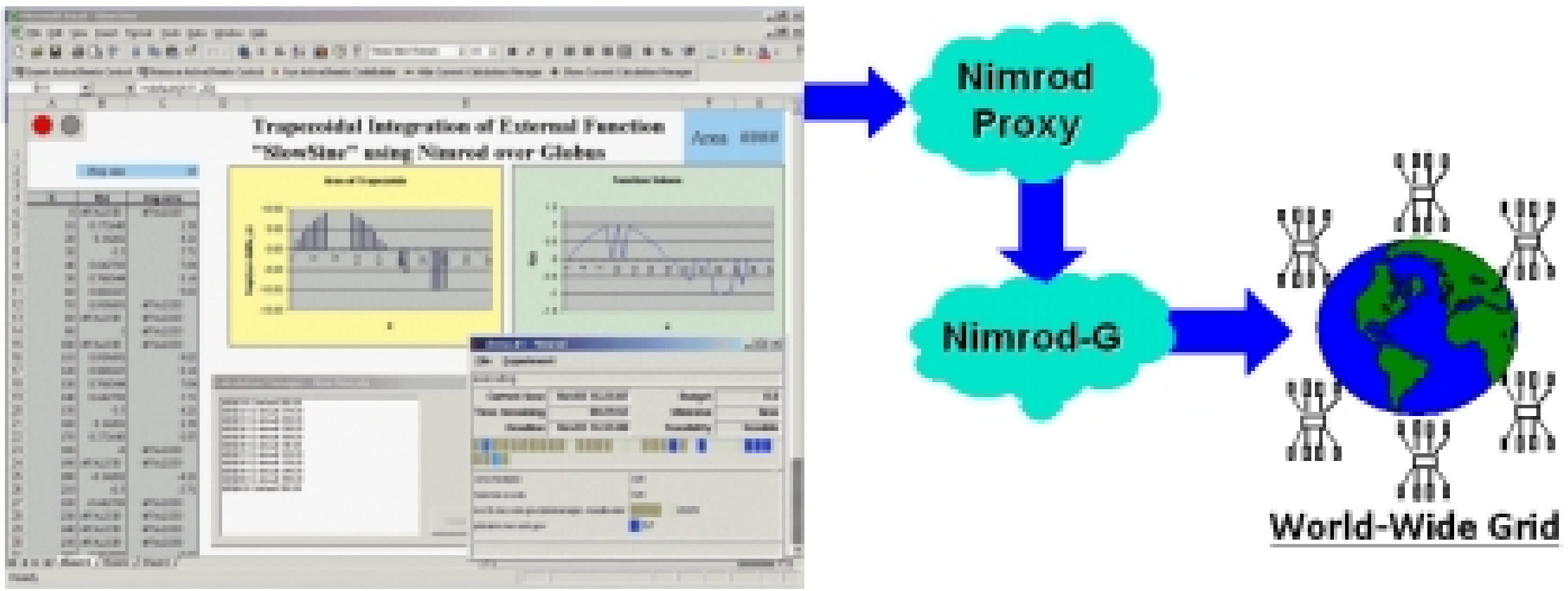

**Figure 4.4: Active Sheet – Spreadsheet processing on the Grid using the Nimrod-G broker.**

### 4.2.5 The Nimrod-G Grid Resource Broker

The Nimrod-G Resource broker is responsible for determining the specific requirements that an experiment places on the Grid and performing resource discovery, scheduling, dispatching jobs to remote Grid nodes, starting and managing job execution, and gathering results back to the home node. The sub-modules of our resource broker are, the task farming engine; the scheduler that consists of a Grid explorer for resource discovery, a schedule advisor backed with scheduling algorithms, and a resource trading manager; a dispatcher and an actuator for deploying agents on Grid resources; and agents for managing execution of Nimrod-G jobs on Grid resources. The interaction between components of the Nimrod-G runtime machinery and Grid services during runtime is shown in Figure 4.5.

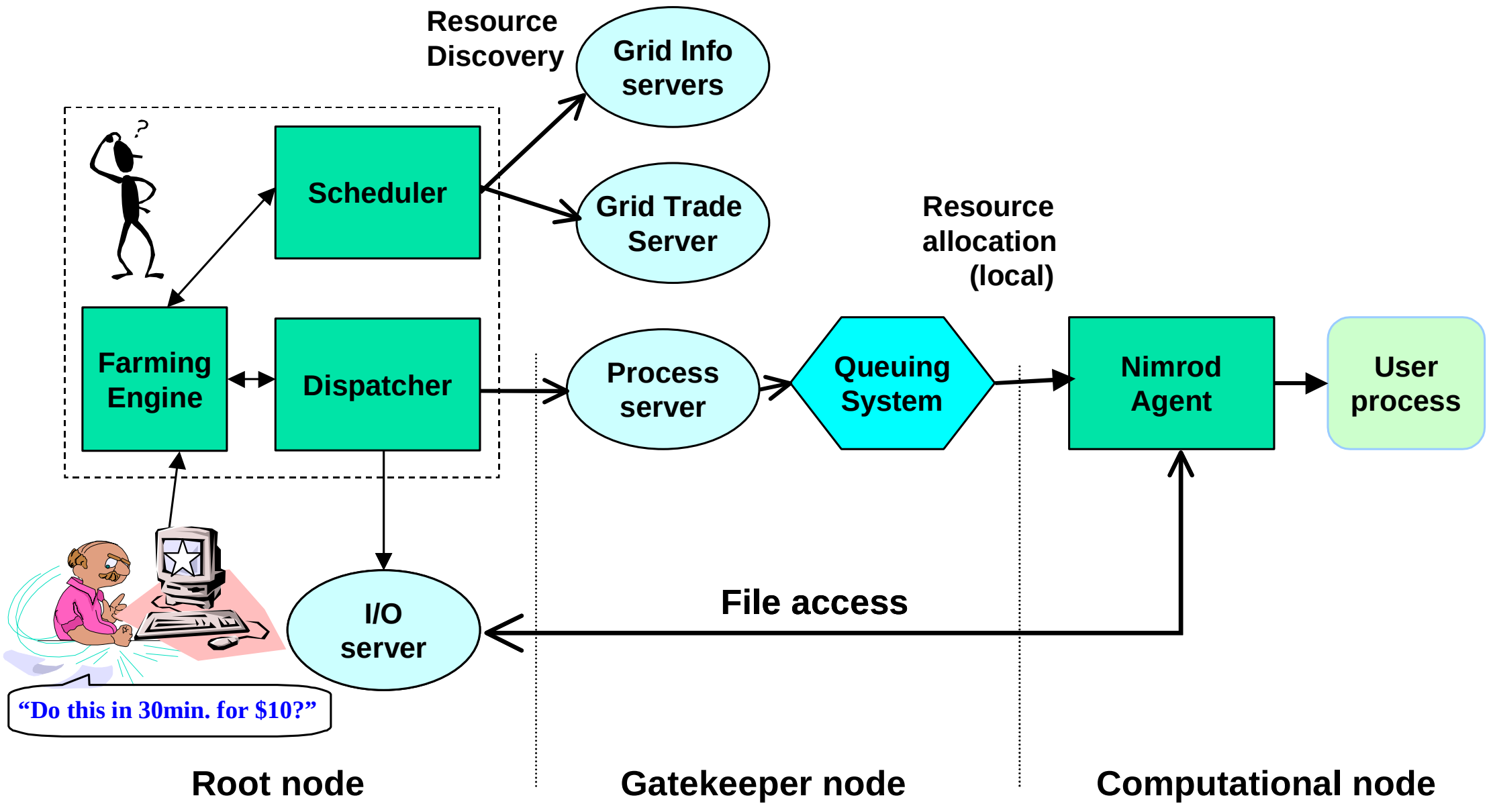

**Figure 4.5: The Flow actions in the Nimrod-G runtime environment.**

### *The Task Farming Engine (TFE)*

The Nimrod-G task-farming engine is a persistent and programmable job control agent that manages and controls an experiment. It consists of a database to provide persistence, which is accessed through a thin management interface. The farming engine is responsible for the parameterization of the experiment, the actual creation of jobs, the maintenance of job status, and providing a means for interaction between the clients, the schedule advisor, and the dispatcher. The TFE interacts with the scheduler and dispatcher in order to process jobs. It manages the experiment under the direction of schedule advisors, and then instructs the dispatcher to allocate an application task to the selected resource.

The TFE maintains the state of an entire experiment and ensures that it is recorded in persistent storage. This allows the experiment to be restarted if root node fails. The TFE exposes interfaces for job, resource, and task management along with the job-to-resource mapping APIs. Accordingly, scheduling policy developers can use these interfaces to implement other schedulers without concern for the complexity of low-level remote execution mechanisms.

The programmable capability of the task-farming engine enables “plugging” of user-defined schedulers and customised clients or problem solving environments (e.g., ActiveSheets) in place of the default components. The task-farming engine is a coordination point for processes performing resource trading, scheduling, data and executable staging, remote execution, and result collation.

### *The Scheduler*

The scheduler is responsible for resource discovery, resource trading, resource selection, and job assignment. The resource discovery algorithm interacts with an information service (the MDS in Globus), identifies the list of authorized and available machines, trades for resource access cost, and keeps track of resource status information. The resource selection algorithm is responsible for selecting those resources that meet the deadline and budget constraints along with optimization requirements. Nimrod-G incorporates three different algorithms (discussed in section 4.4) for deadline and budget constrained scheduling [107].

### *The Dispatcher and Actuators*

The dispatcher triggers appropriate actuators to deploy agents on Grid resources and assign one of the resource-mapped jobs for execution. Even though the schedule advisor creates a schedule for the entire duration based on user requirements, the dispatcher deploys jobs on resources periodically depending on load and number of CPUs that are available. We have implemented different dispatchers and actuators for each different middleware service. For example, a Globus-specific dispatcher is required for Globus resources, and a Legion-specific component for Legion resources.

### *Agents*

Nimrod-G agents are deployed on Grid resources dynamically at runtime depending on the scheduler’s instructions. The agent is submitted as a job to the resource process server (e.g., GRAM gatekeeper in case of a resource running Globus), which then submits to the local resource manager (fork manager in case of time-share resources and queuing system in case of space-shared resource) for starting its execution. The agent is responsible for setting up the execution environment on a given resource for a user job. It is responsible for transporting the code and data to the machine; starting the execution of the task on the assigned resource and sending results back to the TFE. Since the agent operates on the “far side” of the middleware resource management components, it needs to provide error-detection for the user’s job, sending the job termination status information back to the TFE.

The Nimrod-G agent also records the amount of resource consumed during job execution, such as the CPU time and wall clock time. The online measurement of the amount of resource consumed by the job during its execution helps the scheduler evaluate resource performance and change the schedule accordingly. Typically, there is only one type of agent for all mechanisms, irrespective of whether they are fork or queue nodes. However, different agents are required for different middleware systems.

## 4.3 Scheduling and Computational Economy

The integration of computational economy as part of a scheduling system greatly influences the way computational resources are selected to meet the user requirements. The users should be able to submit their application along with their requirements to a scheduling system such as Nimrod-G, which can process the application on the Grid on the user's behalf and try to complete the assigned work within a given deadline and cost. The deadline represents a time by which the user requires the result, and is often imposed by external factors like production schedules or research deadlines.

To arrive at a scheduling decision, the scheduling system needs to take various parameters into consideration including the following:

- Resource Architecture and Configuration
- Resource Capability (clock speed, memory size)
- Resource State (such as CPU load, memory available, disk storage free)
- Resource Requirements of an Application
- Access Speed (such as disk access speed)
- Free or Available Nodes
- Priority (that the user has)
- Queue Type and Length
- Network Bandwidth, Load, and Latency (if jobs need to communicate)
- Reliability of Resource and Connection
- User Preference
- Application Deadline
- User Capacity/Willingness to Pay for Resource Usage
- Resource Cost (in terms of dollars that the user need to pay to the resource owner)
- Resource Cost Variation in terms of Time-scale (like high @ daytime and low @ night)
- Historical Information, including Job Consumption Rate

The important parameters of computational economy that can influence the way resource scheduling is done are:

- Resource Cost (set by its owner)
- Price (that the user is willing to pay)
- Deadline (the period by which an application execution needs to be completed)

The scheduler can use the information gathered by a resource discoverer and also negotiate with resource owners to get the best "value for money". The resource that offers the best price and meets resource requirements can eventually be selected. This can be achieved by resource reservation and bidding. If the user deadline is relaxed, the chances of obtaining low-cost access to resources are high. The cost of resources can vary dynamically from time to time and the resource owner will have the full control over deciding access cost. Further, the cost can vary from one user to another. The scheduler can even solicit bids from resource providers in an open market, and select the feasible service-provider(s). To accomplish this, we need scheduling algorithms that take the application processing requirements, Grid resource dynamics, and the user quality of service (QoS) requirements such as the deadline, budget, and their optimisation preference into consideration. In the next section, we discuss deadline and budget constrained (DBC) algorithms that we developed for scheduling parameter sweep applications on globally distributed Grid resources.

## 4.4 Scheduling Algorithms

The parameter sweep applications, created using a combination of task and data parallel models, contain a large number of independent jobs operating different data sets. A range of scenarios and parameters to be explored are applied to the program input values to generate different data sets. The programming and execution model of such applications resemble the SPMD (Single Program Multiple Data) model. The execution model essentially involves processing N independent jobs (each with the same task specification, but a different dataset) on M distributed computers where N is, typically, much larger than M.

When the user submits a parameter sweep application containing N tasks along with quality of service requirements, the broker performs the following activities:

1. Resource Discovery: Identifying resources and their properties and then selecting resources capable of executing user jobs.
2. Resource Trading: Negotiating and establishing service access cost using a suitable economic

model.

3. Scheduling: Select resources that fit user requirements using *scheduling heuristic/algorithm* and map jobs to them.
4. Deploy jobs on resources [Dispatcher].
5. Monitor and Steer computations
6. Perform load profiling for future usage
7. When the job execution is finished, gather results back to the user home machine [Dispatcher].
8. Record all resource usage details for payment processing purpose.
9. Perform rescheduling: Repeat steps 3-8 until all jobs are processed and the experiment is within the deadline and budget limit.
10. Perform cleanup and post-processing, if required.

High-level steps for adaptive scheduling with deadline and budget constraints are shown in Figure 4.6.

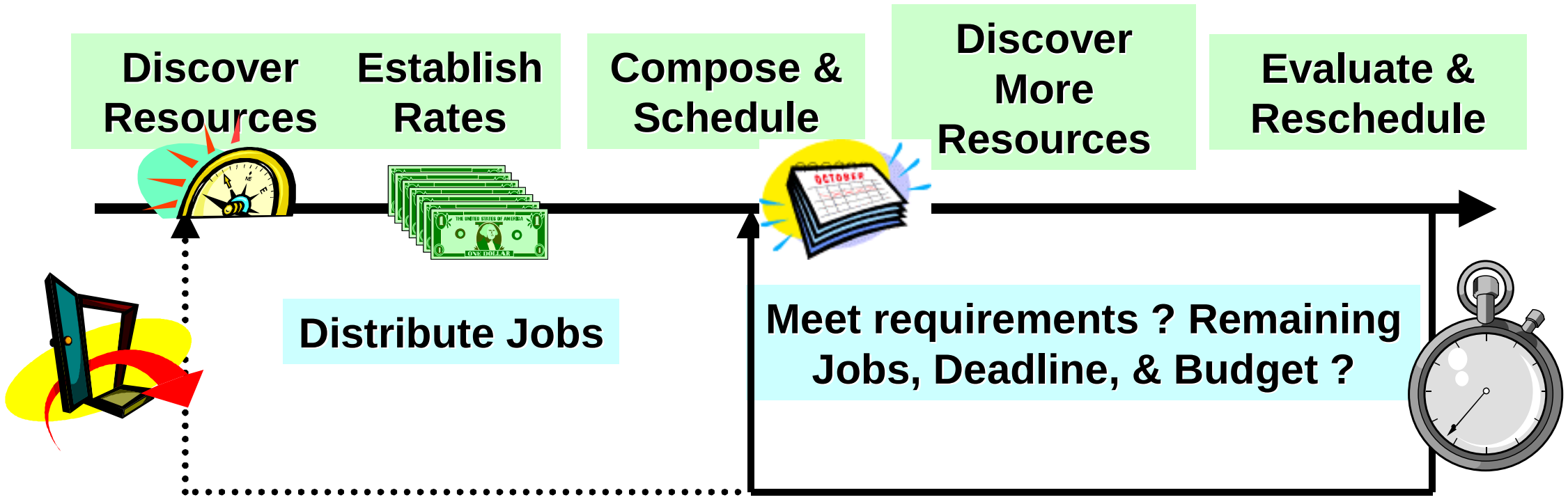

**Figure 4.6: High level steps for adaptive scheduling within the Nimrod-G broker.**

Scheduling and orchestrating the execution of parameter sweep applications on world-wide distributed computers appears simple, but complexity arises when users place QoS constraints like deadline (execution completion time) and computation cost (budget) limitations. Such a guarantee of service is hard to provide in a Grid environment since its resources are shared, heterogeneous, distributed in nature, and owned by different organisations having their own policies and charging mechanisms. In addition, scheduling algorithms need to adapt to the changing load and resource availability conditions in the Grid in order to achieve performance and at the same time meet the deadline and budget constraints. In our Nimrod-G application level resource broker (also called an application level scheduler) for the Grid, we have incorporated three adaptive algorithms for deadline and budget constrained scheduling:

- Cost Optimisation, within time and budget constraints,
- Time Optimisation, within time and budget constraints,
- Conservative Time Optimisation, within time and budget constraints.

We have developed another new algorithm, called cost-time optimisation scheduling, which extends the first two (cost and time optimisation) scheduling algorithms. This new algorithm and the performance evaluation results are discussed in Chapter 6, *Scheduling Simulations*.

The role of deadline and budget constraints in scheduling and objectives of different scheduling algorithms are illustrated in Table 4.2.

**Table 4.2: Adaptive scheduling algorithms constraints and objectives.**

| ***Adaptive Scheduling Algorithms*** | ***Execution Time (not beyond deadline)*** | ***Execution Cost (not beyond budget)*** |
|---|---|---|
| Cost Optimisation | Limited by deadline | Minimise |
| Time Optimisation | Minimise | Limited by budget |
| Conservative Time Optimisation | Limited by deadline | Limited by budget |

The Time Optimisation scheduling algorithm attempts to complete the experiment as quickly as possible, within the budget available. A description of the core of the algorithm follows:

1. For each resource, calculate the next completion time for an assigned job, taking into account previously assigned jobs.
2. Sort resources by next completion time.
3. Assign one job to the first resource for which the cost per job is less than or equal to the remaining budget per job.
4. Repeat all steps until all jobs are assigned.

The Cost Optimisation scheduling algorithm attempts to complete the experiment as economically as possible within the deadline.

1. Sort resources by increasing cost.
2. For each resource in order, assign as many jobs as possible to the resource, without exceeding the deadline.

A final algorithm ("Conservative Time Optimisation scheduling") attempts to complete the experiment within the deadline and cost constraints, minimising the time when higher budget is available. It spends the budget cautiously and ensures that a minimum of "the budget-per-job" from the total budget is available for each unprocessed job.

1. Split resources by whether cost per job is less than or equal to the budget per job.
2. For the cheaper resources, assign jobs in inverse proportion to the job completion time (e.g. a resource with completion time = 5 gets twice as many jobs as a resource with completion time = 10).
3. For the dearer resources, repeat all steps (with a recalculated budget per job) until all jobs are assigned.

Note that the implementations of all the above algorithms contain extra steps for dealing with the initial start-up (when average completion times are unknown), and for when all jobs cannot be assigned to resources (infeasible schedules). Detailed steps of the above scheduling heuristics are described in Chapter 6: "Scheduling Simulations".

## 4.5 Implementation and Technologies Used

The Nimrod-G resource broker follows a modular, extensible, and layered architecture with an "hourglass" principle as applied in the Internet Protocol suite [59]. This architecture enables separation of different Grid middleware systems *mechanisms* for accessing remote resources from the end user applications. The broker provides uniform access to diverse implementations of low-level Grid services. The key components of Nimrod-G, the Task Farming Engine, the scheduler, and the dispatcher are loosely coupled. The interaction among them takes place through network protocols. Apart from the Dispatcher and the Grid Explorer, the Nimrod-G components are independent of low-level middleware used. The modular and extensible architecture of Nimrod-G facilitates a rapid implementation of Nimrod-G support for upcoming peer-to-peer computing infrastructures such as Jxta [68] and Web services [122]. To achieve this, it is only necessary to implement two new components, a dispatcher and an enhanced Grid Explorer. The current implementation of Nimrod-G broker uses low-level Grid services provided by Globus [49] and Legion [123] systems. The Globus toolkit components used in the implementation of Nimrod-G are: GRAM (Globus Resource Allocation Manager), MDS (Metacomputing Directory Service), GSI (Globus Security Infrastructure), and GASS (Global Access to Secondary Storage). We also support Nimrod-G dispatcher implementation for Condor [79] resource management system. The use of various Grid and commodity technologies in implementing Nimrod-G components and functionality is presented in Table 4.3.

While submitting applications to the broker, user requirements such as deadline and budget constraints need to be set and start application execution. These constraints can be changed at any time during execution. The complete details on application parameterization and jobs management information, starting from submission to completion, is maintained in the database. In the past the database was implemented as a file-based hierarchical database. In the latest version of Nimrod-G, the TFE database is implemented using a standard "relational" database management system.

The commodity technologies and software tools used in the Nimrod-G implementation include: the C and Python programming languages, the Perl scripting language, SQL and Embedded C for database management. The PostgreSQL database system is used for the management of the TFE database and its interaction with other components.

**Table 4.3: The Nimrod-G resource broker modules functionality and the role of Grid services.**

| Nimrod-G Module | Implementation and Grid technologies Used |
|---|---|
| Application Model | Coarse Grained Task Farming, Master Worker, and Data Parallelism. |
| Application Composition | We support mechanism for application parameterization through parameterization of input files and command-line inputs for coarse-grained data parallelism. It basically supports coarse-grain, data parallel, task farming application model, which can be expressed using our declarative programming language or GUI tools. |
| Application Interface | The Nimrod-G broker supports protocols and interfaces (described in [136]) that help in job management. Nimrod-G clients or problem solving environments can add, remove, and enquire about job status. They can set user requirements such as deadline and budget; start and stop application execution both at job level and the entire application level. |
| Scheduling Interface | The Nimrod-G broker supports protocols and interfaces (described in [136]) that help in developing schedulers. The schedulers can interact with the TFE to access user constraints and application jobs details to develop a schedule that maps jobs to resources appropriately. |
| Security | Secure access to resources and computations (identification, authentication, computational delegation) is provided by low-level middleware systems (Globus GSI infrastructure). |
| Resource Discovery | Resource discovery involves discovering appropriate resources and their properties that match with the user's requirements. We maintain resource listings for Globus, Legion, and Condor and their static and dynamic properties are discovered using Grid information services. For example, in case of Globus resources, we query Globus LDAP-based GRIS server for resource information. |
| Resource Trading and Market Models | The market-driven resource trading is performed using GRACE trading services. The Nimrod-G broker architecture is generic enough to support various economic models for price negotiation and using the same in developing application schedules. |
| Performance Prediction | The Nimrod-G scheduler performs the user-level resource capability measurement and load profiling by measuring and establishing the job consumption rate. |
| Scheduling Algorithms | Deadline and budget-based constrained (DBC) scheduling performed by Nimrod-G Schedule Advisor. Along with DBC scheduling, we support further optimization of time, cost, or surplus driven divide and conquer in scheduling. |
| Remote Job Submission | The Nimrod-G dispatcher performs deployment of Nimrod-G agents using Globus GRAM, Legion, or Condor commands. The agents are responsible for managing all aspects of job execution. |
| Staging Programs and Data on Remote Resources | In the case of Legion and Condor it is handled by their I/O management systems. On Globus resources, we use http protocols for fetching required files. |
| Accounting (Broker Level) | Nimrod-G agents perform accounting tasks such as measuring resource consumption and the scheduler performs the entire application level accounting. |
| Monitoring and Steering Tools | Nimrod-G Monitoring and Steering Client |
| Problem Solving Environments | ActiveSheets and Nimrod-O are Grid-enabled using the Nimrod-G broker job management services. |
| Execution Testbed | The World Wide Grid (WWG) having resources distributed across five continents. |

## 4.6 Scheduling Evaluation on Nimrod-G Simulated Test Queues

In addition to accessing real computational resources, Nimrod can also simulate the execution of jobs on a test queue. These simulated queues are useful for testing the scheduling algorithms, since their behaviour can be controlled very precisely. A test queue runs each submitted job in succession and the apparent wallclock time and reported CPU usage can be controlled exactly. It simulates job execution by waiting for a job length period in "real time" and it is assumed that each test queue has a single CPU. This feature is meant for a simple testing of scheduling algorithms incorporated into the Nimrod-G broker. For a detailed performance evaluation, discrete-event simulation tools such GridSim are used (discussed in the next two chapters).

For this simulation, we created experiments containing 100 jobs, each with a 90 second running time, giving a total computation time of 9000 seconds. For each experiment, we created 10 test queues with different (but fixed) access costs of 10, 12, 14, 16, 18, 20, 22, 24, 26, and 28 units/CPU-second. The optimal deadline for this experiment is achieved when each queue runs 10 jobs in sequence, giving a running time of 900 seconds for the 100 jobs.

We selected three deadlines: 990 seconds (the optimal deadline plus 10%), 1980 seconds (990 x 2), and 2970 seconds (990 x 3). The 10% allowance allows for the fact that although the queues are simulated, and behave perfectly, the standard scheduler has some delays built in.

We selected three values for the budget. The highest is 252000 units, which is the amount required to run all jobs on the most expensive queue. Effectively, this allows the scheduler full freedom to schedule over the queues with no consideration for the cost. A budget of 171000 units is the budget required to execute 10 jobs on each of the queues. Finally, the lowest budget of 126000 units is the budget required to execute 20 jobs on each of the 5 cheapest queues. Note that for this value, the deadline of 990 seconds is infeasible, and the deadline of 1980 seconds is the optimal deadline plus 10%.

Table 4.4 shows a summary of results for each combination of scheduling algorithm, deadline and budget, and the resulting percentage of completed jobs, the total running time, and the final cost. The jobs marked "infeasible" have no scheduling solution that enables 100% completion of jobs. The jobs marked "hard" have only one scheduling solution.

**Table 4.4: Behaviour of scheduling algorithms for various scenarios on the Grid.**

| Algorithm | Deadline | Budget | Completed | Time | Cost | Remarks |
|---|---|---|---|---|---|---|
| Cost optimise | 990 | 126000 | 85% | 946 | 125820 | Infeasible |
| | 990 | 171000 | 84% | 942 | 139500 | Hard |
| | 990 | 252000 | 94% | 928 | 156420 | Hard |
| | 1980 | 126000 | 97% | 1927 | 124740 | Hard |
| | 1980 | 171000 | 99% | 1918 | 128520 | |
| | 1980 | 252000 | 98% | 1931 | 127620 | |
| | 2970 | 126000 | 98% | 2931 | 116820 | |
| | 2970 | 171000 | 98% | 2925 | 116820 | |
| | 2970 | 252000 | 100% | 2918 | 118800 | |
| Time optimise | 990 | 126000 | 36% | 955 | 50040 | Infeasible |
| | 990 | 171000 | 100% | 913 | 171000 | Hard |
| | 990 | 252000 | 100% | 930 | 171000 | Hard |
| | 1980 | 126000 | 80% | 1968 | 101340 | Hard |
| | 1980 | 171000 | 100% | 909 | 171000 | |
| | 1980 | 252000 | 100% | 949 | 171000 | |
| | 2970 | 126000 | 100% | 2193 | 126000 | |
| | 2970 | 171000 | 100% | 928 | 171000 | |
| | 2970 | 252000 | 100% | 922 | 171000 | |
| Conservative Time optimise | 990 | 126000 | 78% | 919 | 120060 | Infeasible |
| | 990 | 171000 | 99% | 930 | 168480 | Hard |
| | 990 | 252000 | 100% | 941 | 171000 | Hard |
| | 1980 | 126000 | 97% | 1902 | 125100 | Hard |
| | 1980 | 171000 | 100% | 1376 | 160740 | |
| | 1980 | 252000 | 100% | 908 | 171000 | |
| | 2970 | 126000 | 99% | 2928 | 125100 | |
| | 2970 | 171000 | 100% | 1320 | 161460 | |
| | 2970 | 252000 | 100% | 952 | 171000 | |

We analyse the behaviour of the queues by examining the usage of the queues over the period of the experiment. For the Cost Optimisation algorithm, Figure 4.7 shows the node usage for a deadline of 1980 seconds. After an initial spike, during which the scheduler gathers information about the queues, the scheduler calculates that it needs to use the 4-5 cheapest queues only in order to satisfy the deadline. (Actually, it requires exactly 5, but the initial spike reduces the requirements a little.) Note that the schedule is similar, no matter what the allowed budget is. Since we are minimising cost, the budget plays little part in the scheduling, unless the limit is reached. This appears to have happened for the lowest budget, where the completion rate was 97%. The budget of 126000 units is only enough to complete the experiment if the cheapest 5 nodes are used. Because of the initial spike, this experiment appears to have run out of money. The other experiments also did not complete 100% of the jobs, but this is mainly because, in seeking to minimise cost, the algorithm stretches jobs out to the deadline. This indicates the need for a small margin to allow the few remaining jobs to complete close to the deadline.

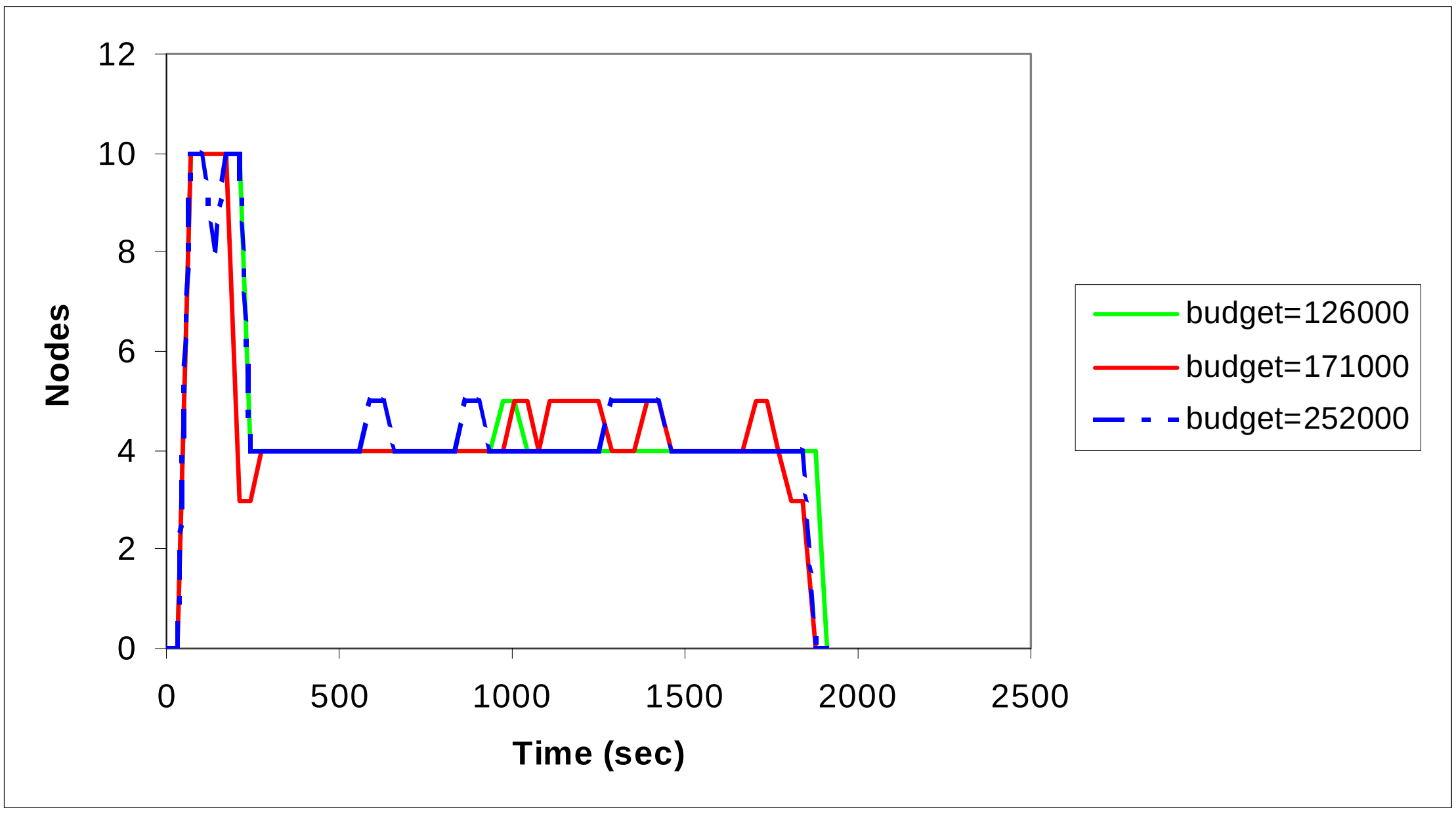

**Figure 4.7: DBC Cost-optimisation scheduling algorithm behaviour for various budgets.**

The equivalent graph for the Time Optimisation algorithm is shown in Figure 4.8. Here we see that except for the case of a limited budget, we get a rectangular shape, indicating the equal mapping of jobs to each resource. Only the experiment with a very limited budget follows the pattern experienced above.

Looking at the equivalent graph for the Conservative Time Optimisation algorithm shown in Figure 4.9, we see a lot more variation in the schedules chosen for different budgets. The schedule with a very large budget is equivalent to the Time Optimisation algorithm. The schedule with the low budget is almost the same as the Cost Optimisation algorithm.

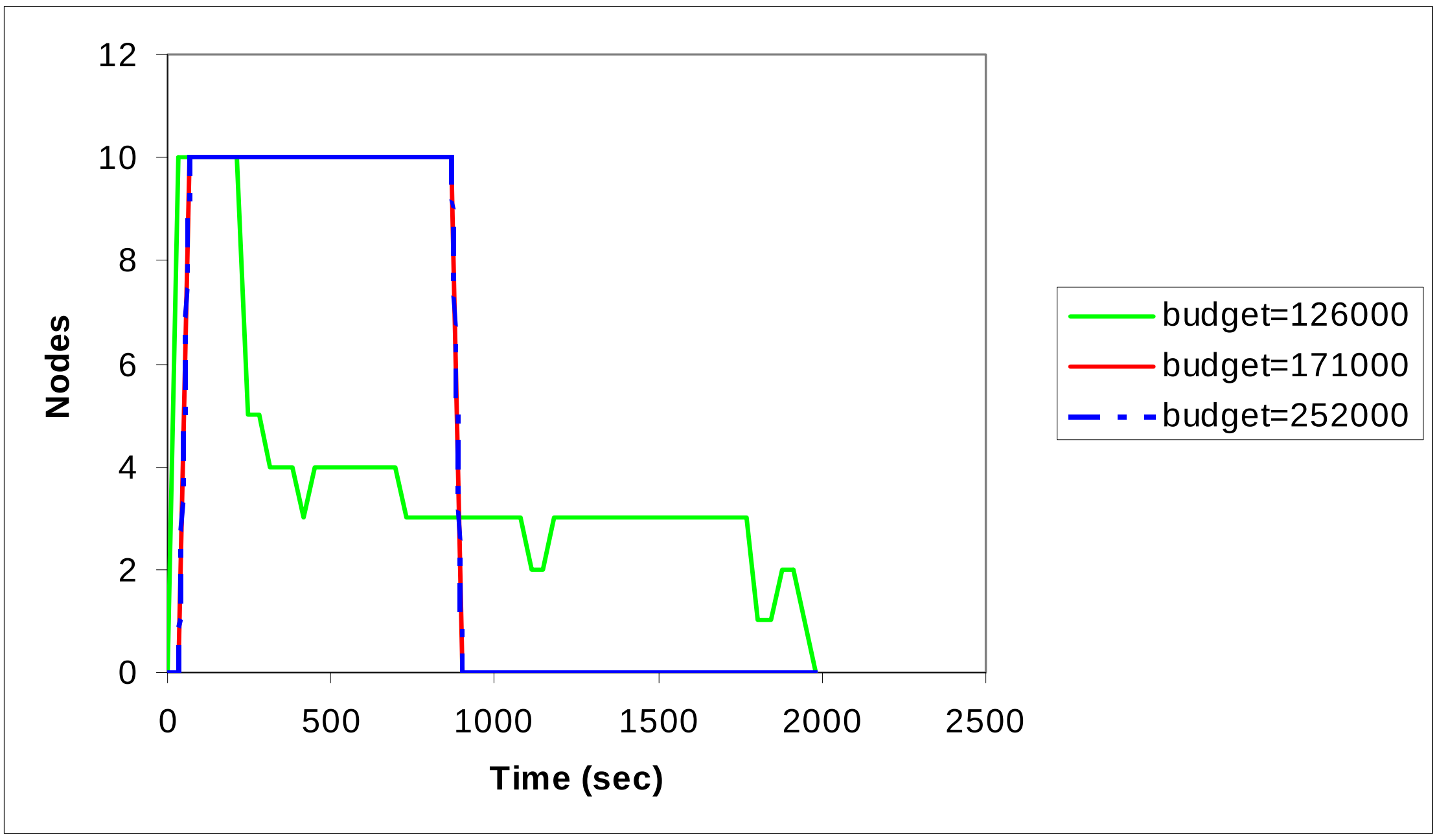

**Figure 4.8: Time optimisation scheduling algorithm behaviour for various budgets.**

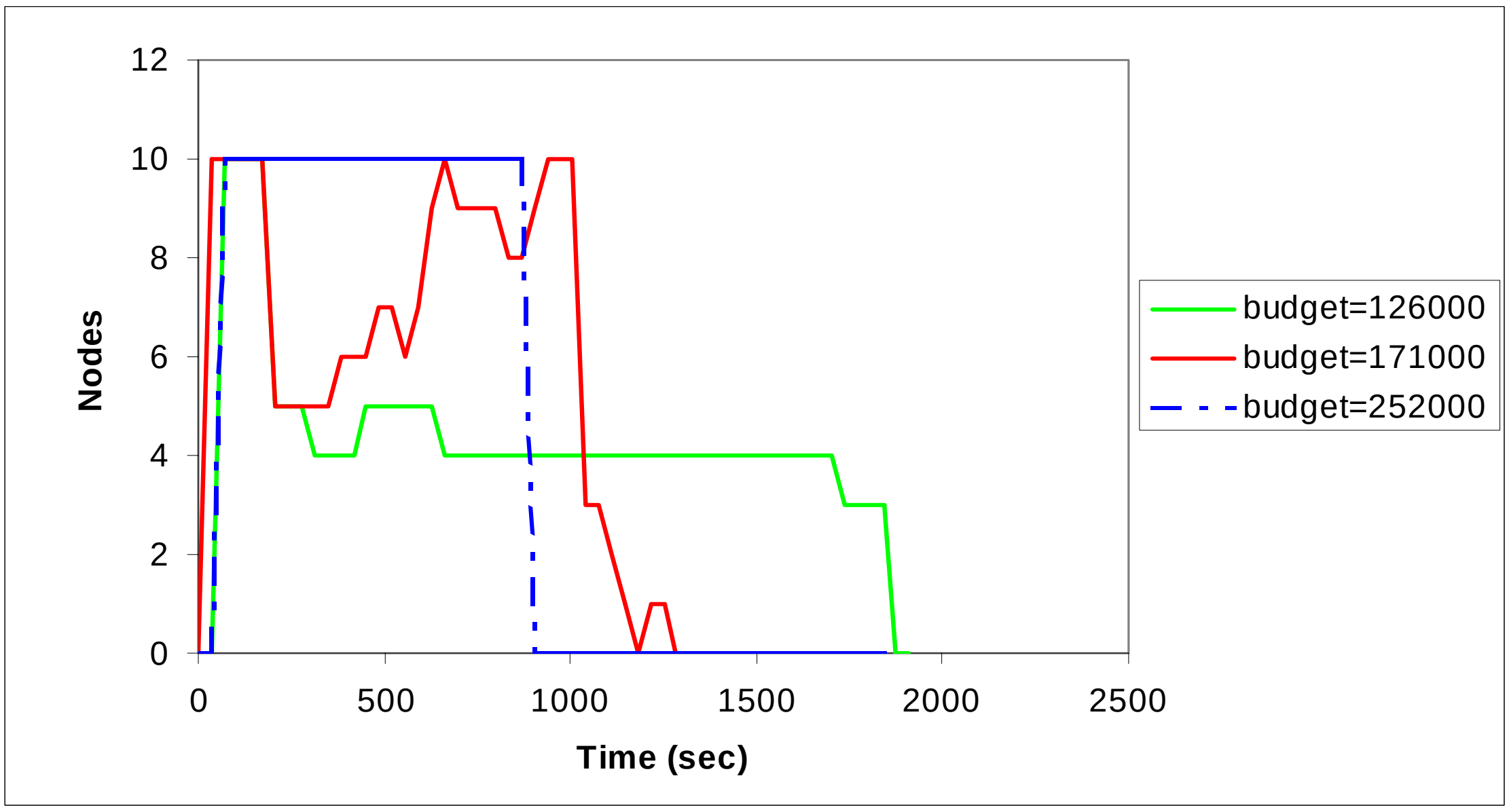

**Figure 4.9: Conservative time optimisation scheduling algorithm behavior for different budgets.**

## 4.7 Scheduling Experiments on the World-Wide Grid

We have performed a number of deadline and budget constrained scheduling experiments with different requirements at different times by selecting different sets of resources available in the World Wide Grid (WWG) [113] testbed during each experiment. They can be categorised into the following scenarios:

- Cost optimisation scheduling during Australian peak and off-peak times,
- Cost and time optimisation scheduling using cheap local and expensive remote resources, and
- Large scale scheduling using cost and time optimisation algorithms.

We briefly discuss the WWG testbed followed by a detailed discussion on these scheduling experiments.

### 4.7.1 The World-Wide Grid (WWG) Testbed

To enable our empirical research and experimentations in distributed computational economy and Grid computing, we created and expanded a testbed called the World-Wide Grid (WWG) in collaboration with colleagues from numerous organizations around the globe. A pictorial view of the WWG testbed depicted in Figure 4.10 shows the name of the organization followed by type of computational resource they have shared. Interestingly, the contributing organizations and the WWG resources themselves are located in five continents: Asia, Australia, Europe, North America, and South America. The organizations whose resources we have used in scheduling experiments reported in this thesis are: Monash University (Melbourne, Australia), Victorian Partnership for Advanced Computing (Melbourne, Australia), Argonne National Laboratories (Chicago, USA), University of Southern California's Information Sciences Institute (Los Angeles, USA), Tokyo Institute of Technology (Tokyo, Japan), National Institute of Advanced Industrial Science and Technology (Tsukuba, Japan), University of Lecce (Italy), and CNUCE-Institute of the Italian National Research Council (Pisa, Italy), Zuse Institute Berlin (Berlin, Germany), Charles University, (Prague, Czech Republic), University of Portsmouth (UK), and University of Manchester (UK).

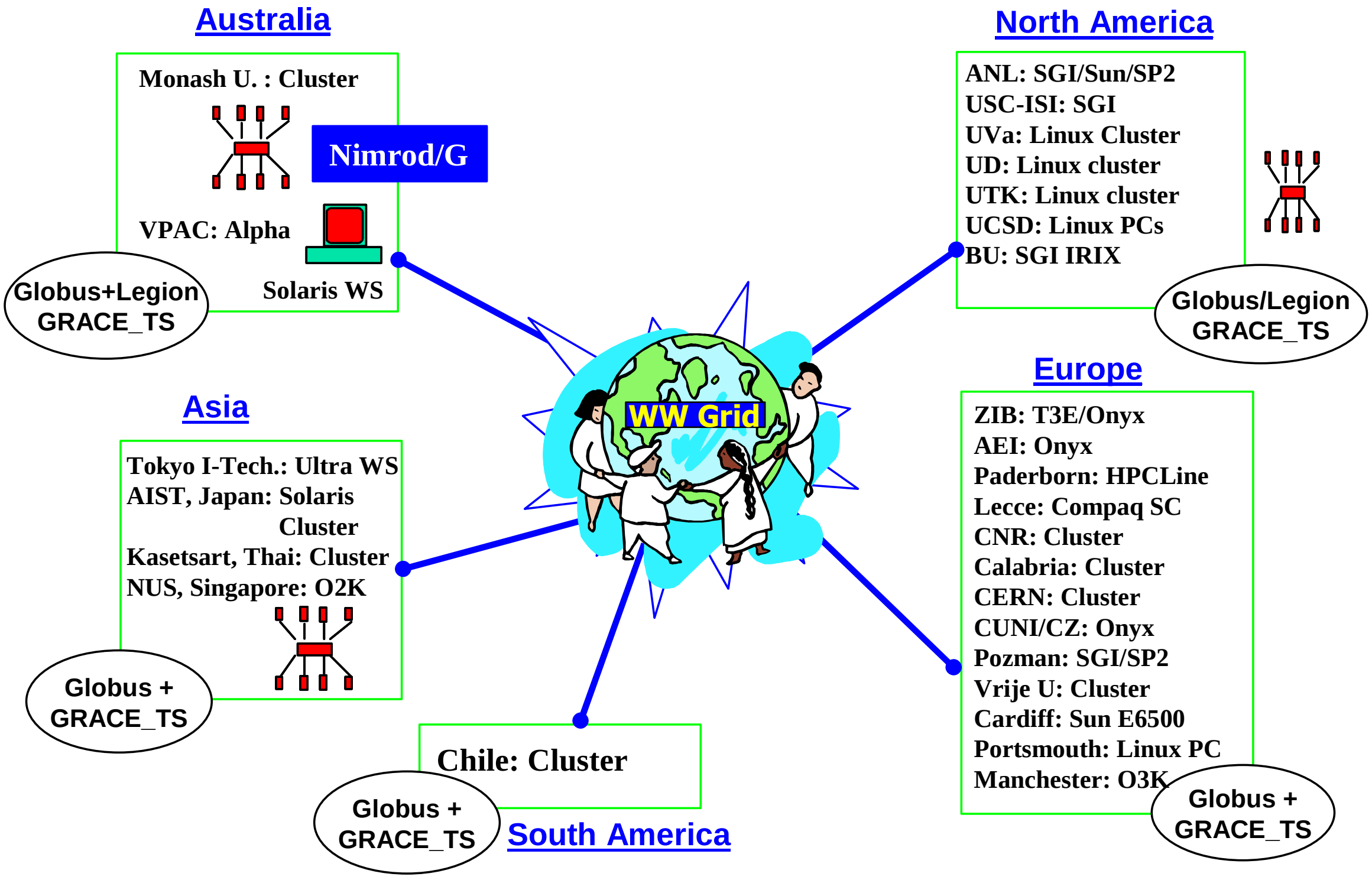

**Figure 4.10: The World Wide Grid (WWG) testbed.**

The WWG testbed contains numerous computers with different architecture, capability, and configuration. They include PCs, workstations, SMPs, clusters, and vector supercomputers running operating systems such as Linux, Sun Solaris, IBM AIX, SGI IRIX, and Compaq Tru64. Further, the systems use a variety of job management systems such as OS-Fork, NQS, Condor, RMS, PBS, and LSF. Most of these resources support secure remote access through the Globus system and a Linux cluster at Virginia is managed using the Legion system. The Solaris workstation from where this scheduling experiment is performed runs Globus, Legion, and Condor systems along with the Nimrod-G resource broker. The Nimrod-G agents are deployed at runtime on resources for managing the execution of jobs.

The properties of WWG testbed resources selected for use in scheduling experiments are discussed in the respective sections. Given that the WWG testbed has been used in numerous scheduling experiments with computational economy and real applications (like molecular modelling for drug design), we believe that it truly represents a blueprint of an emerging scalable Grid computing infrastructure for integrating and aggregating dispersed heterogeneous resources.

### 4.7.2 Cost Optimisation Scheduling – Australian Peak and Off-peak Times

In a competitive commodity-market economy, the resources are priced differently at different times based on the supply and demand. For example, they are priced higher during peak hours and lower during off-peak hours. In this experiment we explore their impact on the processing cost, by scheduling a resource-intensive parameter-sweep application containing a large number of jobs on the World-Wide Grid resources, during Australian peak and off-peak hours.

#### *WWG Computational Resources*

The World-Wide Grid testbed resources selected for use in this experiment and their properties is shown in Table 4.5. To test the Grid Trade Server with the current scheduler, we ran an experiment entirely during peak time and the same experiment entirely during off-peak time. It is important to note access price variations during peak and off-peak times and also time difference between Australia and US. The access price is expressed in Grid units (G$) per CPU second.

**Table 4.5: Wolrd-Wide Grid testbed resources used in the experiment. Price is G$ per CPU sec.**

| Resource Type & Size (No. of nodes) | Organization & Location | Grid Services and Fabric | Price @ AU peak time | Price @ AU off peak time. |
|---|---|---|---|---|
| Linux cluster (60 nodes) | Monash, Australia | Globus/Condor | 20 | 5 |
| IBM SP2 (80 nodes) | ANL, Chicago, USA | Globus/LL | 5 | 10 |
| Sun (8 nodes) | ANL, Chicago, USA | Globus/Fork | 5 | 10 |
| SGI (96 nodes) | ANL, Chicago, USA | Globus/Condor-G | 15 | 15 |
| SGI (10 nodes) | ISI, Los Angeles, USA | Globus/Fork | 10 | 20 |

We selected 5 resources (see Table 4.5) from the testbed, each effectively having 10 nodes available for our experiment. Monash University has a 60-processor Linux cluster running Condor, which was reduced to 10 available processors for the experiment. Similarly, a 96-node SGI at Argonne National Laboratory (ANL) was made to provide 10 nodes by using *Condor glidein* to add 10 processors to the Condor pool. An 8-node Sun at Argonne and a 10-node SGI at the Information Sciences Institute (ISI) of the University of Southern California were accessed using Globus directly. Argonne's 80-node SP2 was also accessed directly through Globus. We relied on its high workload to limit the number of nodes available to us. We assigned artificial-cost (access price per second) for each of those resources depending on their relative capability. This is achieved by setting a resource cost database, which is maintained on each of the resources by their owners. The resource cost database contains access cost (price) that they like to charge to all their Grid users at different times of the day. The access price generally varies from user to user and time to time. The resource broker negotiates with trading servers for establishing access price using the resource trading services provided by the GRACE infrastructure.

#### *Parameter Sweep Application*

We have created a hypothetical parameter sweep application (PSA) that executes a CPU intensive program with 165 different parameter scenarios or values. The program *calc* takes two input parameters and saves results into a file named “output”. The first input parameter `angle_degree` represents the value of angle in degree for processing trigonometric functions. The program *calc* needs to be explored for angular values from 1 to 165 degrees. The second parameter `time_base_value` indicates the expected calculation complexity in minutes plus 0 to 60 seconds positive deviation. That means the program *calc* is expected to run for anywhere between 5 to 6 minutes on resources with some variation depending on resource capability. A plan file modelling this application as a parameter sweep application using the Nimrod-G parameter specification language is shown in Figure 4.11. The first part defines parameters and the second part defines the task that needs to be performed for each job. As the parameter `angle_degree` is defined as a range parameter type with values varying from 1 to 165 in step of 1, it leads to the creation of 165 jobs with 165 different input parameter values. To execute each job on a Grid resource, the Nimrod-G resource broker, depending on its scheduling strategy, first copies the program executable to a Grid node, then executes the program, and finally copies results back to the user home node and stores output with job number as file extension.

```
#Parameters Declaration
parameter angle_degree integer range from 1 to 165 step 1;
parameter time_base_value integer default 5;

#Task Definition
task main
   #Copy necessary executables depending on node type
   copy calc.$OS node:calc
   #Execute program with parameter values on remote node
   node:execute ./calc $angle_degree $time_base_value
   #Copy results file to use home node with jobname as extension
   copy node:output ./output.$jobname
endtask
```

**Figure 4.11: Nimrod-G parameter sweep processing specification.**

### *Scheduling Experiments*

The experiments were run twice, once during the Australian peak time, when the US machines were in their off-peak times, and again during the US peak, when the Australian machine was off-peak. The experiments were configured to *minimise the cost*, within *one-hour deadline*. This requirement instructs the Nimrod-G broker to use “Cost-Optimization Scheduling” algorithm in scheduling jobs for processing on the Grid.

The number of jobs in execution/queued on resources during the Australian peak and off-peak time scheduling experimentations is shown in Figure 4.12 and Figure 4.13 respectively. The results for the Australian peak experiment show the expected typical results. After an initial calibration phase, the jobs were distributed to the cheapest machines for the remainder of the experiment. This characteristic of the scheduler is clearly visible in both experiments. In the Australian peak experiment, after calibration period, the scheduler excluded the usage of Australian resources as they were expensive and the scheduler predicted that it could still meet the deadline using cheaper resources from US resources, which were in off-peak time phase. However, in the Australian off-peak experiment, the scheduler never excluded the usage of Australian resources and excluded the usage of some of the US resources, as they were expensive comparatively at that time (US in peak-time phase). The results for the US peak experiment are somewhat more interesting (see Figure 4.13). When the Sun-ANL machine becomes temporarily unavailable, the SP2, at the same cost, was also busy, so a more expensive SGI is used to keep the experiment on track to complete before the deadline.

When the scheduling algorithm tries to minimize the cost, the total cost Australian peak time experiment is 471205 units and the off-peak time is 427155 units. The result is that costs were quite low in both cases. An experiment using all resources, without the cost optimization algorithm during the Australian peak, costs 686960 units for the same workload. The cost difference indicates a saving in computational cost and it is certainly a successful measure of our budget and deadline-driven scheduling on the Grid.

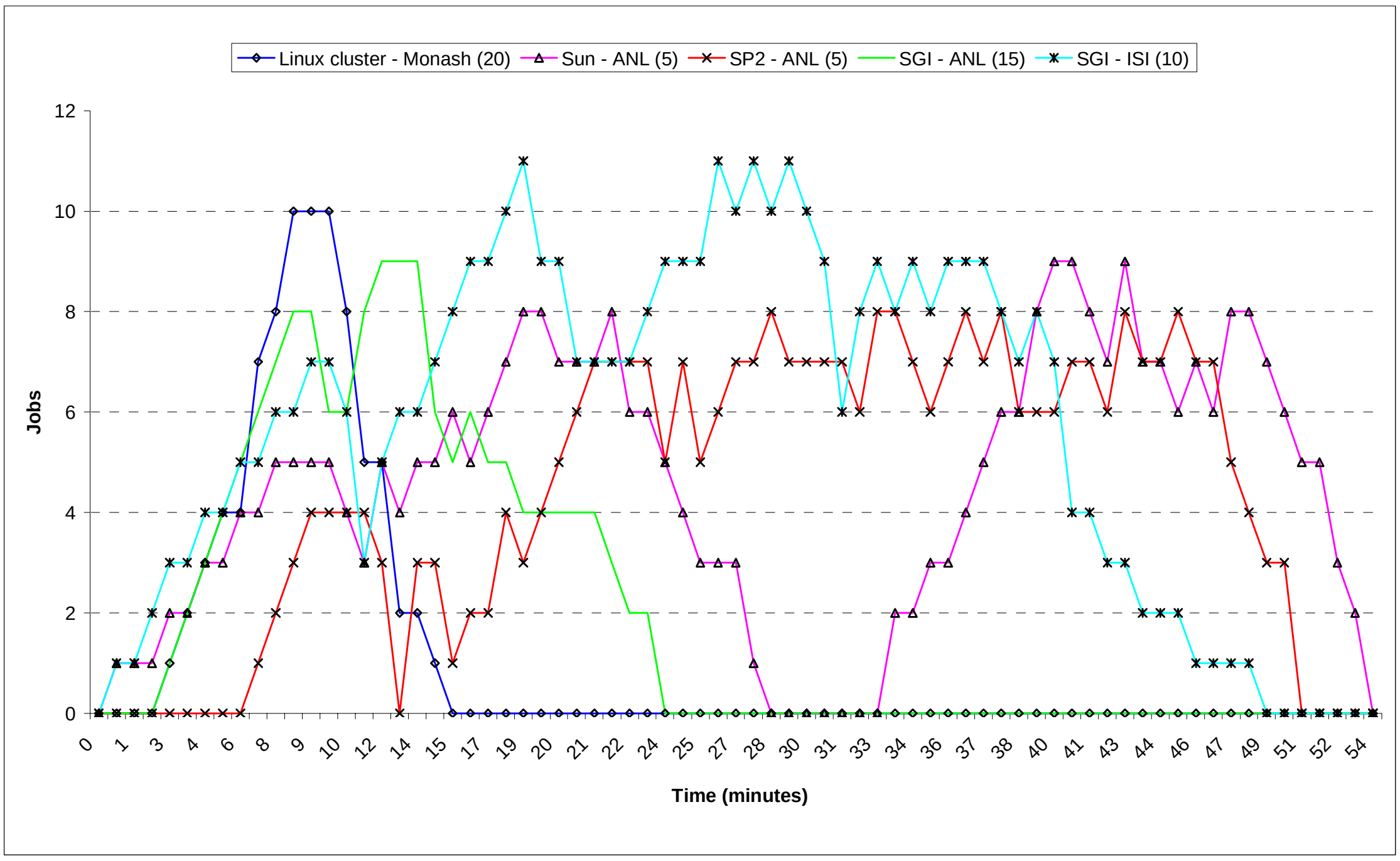

**Figure 4.12: Computational scheduling during Australian peak time (US off-peak time).**

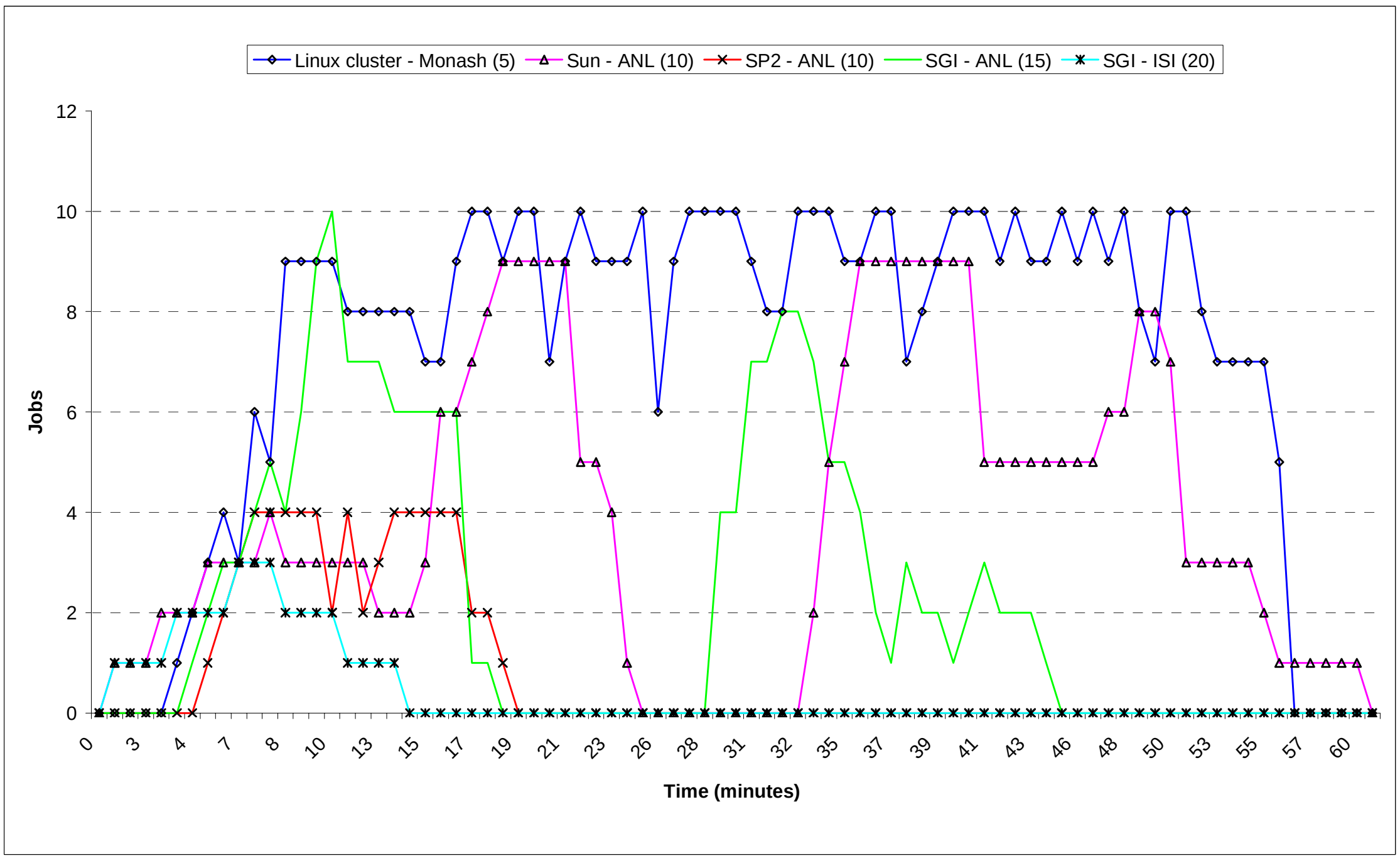

**Figure 4.13: Computational scheduling during Australian off-peak time (US peak time).**

The number of computational nodes (CPUs) in use at different times during the execution of scheduling experimentation at Australian peak-time is shown in Figure 4.14. It can be observed that in the beginning of the experiment (calibration phase), the scheduler had no precise information related to job consumption rate for resources, hence it tried to use as many resources as possible to ensure that it can meet deadline. After calibration phase, scheduler predicted that it could meet the deadline with fewer resources and stopped using more expensive nodes. However, whenever scheduler senses difficulty in meeting the deadline by using the resources currently in use, it includes additional resources. This process continues until deadline is met and at the same time it ensures that the cost of computation is within a given budget.

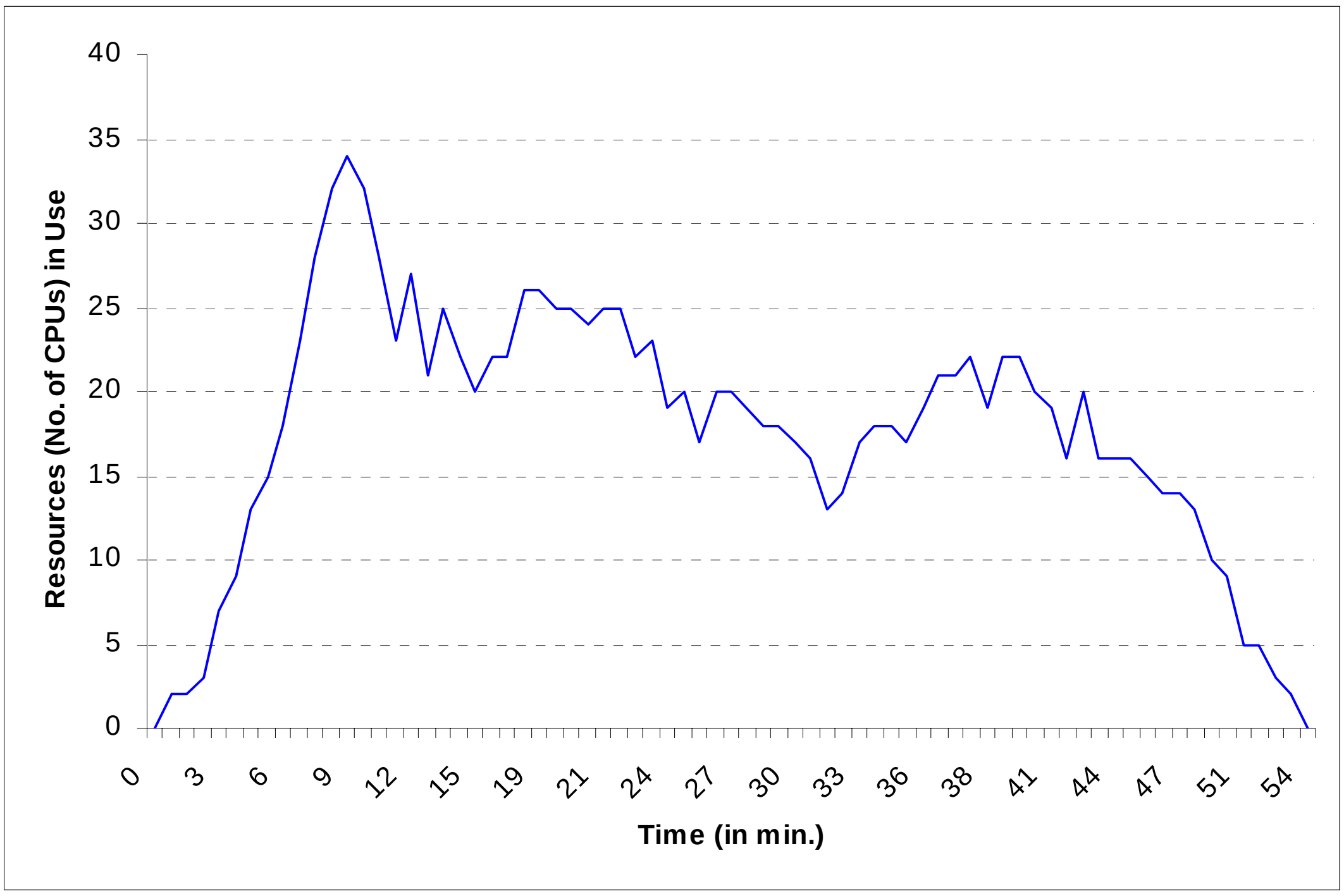

**Figure 4.14: Number of resources in use @ AU peak time scheduling experiment.**

The total cost of resources (sum of the access price for all resources) in use at different times during the execution of scheduling experimentation at Australian peak-time is shown in Figure 4.15. It is interesting to observe the pattern of variation of cost during calibration phase is similar to that of number of resources in use. However, this is not the same as the experiment progresses and in fact the cost of resources decreased almost linearly although the number of resources in use did not decline at the same rate. The reason for this behavior is that a large number of resources that the scheduler selected were from off-peak time zone (i.e., US was in off-peak time when Australia was in peak hours) as they were cheaper. Another reason is that the number of resources used in these experiments contains more US resources compared to Australian resources.

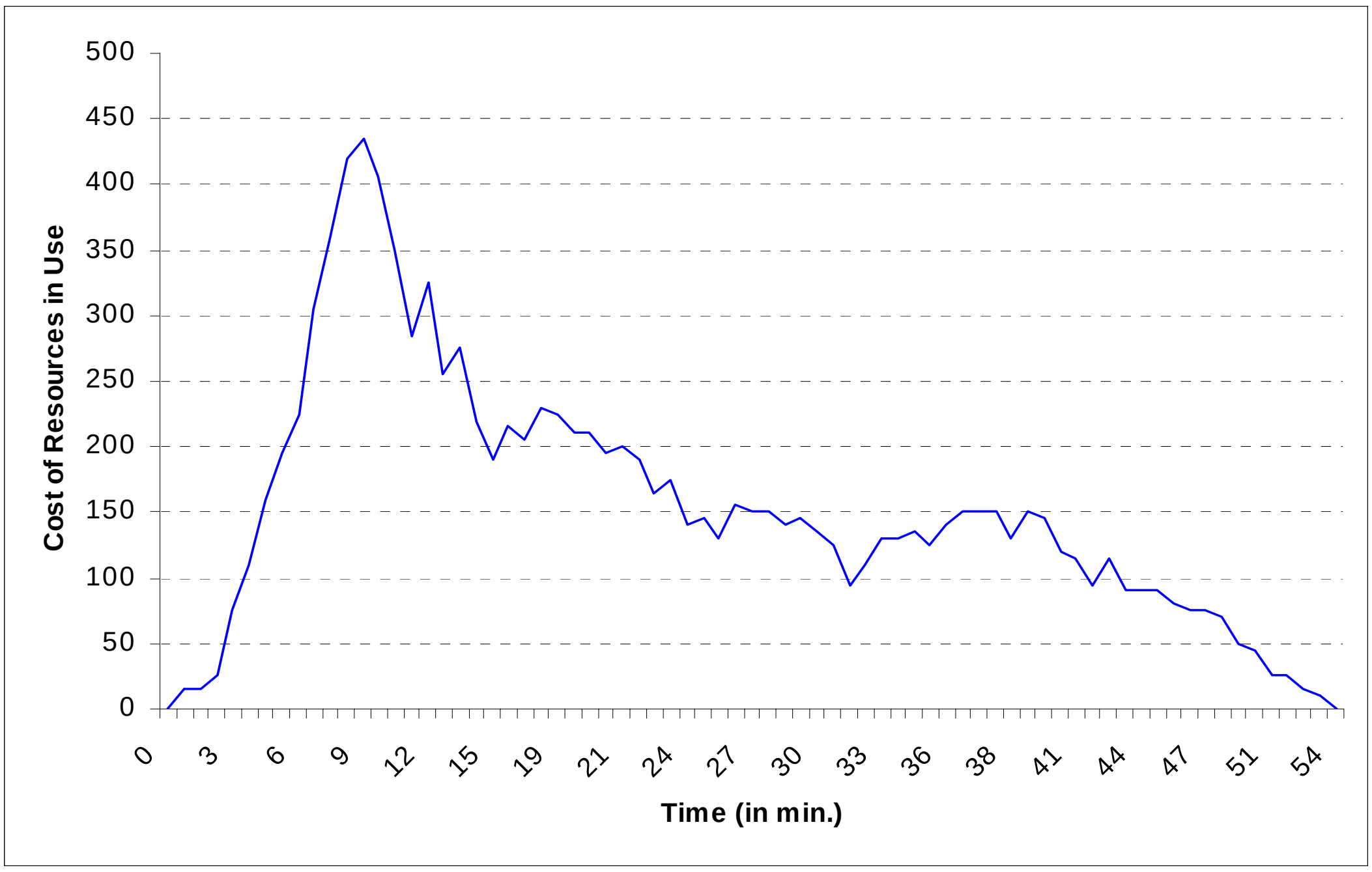

**Figure 4.15: Cost of resources in use @AU peak time scheduling experiment.**

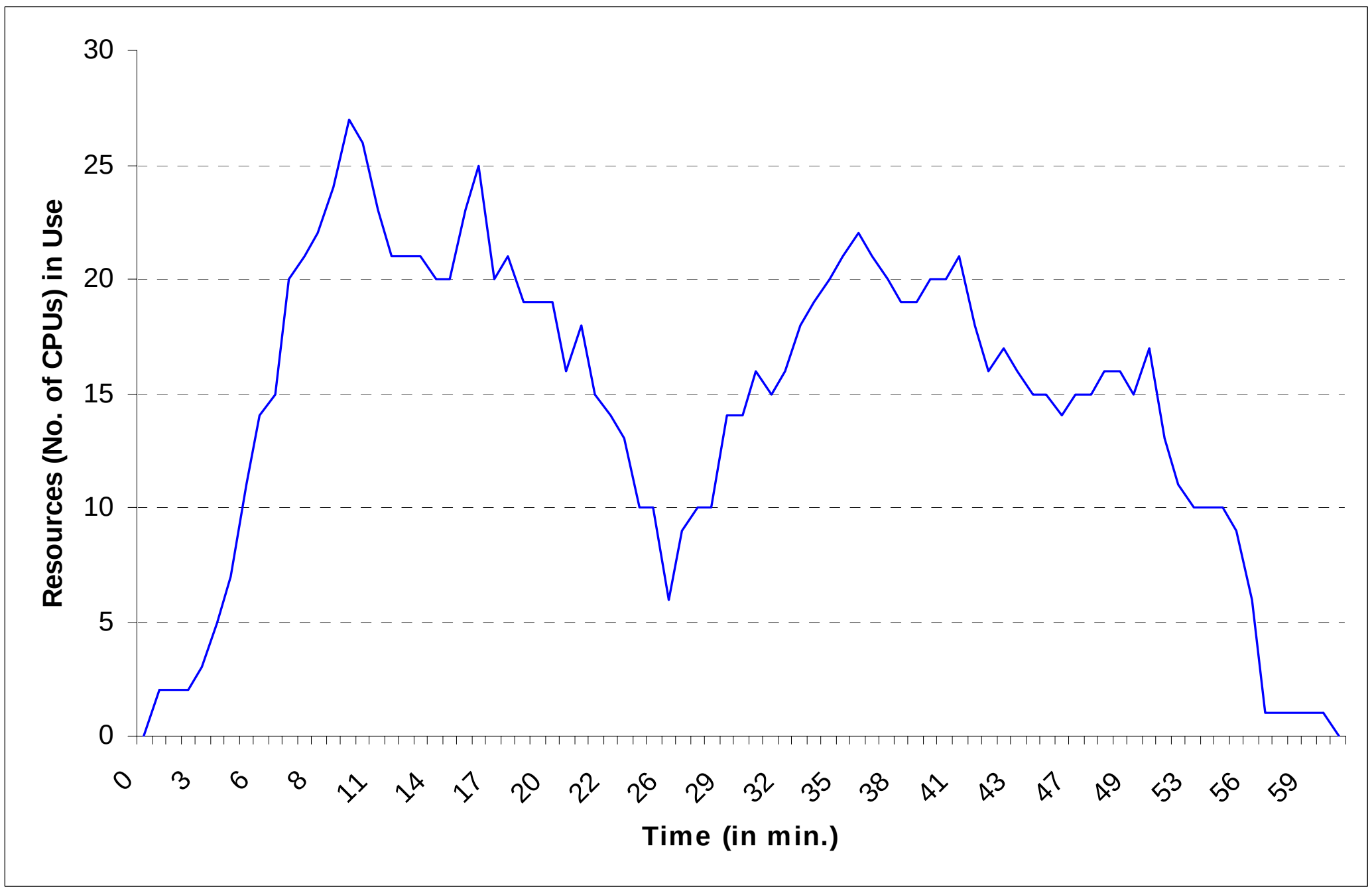

**Figure 4.16: Number of resources in use @ AU off-peak time scheduling experiment.**

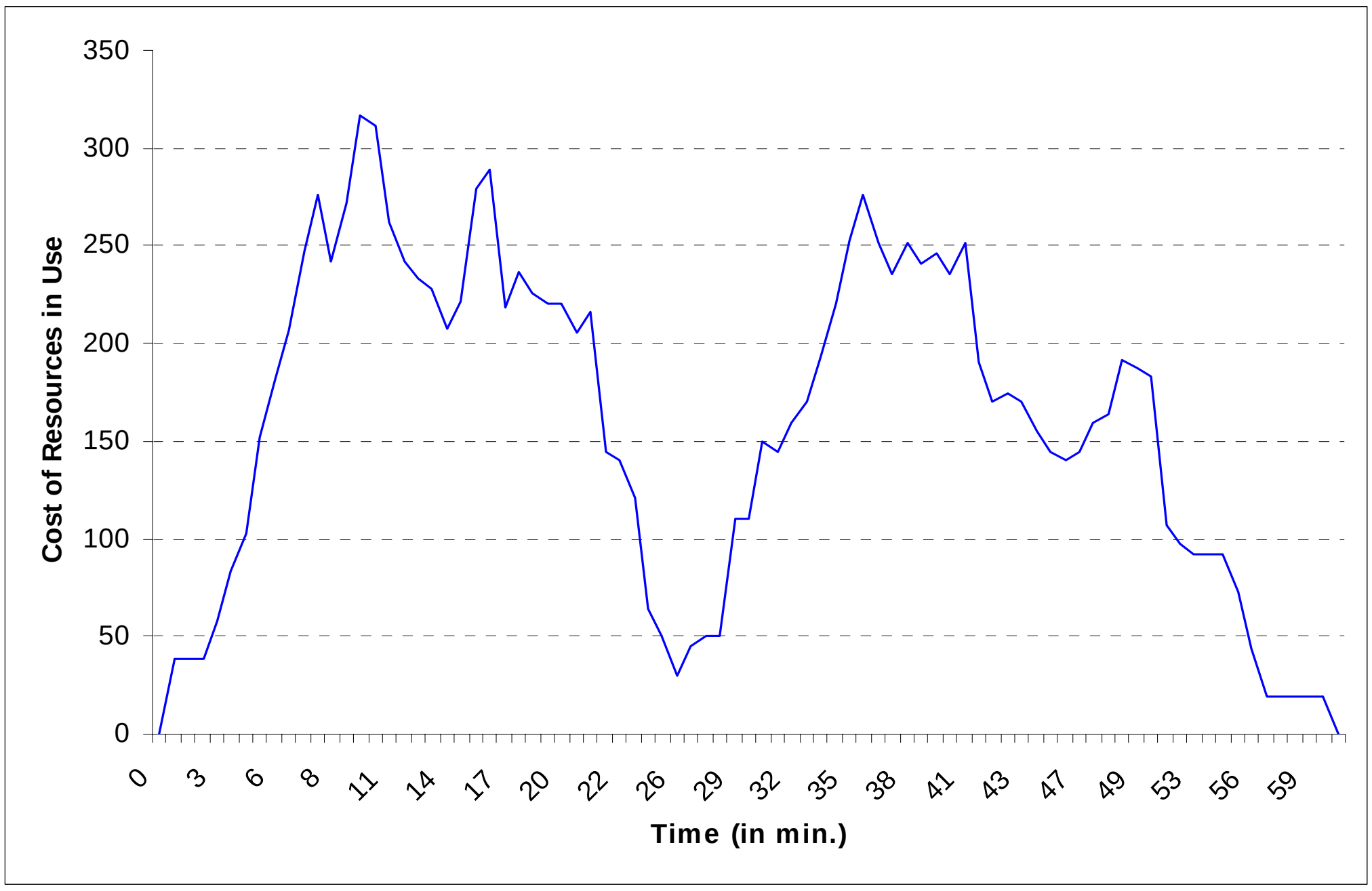

**Figure 4.17: Cost of resources in use @AU off-peak time scheduling experiment.**

Similar behavior did not occur in scheduling experiments conducted during Australian off-peak time (see Figure 4.16 and Figure 4.17). The variation pattern of total number of resources in use and their total cost is similar due to the fact that the larger numbers of US resources were available cheaply. Although the scheduler has used Australian resources throughout the experiment (see Figure 4.13), the scheduler had to depend on US resources to ensure that the deadline is met even if resources were expensive.

### 4.7.3 Cost and Time Optimisation Scheduling using Local and Remote Resources

This experiment demonstrates the use of cheap local resources and expensive remote resources together for processing a parameter sweep application (same as used in the previous scheduling experiment) containing 165 CPU-intensive jobs, each running approximately 5 minutes duration.

We have set the deadline of 2 hours (120 minutes) and budget of 396000 (G$ or tokens) and conducted experiments for two different optimization strategies:

- Optimize for Time - this strategy produces results as early as possible, but before a deadline and within a budget limit.
- Optimize for Cost -this strategy produces results by deadline, but reduces cost within a budget limit.

In these scheduling experiments, the Nimrod-G resource broker employed the *Commodity Market* model for establishing a service access price. It used Grid resource trading services for establishing connection with the Grid Trader running on resource providers' machines and obtained service prices accordingly. The broker architecture is generic enough to use any of the protocols discussed in [104] for negotiating access to resources and choosing appropriate ones. The access price varies for local and remote users: users are encouraged to use local resources since they are available at cheaper price. Depending on the deadline and the specified budget, the broker develops a plan for assigning jobs to resources. While doing so it does dynamic load profiling to learn the ability of resources for executing jobs. Thus, it adapts itself to the changing resource conditions including failure of resources or jobs on the resource.

We have used a subset of resources of the WWG testbed in these scheduling experiments. Table 4.6 shows resources details such as architecture, location, and access price along with type of Grid middleware systems used in making them Grid enabled. These are shared resources and hence they were not fully available to us. The access price indicated in the table is being established dynamically using GRACE resource trading protocols (commodity market model), but is based on an arbitrary assignment by us for demonstration purposes only.

**Table 4.6: The WWG testbed resources used in scheduling experiments, job execution and costing.**

| Resource Type & Size (No. of nodes) | Organization & Location | Grid Services and Fabric | Price (G$ per CPU sec.) | Jobs Executed on Resources | |
|---|---|---|---|---|---|
| | | | | Time_Opt | Cost_Opt |
| Linux cluster (60 nodes) | Monash, Australia | Globus, GTS, Condor | 2 | 64 | 153 |
| Solaris (Ultra-2) | Tokyo Institute of Technology, Japan. | Globus, GTS, Fork | 3 | 9 | 1 |
| Linux PC (Prosecco) | CNUCE, Pisa, Italy | Globus, GTS, Fork | 3 | 7 | 1 |
| Linux PC (Barbera) | CNUCE, Pisa, Italy | Globus, GTS, Fork | 4 | 6 | 1 |
| Sun (8 nodes) | ANL, Chicago, USA | Globus, GTS, Fork | 7 | 42 | 4 |
| SGI (10 nodes) | ISI, Los Angeles, USA | Globus, GTS, Fork | 8 | 37 | 5 |
| | | Total Experiment Cost (G$) | | 237000 | 115200 |
| | | Time to Complete Experiment (Min.) | | 70 | 119 |

The number of jobs in execution on resources (Y-axis) at different times (X-axis) during the experimentation is shown in Figure 4.18 and Figure 4.19 for the time and cost optimization scheduling strategies respectively. In the first (time minimization) experiment, the broker selected resources in such a way that the whole application execution is completed at the earliest time for a given budget. In this experiment, it completed execution of all jobs within *70 minutes* and spent *237000 G$*. In the second experiment (cost minimization), the broker selected cheap resources as much as possible to minimize the execution cost whilst still trying to meet the deadline (completed in *119 minutes*) and spent *115200 G$*. After the initial *calibration phase*, the jobs were distributed to the cheapest machines for the remainder of the experiment. The processing expense of the time-optimization scheduling experiment is much larger than the cost-optimization scheduling experiment due to the use of expensive resources to complete the experiment early. The results show that our Grid brokering system can take advantage of economic models and user input parameters to meet their requirements.

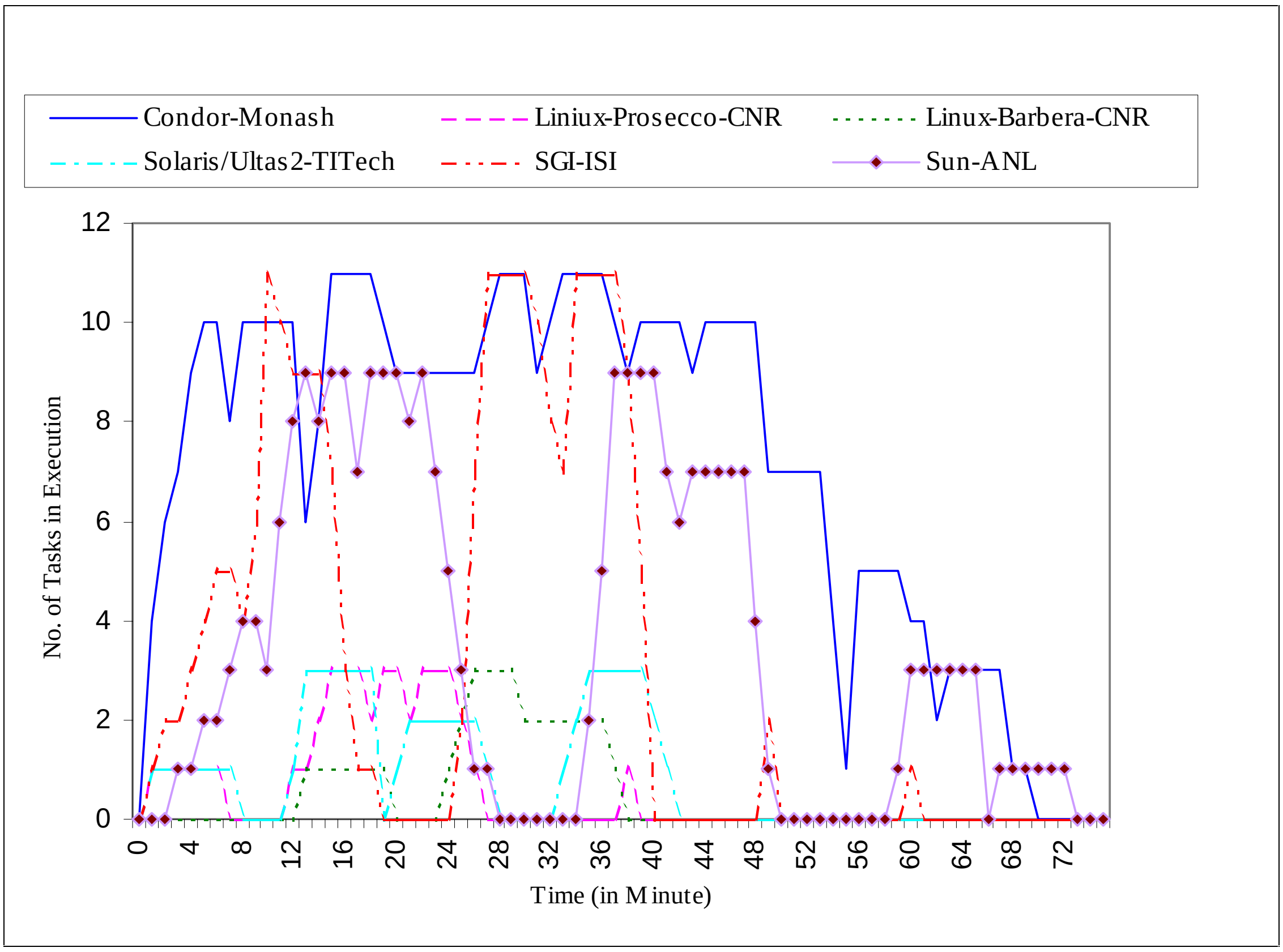

**Figure 4.18: Resource selection in deadline and budget constrained time optimization scheduling.**

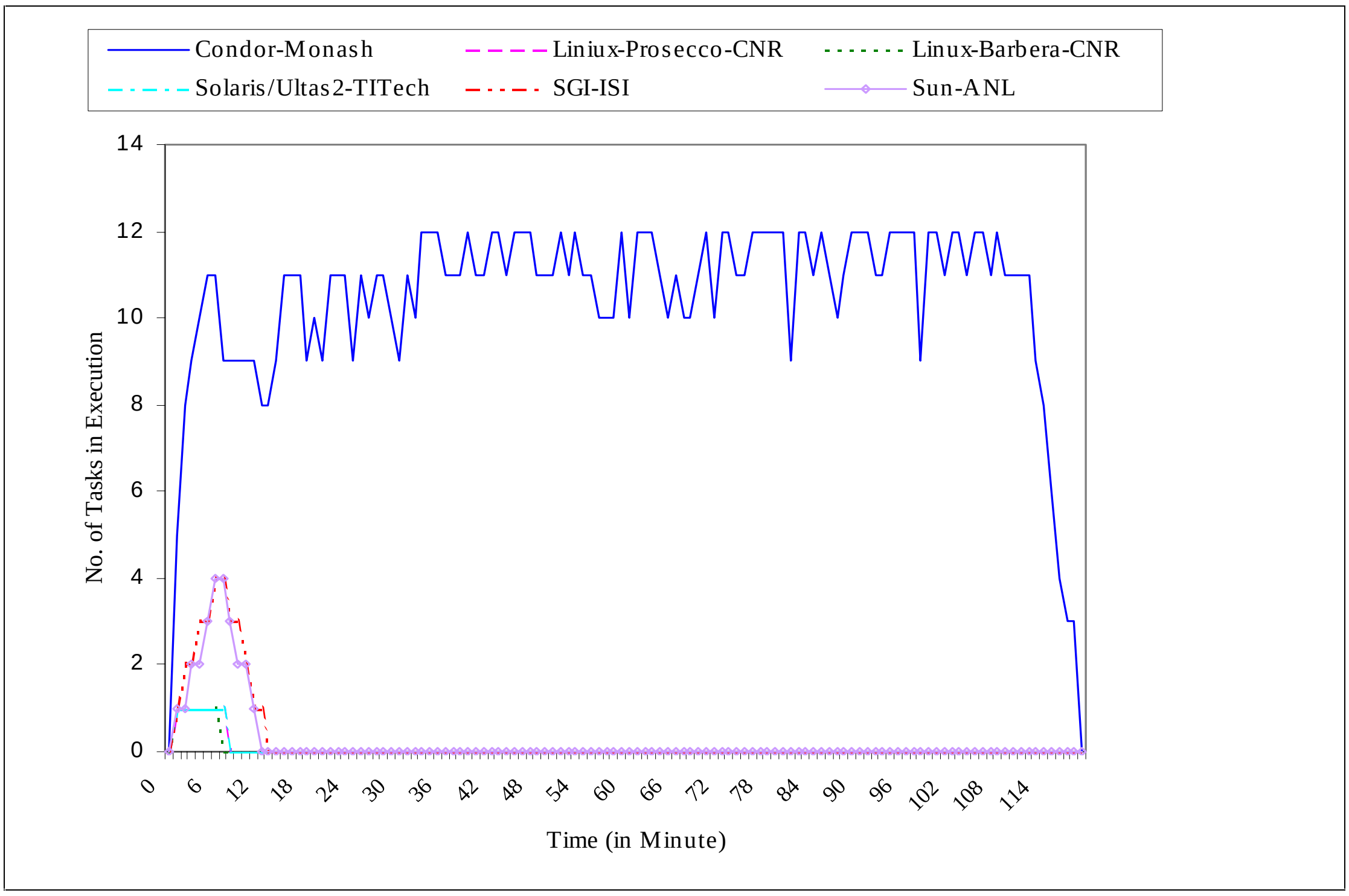

**Figure 4.19: Resource selection in deadline and budget constrained cost optimization scheduling.**

### 4.7.4 Large Scale Scheduling with Cost and Time Optimisation

We have created a hypothetical parameter sweep application (PSA) that executes a CPU intensive program with 200 different parameter scenarios or values. The program *calc* takes two input parameters and saves results into a file named "output". The first input parameter `angle_degree` represents the value of angle in degree for processing trigonometric functions. The program *calc* needs to be explored for angular values from 1 to 200 degrees. The second parameter `time_base_value` indicates the expected calculation complexity in minutes plus 0 to 60 seconds positive deviation. That means, the program *calc* is modeled to execute for anywhere between 10 to 11 minutes on resources randomly. A plan for modelling this application as a parameter sweep application using the Nimrod-G parameter specification language is shown in Figure 4.20. The first part defines parameters and the second part defines the task that needs to be performed for each job. As the parameter `angle_degree` is defined as a range parameter type with values varying from 1 to 200 in step of 1, it leads to the creation of 200 jobs with 200 different input parameter values. To execute each job on a Grid resource, the Nimrod-G resource broker, depending on its scheduling strategy, first copies the program executable to a Grid node, then executes the program, and finally copies results back to the user home node and stores output with job number as file extension.

```
#Parameters Declaration
parameter angle_degree integer range from 1 to 200 step 1;
parameter time_base_value integer default 10;

#Task Definition
task main
   #Copy necessary executables depending on node type
   copy calc.$OS node:calc
   #Execute program with parameter values on remote node
   node:execute ./calc $angle_degree $time_base_value
   #Copy results file to use home node with jobname as extension
   copy node:output ./output.$jobname
endtask
```

**Figure 4.20: Nimrod-G parameter sweep processing specification.**

We have conducted two scheduling experiments for a given deadline of 4 hours (with 18 minutes of extension for the second experiment) and budget of 250000 (G$ or tokens) with different optimization strategies [107]:

- Optimize for Time: This strategy produces results as early as possible, but before a deadline and within a budget limit.
- Optimize for Cost: This strategy produces results by deadline, but reduces cost within a budget limit.

In these scheduling experiments, the Nimrod-G resource broker employed the c*ommodity market* model for establishing a service access price. It used Grid resource trading services for establishing connection with the Grid Trader running on resource providers' machines and obtained service prices accordingly. The broker architecture is generic enough to use any of the protocols discussed above for negotiating access to resources and choosing appropriate ones. The access price varies from one consumer to another and from time to time, as defined by the resource owners. Depending on the deadline and the specified budget, the broker develops a plan for assigning jobs to resources. While doing so it does dynamic load profiling to learn the ability of resources to execute jobs. Thus, it adapts itself to the changing resource conditions including failure of resources or jobs on the resource.

We have used a subset of resources of the WWG testbed in these scheduling experimentations. Table 4.7 shows resources properties such as architecture, location, and access price, and middleware systems loaded on them. A snapshot of the Nimrod-G monitoring and steering clients taken immediately after the completion of application processing is shown in Figure 4.21 and Figure 4.22. The resources used in both experiments are (time/space) shared resources with many other independent users. Hence, they were partially available to us, which changed dynamically depending on other users requirements and priorities. The access price indicated in the table is being established dynamically using GRACE resource trading protocols (commodity market model), but is based on an arbitrary assignment by us for demonstration purposes only.

**Table 4.7: The WWG testbed resources used in scheduling experiments, job execution and costing.**

| Organization & Location | Vendor, Resource Type, # CPU, OS, hostname | Grid Services, Fabric, and Role | Price (G$ per CPU sec.) | Number of Jobs Executed | |
|---|---|---|---|---|---|
| | | | | TimeOpt | CostOpt |
| Monash University, Melbourne, Australia | Sun: Ultra-1, 1 node, bezek.dstc.monash.edu.au | Globus, Nimrod-G, CDB Server, Fork (Master node) | -- | -- | -- |
| VPAC, Melbourne, Australia | Compaq: Alpha, 4 CPU, OSF1, grendel.vpac.org | Globus, GTS, Fork (Worker node) | 1 | 7 | 59 |
| AIST, Tokyo, Japan | Sun: Ultra-4, 4 nodes, Solaris, hpc420.hpcc.jp | Globus, GTS, Fork (Worker node) | 2 | 14 | 2 |
| AIST, Tokyo, Japan | Sun: Ultra-4, 4 nodes, Solaris, hpc420-1.hpcc.jp | Globus, GTS, Fork (Worker node) | 1 | 7 | 3 |
| AIST, Tokyo, Japan | Sun: Ultra-2, 2 nodes, Solaris, hpc420-2.hpcc.jp | Globus, GTS, Fork (Worker node) | 1 | 8 | 50 |
| University of Lecce, Italy | Compaq: Alpa cluster, OSF1, sierra0.unile.it | Globus, GTS, RMS (Worker node) | 2 | 0 | 0 |
| Institute of the Italian National Research Council, Pisa, Italy | Unknown: Dual CPU PC, Linux, barbera.cnuce.cnr.it | Globus, GTS, Fork (Worker node) | 1 | 9 | 1 |
| Institute of the Italian National Research Council, Pisa, Italy | Unknown: Dual CPU PC, Linux, novello.cnuce.cnr.it | Globus, GTS, Fork (Worker node) | 1 | 0 | 0 |
| Konrad-Zuse-Zentrum Berlin, Berlin, Germany | SGI: Onyx2K, IRIX, 6, onyx1.zib.de | Globus, GTS, Fork (Worker node) | 2 | 38 | 5 |
| Konrad-Zuse-Zentrum Berlin, Berlin, Germany | SGI: Onyx2K, IRIX, 16 onyx3.zib.de | Globus, GTS, Fork (Worker node) | 3 | 32 | 7 |
| Charles University, Prague, Czech Republic | SGI: Onyx2K, IRIX, mat.ruk.cuni.cz | Globus, GTS, Fork (Worker node) | 2 | 20 | 11 |
| University of Portsmouth, UK | Unknown: Dual CPU PC, Linux, marge.csm.port.ac.uk | Globus, GTS, Fork (Worker node) | 1 | 1 | 25 |
| University of Manchester, UK | SGI: Onyx3K, 512 node, IRIX, green.cfs.ac.uk | Globus, GTS, NQS, (Worker node) | 2 | 15 | 12 |
| Argonne National Lab, Chicago, USA | SGI: IRIX lemon.mcs.anl.gov | Globus, GTS, Fork (Worker node) | 2 | 0 | 0 |
| Argonne National Lab, Chicago, USA | Sun: Ultra –8, Solaris, 8, pitcairn.mcs.anl.gov | Globus, GTS, Fork (Worker node) | 1 | 49 | 25 |
| | | Total Experiment Cost (G$) | | 199968 | 141869 |
| | | Time to Finish Expt. (Min.) | | 150 | 258 |

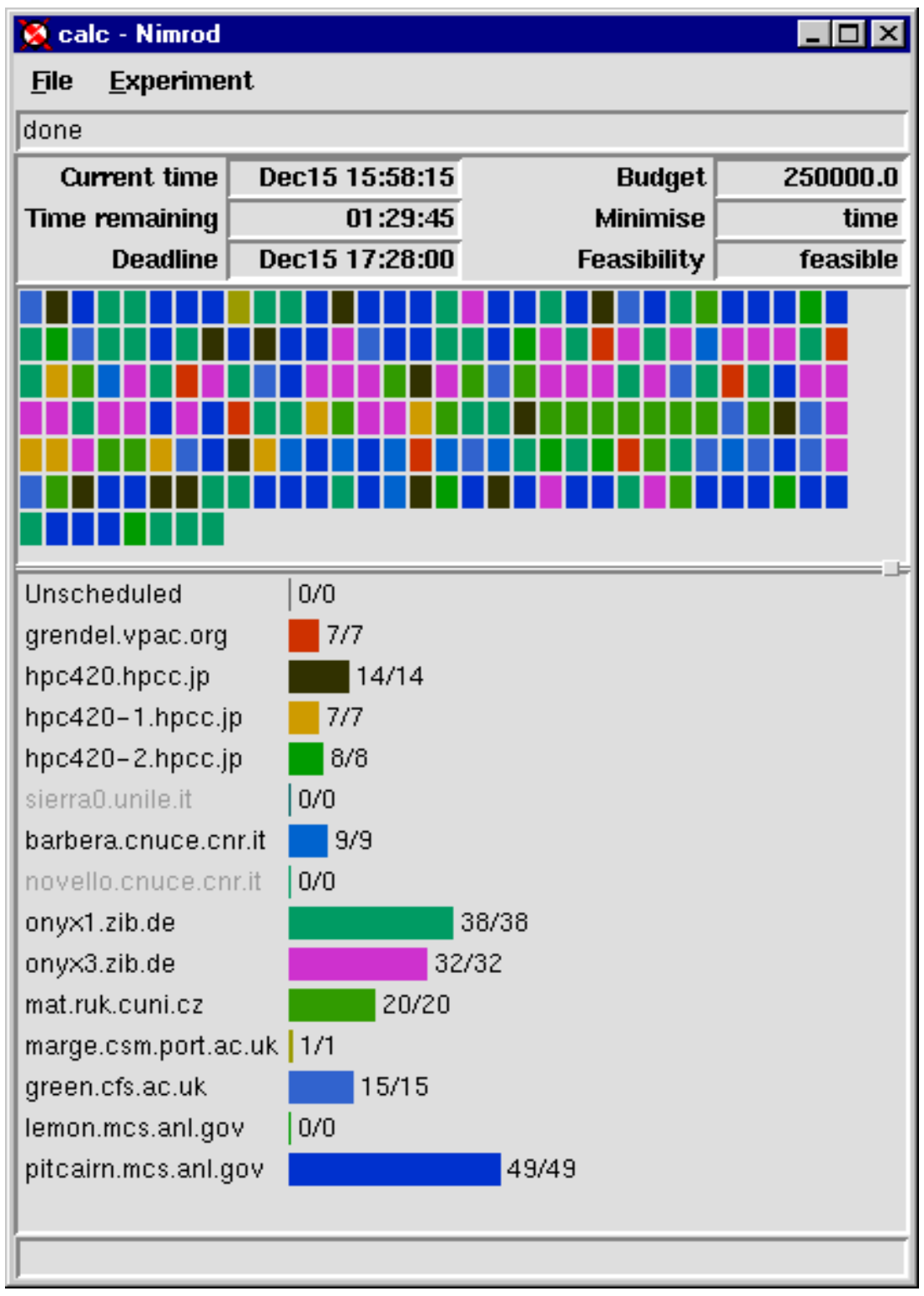

**Figure 4.21: A snapshot of the Nimrod-G monitor during "optimize for time" scheduling experiment.**

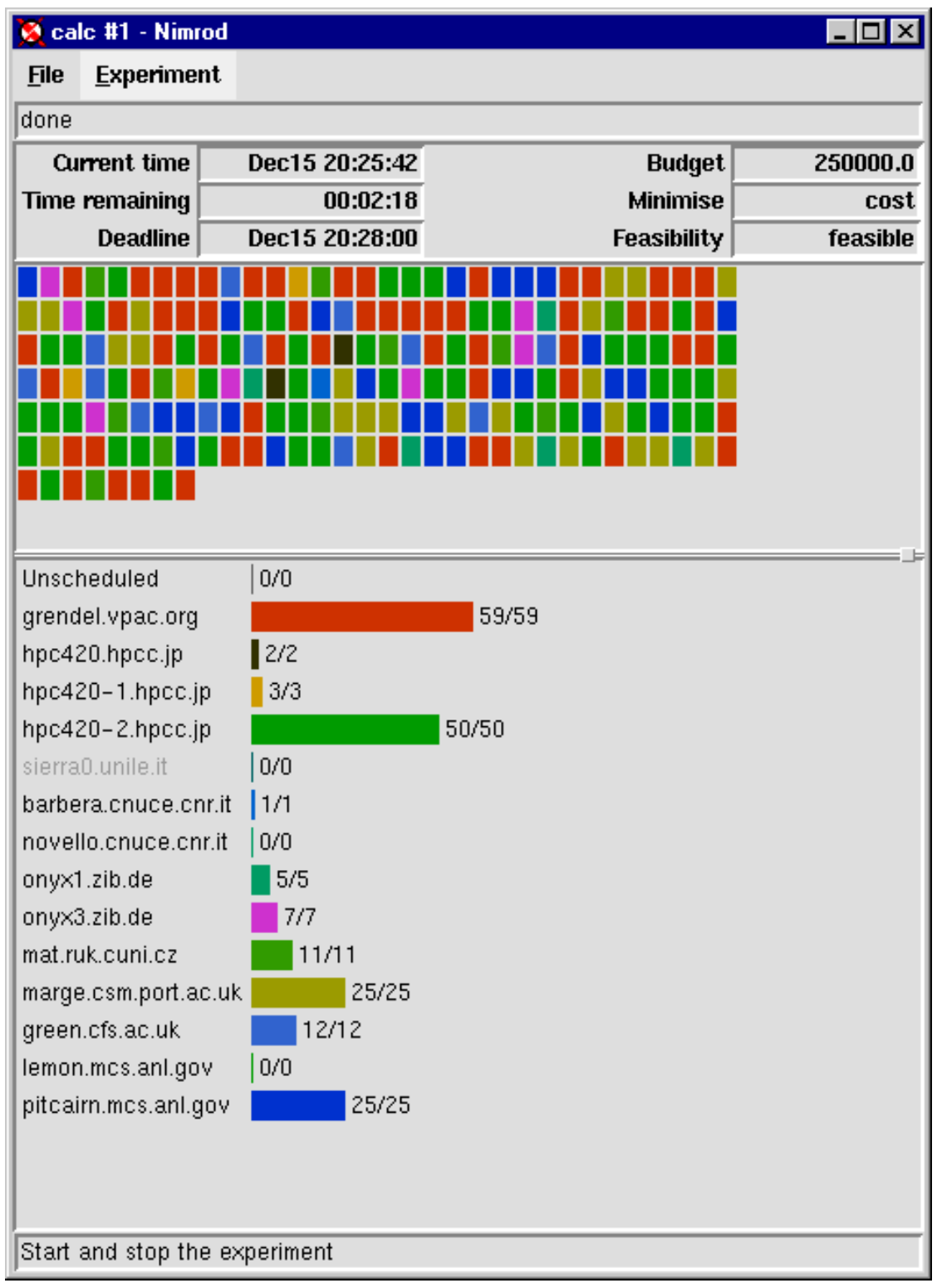

**Figure 4.22: A snapshot of the Nimrod-G monitor during "optimize for cost" scheduling experiment.**

### *DBC Constrained Time Optimization Scheduling*

The first experiment, *Optimize for Time* scheduling, we performed on December 15, 2001 at 13.28:00, Australian Eastern Daylight Saving Time (AEDT), with 4 hours deadline and finished on the same day by 15:58:15. A snapshot of the Nimrod-G monitoring and steering client, taken immediately after the completion of experiment, is shown in Figure 4.21. This experiment took 2½ hours to finish the processing of all jobs using resources available at that time with an expense of 199968 G$. Figure 4.23 shows the number of jobs in execution on different resources and Figure 4.24 shows the total number of jobs in execution on the Grid during the experiment execution period. It can be observed during the first hour of deadline, called the *calibration phase*, the broker aggressively consumed resources for processing jobs to bring the experiment to a feasible state.

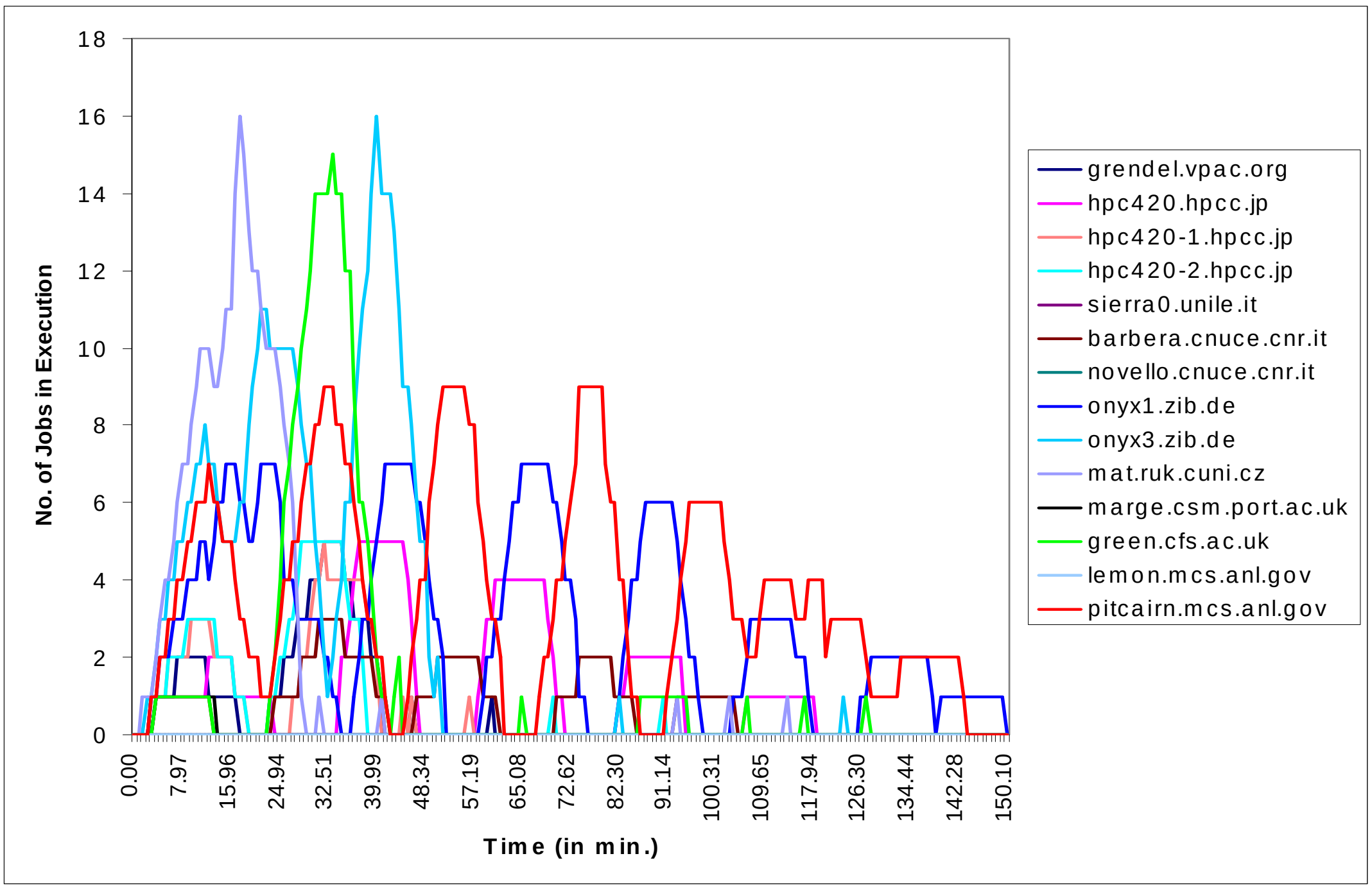

**Figure 4.23: No. of jobs in execution on different Grid resources during DBC time optimization scheduling.**

Throughout the experiment, two resources, marked in *gray* color in Figure 4.21, were unavailable (either they were shutdown or their resource information servers had failed). We were unable to schedule any jobs on the first ANL node, lemon.mcs.anl.gov, as that was overloaded with jobs from other users. Figure 4.25 shows the number of jobs processed on different resources selected depending on their cost and availability. Figure 4.26 shows the total number of jobs processed on the Grid. Figure 4.27 shows the corresponding expenses of processing on different resources and Figure 4.28 shows the aggregated processing expenses. From the graphs it can be observed that the broker selected resources to ensure that the experiment was completed at the earliest possible time given the current availability of resources and the budget limitations. It continued to use expensive resources even after the calibration phase depending on the amount of remaining budget. Another interesting pattern to be observed in Figure 4.25 and Figure 4.27 is that the amount of budget consumed by a resource is not always in proportion to the consumption amount and jobs processed. The most expensive machine (3G$/sec.), onyx3.zib.de, was used to process less jobs compared to the other two machines, onyx1.zib.de and pitcairn.mcs.anl.gov, but we had to spend more budget for processing on it.

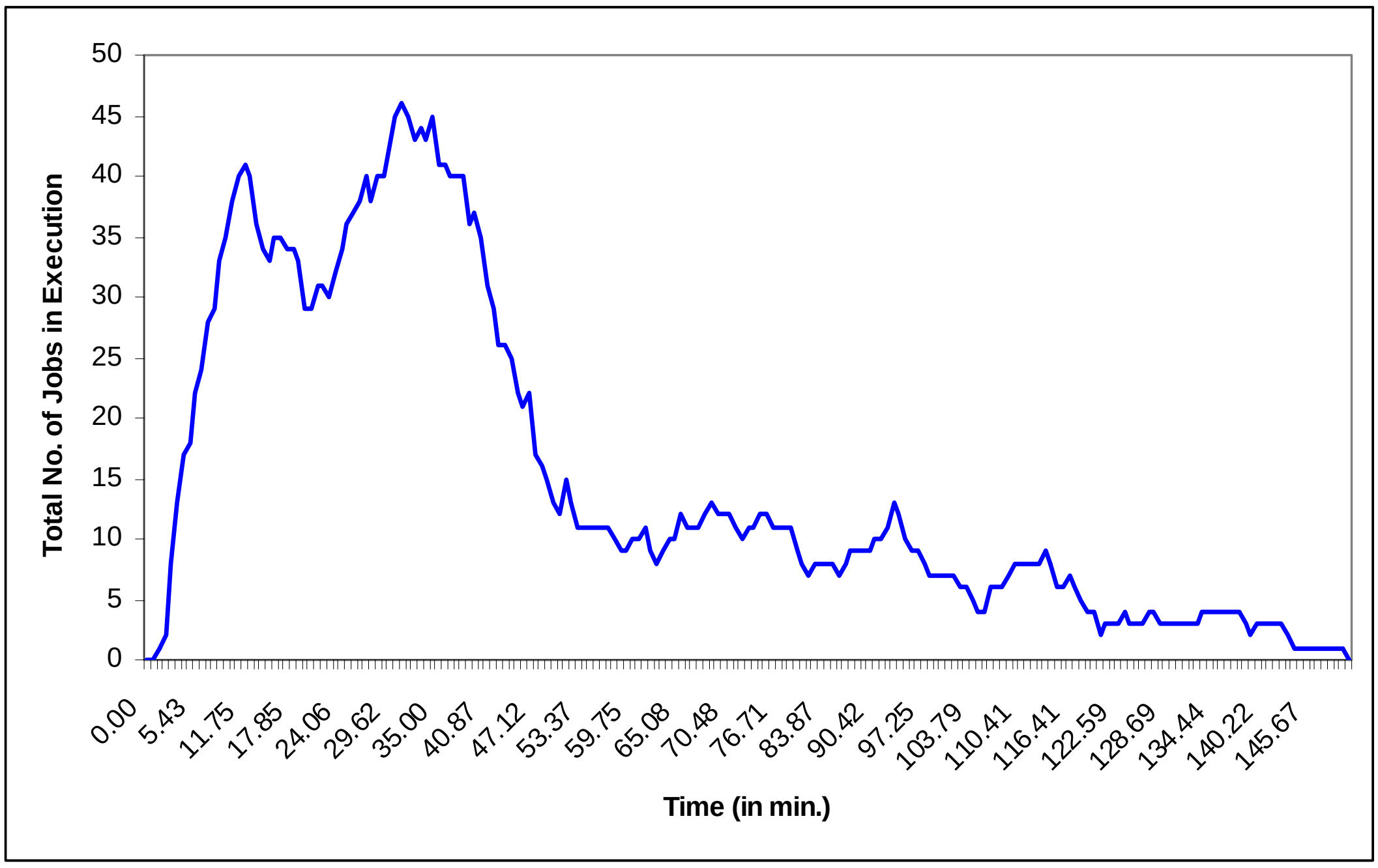

**Figure 4.24: Total No. of jobs in execution on Grid during DBC time optimization scheduling.**

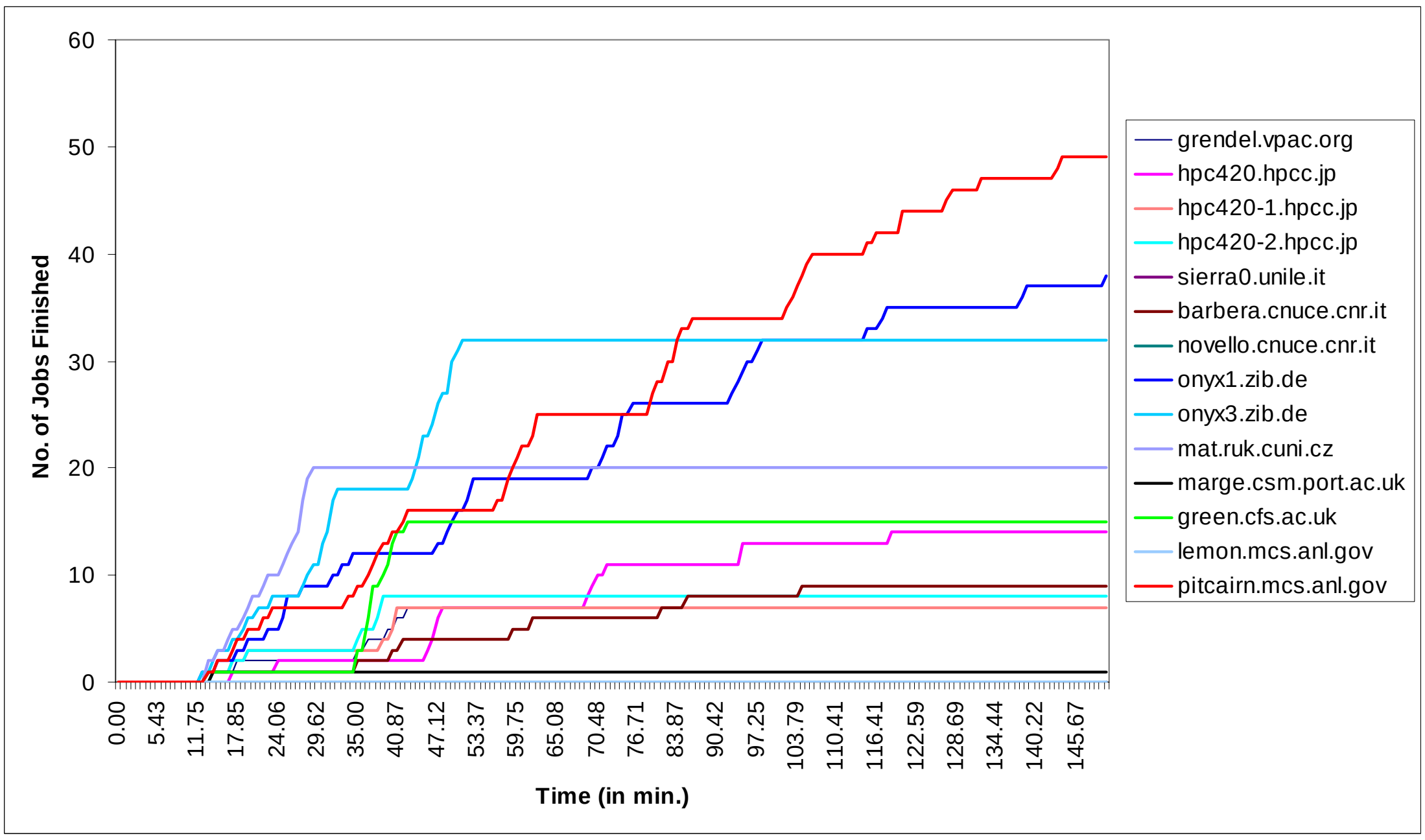

**Figure 4.25: No. of jobs processed on different Grid resources during DBC time optimization scheduling.**

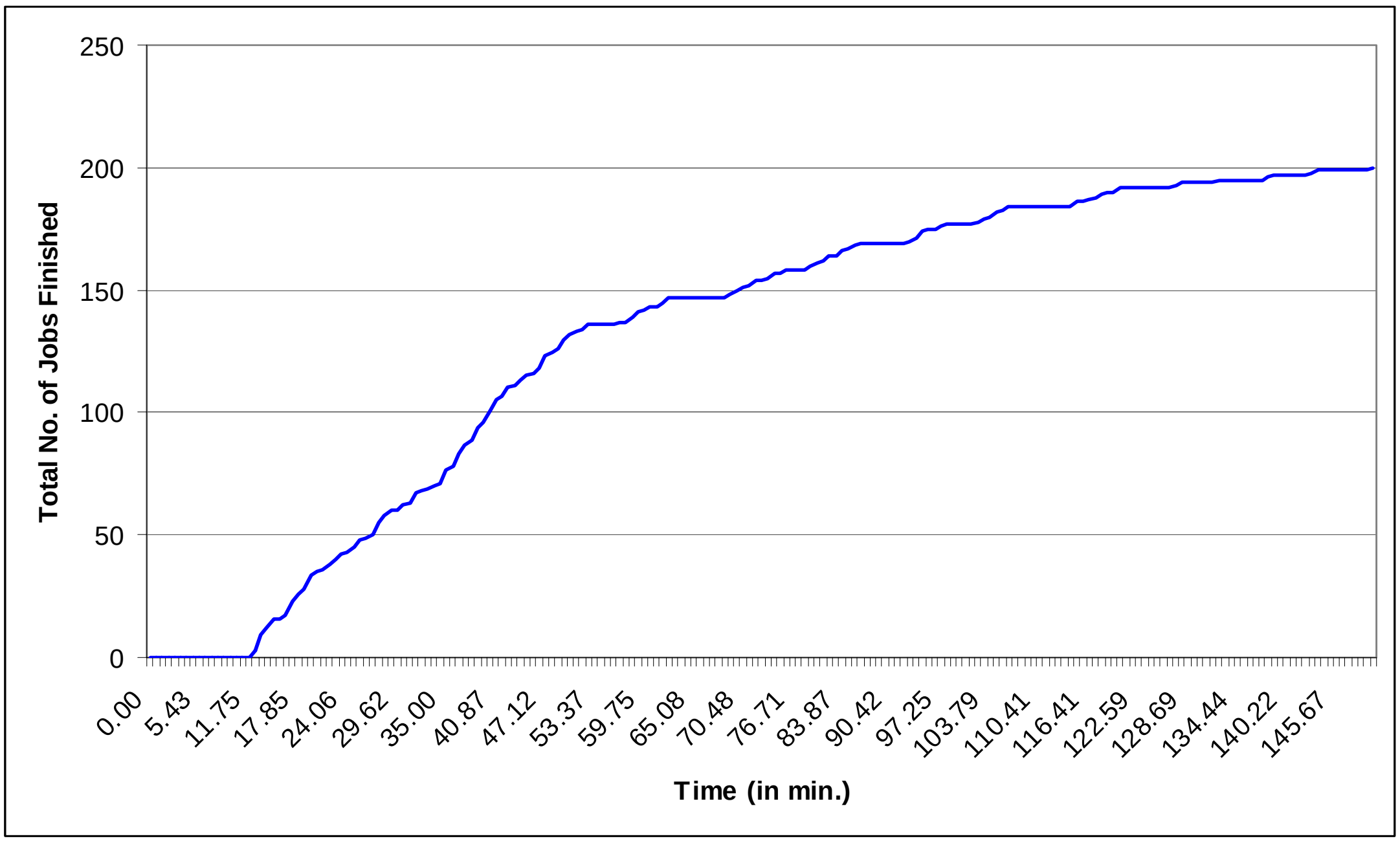

**Figure 4.26: Total no. of jobs processed on Grid resources during DBC time optimization scheduling.**

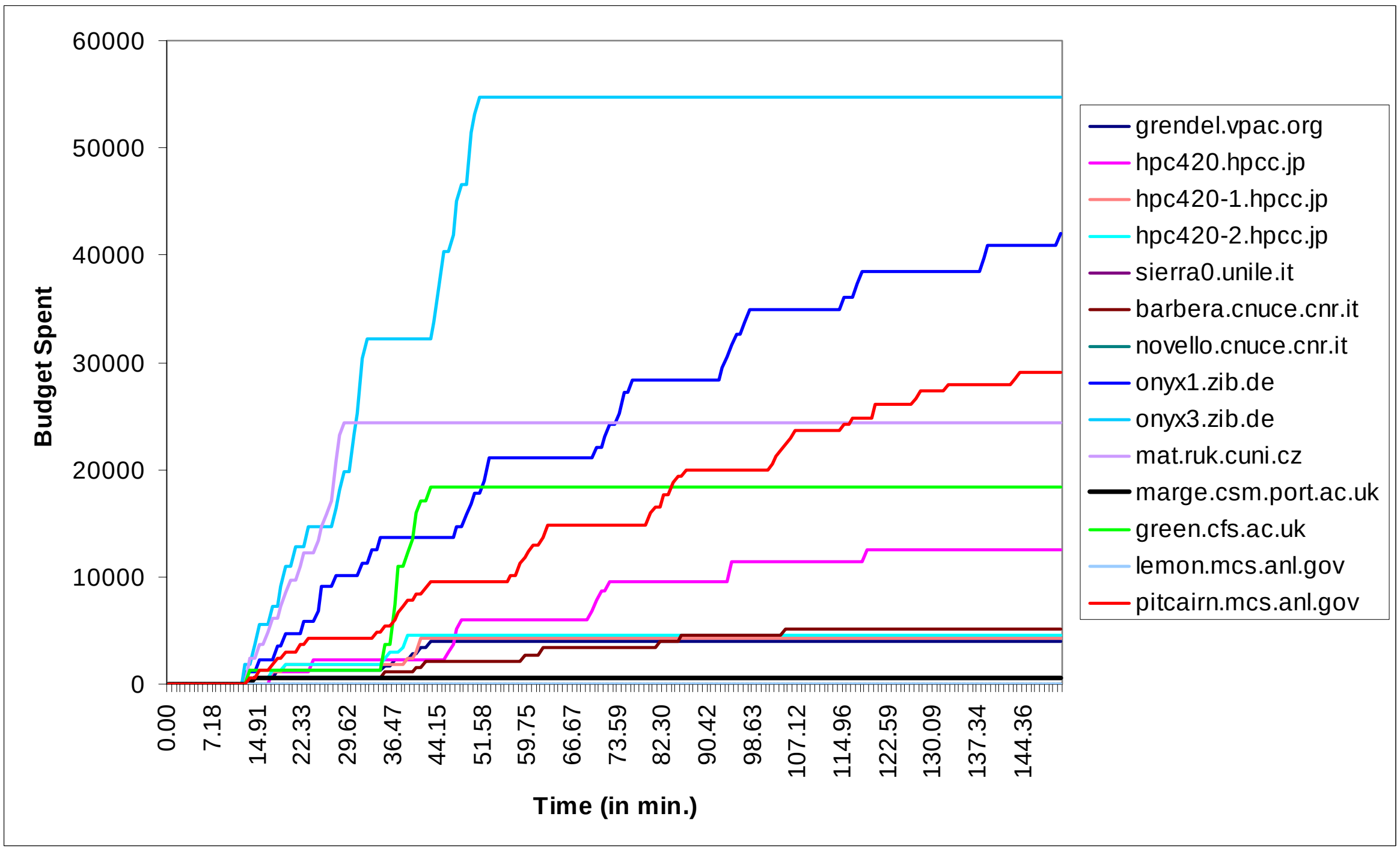

**Figure 4.27: The amount spent on different Grid resources during DBC time optimization scheduling.**

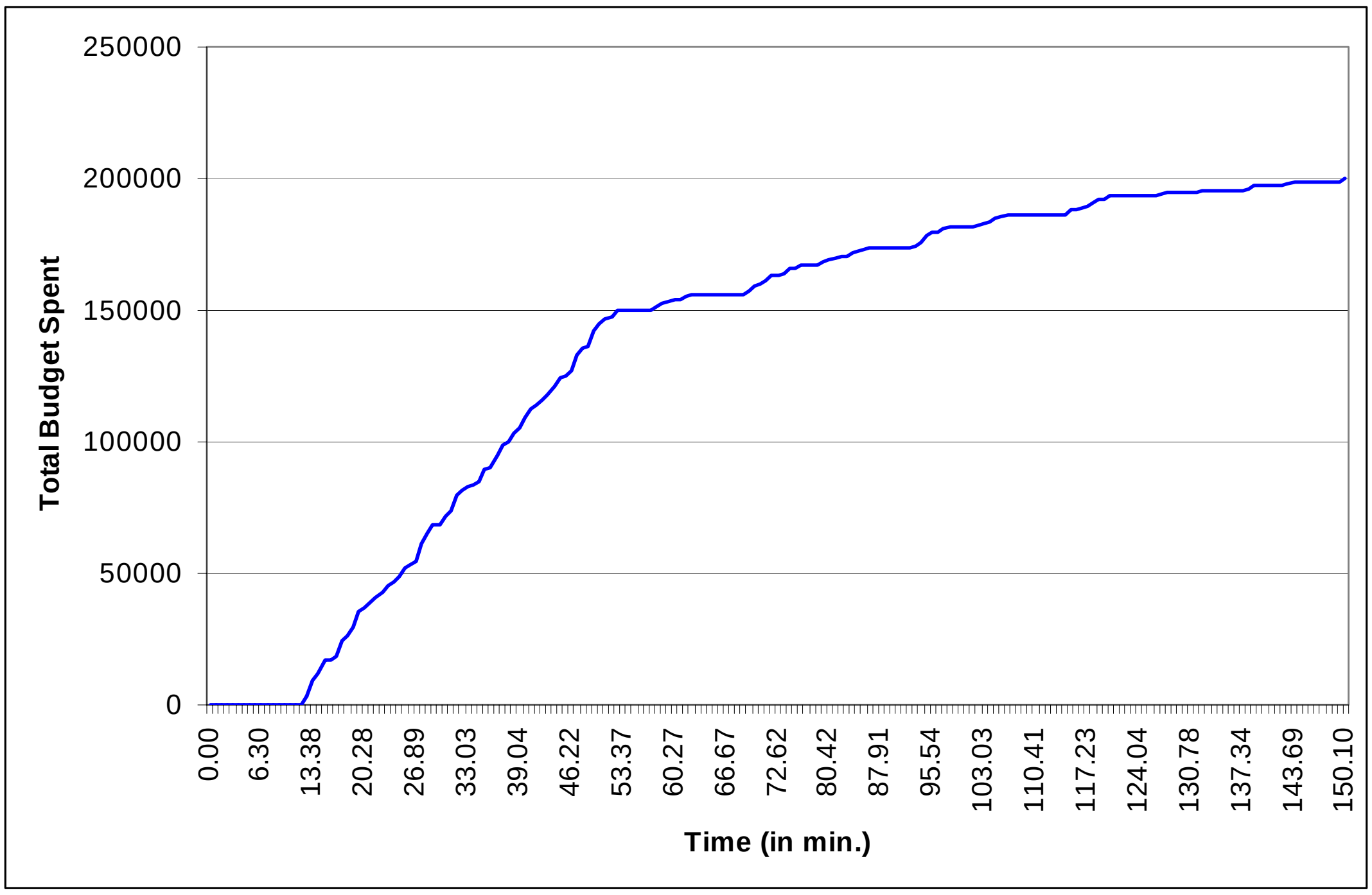

**Figure 4.28: The total amount spent on Grid during DBC time optimization scheduling.**

### *DBC Constrained Cost Optimization Scheduling*

The second experiment, *Optimize for Cost* scheduling, was performed on December 15, 2001 at 16.10:00, Australian Eastern Daylight Saving Time (AEDT), with 4 hours deadline, which had been extended by 20 minutes after 3 hours of execution time. It finished on the same day by 20:25:52. A snapshot of the Nimrod-G monitoring and steering client, taken immediately after the completion of experiment, is shown in Figure 4.22. This experiment took 4 hours and 18 minutes to finish the processing of all jobs using resources available at that time with an expense of 141869 G$. Even though it took more time compared to the first experiment, it saved 58099 G$ from expenses.

Figure 4.29 shows the number of jobs in execution on different resources and Figure 4.30 shows the total number of jobs in execution on the Grid during the experiment execution period. It can be observed during the first half-hour, called the *calibration phase*, the broker aggressively consumed resources for processing jobs to bring the experiment to a feasible state. During the experiment, one Italian resource, marked in *gray* color in Figure 4.22, was unavailable. As in the first experiment, we were unable to schedule any jobs on the first ANL node, lemon.mcs.anl.gov, as that was overloaded with jobs from other users. Figure 4.31 shows the number of jobs processed on different resources selected depending on their cost and availability and Figure 4.32 shows the aggregation of jobs processing on the Grid at different times during the experimentation. The graphs (Figure 4.33 and Figure 4.34) show the corresponding amount of budget spent for processing on individual resources and the Grid respectively.

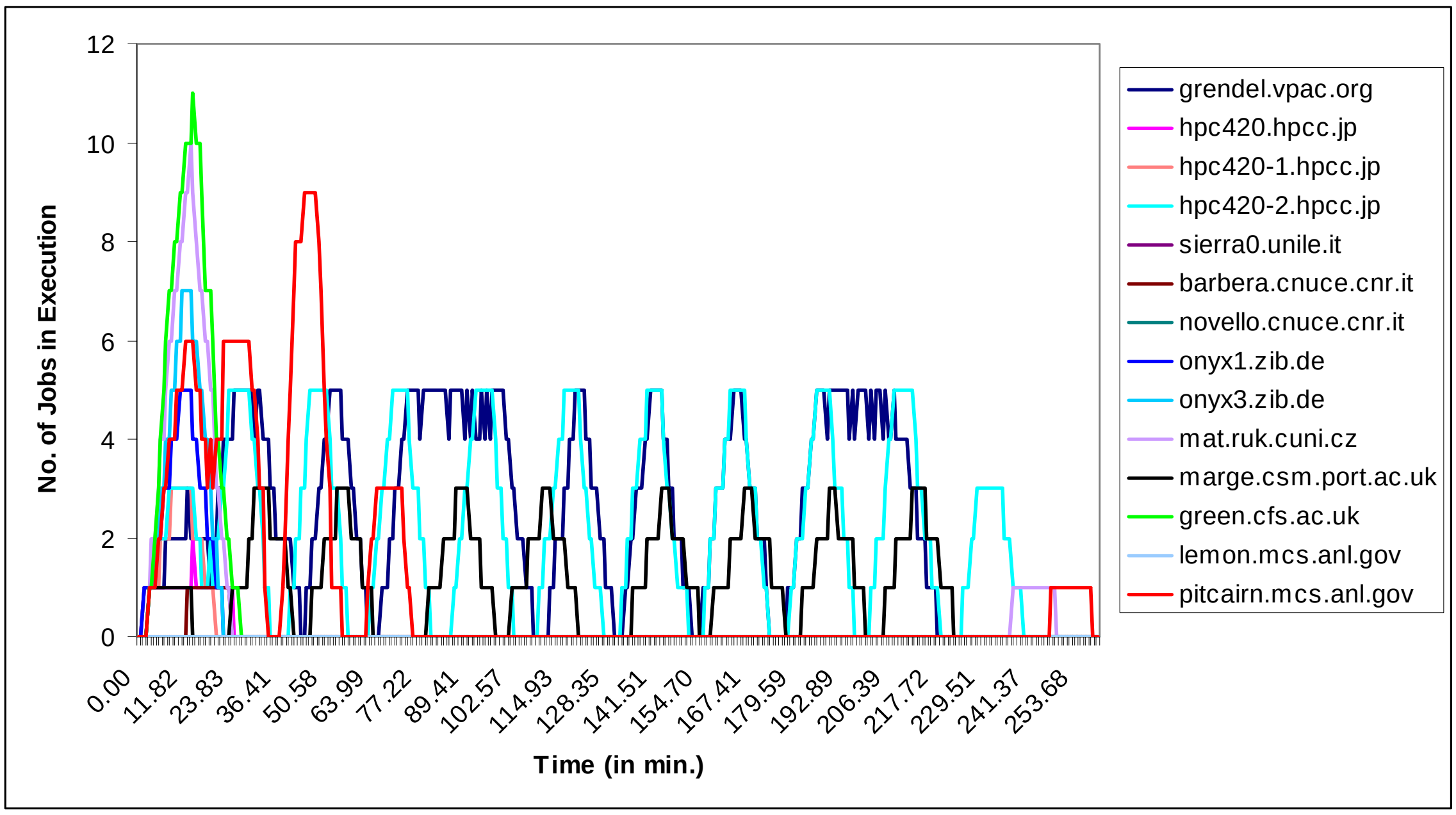

**Figure 4.29: No. of jobs in execution on different Grid resources during DBC cost optimization scheduling.**

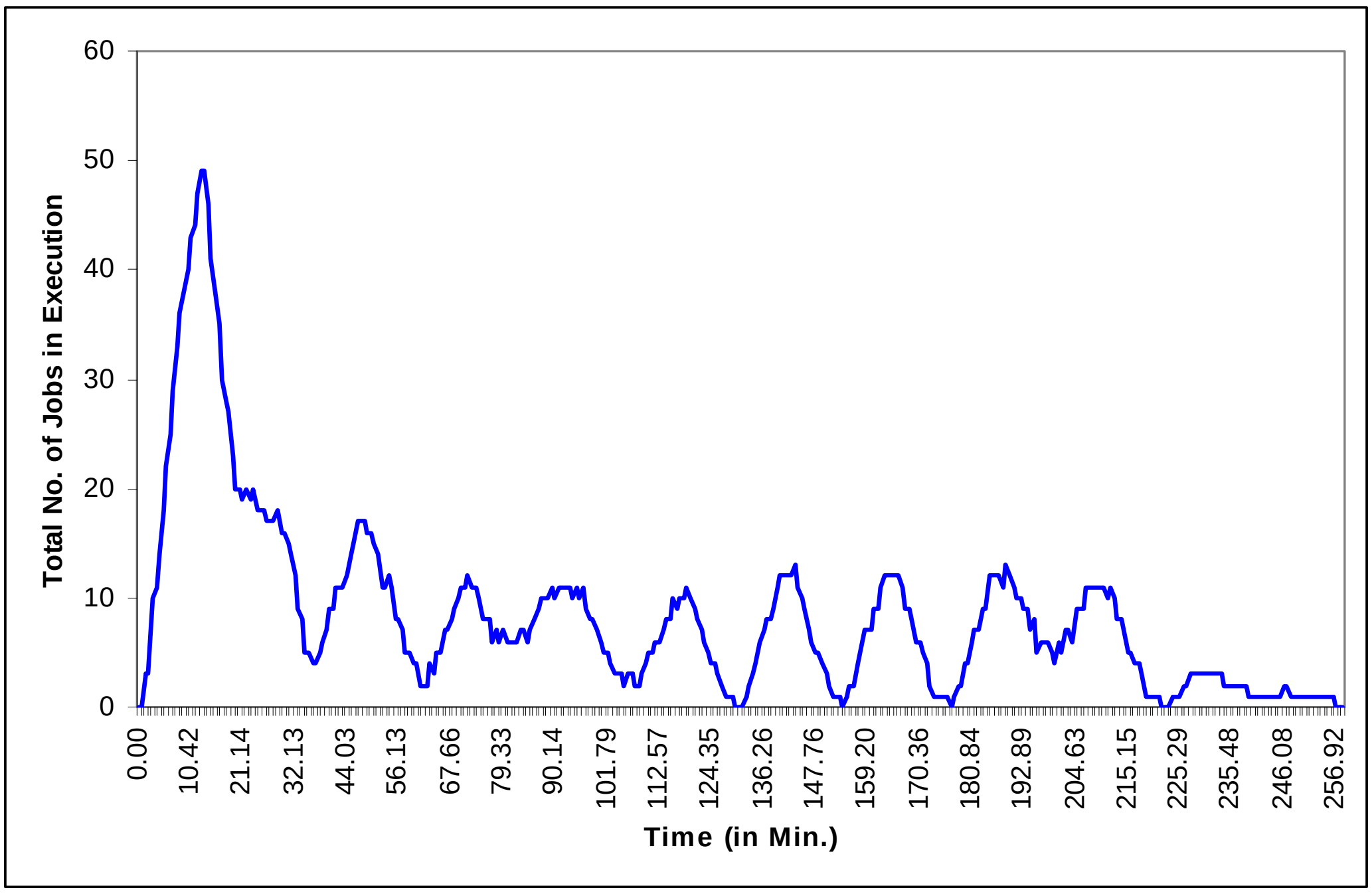

**Figure 4.30: Total No. of jobs in execution on Grid during DBC cost optimization scheduling.**

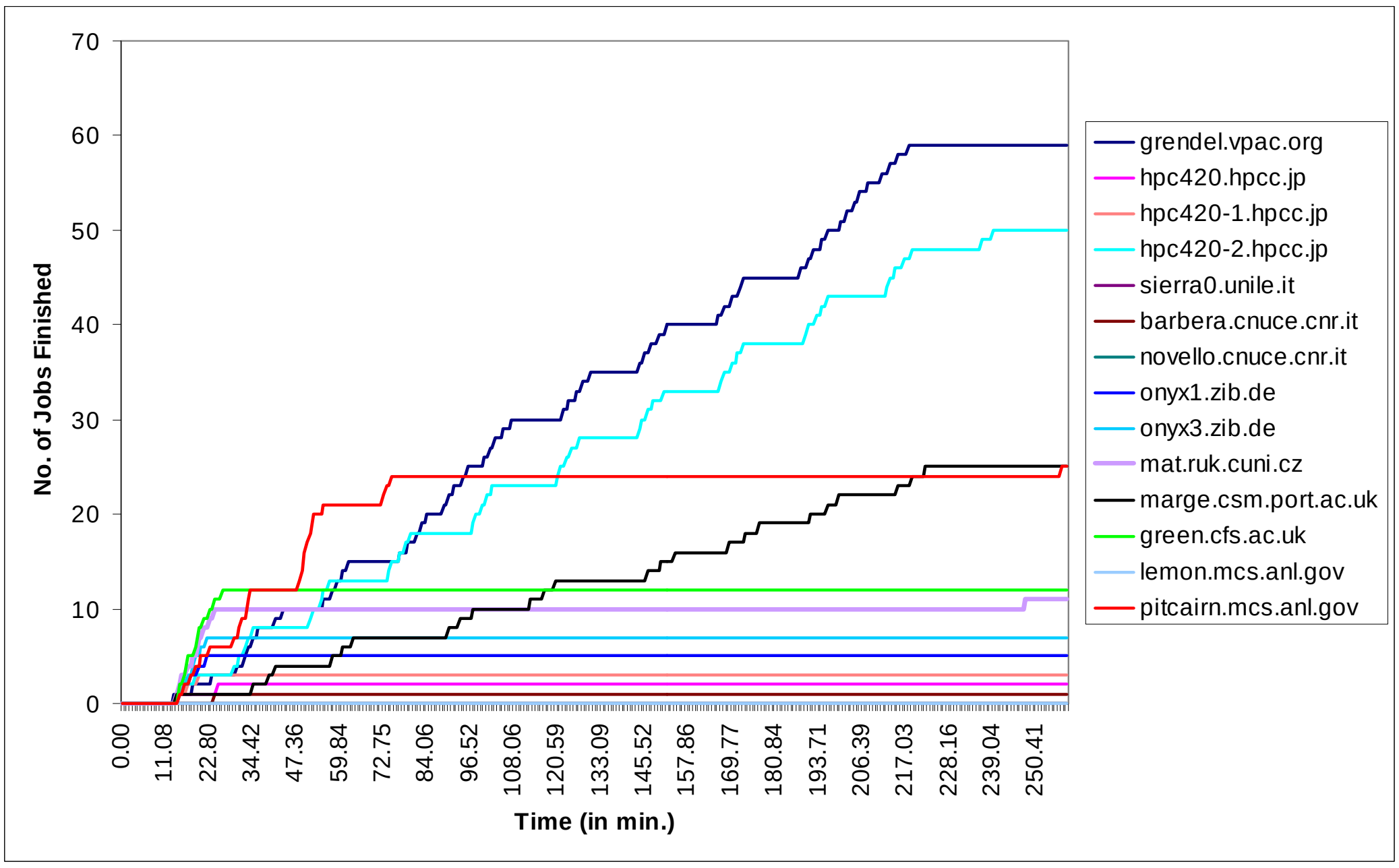

**Figure 4.31: No. of jobs processed on different Grid resources during DBC cost optimization scheduling.**

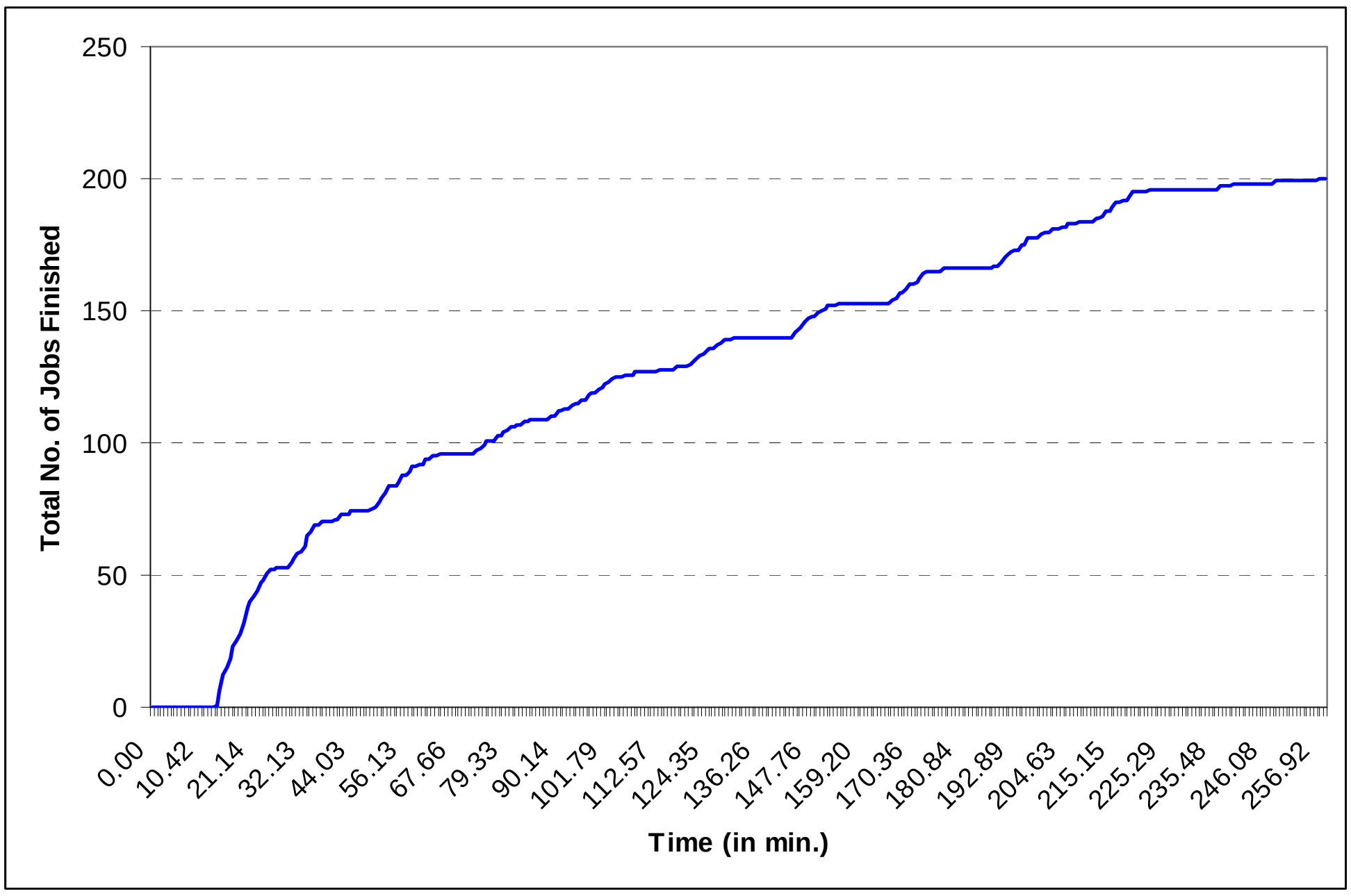

**Figure 4.32: Total No. of jobs processed on Grid during DBC cost optimization scheduling.**

From the graphs it can be observed that, after the calibration phase, the broker selected the cheapest and most powerful resources extensively to process jobs, as cost minimization was the top priority as long as the deadline could be met. However, it did use a moderately expensive resource, for example, resource mat.ruk.cuni.cz costing 2G$/CPU-sec., to process one job to ensure that the deadline can be met and this was essential as the availability of the cheapest resources had changed from the forecasted availability. As in the first experiment, it can be observed that the amount of budget consumed by a resource was not always in proportion to the number of jobs processed (see Figure 4.31 and Figure 4.33) since we had to

spend the higher amount for processing on expensive resources.

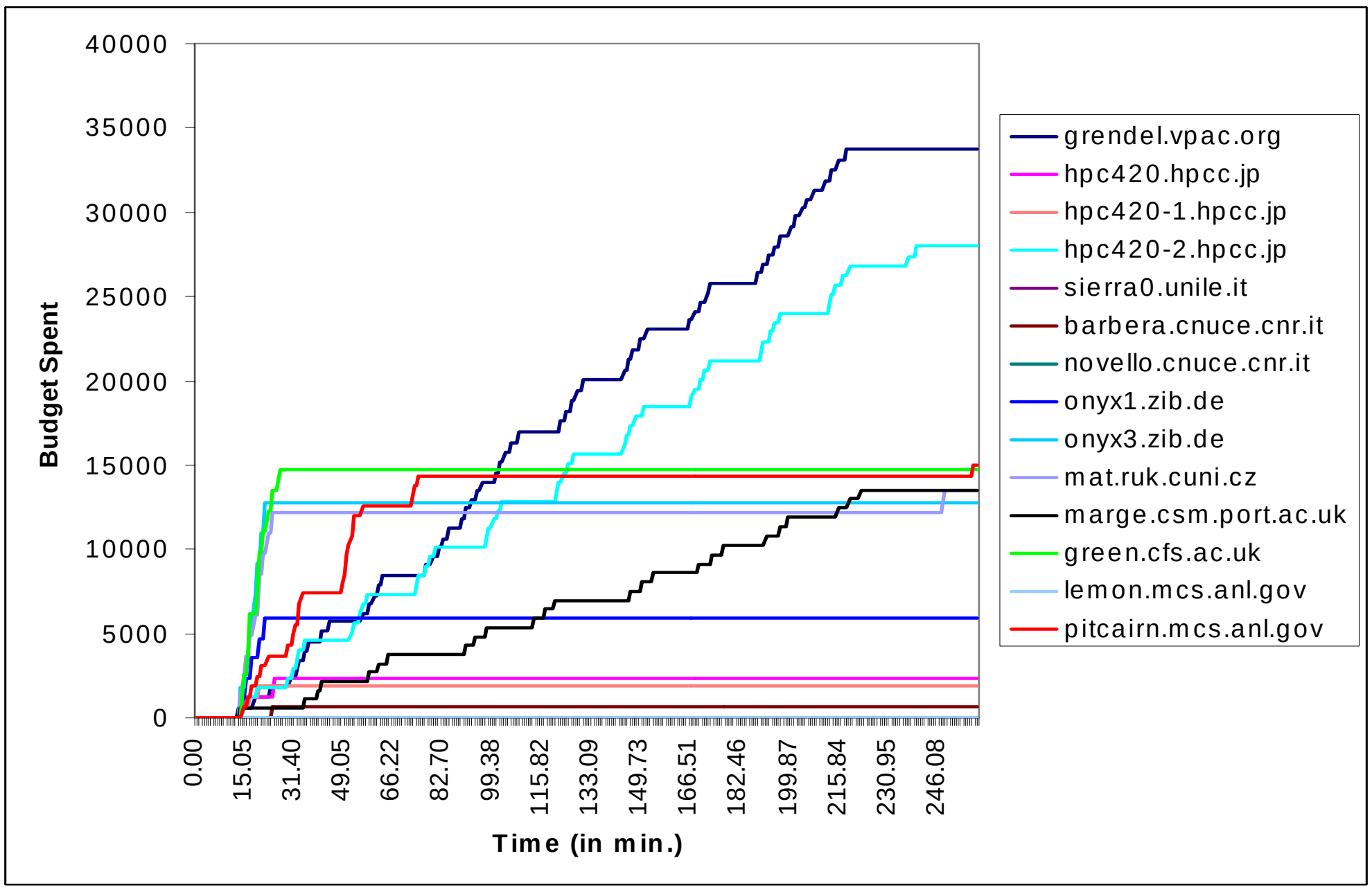

**Figure 4.33: The amount spent on different Grid resources during DBC cost optimization scheduling.**

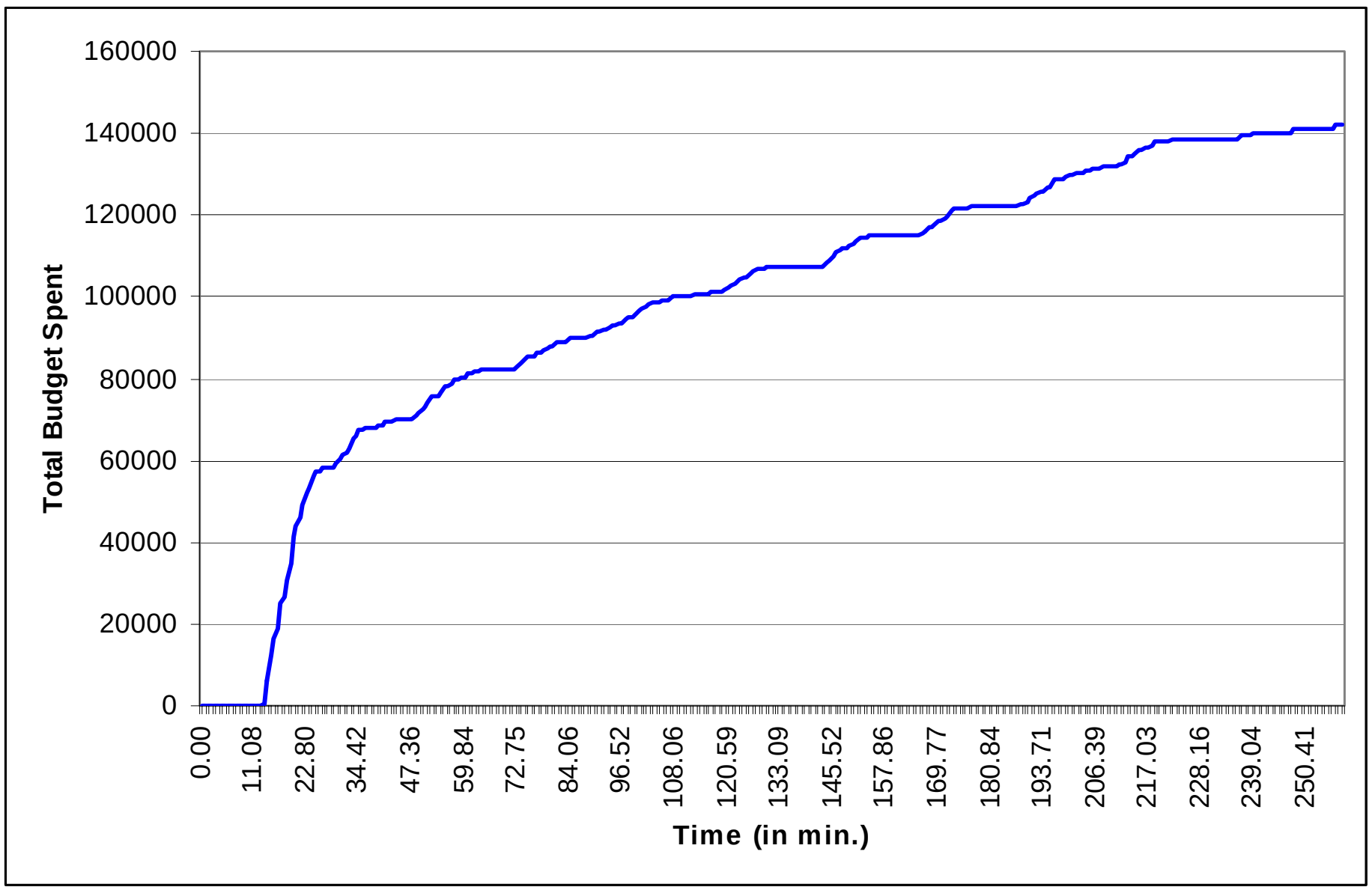

**Figure 4.34: The total amount spent on Grid during DBC cost optimization scheduling.**

## 4.8 Summary and Comments

We have discussed the design and development of the Nimrod-G Grid resource broker, that supports deadline and budget constrained and quality of service requirements-driven scheduling, of parameter sweep applications on world-wide distributed resources. It has an ability of dynamically adapting to the changes in availability of resources and user requirements at runtime. It also provides scalable, controllable, measurable, and easily enforceable policies for allocation of resources to user applications. We believe that the computational economy approach for Grid computing provides one of the essential ingredients for pushing Grid into mainstream computing.

The Nimrod tools for modeling parametric experiments are quite mature and in production use for cluster computing. A prototype version of Grid enabled tools and Nimrod-G resource broker have been implemented and they are available for download from our project web page. The Nimrod-G task farming engine (TFE) services have been used in developing customized clients and applications. An associated dispatcher is capable of deploying computations (jobs) on Grid resources enabled by Globus, Legion, and Condor. The TFE jobs management protocols and services can be used for developing new scheduling policies. We have built a number of market-driven deadline and budget constrained scheduling algorithms, namely, time and cost optimizations with deadline and budget constraints. The results of scheduling experiments with different QoS requirements on the World-Wide Grid resource show promising insights into the effectiveness of an economic paradigm for management of resources, and their usefulness in application scheduling with optimizations. The results demonstrate that the users have choice and they can indeed trade-off between the deadline and budget depending on their requirements; thus encouraging them to reveal their true requirements to increase the value delivered by the utility.

Although our scheduling experiments on the WWG testbed resources demonstrated the capability of Nimrod-G broker and provided worthwhile insights into our economic scheduling algorithms, we were unable to perform truly comparable and repeatable exhaustive evaluation. For example, with empirical studies in Grid environment, it is impossible to provide answers to questions such as: what if the cost-optimisation scheduling is selected instead of the time optimisation?; what if the deadline is changed from A to B and the budget from X to Y?; what if the parameter range is changed from M to N?; what if resource R failed?; and what if the number of available resources drastically changed?. It is impossible to repeat the scheduling experiment for different requirements with the same resource scenario as in the previous experiment – the availability of Grid resources and load continuously varies from time to time and it is impossible for an individual user/domain to control activities of other users in different administrative domains.

In order to answer the above questions and overcome the limitations of scheduling studies using a real Grid testbed, we turned to simulation. While it is absolutely worthwhile to perform empirical evaluation of scheduling on the Grid, simulation-based investigations are valuable since they allow a controlled and repeatable evaluation for a range of scenarios: varying number of users, requirements, parameters, resources, and workload, etc. Towards this end, the next two chapters of this thesis discuss the GridSim toolkit we developed and the results of an extensive series of scheduling simulations. We also present comparative analysis of scheduling algorithms presented in this chapter along with a new cost-time optimisation algorithm that evolved during our simulation studies.

## Software Availability

The Nimrod-G resource broker software with source code can be downloaded from the project website:

`http://www.csse.monash.edu.au/~davida/nimrod/`

## Chapter 5

# GridSim: A Toolkit for Modeling and Simulation of Grid Resource Management and Scheduling

This chapter presents the design and implementation of GridSim, a toolkit for modelling and simulation of resources and application scheduling in large-scale parallel and distributed computing environments. We identify the requirements for simulating complex systems and present the mechanisms that GridSim uses to support the modelling and simulation of various Grid entities—resources, users, application tasks, and schedulers/brokers—and their characteristics using discrete events. GridSim supports the creation of *repeatable* and *controllable* Grid environments for quicker performance evaluation of scheduling strategies under different scenarios such as varying number of resources and users with different requirements. A recipe for developing application scheduling simulators using the GridSim toolkit is presented at the end.

## 5.1 Introduction

In order to prove the effectiveness of resource brokers and associated scheduling algorithms, their performance needs to be evaluated under different scenarios such as varying the number of resources and users with different requirements. In a real Grid environment, it is hard and perhaps even impossible to perform scheduler performance evaluation in a *repeatable* and *controllable* manner for different scenarios—the availability of resources and their load continuously varies from time to time and it is impossible for an individual user/domain to control activities of other users in different administrative domains.

Those investigating resource management and scheduling strategies for large-scale distributed computing systems need a simple framework for deterministic modeling and simulation of resources and applications to evaluate scheduling strategies. For most who do not have access to ready-to-use testbed infrastructures, building them is expensive and time consuming. Also, even for those who have access, the testbed size is limited to a few resources and domains; and testing scheduling algorithms for scalability and adaptability, and evaluating scheduler performance for various applications and resource scenarios is harder to trace and resource intensive. The Grid computing researchers and educators also recognized the importance and the need for such a toolkit for modeling and simulation environments [61]. We have developed a Java-based discrete-event Grid simulation toolkit called *GridSim*. The toolkit supports modeling and simulation of heterogeneous Grid resources (both time- and space-shared), users and application models. It provides primitives for creation of application tasks, mapping of tasks to resources, and their management. To demonstrate suitability of the GridSim toolkit, we have simulated a Nimrod-G like Grid resource broker and evaluated the performance of deadline and budget constrained cost- and time-minimization scheduling algorithms.

Our interest in building a simulation environment arose from the need for performing a detailed evaluation of deadline and budget constraint scheduling algorithms implemented within the Nimrod-G broker [100]. We performed many experiments using the Nimrod-G broker for scheduling task farming applications on the WWG (World-Wide Grid) [113] testbed resources with small configuration (like 2 hours deadline and 10 machines for a single user). The ability to experiment with a large number of Grid scenarios was limited by the number of resources that were available in the WWG testbed. Also, it was impossible perform repeatable evaluation of scheduling strategies as the availability, allocation, and usage of resources changed from time to time. Also conducting performance evaluation on a real Grid tested for a number of different scenarios is resource intensive and time consuming task, which can be drastically minimized by using discrete event simulation techniques.

The GridSim toolkit supports modeling and simulation of a wide range of heterogeneous resources, such as single or multiprocessors, shared and distributed memory machines such as PCs, workstations, SMPs, and clusters managed by time or space-shared schedulers. That means, GridSim can be used for modeling and simulation of application scheduling on various classes of parallel and distributed computing systems such as clusters, Grids, and P2P networks. The resources in clusters are located in a single administrative domain and managed by a single entity whereas, in Grid and P2P systems, resources are geographically distributed across multiple administrative domains with their own management policies and goals. Another key difference between cluster and Grid/P2P systems arises from the way application scheduling is performed. The *schedulers* in cluster systems focus on enhancing overall system performance and utility, as they are responsible for the whole system. Whereas, schedulers in Grid/P2P systems called *resource brokers*, focus on enhancing performance of a specific application in such a way that its end-users requirements are met.

The rest of this chapter is organized as follows. Section 5.2 discusses related work with highlights on unique features that distinguish our toolkit from other packages. The GridSim architecture and internal components that make up GridSim simulations are discussed in Section 5.3. Section 5.4, discusses how to build GridSim based scheduling simulations. The final section summarizes the chapter along with comments on adoption and usage of the GridSim toolkit.

## 5.2 Related Work

Simulation has been used extensively for modeling and evaluation of real world systems, from business process and factory assembly line to computer systems design. Accordingly, over the years, modeling and simulation has emerged as an important discipline and many standard and application-specific tools and technologies have been built. They include simulation languages (e.g., Simscript [15]), simulation environments (e.g., Parsec [90]), simulation libraries (SimJava [29]), and application specific simulators (e.g., OMNet++ network simulator [5]). While there exists a large body of knowledge and tools, there are very few tools available for application scheduling simulation in Grid computing environments. The notable ones are: Bricks [62], MicroGrid [46], Simgrid [43], and our GridSim toolkit.

The Bricks simulation system [62], developed at the Tokyo Institute of Technology in Japan, helps in simulating client-server like global computing systems that provide remote access to scientific libraries and packages running on high performance computers. It follows centralized global scheduling methodology as opposed to our work in which each application scheduling is managed by the users' own resource broker.

The MicroGrid emulator [46], undertaken in the University of California at San Diego (UCSD), is modeled after Globus. It allows execution of applications constructed using Globus toolkit in a controlled virtual Grid emulated environment. The results produced by emulation can be precise, but modeling numerous applications, Grid environments, and scheduling scenarios for realistic statistical analysis of scheduling algorithms is time consuming as applications run on emulated resources. Also, scheduling algorithms designers generally work with application models instead of constructing actual applications. Therefore, MicroGrid's need for an application constructed using Globus imposes significant development overhead. However, when an actual system is implemented by incorporating scheduling strategies that are evaluated using simulation, the MicroGrid emulator can be used as a complementary tool for verifying simulation results with real applications.

The Simgrid toolkit [43], developed in the University of California at San Diego (UCSD), is a C language based toolkit for the simulation of application scheduling. It supports modeling of resources that are *time-shared* and the load can be injected as constants or from real traces. It is a powerful system that allows creation of tasks in terms of their execution time and resources with respect to a standard machine capability. Using Simgrid APIs, tasks can be assigned to resources depending on the scheduling policy being simulated. It has been used for a number of real studies, and demonstrates the power of simulation. However, because Simgrid is restricted to a single scheduling entity and time-shared systems, it is difficult to simulate multiple competing users, applications, and schedulers, each with their own policies when operating under market like Grid computing environment, without extending the toolkit substantially.

Also, many large-scale resources in the Grid environment are space-shared machines and they need to be supported in simulation. Hence, our GridSim toolkit extends the ideas in existing systems and overcomes their limitations accordingly.

Finally, we have chosen to implement GridSim in Java by leveraging SimJava's [29] basic discrete event simulation infrastructure. This feature is likely to appeal to educators and students since Java has emerged as a popular programming language for network computing.

## 5.3 GridSim: Grid Modeling and Simulation Toolkit

The GridSim toolkit provides a comprehensive facility for simulation of different classes of heterogeneous resources, users, applications, resource brokers, and schedulers. It can be used to simulate application schedulers for single or multiple administrative domain(s) distributed computing systems such as clusters and Grids. Application schedulers in Grid environment, called resource brokers, perform resource discovery, selection, and aggregation of a diverse set of distributed resources for an individual user. That means, each user has his own private resource broker and hence, it can be targeted to optimize for the requirements and objectives of its owner. Whereas schedulers, managing resources such as clusters in a single administrative domain, have complete control over the policy used for allocation of resources. That means, all users need to submit their jobs to the *central* scheduler, which can be targeted to perform global optimization such as higher system utilization and overall user satisfaction depending on resource allocation policy or optimize for high priority users.

### 5.3.1 Features

Salient features of the GridSim toolkit include the following:

- It allows modeling of heterogeneous types of resources.
- Resources can be modeled operating under space- or time-shared mode.
- Resource capability can be defined (in the form of MIPS as per SPEC benchmark).
- Resources can be located in any time zone.
- Weekends and holidays can be mapped depending on resource's local time to model non-Grid (local) workload.
- Resources can be booked for advance reservation.
- Applications with different parallel application models can be simulated.
- Application tasks can be heterogeneous and they can be CPU or I/O intensive.
- There is no limit on the number of application jobs that can be submitted to a resource.
- Multiple user entities can submit tasks for execution simultaneously in the same resource, which may be time-shared or space-shared. This feature helps in building schedulers that can use different market-driven economic models for selecting services competitively.
- Network speed between resources can be specified.
- It supports simulation of both static and dynamic schedulers.
- Statistics of all or selected operations can be recorded and they can be analyzed using GridSim statistics analysis methods.

### 5.3.2 System Architecture

We employed a layered and modular architecture for Grid simulation to leverage existing technologies and manage them as separate components. A multi-layer architecture and abstraction for the development of GridSim platform and its applications is shown in Figure 5.1. The first layer is concerned with the scalable Java's interface and the runtime machinery, called JVM (Java Virtual Machine), whose implementation is available for single and multiprocessor systems including clusters [148]. The second layer is concerned with a basic discrete-event infrastructure built using the interfaces provided by the first layer. One of the popular discrete-event infrastructure implementations available in Java is SimJava [29]. Recently a distributed implementation of SimJava is also made available. The third layer is concerned with modeling and simulation of core Grid entities such as resources, information services, and so on; application model, uniform access interface, and primitives application modeling and framework for creating higher level entities. The GridSim toolkit focuses on this layer that simulates system entities using the discrete-event services offered by the lower-level infrastructure. The fourth layer is concerned with the simulation of resource aggregators called Grid resource brokers or schedulers. The final layer focuses on application and resource modeling with different scenarios using the services provided by the two lower-level layers for evaluating scheduling and resource management policies, heuristics, and algorithms. In this section, we briefly discuss SimJava model for discrete events (a second-layer component) and focus mainly on the GridSim (the third-layer) design and implementation. The resource broker simulation and performance evaluation is highlighted in the next two sections.

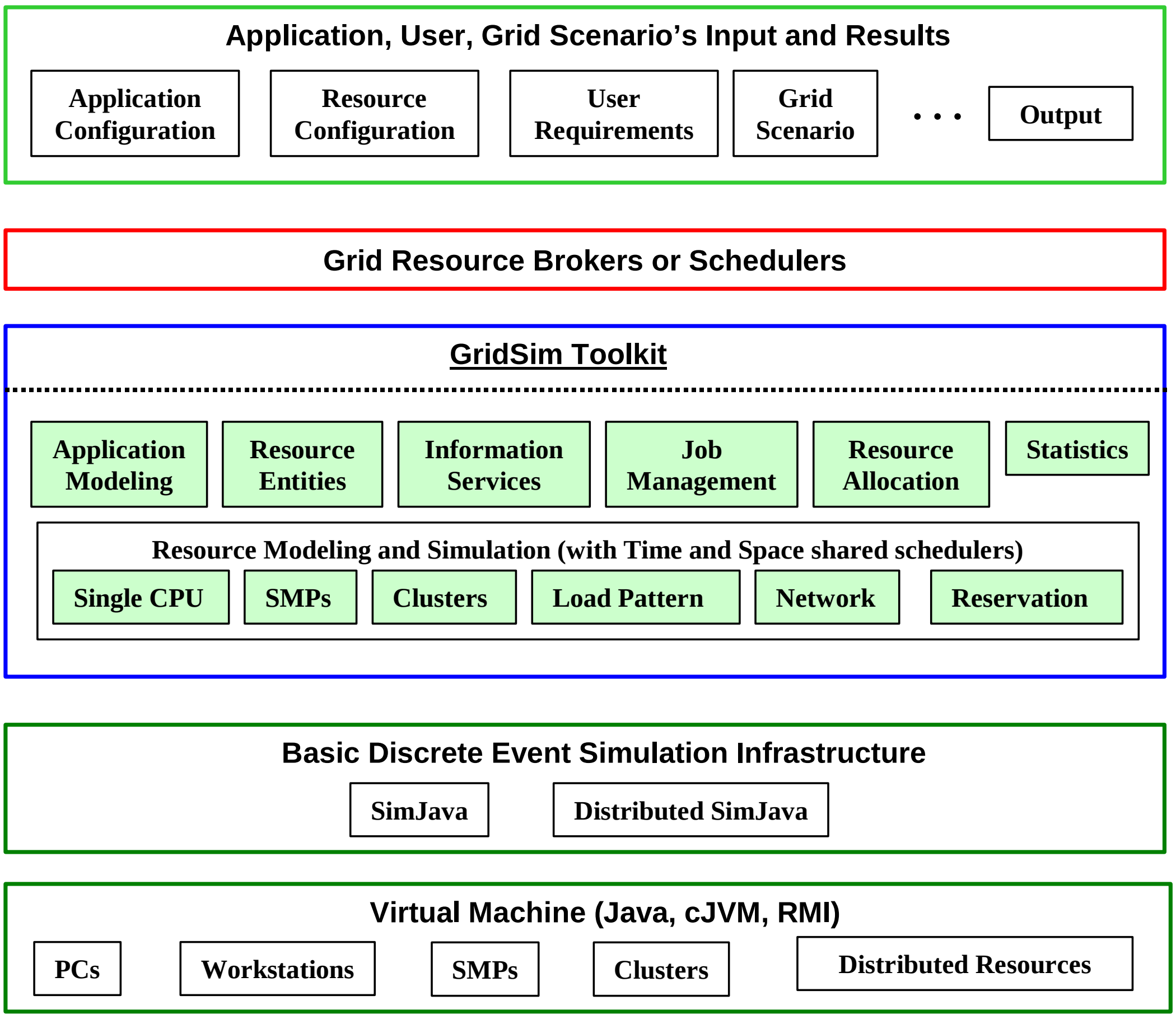

**Figure 5.1**: **A modular architecture for GridSim platform and components.**

### 5.3.3 SimJava Discrete Event Model

SimJava [29] is a general-purpose discrete event simulation package implemented in Java. Simulations in SimJava contain a number of entities each of which runs in parallel in its own thread. An entity's behaviour is encoded in Java using its body() method. Entities have access to a small number of simulation primitives:

- sim_schedule() sends event objects to other entities via ports;
- sim_hold() holds for some simulation time;
- sim_wait() waits for an event object to arrive.

These features help in constructing a network of active entities that communicate by sending and receiving passive event objects efficiently.

The sequential discrete event simulation algorithm, in SimJava, is as follows. A central object Sim_system maintains a timestamp ordered queue of future events. Initially all entities are created and their body() methods are put in run state. When an entity invokes a simulation function, the *Sim_system* object halts that entity's thread and places an event on the future queue to signify processing the function. When all entities have halted, Sim_system pops the next event off the queue, advances the simulation time accordingly, and restarts entities as appropriate. This continues until no more events are generated. If the Java virtual machine supports native threads, then all entities starting at exactly the same simulation time may run concurrently.

### 5.3.4 GridSim Entities

GridSim supports entities for simulation of single processor and multiprocessor, heterogeneous resources that can be configured as time or space shared systems. It allows setting their clock to different time zones to simulate geographic distribution of resources. It supports entities that simulate networks used for communication among resources. During simulation, GridSim creates a number of multi-threaded entities, each of which runs in parallel in its own thread. An entity's behavior needs to be simulated within its body() method, as dictated by SimJava.

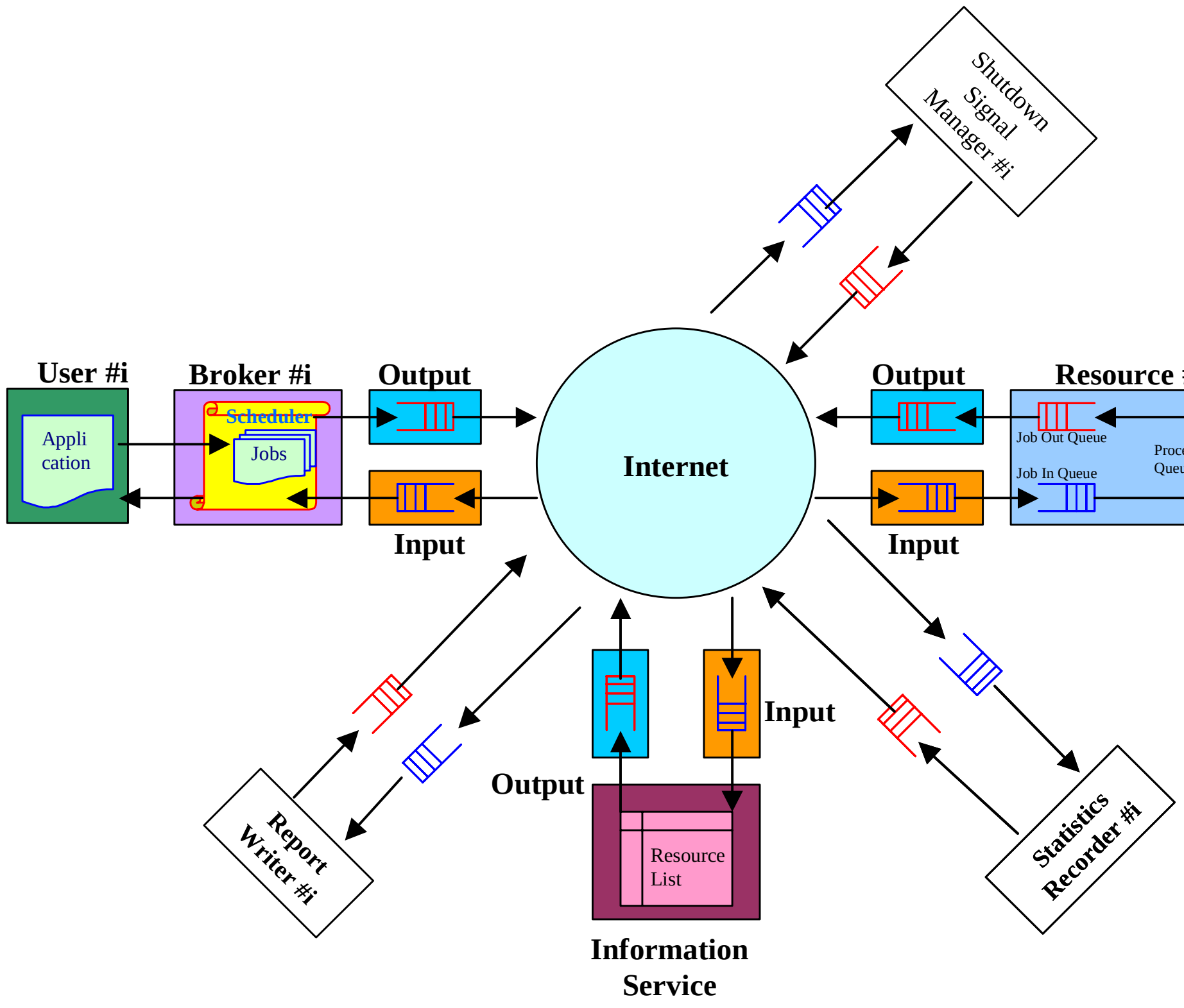

**Figure 5.2**: **A flow diagram in GridSim based simulations.**

A simulation environment needs to abstract all the entities and their time dependent interactions in the real system. It needs to support the creation of user-defined time dependent response functions for the interacting entities. The response function can be a function of the past, current, or both states of entities. GridSim based simulations contain entities for the users, brokers, resources, information service, statistics, and network based I/O as shown in Figure 5.2. The design and implementation issues of these GridSim entities are discussed below:

**User –** Each instance of the User entity represents a Grid user. Each user may differ from the rest of the users with respect to the following characteristics:

- Types of job created e.g., job execution time, number of parametric replications, etc.,
- Scheduling optimization strategy e.g., minimization of cost, time, or both,
- Activity rate e.g., how often it creates new job,
- Time zone, and

- Absolute deadline and budget, or
- D-and B-factors, deadline and budget relaxation parameters, measured in the range [0,1] express deadline and budget affordability of the user relative to the application processing requirements and available resources.

**Broker –** Each user is connected to an instance of the Broker entity. Every job of a user is first submitted to its broker and the broker then schedules the parametric tasks according to the user's scheduling policy. Before scheduling the tasks, the broker dynamically gets a list of available resources from the global directory entity. Every broker tries to optimize the policy of its user and therefore, brokers are expected to face extreme competition while gaining access to resources. The scheduling algorithms used by the brokers must be highly adaptable to the market's supply and demand situation.

**Resource –** Each instance of the Resource entity represents a Grid resource. Each resource may differ from the rest of resources with respect to the following characteristics:

- Number of processors;
- Cost of processing;
- Speed of processing;
- Internal process scheduling policy e.g., time shared or space shared;
- Local load factor; and
- Time zone.

The resource speed and the job execution time can be defined in terms of the ratings of standard benchmarks such as MIPS and SPEC. They can also be defined with respect to the standard machine. Upon obtaining the resource contact details from the Grid information service, brokers can query resources directly for their static and dynamic properties.

**Grid Information Service –** It provides resource registration services and keeps track of a list of resources available in the Grid. The brokers can query this for resource contact, configuration, and status information.

**Input and Output –**The flow of information among the GridSim entities happen via their Input and Output entities. Every networked GridSim entity has I/O channels or ports, which are used for establishing a link between the entity and its own Input and Output entities.  Note that the GridSim entity and its Input and Output entities are threaded entities i.e., they have their own execution thread with body() method that handle the events. The architecture for entity communication model in GridSim is illustrated in Figure 5.3. The use of separate entities for input and output enables a networked entity to model full duplex and multi-user parallel communications. The support for buffered input and output channels associated with every GridSim entity provides a simple mechanism for an entity to communicate with other entities and at the same time enables the modeling of a communication delay transparently.

### 5.3.5 Application Model

GridSim does not explicitly define any specific application model. It is up to the developers (of schedulers and resource brokers) to define them. We have experimented with a task-farming application model and we believe that other parallel application models such as process parallelism, DAGs (Directed Acyclic Graphs), divide and conquer etc., described in [70], can also be modeled and simulated using GridSim.

In GridSim, each independent task may require varying processing time and input files size. Such tasks can be created and their requirements are defined through *Gridlet* objects.  A *Gridlet* is a package that contains all the information related to the job and its execution management details such as job length expressed in MIPS, disk I/O operations, the size of input and output files, and the job originator. These basic parameters help in determining execution time, the time required to transport input and output files between users and remote resources, and returning the processed Gridlets back to the originator along with the results. The GridSim toolkit supports a wide range of Gridlet management protocols and services that allow schedulers to map a Gridlet to a resource and manage it through out the life cycle.

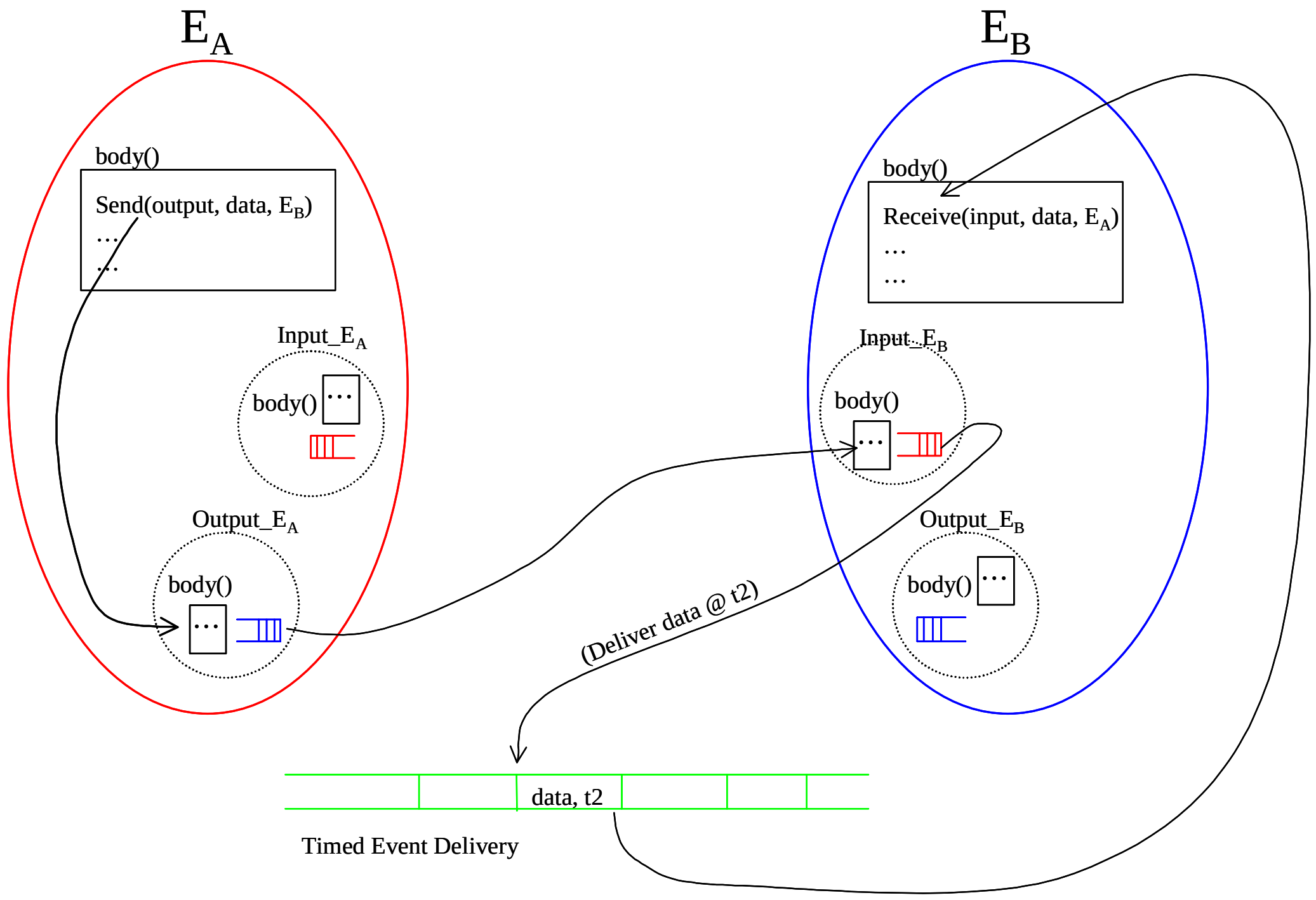

**Figure 5.3: Entity communication model via its Input and Output entities.**

### 5.3.6 Interaction Protocols Model

The protocols for interaction between GridSim entities are implemented using events. In GridSim, entities use events for both service requests and service deliveries. The events can be raised by any entity to be delivered immediately or with specified delay to other entities or itself. The events that are originated from the same entity are called *internal events* and those originated from the external entities are called *external events*. Entities can distinguish these events based on the source identification associated with them. The GridSim protocols are used for defining entity services. Depending on the service protocols, the GridSim events can be further classified into *synchronous* and *asynchronous* events. An event is called *synchronous* when the event source entity waits until the event destination entity performs all the actions associated with the event (i.e., the delivery of full service). An event is called *asynchronous* when the event source entity raises an event and continues with other activities without waiting for its completion. When the destination entity receives such events or service requests, it responds back with results by sending one or more events, which can then take appropriate actions. It should be noted that external events could be synchronous or asynchronous, but internal events need to be raised as asynchronous events only to avoid deadlocks.

A complete set of entities in a typical GridSim simulation and the use of events for simulating interaction between them are shown in Figure 5.4 and Figure 5.5. Figure 5.4 emphasizes the interaction between a resource entity that simulates time-shared scheduling and other entities. Figure 5.5 emphasizes the interaction between a resource entity that simulates space-shared system and other entities. In this section we briefly discuss the use of the events for simulating Grid activities.

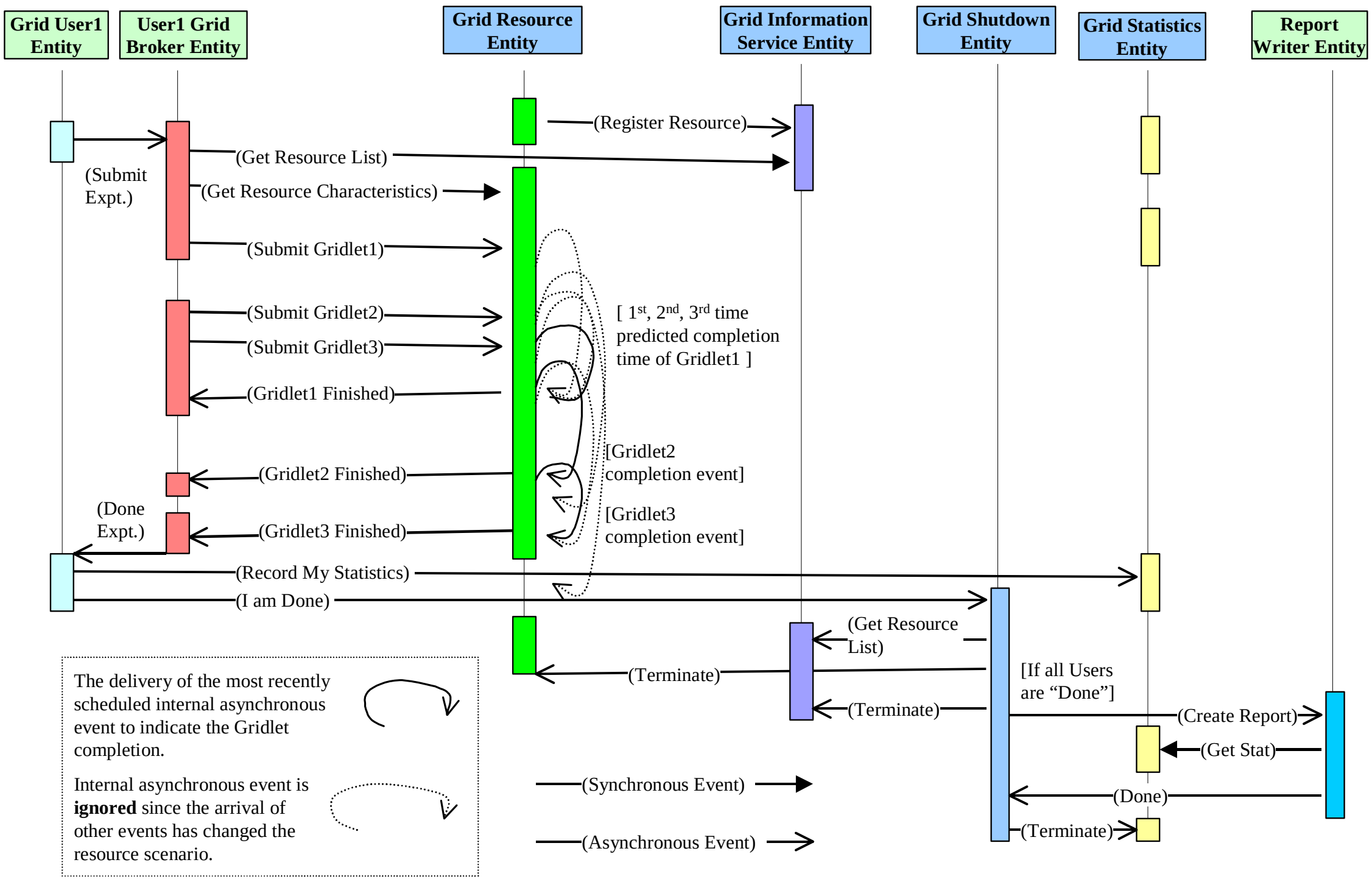

**Figure 5.4: An event diagram for interaction between a time-shared resource and other entities.**

The GridSim entities (user, broker, resource, information service, statistics, shutdown, and report writer) send events to other entities to signify the request for service, deliver results, or raise internal actions. Note that GridSim implements core entities that simulate resource, information service, statistics, and shutdown services. These services are used to simulate user with application, broker for scheduling, and an optional report writer for creating statistical reports at the end of a simulation. The event source and destination entities must agree upon the protocols for service request and delivery. The protocols for interaction between the user-defined and core entities are pre-defined.

When GridSim starts, the resource entities register themselves with the Grid Information Service (GIS) entity, by sending events. This resource registration process is similar to GRIS (Grid Resource Information Server) registering with GIIS (Grid Index Information Server) in Globus system. Depending on the user entity's request, the broker entity sends an event to the GIS entity, to signify a query for resource discovery. The GIS entity returns a list of registered resources and their contact details. The broker entity sends events to resources with request for resource configuration and properties. They respond with dynamic information such as resources cost, capability, availability, load, and other configuration parameters. These events involving the GIS entity are synchronous in nature.

Depending on the resource selection and scheduling strategy, the broker entity places asynchronous events for resource entities in order to dispatch Gridlets for execution—the broker need not wait for a resource to complete the assigned work. When the Gridlet processing is finished, the resource entity updates the Gridlet status and processing time and sends it back to the broker by raising an event to signify its completion.

The GridSim resources use internal events to simulate resource behavior and resource allocation. The entity needs to be modeled in such a way that it is able to receive all events meant for it. However, it is up to the entity to decide on the associated actions. For example, in time-shared resource simulations (see Figure 5.4) internal events are scheduled to signify the completion time of a Gridlet, which has the smallest remaining processing time requirement. Meanwhile, if an external event arrives, it changes the share resource availability for each Gridlet. That means the most recently scheduled event may not necessarily signify the completion of a Gridlet. The resource entity can discard such internal events without processing. The use of internal events for simulating resources is discussed in detail in Section 5.3.7.

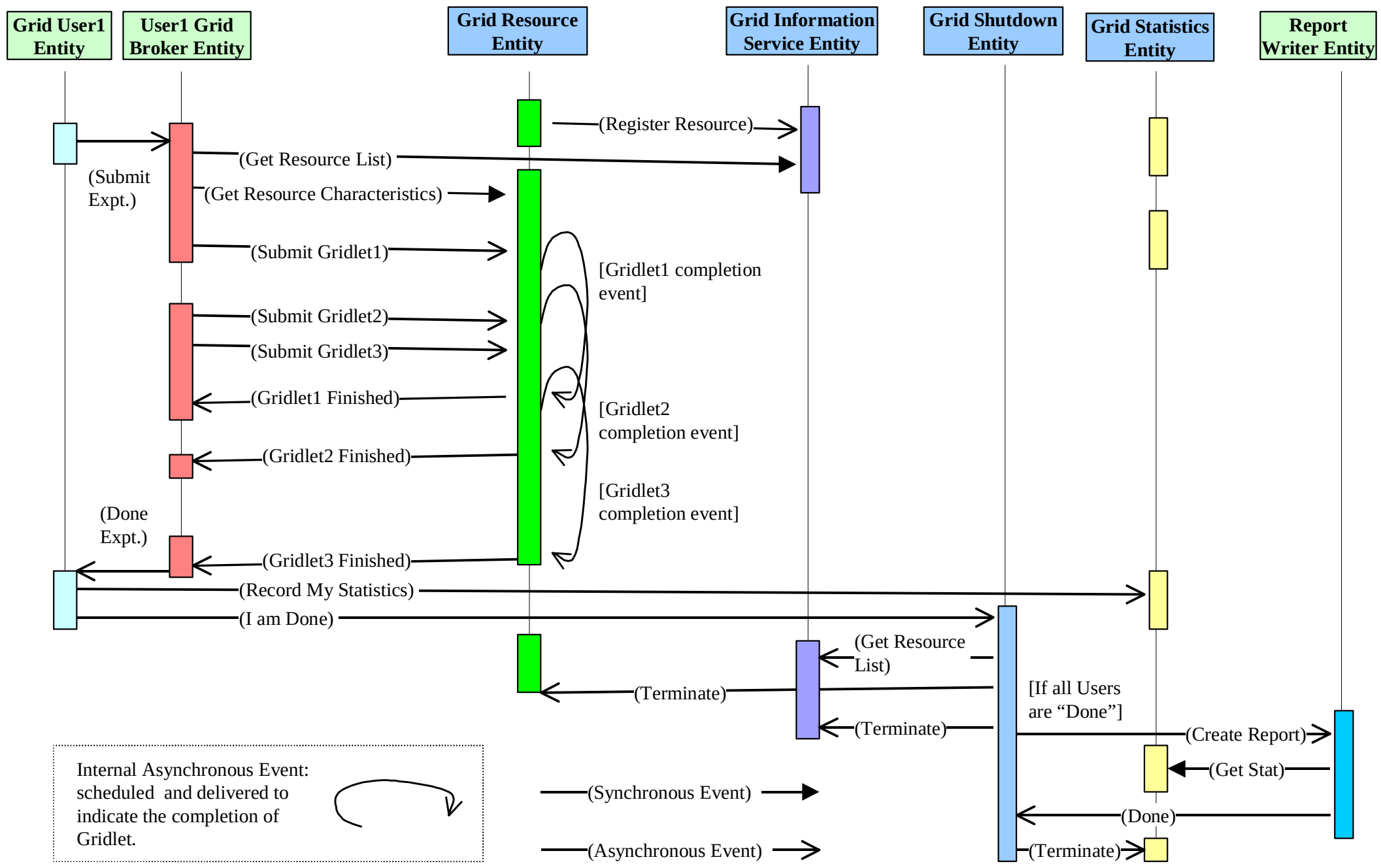

**Figure 5.5: An event diagram for interaction between a space-shared resource and other entities.**

### 5.3.7 Resource Model – Simulating Multitasking and Multiprocessing

In the GridSim toolkit, we can create Processing Elements (PEs) with different speeds (measured in either MIPS or SPEC-like ratings). Then, one or more PEs can be put together to create a machine. Similarly, one or more machines can be put together to create a Grid resource. Thus, the resulting Grid resource can be a single processor, shared memory multiprocessors (SMP), or a distributed memory cluster of computers. These Grid resources can simulate time- or space-shared scheduling depending on the allocation policy. A single PE or SMP type Grid resource is typically managed by time-shared operating systems that use round-robin scheduling policy (see Figure 5.8) for multitasking. The distributed memory multiprocessing systems (such as clusters) are managed by queuing systems, called space-shared schedulers, that execute a Gridlet by running it on a dedicated PE (see Figure 5.11) when allocated. The space-shared systems use resource allocation policies such as first-come-first-served (FCFS), back filling, shortest-job-first served (SJFS), and so on. It should also be noted that resource allocation within high-end SMPs could also be performed using the space-shared schedulers.

Multitasking and multiprocessing systems allow concurrently running tasks to share system resources such as processors, memory, storage, I/O, and network by scheduling their use for very short time intervals. A detailed simulation of scheduling tasks in the real systems would be complex and time consuming. Hence, in GridSim, we abstract these physical entities and simulate their behavior using process oriented, discrete event "interrupts" with time interval as large as the time required for the completion of a smallest remaining-time job. The GridSim resources can send, receive, or schedule events to simulate the execution of jobs. It schedules self-events for simulating resource allocation depending on the scheduling policy and the number of jobs in queue or in execution.

Let us consider the following scenario to illustrate the simulation of Gridlets execution and scheduling within a GridSim resource. A resource consists of two shared or distributed memory PEs each with MIPS rating of 1, for simplicity. Three Gridlets that represent jobs with processing requirements equivalent to 10, 8.5, and 9.5 MI (million instructions) arrive in simulation times 0, 4, and 7 respectively. The way GridSim schedules jobs to PEs is shown schematically in Figure 5.8 for time-shared resources and Figure 5.11 for space-shared resources.

### *Simulation of Scheduling in Time-Shared Resources*

The GridSim resource simulator uses internal events to simulate the execution and allocation of PEs share to Gridlet jobs. When jobs arrive, time-shared systems start their execution immediately and share resources among all jobs. Whenever a new Gridlet job arrives, we update the processing time of existing Gridlets and then add this newly arrived job to the execution set. We schedule an internal event to be delivered at the earliest completion time of smallest job in the execution set. It then waits for the arrival of events.

A complete algorithm for simulation of time-share scheduling and execution is shown in Figure 5.6. If a newly arrived event happens to be an internal event whose tag number is the same as the most recently scheduled event, then it is recognized as a job completion event. Depending on the number of Gridlets in execution and the number of PEs in a resource, GridSim allocates appropriate amount of PE share to all Gridlets for the event duration using the algorithm shown in Figure 5.7. It should be noted that Gridlets sharing the same PE would get an equal amount of PE share. The completed Gridlet is sent back to its originator (broker or user) and removed from the execution set. GridSim schedules a new internal event to be delivered at the forecasted earliest completion time of the remaining Gridlets.

*Algorithm: Time-Shared Grid Resource Event Handler ()*

1. Wait for an event
2. If the external and Gridlet arrival event, then:
   BEGIN /* a new job has arrived */
   a. Allocate PE Share for Gridlets Processed so far
   b. Add arrived Gridlet to Execution_Set
   c. Forecast completion time of all Gridlets in Execution_Set
   d. Schedule an event to be delivered at the smallest completion time
   
   END
3. If event is internal and its tag value is the same as the recently scheduled internal event tag,
   BEGIN /* a job finish event */
   a. Allocate PE Share of all Gridlets processed so far
   b. Update finished Gridlet's PE and Wall clock time parameters and send it back to the broker
   c. Remove finished Gridlet from the Execution_Set and add to Finished_Set
   d. Forecast completion time of all Gridlets in Execution_Set
   e. Schedule an event to be delivered at the smallest completion time
   
   END
4. Repeat the above steps until the end of simulation event is received

**Figure 5.6: An event handler for simulating time-shared resource scheduling.**

Figure 5.8 illustrates the simulation of time-share scheduling algorithm and the Gridlets' execution. When Gridlet1 arrives at time 0, it is mapped to PE1 and an internal event to be delivered at the time 10 is scheduled since the predicted completion time is still 10. At time 4, Gridlet2 arrives and it is mapped to the PE2. The completion time of Gridlet2 was predicted as 12.5 and the completion time of Gridlet1 is still 10 since both of them are executing on different PEs. A new internal event is scheduled, which will still be delivered at time 10. At time 7, Gridlet3 arrives, which is mapped to the PE2. It shares the PE time with Gridlet2. At time 10, an internal event is delivered to the resource to signify the completion of the Gridlet1, which is then sent back to the broker. At this moment, as the number of Gridlets equal the number of PEs, they are mapped to different PEs. An internal event to be delivered at time 14 is scheduled to indicate the predicted completion time of Gridlet2. As simulation proceeds, an internal event is delivered at time 14 and Gridlet2 is sent back to the broker. An internal event to be delivered at time 18 is scheduled to indicate the predicted completion time of Gridlet3. Since there were no other Gridlets submitted before this time, the resource receives an internal interrupt at time 18, which signifies the completion of Gridlet3. A schematic representation of Gridlets arrival, internal events delivery, and sending them back to the broker is shown in Figure 5.4. A detailed statistical data on the arrival, execution start, finish, and elapsed time of all Gridlets is shown in Table 5.1.

*Algorithm: PE_Share_Allocation(Duration)*

1. Identify total MI per PE for the duration and the number of PE that process one extra Gridlet
   TotalMIperPE = MIPSRatingOfOnePE()*Duration
   MinNoOfGridletsPerPE = NoOfGridletsInExec / NoOfPEs
   NoOfPEsRunningOneExtraGridlet = NoOfGridletsInExec % NoOfPEs
2. Identify maximum and minimum MI share that Gridlet get in the Duration
   If(NoOfGridletsInExec <= NoOfPEs), then:
      MaxSharePerGridlet = MinSharePerGridlet  = TotalMIperPE
      MaxShareNoOfGridlets = NoOfGridletsInExec
   Else /*  NoOfGridletsInExec > NoOfPEs */
      MaxSharePerGridlet = TotalMIperPE/ MinNoOfGridletsPerPE
      MinSharePerGridlet = TotalMIperPE/(MinNoOfGridletsPerPE+1)
      MaxShareNoOfGridlets = (NoOfPEs - NoOfPEsRunningOneExtraGridlet)* MinNoOfGridletsPerPE

**Figure 5.7: PE share allocation to Gridlet in time-shared GridSim resource.**

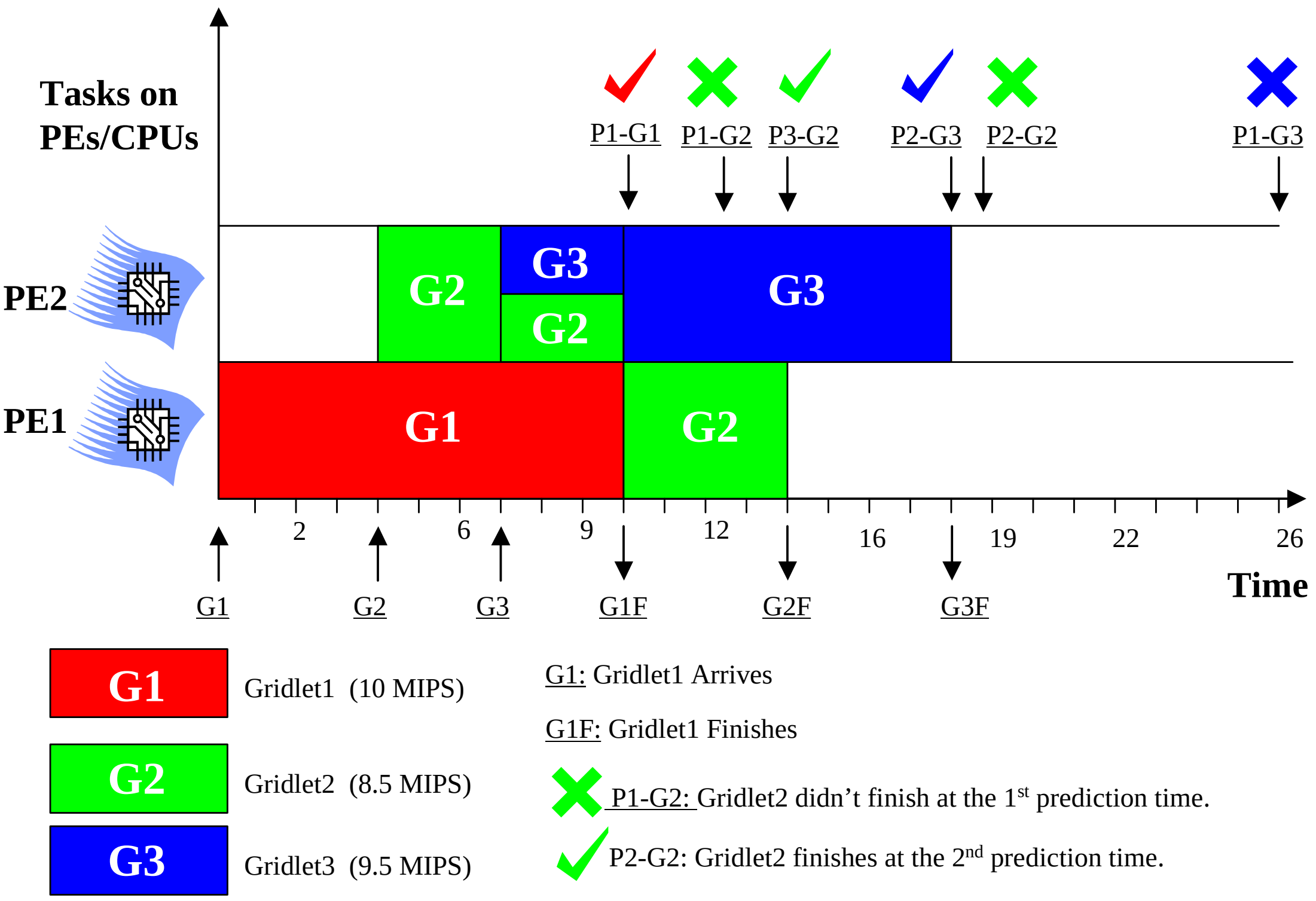

**Figure 5.8: Modeling time-shared multitasking and multiprocessing based on an event scheme.**

**Table 5.1: A scheduling statistics scenario for time- and space-shared resources in GridSim.**

| | | | Time-Shared Resource | | | Space-Shared Resource | | |
|---|---|---|---|---|---|---|---|---|
| **Grilets Number** | **Length (MI)** | **Arrival Time (a)** | **Start Time (s)** | **Finish Time (f)** | **Elapsed Time (f-a)** | **Start Time (s)** | **Finish Time (f)** | **Elapsed Time (f-a)** |
| G1 | 10 | 0 | 0 | 10 | 10 | 0 | 10 | 10 |
| G2 | 8.5 | 4 | 4 | 14 | 10 | 4 | 12.5 | 8.5 |
| G3 | 9.5 | 7 | 7 | 18 | 11 | 10 | 19.5 | 12.5 |

### *Simulation of Scheduling in Space-Shared Resources*

The GridSim resource simulator uses internal events to simulate the execution and allocation of PEs to Gridlet jobs. When a job arrives, space-shared systems start its execution immediately if there is a free PE available, otherwise, it is queued. During the Gridlet assignment, job-processing time is determined and event is scheduled for delivery at the end of execution time. Whenever the Gridlet job finishes and the internal event is delivered to signify the completion of scheduled Gridlet job, the resource simulator frees the PE allocated to it and then checks if there are any other jobs waiting in the queue. If there are jobs waiting in the queue, then it selects a suitable job depending on the policy and assigns to the PE, which is free.

A complete algorithm for simulation of space-share scheduling and execution is shown in Figure 5.9. If a newly arrived event happens to be an internal event whose tag number is the same as the most recently scheduled event, then it is recognized as a Gridlet completion event. If there are Gridlets in the submission queue, then depending on the allocation policy (e.g., the first Gridlet in the queue if FCFS policy is used), GridSim selects suitable Gridlets from the queue and assigns it to the PE or a suitable PE if more than one PE is free. See Figure 5.11 for illustration of the allocation of PE to Gridlets. The completed Gridlet is sent back to its originator (broker or user) and removed from the execution set. GridSim schedules a new internal event to be delivered at the completion time of the scheduled Gridlet job.

*Algorithm: Space-Shared Grid Resource Event Handler ()*

1. Wait for event and Identity Type of Event received
2. If it external and Gridlet arrival event, then:
   BEGIN /* a new job arrived */
   - If the number of Gridlets in execution are less than the number of PEs in the resource, then, Allocate_PE_to_the_Gridlet() /* It should schedule an Gridlet completion event */
   - If not, Add Gridlet to the Gridlet_Submitted_Queue

   END
3. If event is internal and its tag value is the same recently scheduled internal event tag,
   BEGIN /* a job finish event */
   - Update finished Gridlet's PE and Wall clock time parameters and send it back to the broker
   - Set the status of PE to FREE
   - Remove finished Gridlet from the Execution_Set and add to Finished_Set
   - If Gridlet_Submitted_Queue has Gridlets in waiting, then
     Choose the Gridlet to be Processed() /* e.g., first one in Q if FCFS policy is used */
     Allocate_PE_to_the_Gridlet() /* It should schedule an Gridlet completion event */

   END
4. Repeat the above steps until the end of simulation event is received

**Figure 5.9: An event handler for simulating space-shared resource scheduling.**

*Algorithm: Allocate_PE_to_the_Gridlet(Gridlet gl)*

1. Identify a suitable Machine with Free PE
2. Identify a suitable PE in the machine and Assign to the Gridlet
3. Set Status of the Allocated PE to BUSY
4. Determine the Completion Time of Gridlet and Set an internal event to be delivered at the completion time

**Figure 5.10: PE allocation to the Gridlets in space-shared GridSim resource.**

Figure 5.11 illustrates the simulation of space-share scheduling algorithm and Gridlets' execution. When Gridlet1 arrives at time 0, it is mapped to PE1 and an internal event to be delivered at the time 10 is scheduled since the predicted completion time is still 10. At time 4, Gridlet2 arrives and it is mapped to the PE2. The completion time of Gridlet2 is predicted as 12.5 and the completion time of Gridlet1 is still 10 since both of them are executing on different PEs. A new internal event to be delivered at time 12.5 is scheduled to signify the completion of Gridlet2. At time 7, Gridlet3 arrives. Since there is no free PE available on the resource, it is put into the queue. The simulation continues i.e., GridSim resource waits for the arrival of a new event. At time 10 a new event is delivered which happens to signify the completion of Gridlet1, which is then sent back to the broker. It then checks to see if there are any Gridlets waiting in the queue and chooses a suitable Gridlet (in this case as Gridlet2 is based on FCFS policy) and assign the available PE to it. An internal event to be delivered at time 19.5 is scheduled to indicate the completion time of Gridlet3 and then waits for the arrival of new events. A new event is delivered at the simulation time 12.5, which signifies the completion of the Gridlet2, which is then sent back to the broker. There is no Gridlet waiting in the queue, so it proceeds without scheduling any events and waits for the arrival of the next event. A new internal event arrives at the simulation time 19.5, which signifies the completion of Gridlet3. This process continues until resources receive an external event indicating the termination of simulation. A schematic representation of Gridlets arrival, internal events delivery, and sending them back to the broker is shown in Figure 5.5. A detailed statistical data on the arrival, execution start, finish, and elapsed time of all Gridlets is shown in Table 5.1.

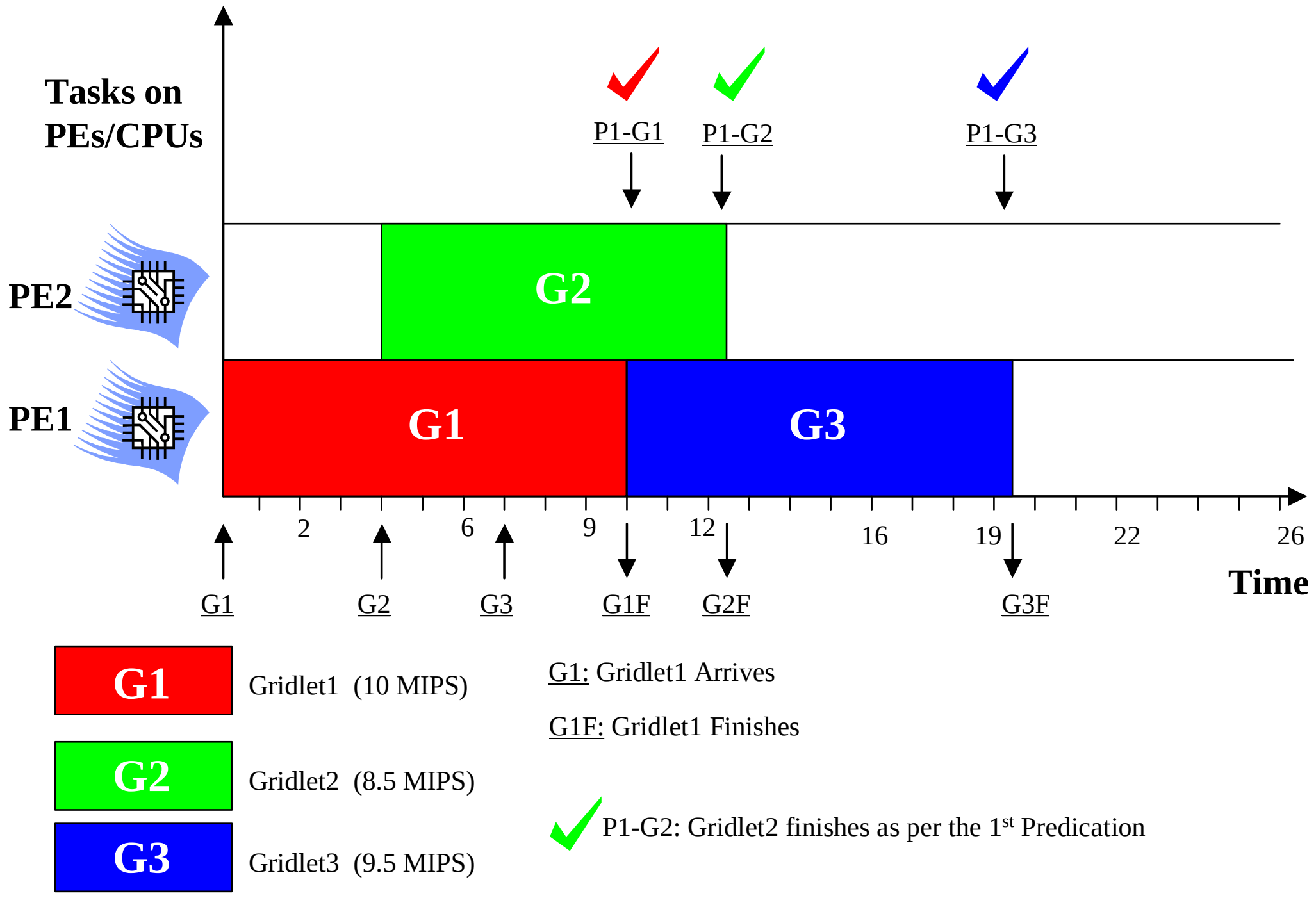

**Figure 5.11: Modeling space-shared multiprocessing based on an event scheme.**

For every Grid resource, the non-Grid (local) workload is estimated based on typically observed load conditions depending on the time zone of the resource. The network communication speed between a user and the resources is defined in terms of a data transfer speed (baud rate).

### 5.3.8 GridSim Java Package Design

A class diagram hierarchy of the *gridsim* package, represented using unified modeling language (UML) notation, is shown in Figure 5.12. The specification of each class contains up to three parts: attributes, methods, and internal classes. In the class diagram, attributes and methods are prefixed with characters "+", "-", and "#" indicating access modifiers public, private, and protected respectively. The gridsim package implements the following classes:

`class gridsim.Input` – This class extends the eduni.simjava.Sim_entity class. This class defines a port through which a simulation entity receives data from the simulated network. It maintains an event queue to serialize the data-in-flow and delivers to its parent entity. Simultaneous inputs can be modeled using multiple instances of this class.

`class gridsim.Output` – This class is very similar to the *gridsim.Input* class and it defines a port through which a simulation entity sends data to the simulated network. It maintains an event queue to serialize the data-out-flow and delivers to the destination entity. Simultaneous outputs can be modeled by using multiple instances of this class.

`class gridsim.GridSim` – This is the main class of Gridsim package that must be extended by GridSim entities. It inherits event management and threaded entity features from the eduni.simjava.Sim_entity class. The GridSim class adds networking and event delivery features, which allows synchronous or asynchronous communication for service access or delivery. All classes that extend the GridSim class must implement a method called "body()", which is automatically invoked since it is expected to be responsible for simulating entity behavior. The entities that extend the GridSim class can be instantiated with or without networked I/O ports. A networked GridSim entity gains communication capability via the objects of GridSim's I/O entity classes gridsim.Input and gridsim.Output classes. Each I/O entity will have a unique name assuming each GridSim entity that the user creates has unique name. For example, a resource entity with name "Resource2" will have an input entity whose name is prefixed with "Input_", making input entity full name as "Input_Resource2", which is expected to be unique. The I/O entities are concurrent entities, but they are visible within GridSim entity and are able to communicate with other GridSim entities by sending messages.

The GridSim class supports methods for simulation initialization, management, and flow control. The GridSim environment must be initialized to setup simulation environment before creating any other GridSim entities at the user level. This method also prepares the system for simulation by creating three GridSim internal entities—GridInformationService, GridSimShutdown, and GridStatistics. As explained in Section 5.3.2, the GridInformationService entity simulates the directory that dynamically keeps a list of resources available in the Grid. The GridSimShutdown entity helps in wrapping up a simulation by systematically closing all the opened GridSim entities. The GridStatistics entity provides standard services during the simulation to accumulate statistical data. Invoking the GridSim.Start () method starts the Grid simulation. All the resource and user entities must be instantiated in between invoking the above two methods.

The GridSim class supports static methods for sending and receiving messages between entities directly or via network entities, managing and accessing handle to various GridSim core entities, and recording statistics.

`class gridsim.PE` – It is used to represent CPU/*Processing Element* (PE) whose capability is defined in terms of MIPS rating.

`class gridsim.PEList` – It maintains a list of PEs that make up a machine.

`class gridsim.Machine` – It represents a uniprocessor or shared memory multiprocessor machine.

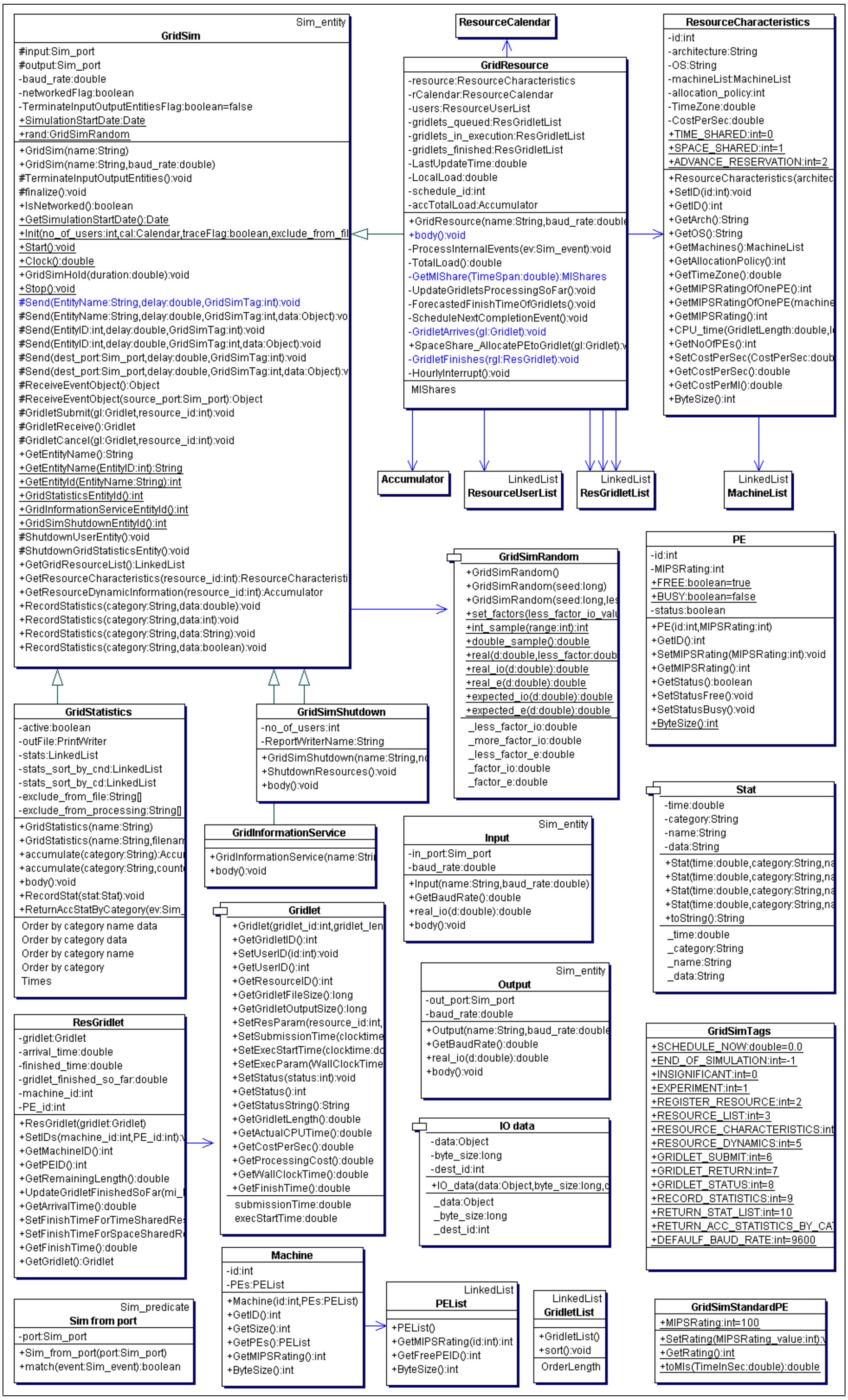

**Figure 5.12: A class hierarchy diagram of GridSim package.**

`class gridsim.MachineList` – An instance of this class simulates a collection of machines. It is up to the GridSim users to define the connectivity among the machines in a collection. Therefore, this class can be instantiated to model simple LAN to cluster to WAN.

`class gridsim.ResourceCharacteristics` –It represents static properties of a resource such as resource architecture, OS, management policy (time or space shared), cost, and time zone at which the resource is located along resource configuration.

`class gridsim.GridResource` – It extends the GridSim class and gains communication and concurrent entity capability. An instance of this class simulates a resource with properties defined in an object of the `gridsim.ResourceCharacteristics` class. The process of creating a Grid resource is as follows: first create PE objects with a suitable MIPS/SPEC rating, second assemble them together to create a machine. Finally, group one or more objects of the Machine to form a resource. A resource having a single machine with one or more PEs is managed as a time-shared system using round robin scheduling algorithm. A resource with multiple machines is treated as a distributed memory cluster and is managed as a space-shared system using first-come first served scheduling policy or its variants.

`class gridsim.GridSimStandardPE` – It defines MIPS rating for a standard PE or enables the users to define their own MIPS/SPEC rating for a standard PE. This value can be used for creating PEs with relative MIPS/SPEC rating for GridSim resources and creating Gridlets with relative processing requirements.

`class gridsim.ResourceCalendar` – This class implements a mechanism to support modeling local load on Grid resources that may vary according to the time zone, time, weekends, and holidays.

`class gridsim.GridInformationService` – A GridSim entity that provides Grid resource registration, indexing and discovery services. The Grid resources register their readiness to process Gridlets by registering themselves with this entity. GridSim entities such as the resource broker can contact this entity for resource discovery service, which returns a list of registered resource entities and their contact address. For example, scheduling entities use this service for resource discovery.

`class gridsim.Gridlet` – This class acts as job package that contains job length in MI, the length of input and out data in bytes, execution start and end time, and the originator of job. Individual users model their application by creating Gridlets for processing them on Grid resources assigned by scheduling entities (resource brokers).

`class gridsim.GridletList` – It can be used to maintain a list of Gridlets and supports methods for organizing them.

`class gridsim.GridSimTags` – It contains various static command tags that indicate a type of action that needs to be undertaken by GridSim entities when they receive events. The different types of tags supported in GridSim along with comments indicating possible purpose are shown in Figure 5.13.

```
public class GridSimTags {
  public static final double SCHEDULE_NOW = 0.0;  // 0.0 indicates NO delay
  public static final int END_OF_SIMULATION = -1;
  public static final int INSIGNIFICANT = 0; // ignore tag
  public static final int EXPERIMENT = 1; // User <-> Broker
  public static final int REGISTER_RESOURCE = 2; // GIS -> ResourceEntity
  public static final int RESOURCE_LIST = 3; // GIS <-> Broker
  public static final int RESOURCE_CHARACTERISTICS = 4; // Broker <-> ResourceEntity
  public static final int RESOURCE_DYNAMICS = 5; // Broker <-> ResourceEntity
  public static final int GRIDLET_SUBMIT = 6; // Broker -> ResourceEntity
  public static final int GRIDLET_RETURN = 7; // Broker <- ResourceEntity
  public static final int GRIDLET_STATUS = 8; // Broker <-> ResourceEntity
  public static final int RECORD_STATISTICS = 9; // Entity -> GridStatistics
  public static final int RETURN_STAT_LIST = 10; // Entity <- GridStatistics
  public static final int RETURN_ACC_STATISTICS_BY_CATEGORY = 11;
  public static final int DEFAULF_BAUD_RATE = 9600; // Default Baud Rate for entities
}
```

**Figure 5.13: Global tags in GridSim package.**

`class gridsim.ResGridlet` – It represents a Gridlet submitted to the resource for processing. It contains Gridlet object along with its arrival time and the ID of machine and PE allocated to it. It acts as a placeholder for maintaining the amount of resource share allocated at various times for simulating time-shared scheduling using internal events.

`class gridsim.GridStatistics` – This is a GridSim entity that records statistical data reported by other entities. It stores data objects with their label and timestamp. At the end of simulation, the user-defined report-writer entity can query recorded statistics of interest for report generation.

`class gridsim.Accumulator` – The objects of this class provide a placeholder for maintaining statistical values of a series of data added to it. It can be queried for mean, sum, standard deviation, and the largest and smallest values in the data series.

`class gridsim.GridSimShutdown` – This is a GridSim entity that waits for termination of all User entities to determine the end of simulation. It then signals the user-defined report-writer entity to interact with GridStatistics entity to generate report. Finally, it signals the end of simulation to other GridSim core entities.

`class gridsim.GridSimRandom` – This class provides static methods for incorporating randomness in data used for any simulation. Any predicted/estimated data, e.g., number of Gridlets used by an experiment, execution time and output size of a Gridlet etc., need to be mapped to real-world data by introducing randomness to reflect the uncertainty that is present in the prediction/estimation process and the randomness that exists in the nature itself. The execution time of a Gridlet on a particular resource, for example, can vary depending on the local load, which is not covered by the scope of GridSim to simulate.

The real($d$, $f_L$, $f_M$) method of this class maps the predicted/estimated value $d$ to a random real-world value between $(1- f_L)\times d$ to $(1+ f_M)\times d$, using the formula $d\times(1 - f_L + (f_L + f_M)\times rd)$ where $0.0 \le f_L, f_M \le 1.0$ and $rd$ is a uniformly distributed `double` value between 0.0 and 1.0. This class also maintains different values of $f_L$ and $f_M$ factors for different situations to represent different level of uncertainty involved.

## 5.4 Building Simulations with GridSim

To simulate Grid resource brokers using the GridSim toolkit, the developers need to create new entities that exhibit the behavior of Grid users and scheduling systems. The user-defined entities extend the GridSim base class to inherit the properties of concurrent entities capable of communicating with other entities using events. The high-level steps involved in modeling resources and applications, and simulating brokers using the GridSim toolkit are discussed below. The simulation of a Nimrod-G like economic Grid broker and evaluation of deadline and budget constrained scheduling algorithms are presented in the next chapter.

In this section we present a recipe for simulating application scheduling, with sample code clips, to demonstrate how GridSim can be used to simulate a Grid environment to evaluate schedulers:

- First, we need to create Grid resources of different capability and configuration (a single or multiprocessor with time/space-shared resource manager) similar to those used in application scheduling on the World-Wide Grid (WWG) testbed. We also need to create users with different requirements (application and quality of service requirements). A sample code for creating a Grid environment is given in Figure 5.14.

```
public static void CreateSampleGridEnvironement(int no_of_users, int no_of_resources,
  double B_factor, double D_factor, int policy, double how_long, double seed) {
  Calendar now = Calendar.getInstance();

  String ReportWriterName = "MyReportWriter";
  GridSim.Init(no_of_users, calender, true, eff, efp, ReportWriterName);

  String[] category = {"*.USER.TimeUtilization", "*.USER.GridletCompletionFactor",
"*.USER.BudgetUtilization"};

// Create Report Writer Entity and category indicates types of information to be recorded.
  new ReportWriter(ReportWriterName, no_of_users, no_of_resources, ReportFile, category,
report_on_next_row_flag);

  // Create Resources
  for(int i=0; i<no_of_resources; i++) {
     // Create PEs
     PEList peList = new PEList();
     for(int j=0; j<(i*1+1); j++)
        peList.add(new PE(0, 100));

     // Create machine list
     MachineList mList = new MachineList();
     mList.add(new Machine(0, peList));

     // Create a resource containing machines
     ResourceCharacteristics resource = new ResourceCharacteristics("INTEL", "Linux",
         mList, ResourceCharacteristics.TIME_SHARED, 0.0, i*0.5+1.0);
     LinkedList Weekends = new LinkedList();
     Weekends.add(new Integer(Calendar.SATURDAY));
     Weekends.add(new Integer(Calendar.SUNDAY));
     LinkedList Holidays = new LinkedList(); // no holiday is set!

     // Setup resource as simulated entity with a name (e.g. "Resource_1").
     new GridResource("Resource_"+i, 28000.0, seed, resource,
                                       0.0, 0.0, 0.0, Weekends, Holidays);
  }
  Random r = new Random(seed);
  // Create Application, Experiment, and Users
  for(int i=0; i<no_of_users; i++)
  {
    Random r = new Random(seed*997*(1+i)+1);
    GridletList glList = Application1(r);  // it creates Gridlets and returns their list
    Experiment expt = new Experiment(0, glList, policy, true, B_factor, D_factor);
    new UserEntity("U"+i, expt, 28000.0, how_long, seed*997*(1+i)+1, i, user_entity_report);
  }
  // Perform Simulation
  GridSim.Start();
}
```

**Figure 5.14: A sample code segment for creating Grid resource and user entities in GridSim.**

- Second, we need to model applications by creating a number of Gridlets (that appear similar to Nimrod-G jobs) and define all parameters associated with jobs as shown in Figure 5.15. The Gridlets need to be grouped together depending on the application model.

```
Gridlet gl = new Gridlet(Gridlet_id, Gridlet_length, GridletFileSize,
                       GridletOutputSize);
```

**Figure 5.15: The Gridlet method in GridSim.**

- Then, we need to create a GridSim User entity that creates and interacts with the resource broker scheduling entity to coordinate execution experiment. It can also directly interact with GIS and resource entities for Grid information and submitting or receiving processed Gridlets, however, for modularity sake, we encourage the implementation of a separate resource broker entity by extending the GridSim class.
- Finally, we need to implement a resource broker entity that performs application scheduling on Grid resources. A sample code for implementing the broker is shown in Figure 5.16. First, it accesses the Grid Information Service (GIS), and then inquires the resource for its capability including cost. Depending on processing requirements, it develops schedule for assigning Gridlets to resources and coordinates the execution. The scheduling policies can be systems-centric like those implemented in many Grid systems such as Condor or user-centric like the Nimrod-G broker's quality of service (QoS) driven application scheduling algorithms [107].

```
class Broker extends GridSim {
  private Experiment experiment;
  private LinkedList ResIDList;
  private LinkedList BrokerResourceList;

  public Broker(String name, double baud_rate)
  {
    super(name, baud_rate);
    GridletDispatched = 0;
    GridletReturned = 0;
    Expenses = 0.0;
    MaxGridletPerPE = 2;
  }

  ... // Gridlet scheduling flow code at the Grid Resource Broker level

  public void body() {

    Sim_event ev = new Sim_event();
    // Accept User Commands and Process
    for( sim_get_next(ev); ev.get_tag()!=GridSimTags.END_OF_SIMULATION; sim_get_next(ev))
    {
      experiment = (Experiment) ev.get_data();
      int UserEntityID = ev.get_src();

      // Record Experiment Start Time.
      experiment.SetStartTime();

     // Set Gridlets' OwnerID as this BrokerID so that Resources knows where to return them.
      for(int i=0; i<experiment.GetGridletList().size(); i++)
        ((Gridlet) experiment.GetGridletList().get(i)).SetUserID(get_id());

      // RESOURCE DISCOVERY
      ResIDList = (LinkedList) GetGridResourceList();

      // RESOURCE TRADING and SORTING
      // SCHEDULING
      while (glFinishedList.size() < experiment.GetGridletList().size())
      {
        if((GridSim.Clock()>=experiment.GetDeadline())||(Expenses>=experiment.GetBudget()) )
          break;

        scheduled_count = ScheduleAdviser();
        dispatched_count = Dispatcher();
        received_count = Receiver();

        // Heurisitics for deciding hold condition
        if(dispatched<=0 && received<=0 && glUnfinishedList.size()>0)
        {
          double deadline_left = experiment.GetDeadline()-GridSim.Clock();
          GridSimHold(Math.max(deadline_left*0.01, 1.0));
        }
      }
    }
  ...  // Code for actual scheduling policy
  ...  // Code for dispatch policy
  }
 }
```

**Figure 5.16: A sample code segment for creating a Grid resource broker in GridSim.**

## 5.5 Summary and Comments

We discussed an object-oriented toolkit, called GridSim, for distributed resource modeling and scheduling simulation. GridSim simulates time- and space-shared resources with different capabilities, time zones, and configurations. It supports different application models that can be mapped to resources for execution by developing simulated application schedulers. We have discussed the architecture and components of the GridSim toolkit along with steps involved in creating GridSim based application-scheduling simulators.

The implementation of GridSim toolkit in Java is an important contribution since Java provides a rich set of tools that enhance programming productivity, application portability, and a scalable runtime environment. As the JVM (Java Virtual Machine) is available for single, multiprocessor shared or distributed machines such as clusters, GridSim scales with them due to its concurrent implementation. Also, we were able to leverage the existing basic discrete-event infrastructure from SimJava while implementing the GridSim toolkit.

We have used the GridSim toolkit to develop a Nimrod-G like economic Grid resource broker simulator and evaluated the performance of a number of scheduling algorithms based on deadline and budget based constraints (see the next chapter). The results are promising and demonstrate the suitability of GridSim for developing simulators for scheduling in parallel and distributed systems. Furthermore, GridSim is gaining rapid acceptance as a tool for simulation and performance evaluation of computational Grids and Grid schedulers. It is in use at several academic institutions and commercial enterprises all over the world including California Institute of Technology, University of Southern California, University of Illinois at Urbana-Champaign, San Diego Supercomputing Centre, Carnegie Melon University, University of Adelaide, Manchester University, CERN (European Organisation for Nuclear Research), University of Paderborn, Hong Kong University, National University of Singapore, Sun Microsystems, C-DOT (Centre for Development of Telematics), IBM (International Business Machines), Unisys, HP (Hewlett and Packard), Compaq, British Telecom, and WorldCom.

## Software Availability

The GridSim toolkit software with source code can be downloaded from the project website:

`http://www.buyya.com/gridsim/`

# Chapter 6

# Scheduling Simulations

This chapter presents a performance evaluation of economic-based Grid resource management and scheduling. The GridSim toolkit is used to develop an economic Grid resource broker that supports the deadline and budget constrained (DBC) scheduling strategies and to quantify the broker's ability to dynamically select resources at runtime depending on their availability, capability, cost, and user quality of service requirements (QoS). The broker supports DBC algorithms with the four different optimisation strategies—cost, time, cost-time, and conservative time. The detailed performance evaluation of economic-driven scheduling algorithms is carried out through a series of simulations by varying the number of users, deadline, budget, and optimisation strategies and simulating geographically distributed Grid resources.

## 6.1 Economic Grid Resource Broker Simulation

We used the GridSim toolkit to simulate a Grid environment and a Nimrod-G like deadline and budget constrained scheduling system called economic Grid resource broker. The simulated Grid environment contains multiple resources and user entities with different requirements. The users create an experiment that contains an application specification (a set of Gridlets that represent application jobs with different processing) and quality of service requirements (deadline and budget constraints with optimization strategy). We created two entities that simulate users and the brokers by extending the GridSim class. When simulated, each user entity having its own application and quality of service requirements, creates its own instance of the broker entity for scheduling Gridlets on resources.

### 6.1.1 Broker Architecture

The broker entity architecture along with its interaction flow diagram with other entities is shown in Figure 6.1. The key components of the broker are: experiment interface, resource discovery and trading, scheduling flow manager backed with scheduling heuristics and algorithms, Gridlets dispatcher, and Gridlets receptor. The following high-level steps describe functionality of the broker components and their interaction:

1. The user entity creates an experiment that contains an application description (a list of Gridlets to be processed) and user requirements to the broker via the experiment interface.
2. The broker resource discovery and trading module interacts with the GridSim GIS entity to identify contact information of resources and then interacts with resources to establish their configuration and access cost. It creates a Broker Resource list that acts as placeholder for maintaining resource properties, a list of Gridlets committed for execution on the resource, and the resource performance data as predicted through the measurement and extrapolation methodology.
3. The scheduling flow manager selects an appropriate scheduling algorithm for mapping Gridlets to resources depending on the user requirements (deadline and budget limits; and optimisation strategy—cost, cost-time, time, or time variant). Gridlets that are mapped to a specific resource are added to the Gridlets list in the Broker Resource.
4. For each of the resources, the dispatcher selects the number of Gridlets that can be staged for execution according to the usage policy to avoid overloading resources with single user jobs.

5. The dispatcher then submits Gridlets to resources using the GridSim's asynchronous service.
6. When the Gridlet processing completes, the resource returns it to the broker's Gridlet receptor module, which then measures and updates the runtime parameter, *resource or MI share available to the user*. It aids in predicting the job consumption rate for making scheduling decisions.
7. The steps, 3–6, continue until all the Gridlets are processed or the broker exceeds deadline or budget limits. The broker then returns the updated experiment data along with processed Gridlets back to the user entity.

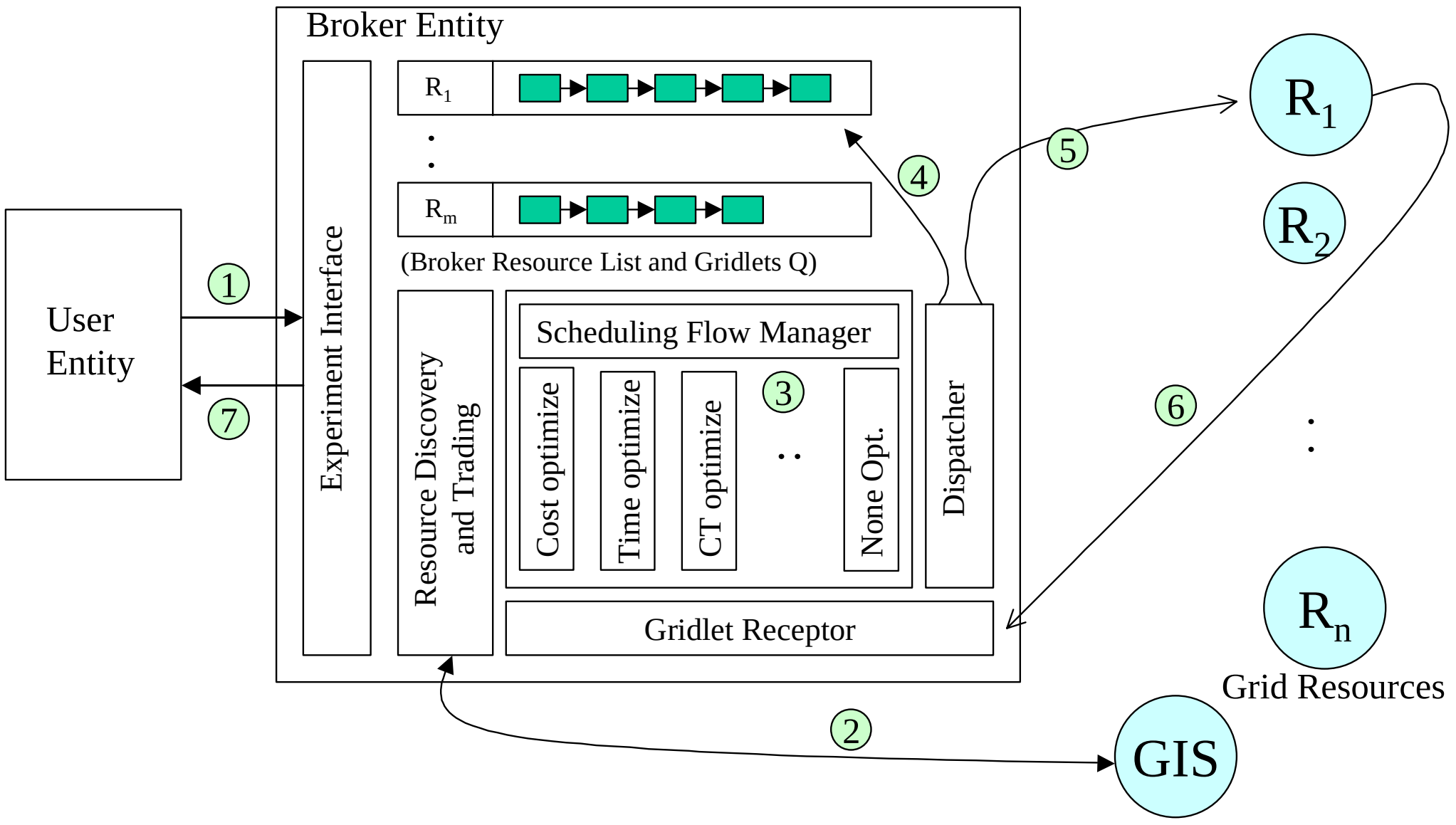

**Figure 6.1: Economic Grid resource broker architecture and its interaction with other entities.**

A class diagram hierarchy of the Grid broker package built using the GridSim toolkit is shown in Figure 6.2. The Grid broker package implements the following key classes:

`class Experiment` – It acts as a placeholder for representing simulation experiment configuration that includes synthesized application (a set of Gridlets stored in GridletList) and user requirements such as D and B-factors or deadline and budget constraints, and optimization strategy. It provides methods for updating and querying the experiment parameters and status. The user entity invokes the broker entity and passes its requirements via experiment object. On receiving an experiment from its user, the broker schedules Gridlets according to the optimization policy set for the experiment.

`class UserEntity` – A GridSim entity that simulates the user. It invokes the broker and passes the user requirements. When it receives the results of application processing, it records parameters of interest with the gridsim.Statistics entity. When it has no more processing requirements, it sends `END_OF_SIMULATION` event to the broker and gridsim.GridSimShutdown entities.

`class Broker` – A GridSim entity that simulates the Grid resource broker. On receiving an experiment from the user entity, it does resource discovery, and determines deadline and budget values based on D and B factors, and then proceeds with scheduling. It schedules Gridlets on resources depending on user constraints, optimization strategy, and cost of resources and their availability. When it receives the results of application processing, it records parameters of interest with the gridsim.Statistics entity. When it has no more processing requirements, it sends `END_OF_SIMULATION` event to the gridsim.GridSimShutdown entity.

`class BrokerResource` – It acts as placeholder for the broker to maintain a detailed record of the resources it uses for processing user application. It maintains resource characteristics, a list of Gridlets assigned to the resource, the actual amount of MIPS available to the user, and a report on the Gridlets processed. These measurements help in extrapolating and predicting the resource performance from the user point of view and aid in scheduling jobs dynamically at runtime.

`class ReportWriter` – A user-defined, optional GridSim entity which is meant for creating a report at the end of each simulation by interacting with the gridsim.Statistics entity. If the user does not want to create a report, then it can pass "`null`" as the name of the ReportWriter entity. Note that the users can choose any name for the ReportWriter entity and for the class name since all entities are identified by their name defined at the runtime.

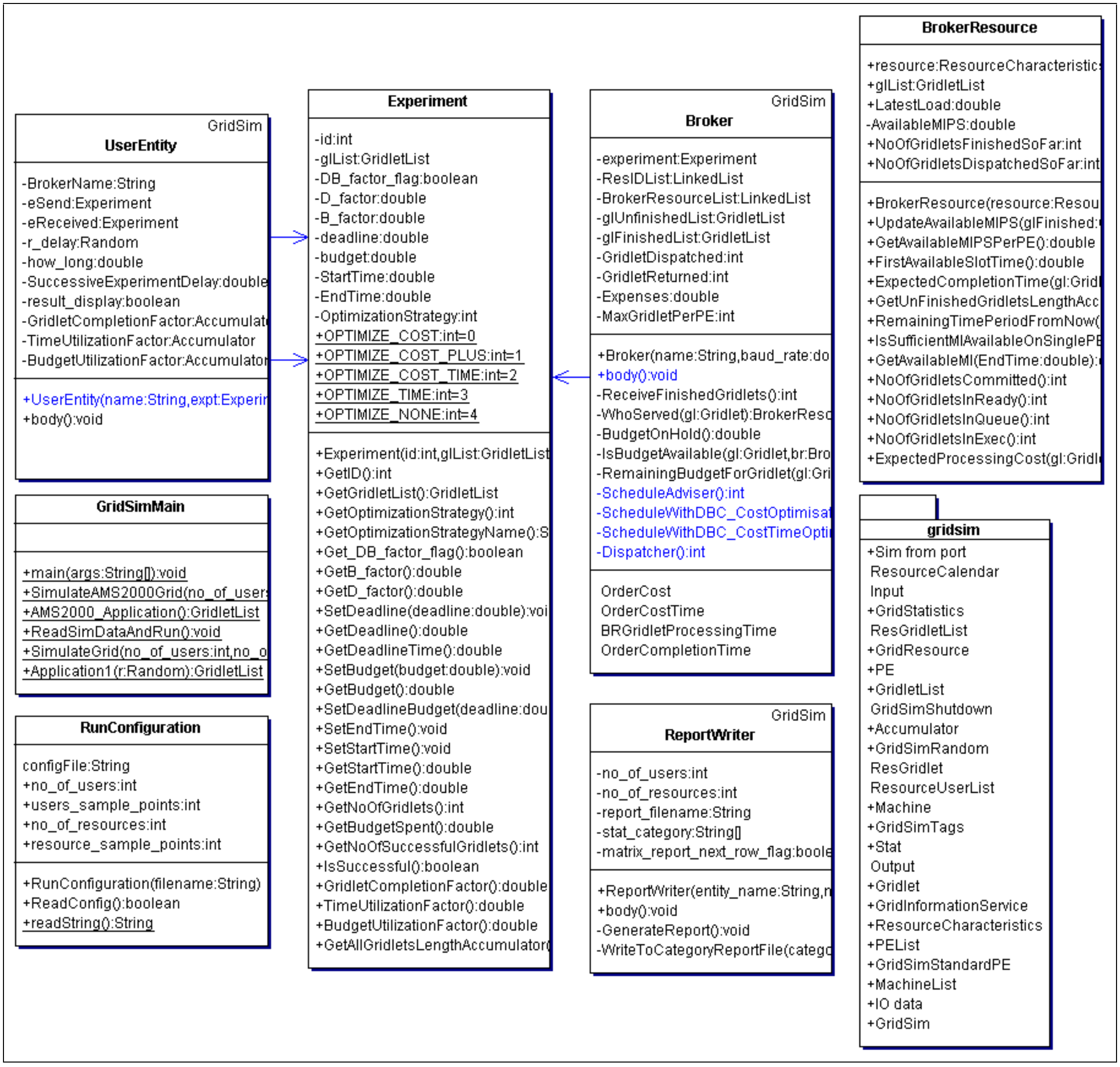

**Figure 6.2: A class hierarchy diagram of Grid broker using the gridsim package.**

An interactive class hierarchy diagram of the economic Grid resource broker (accessible from [105]) provides syntax and semantic information of data members and methods of each class discussed above.

### 6.1.2 Determining the Deadline and Budget

A D-factor close to 1 signifies the user's willingness to set a highly relaxed deadline, which is sufficient to process applications even when only the slowest resources are available. Similarly a B-factor close to 1 signifies that the user is willing to spend as much money as required even when only the most expensive resource is used. The jobs are scheduled on the Grid through user's broker. The broker uses these factors in determining the absolute deadline (see Equation 6.1) and budget (see Equation 6.2) values for a given execution scenario at runtime as follows:

***Determining the Absolute Deadline Value:***

$$Deadline = T_{MIN} + D_{FACTOR} * (T_{MAX} - T_{MIN})$$

**Equation 6.1**

where,

- $T_{MIN}$ = the time required to process all the jobs, in parallel, giving the *fastest* resource the highest priority.
- $T_{MAX}$ = the time required to process all the jobs, serially, using the slowest resource.
- An application with $D_{FACTOR}$ < **0** would **never** be completed.
- An application with $D_{FACTOR}$ $\geq$ **1** would **always** be completed as long as some resources are available with minimal user-share throughout the deadline.

***Determining the Absolute Budget Value:***

$$1. \quad Budget = C_{MIN} + B_{FACTOR} * (C_{MAX} - C_{MIN}$$

**Equation 6.2**

where,

- $C_{MIN}$ = the cost of processing all the jobs, in parallel within deadline, giving the *cheapest* resource the highest priority.
- $C_{MAX}$ = the cost of processing all the jobs, in parallel within deadline, giving the *costliest* resource the highest priority.
- An application with $B_{FACTOR}$ < **0** would **never** be completed.
- An application with $B_{FACTOR}$ $\geq$ **1** would **always** be completed as long as some resources are available with minimal user-share throughout the deadline.

### 6.1.3 Scheduling Algorithms

We propose deadline and budget constrained (DBC) algorithms with four different optimisation strategies—cost optimisation, cost-time optimisation, time optimisation, and conservative-time optimisation—for scheduling task-farming applications on geographically distributed resources. The properties of DBC scheduling algorithms are shown in Table 6.1.

**Table 6.1: Deadline and budget constrained adaptive scheduling algorithms.**

| Algorithm/ Strategy | Execution Time (Deadline, D) | Execution Cost (Budget, B) |
|---|---|---|
| Cost Opt | Limited by D | Minimize |
| Cost-Time Opt | Minimize when possible | Minimize |
| Time Opt | Minimize | Limited by B |
| Conservative-Time Opt | Minimize | Limited by B, but all unprocessed jobs have guaranteed minimum budget |

The *cost-optimisation* scheduling algorithm uses the cheapest resources to ensure that the deadline can be met and the computational cost is minimized. The *time-optimisation* scheduling algorithm uses all the affordable resources to process jobs in parallel as early as possible. The *cost-time optimisation* scheduling is similar to cost optimisation, but if there are multiple resources with the same cost, it applies time optimisation strategy while scheduling jobs on them. The *conservative-time optimisation* scheduling strategy is similar to the time-optimisation scheduling strategy, but it guarantees that each unprocessed job has a minimum budget-per-job.

We have incorporated DBC cost, time, and conservative time optimisation scheduling algorithms into the Nimrod-G broker and explored their ability in scheduling parameter sweep applications on the World-Wide Grid (WWG) testbed. A detailed evaluation of DBC cost, time, and cost-time scheduling algorithms by simulation for various scenarios is presented in the next sections.

## 6.2 Simulation Experiment Setup

To simulate application scheduling in GridSim environment using the economic Grid broker requires the modeling and creation of GridSim resources and applications that model jobs as Gridlets. In this section, we present resource and application modeling along with the results of experiments with quality of services driven application processing.

**Table 6.2: WWG testbed resources simulated using GridSim.**

| Resource Name in Simulation | Simulated Resource Characteristics<br>Vendor, Resource Type, Node OS, No of PEs | Equivalent Resource in Worldwide Grid<br>(Hostname, Location) | A PE SPEC/ MIPS Rating | Resource Manager Type | Price (G$/PE time unit) | MIPS per G$ |
|---|---|---|---|---|---|---|
| R0 | Compaq, AlphaServer, CPU, OSF1, 4 | grendel.vpac.org, VPAC, Australia | 515 | Time-shared | 8 | 64.37 |
| R1 | Sun, Ultra, Solaris, 4 | hpc420.hpcc.jp, AIST, Tokyo, Japan | 377 | Time-shared | 4 | 94.25 |
| R2 | Sun, Ultra, Solaris, 4 | hpc420-1.hpcc.jp, AIST, Tokyo, Japan | 377 | Time-shared | 3 | 125.66 |
| R3 | Sun, Ultra, Solaris, 2 | hpc420-2.hpcc.jp, AIST, Tokyo, Japan | 377 | Time-shared | 3 | 125.66 |
| R4 | Intel, Pentium/VC820, Linux, 2 | barbera.cnuce.cnr.it, CNR, Pisa, Italy | 380 | Time-shared | 2 | 190.0 |
| R5 | SGI, Origin 3200, IRIX, 6 | onyx1.zib.de, ZIB, Berlin, Germany | 410 | Time-shared | 5 | 82.0 |
| R6 | SGI, Origin 3200, IRIX, 16 | Onyx3.zib.de, ZIB, Berlin, Germany | 410 | Time-shared | 5 | 82.0 |
| R7 | SGI, Origin 3200, IRIX, 16 | mat.ruk.cuni.cz, Charles U., Prague, Czech Republic | 410 | Space-shared | 4 | 102.5 |
| R8 | Intel, Pentium/VC820, Linux, 2 | marge.csm.port.ac.uk, Portsmouth, UK | 380 | Time-shared | 1 | 380.0 |
| R9 | SGI, Origin 3200, IRIX, 4 (accessible) | green.cfs.ac.uk, Manchester, UK | 410 | Time-shared | 6 | 68.33 |
| R10 | Sun, Ultra, Solaris, 8, | pitcairn.mcs.anl.gov, ANL, Chicago, USA | 377 | Time-shared | 3 | 125.66 |

### 6.2.1 Resource Modeling

We modeled and simulated a number of time- and space-shared resources with different characteristics, configuration, and capability as those in the WWG testbed. We have selected the latest CPUs models AlphaServer ES40, Sun Netra 20, Intel VC820 (800EB MHz, Pentium III), and SGI Origin 3200 1X 500MHz R14k released by their manufacturers Compaq, Sun, Intel, and SGI respectively. The processing capability of these PEs in simulation time-unit is modeled after the base value of SPEC CPU (INT) 2000 benchmark ratings published in [130]. To enable the users to model and express their application processing requirements in terms of MI (million instructions) or MIPS (million instructions per second), we *assume the MIPS rating of PEs is same as the SPEC rating*.

Table 6.2 shows characteristics of resources simulated and their PE cost per time unit in G$ (Grid dollar). These simulated resources resemble the WWG testbed resources used in processing a parameter sweep application using the Nimrod-G broker [102]. The PE cost in G$/unit time not necessarily reflects the cost of processing when PEs have different capability. The brokers need to translate it into the G$ per MI (million instructions) for each resource. Such translation helps in identifying the relative cost of resources for processing Gridlets on them.

### 6.2.2 Application Modeling

We have modeled a task farming application that consists of 200 jobs. In GridSim, these jobs are packaged as Gridlets whose contents include the job length in MI, the size of job input and output data in bytes along with various other execution related parameters when they move between the broker and resources. The job length is expressed in terms of the time it takes to run on a standard resource PE with SPEC/MIPS rating of 100. Gridlets processing time is expressed in such a way that they are expected to take at least 100 time-units with a random variation of 0 to 10% on the positive side of the standard resource. That means, Gridlets' job length (processing requirements) can be at least 10,000 MI with a random variation of 0 to 10% on the positive side. This 0 to 10% random variation in Gridlets' job length is introduced to model heterogeneous tasks similar to those present in the real world parameter sweep applications.

*Algorithm: DBC_Scheduling_with_Cost_Optimisation()*

1. RESOURCE DISCOVERY: Identify resources that can be used in this execution with their capability through the Grid Information Service.
2. RESOURCE TRADING: Identify cost of each of the resources in terms of CPU cost per second and capability to be delivered per cost-unit.
3. If the user supplies D and B factors, then determine the absolute deadline and budget based on the capability and cost of resources and user's requirements.
4. SORT resources by increasing order of cost.
5. SCHEDULING: Repeat while there exist unprocessed jobs in application job list with a delay of scheduling event period or occurrence of an event AND the time and process expenses are within deadline and budget limits:

[SCHEDULE ADVISOR with Policy]

   a. For each resource predict and establish the job consumption rate or the available resource share through measure and extrapolation.
   b. For each resource based on its job consumption rate or available resource share, predict and establish the number of jobs a resource can process by the deadline.
   c. For each resource in order:
      i. If the number of jobs currently assigned to a resource is less than the predicted number of jobs that a resource can consume, assign more jobs from unassigned job queue or from the most expensive machines based on job state and feasibility. Assign job to a resource only when there is enough budget available.
      ii. Alternatively, if a resource has more jobs than it can complete by the deadline, move those extra jobs to unassigned job queue.

6. [DISPATCHER with Policy]

*Repeat the following steps for each resource if it has jobs to be dispatched:*

- Identify the number of jobs that can be submitted without overloading the resource. Our default policy is to dispatch jobs as long as the number of user jobs deployed (active or in queue) is less than the number of PEs in the resource.

**Figure 6.3: Deadline and budget constrained (DBC) scheduling with cost-optimization.**

## 6.3 Deadline and Budget Constrained Cost Optimisation Scheduling

The steps for implementing DBC cost-optimisation scheduling algorithms within economic broker simulator are shown in Figure 6.3. This algorithm attempts to process jobs as economically as possible within the deadline and budget.

### 6.3.1 Scheduling Experiments with a Single User

In this experiment, we perform scheduling experiments with different values of deadline and budget constraints (DBC) for a single user. The deadline is varied in simulation time from 100 to 3600 in steps of 500. The budget is varied from G$ 5000 to 22000 in steps of 1000. For this scenario, we performed scheduling simulation for DBC *cost-optimization* algorithm. The number of Gridlets processed, deadline utilized, and budget spent for different scheduling scenario is shown in Figure 6.4–Figure 6.7. From Figure 6.4, it can be observed that for a tight deadline (e.g., 100 time unit), the number of Gridlets processed increased with the increase in budget value. Because, when a higher budget is available, the broker leases expensive resources to process more jobs within the deadline. Alternatively, when scheduling with a low budget value, the number of Gridlets processed increases as the deadline is relaxed (see Figure 6.5).

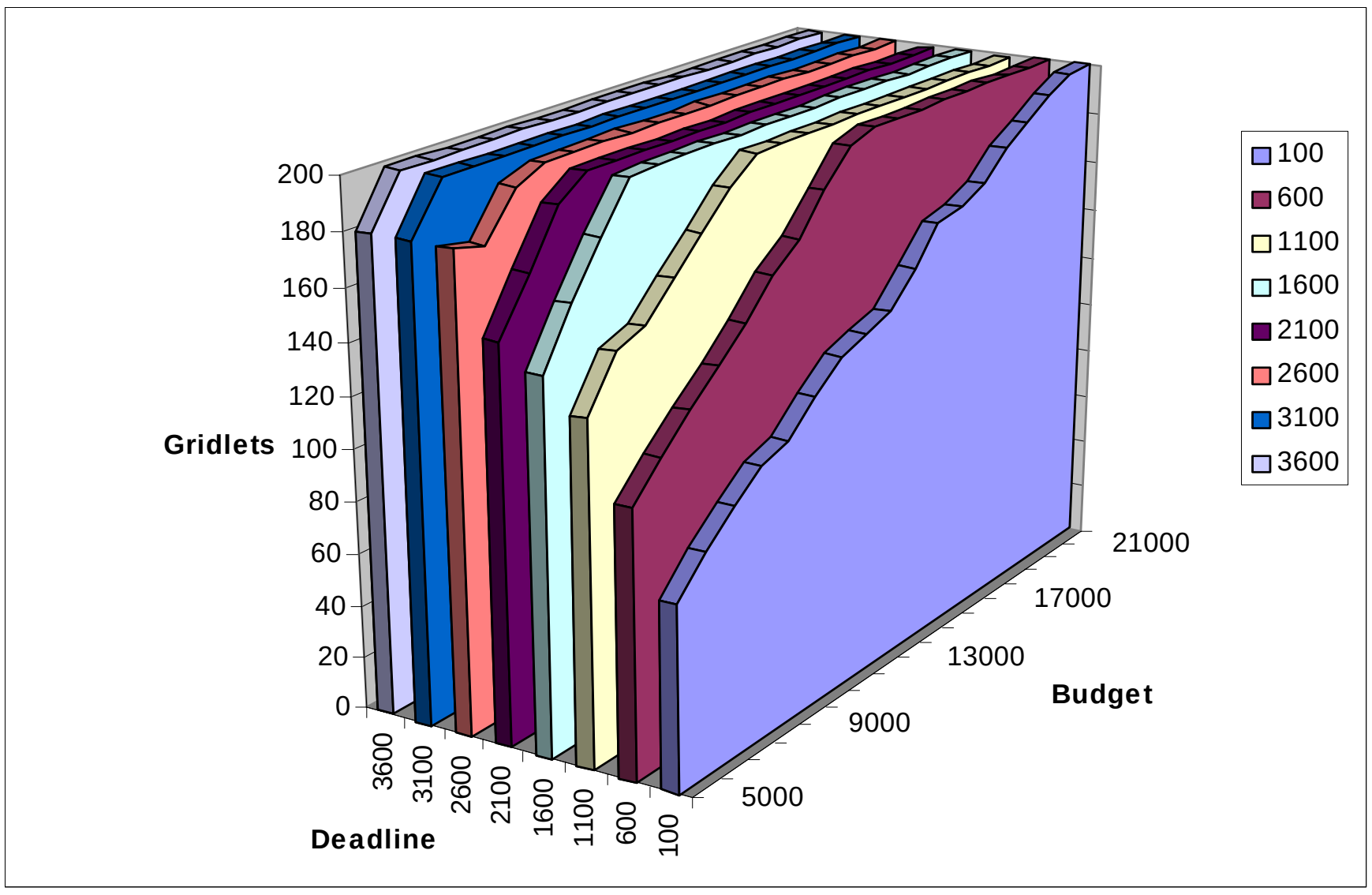

**Figure 6.4: No. of Gridlets processed for different budget limits with a fixed deadline for each.**

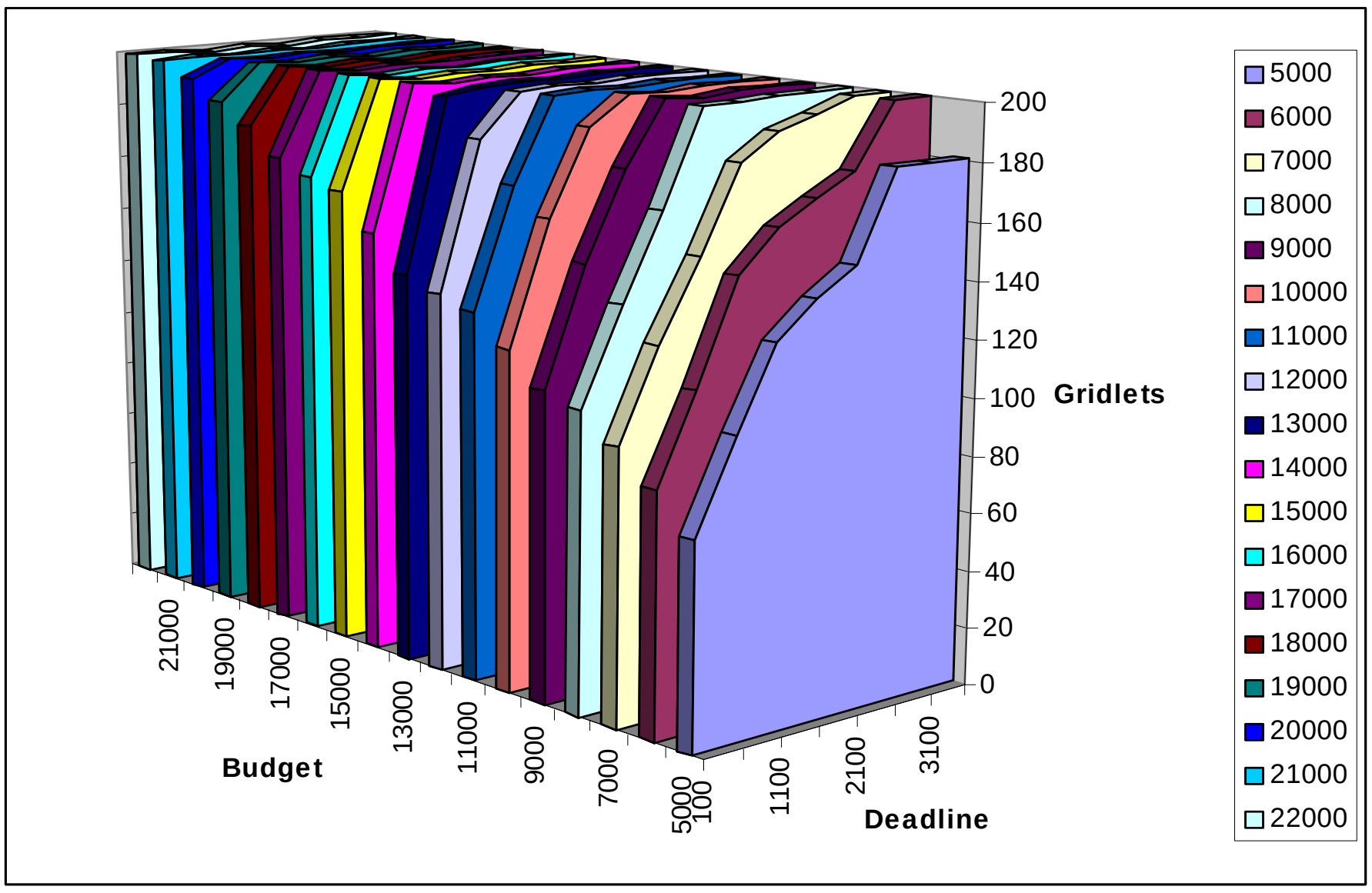

**Figure 6.5: No. of Gridlets processed for different deadline limits with a fixed budget for each.**

The impact of budget for different values of deadline is shown in Figure 6.6. In cost-optimization scheduling, for a larger deadline value (see time utilization for deadline of 3600), the increase in budget value does not have much impact on resource selection. This trend can also be observed from the budget spent for processing Gridlets with different deadline constraints (see Figure 6.7). When the deadline is too tight (e.g., 100), it is likely that the complete budget is spent for processing Gridlets within the deadline.

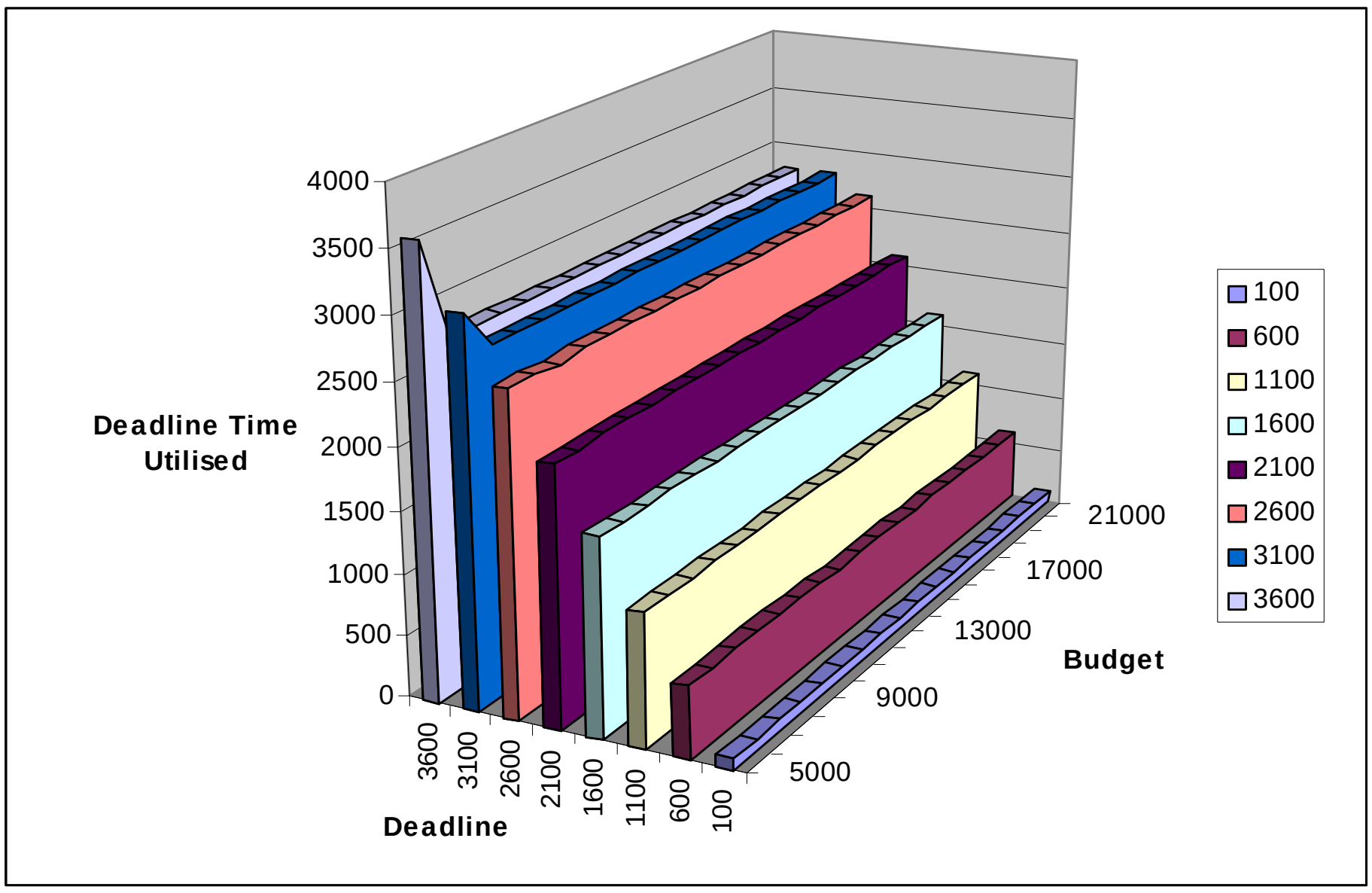

**Figure 6.6: Deadline time utilized for processing Gridlets for different values of deadline and budget.**

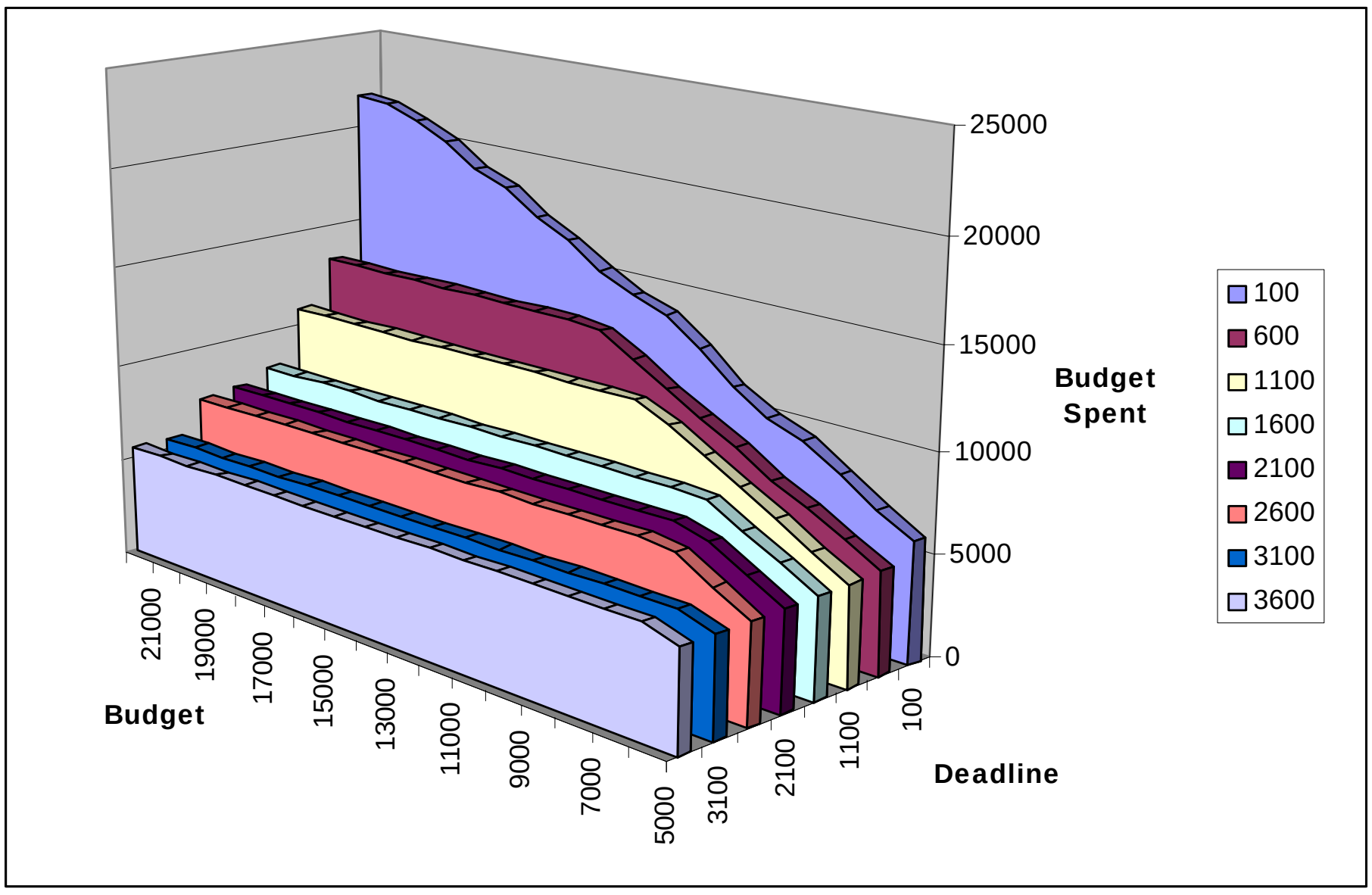

**Figure 6.7: Budget spent for processing Gridlets for different values of deadline and budget.**

Three diagrams (Figure 6.8–Figure 6.10) show the selection of resources for processing Gridlets for different budget values with a fixed deadline of 100, 1100, and 3100 (low, medium, and high deadline value) respectively. It can be observed that when the deadline is low, the economic broker also leases expensive resources to process Gridlets whenever the budget permits (see Figure 6.8). In this, all resources have been used depending on the budget availability. When the deadline is increased to a high value (a medium deadline of 1100), the broker processes as many Gridlets as possible on cheaper resources by the deadline (see Figure 6.9) and utilizes expensive resources if required. When the deadline is highly relaxed (a high deadline of 3100), the broker allocates Gridlets to the cheapest resource since it was able to process all Gridlets within this deadline (see Figure 6.10). In all three diagrams (Figure 6.8 –Figure 6.10), the left most solid curve marked with the label “All” in the resources axis represents the aggregation of all resources and shows the total number of Gridlets processed for the different budgets.

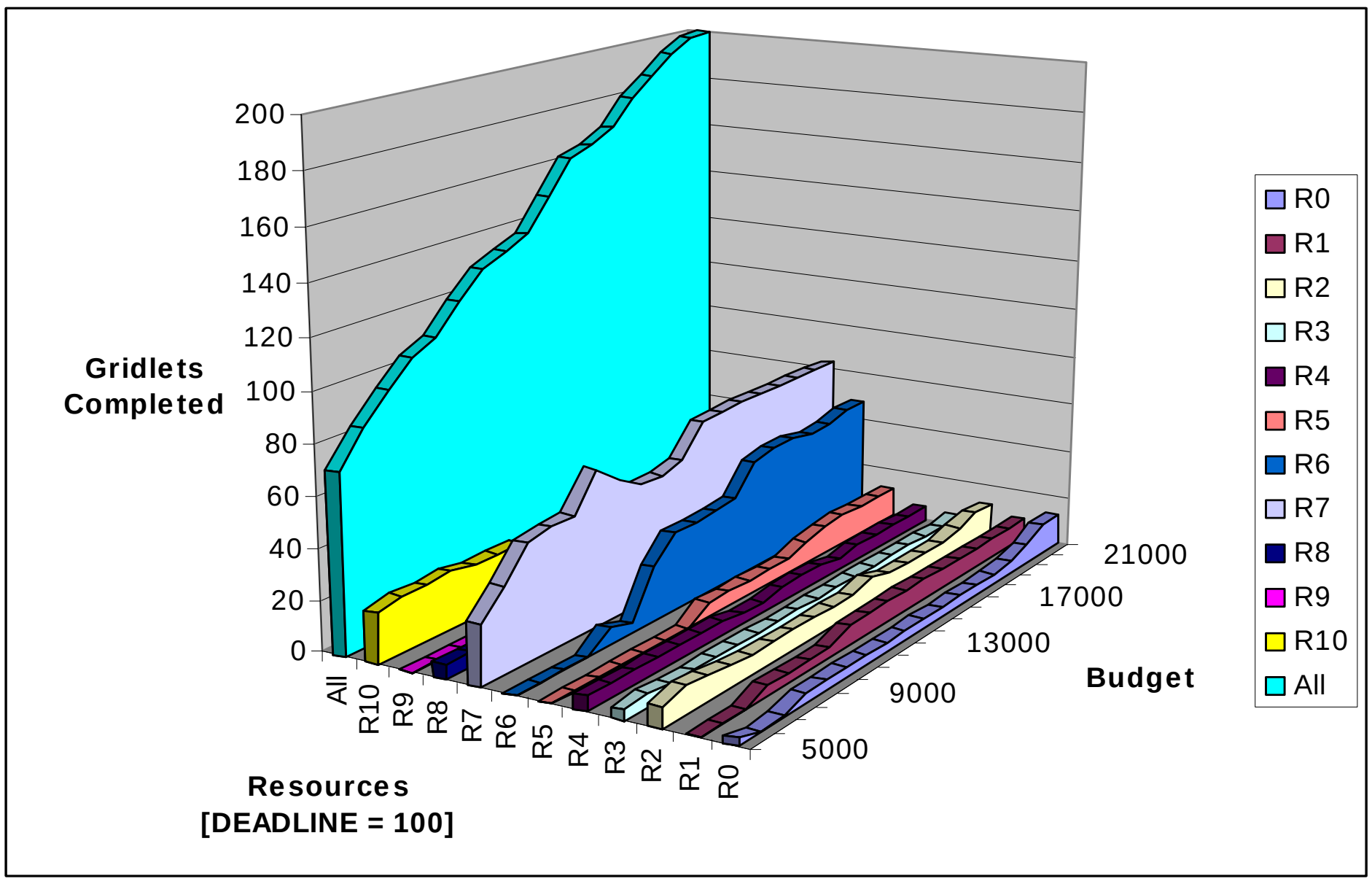

**Figure 6.8: Gridlets processed on resources for different budget values with low deadline.**

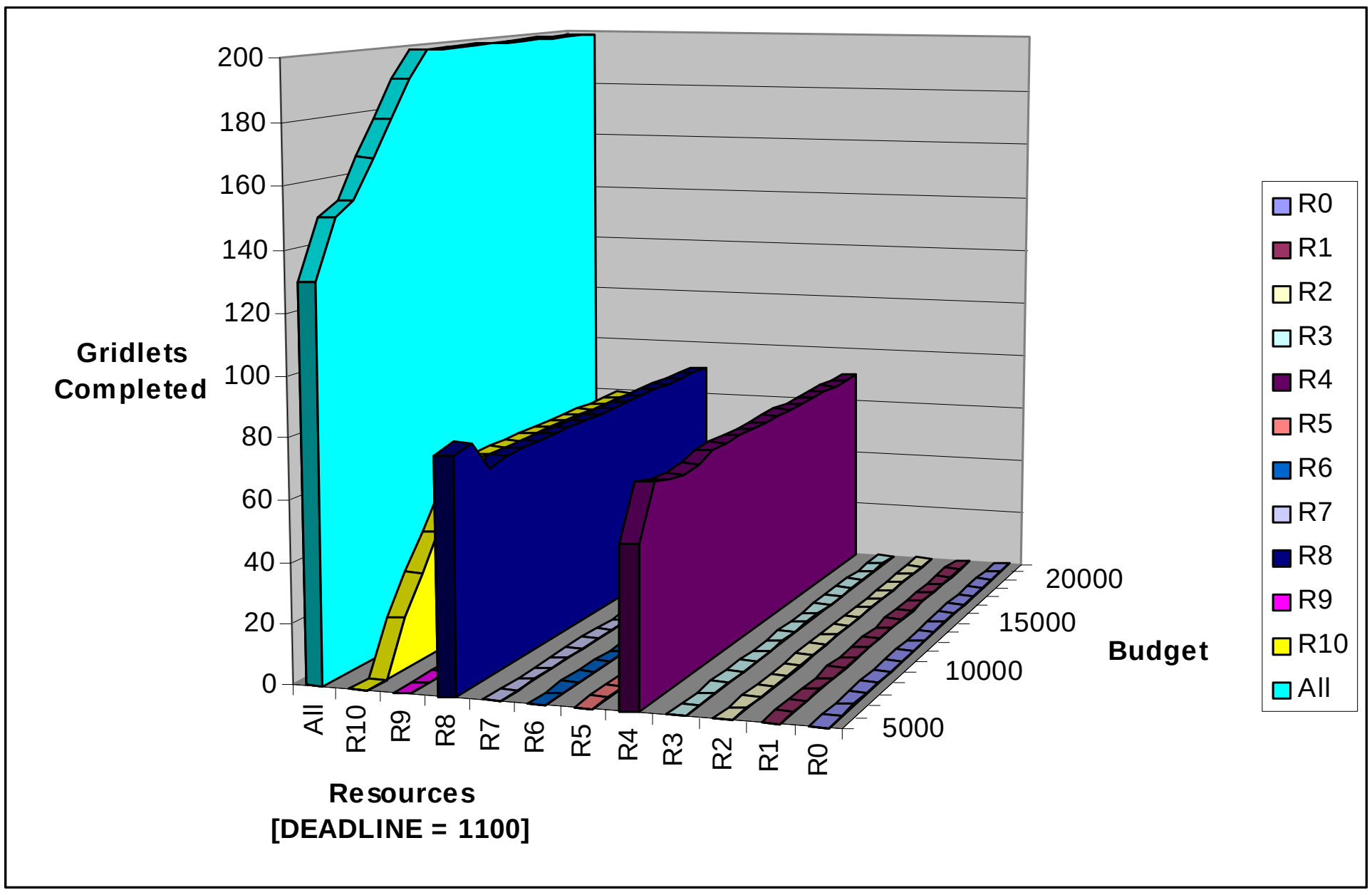

**Figure 6.9: Gridlets processed on resources for different budget values with medium deadline.**

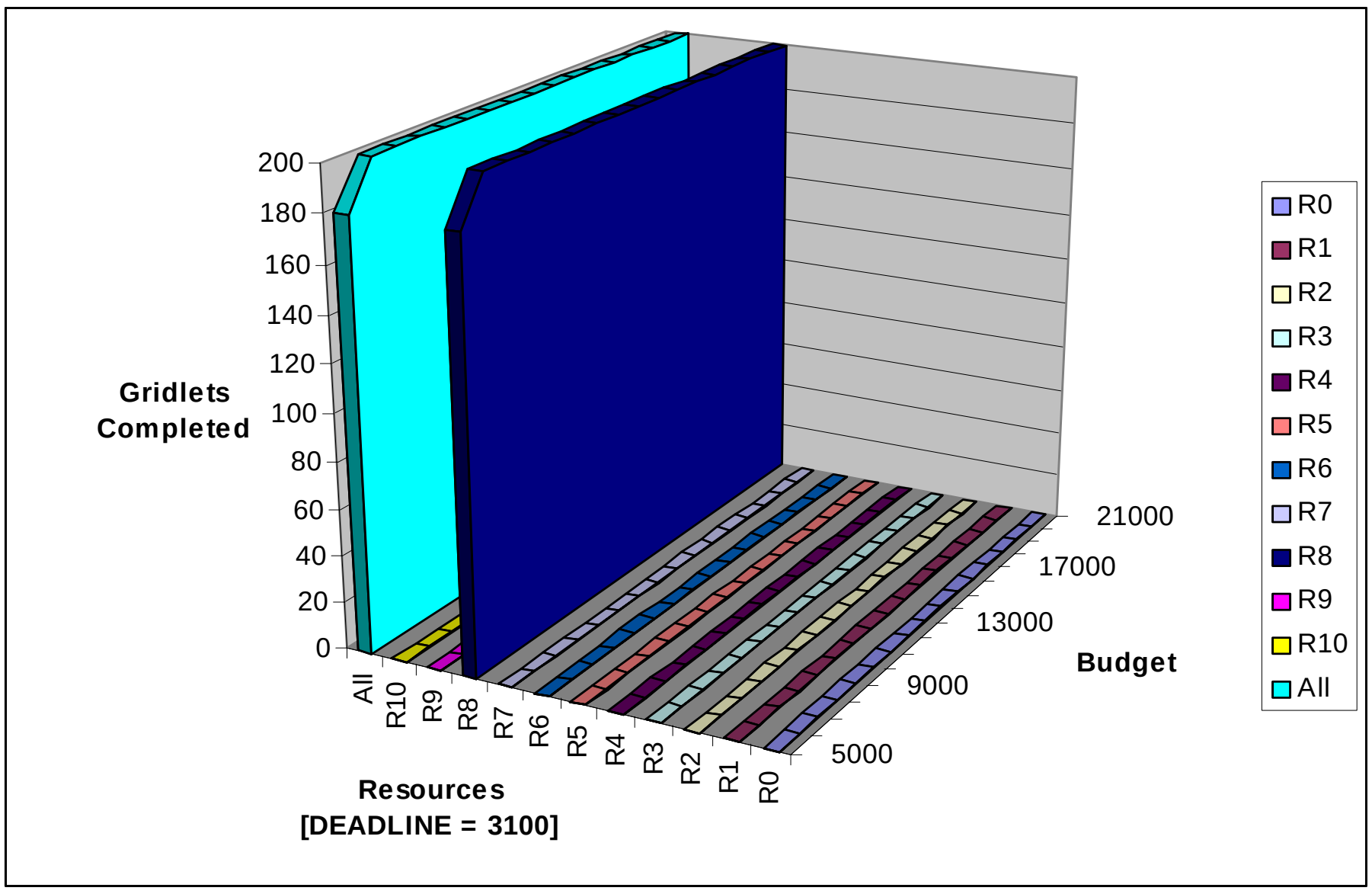

**Figure 6.10: Gridlets processed on resources for different budget values with high deadline.**

Let us now take a microscopic look at the allocation of resources at different times during the scheduling experimentation. The two graphs (Figure 6.11, Figure 6.13, and Figure 6.14) show a trace of leasing resources at different times during the scheduling experiment for processing Gridlets for different budget values with a fixed deadline of 100, 1100, and 3100 (low, medium, and high deadline value) respectively. It can be observed that when the deadline value is low, the economic broker also leases expensive resources to process Gridlets whenever the budget permits. The broker had to allocate powerful resources even if they are expensive since the deadline is too tight (see Figure 6.11 for Gridlets completed and Figure 6.12 for budget spent in processing). But this is not the case when the deadline is highly relaxed (see Figure 6.14)— the broker leased just one resource, which happened to process all Gridlets within the given deadline. From the diagrams (Figure 6.11 and Figure 6.12), it can be observed that the resource R7 has processed more Gridlets than the resource R6, but had to spend more budget on the resource R6 since it is more expensive than the resource R7.

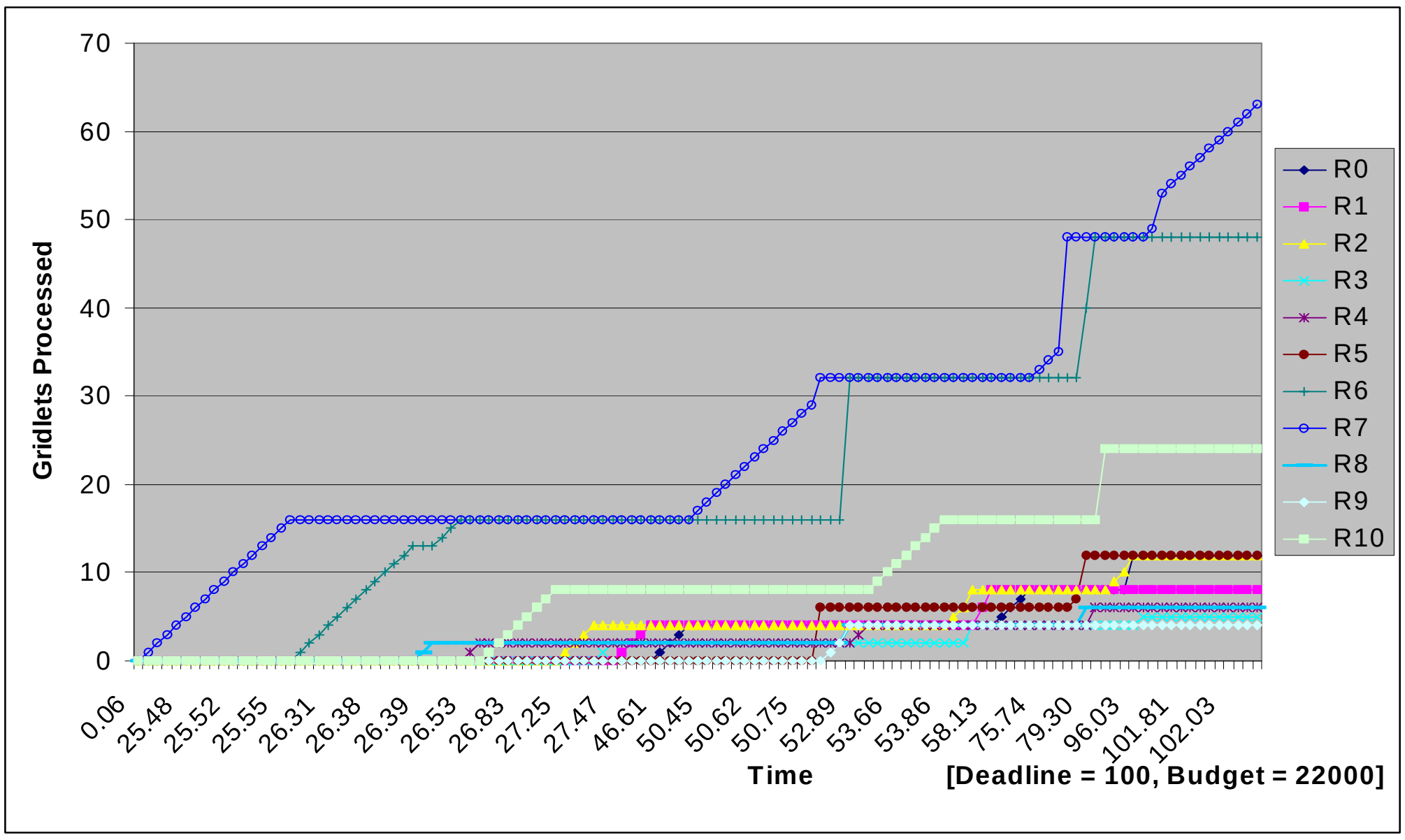

**Figure 6.11: Trace of No. of Gridlets processed on resources for a low deadline and high budget constraints.**

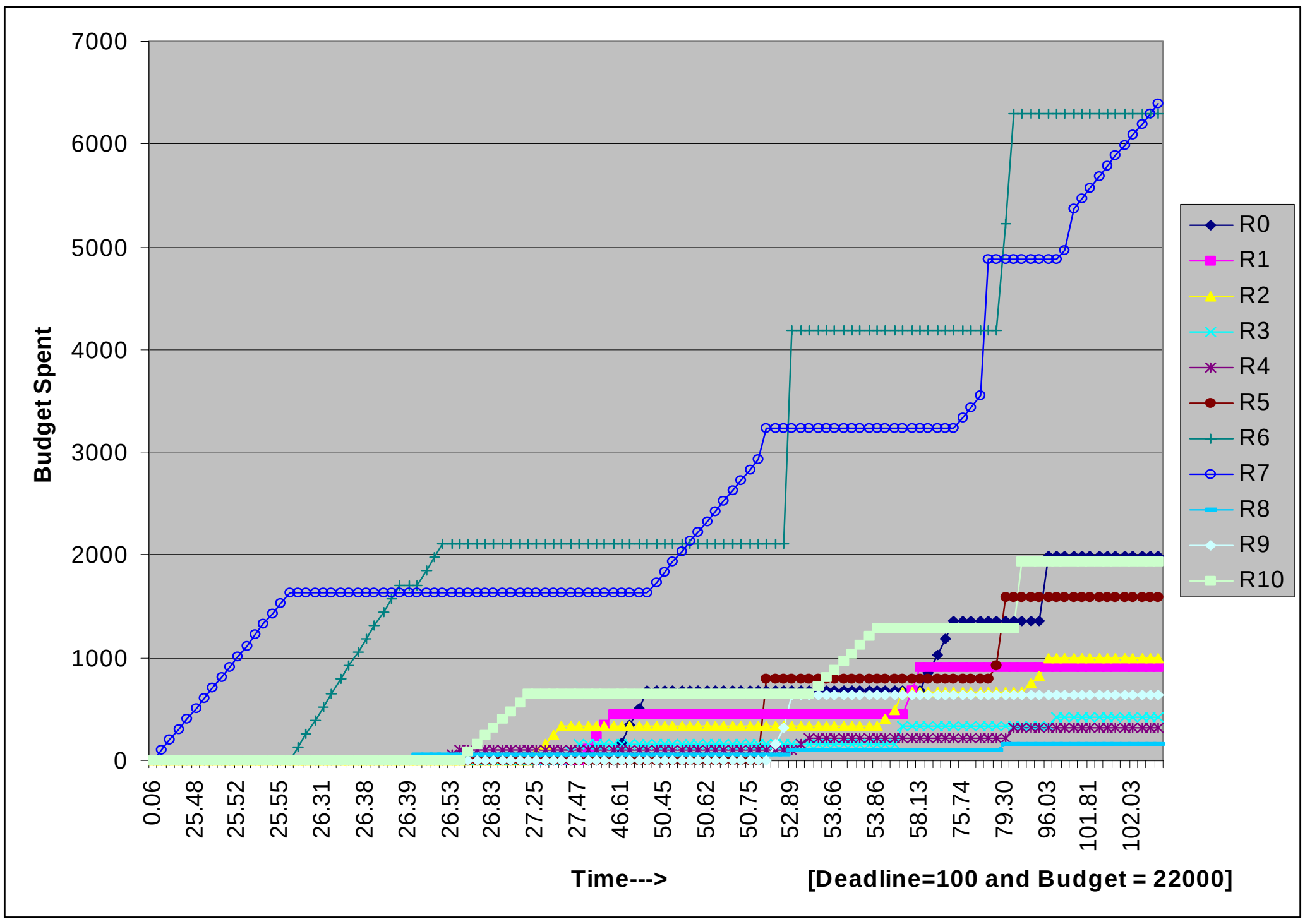

**Figure 6.12: Trace of budget spent for low deadline and high budget constraints.**

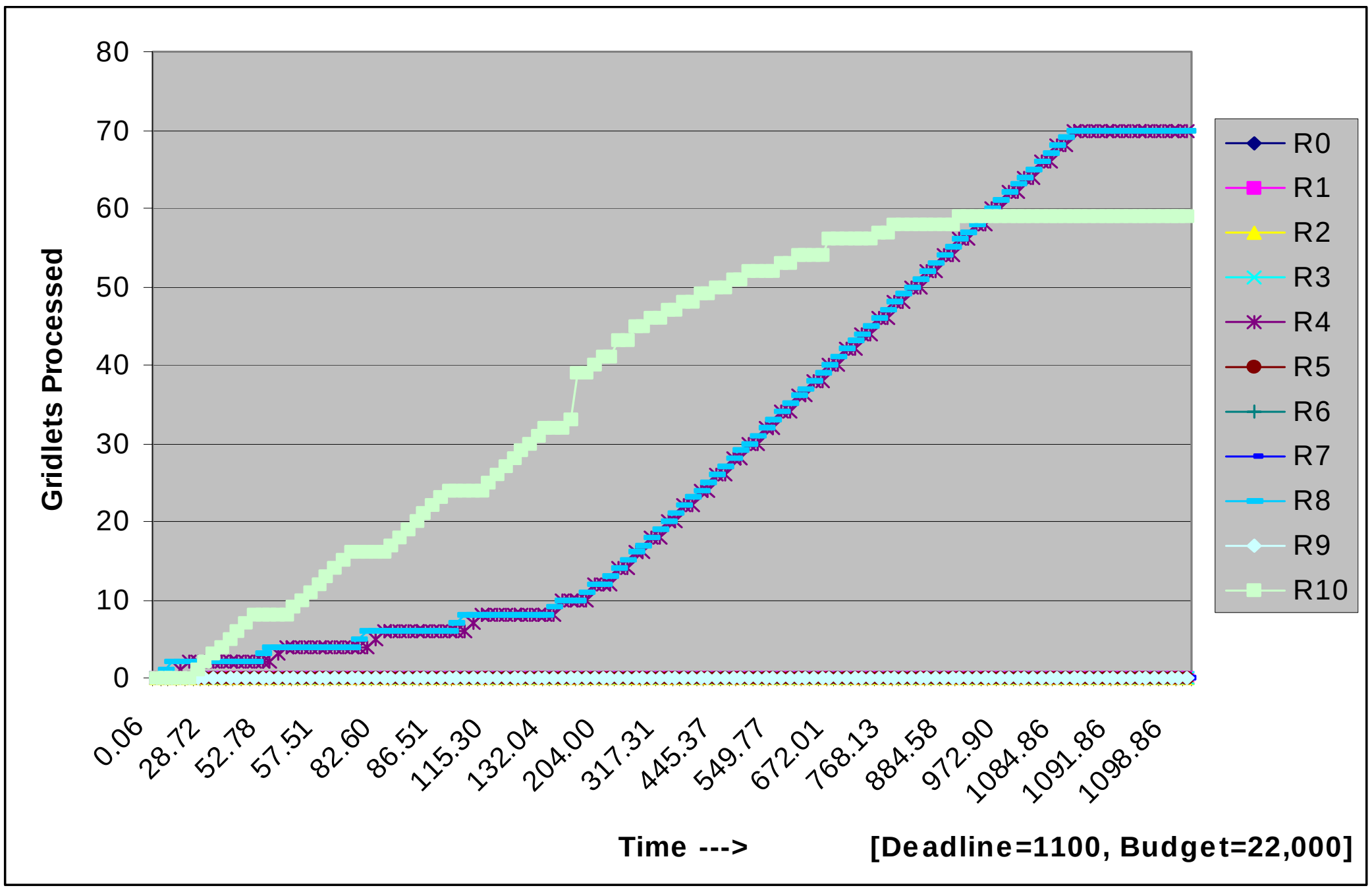

**Figure 6.13: Trace of No. of Gridlets processed for a medium deadline and low budget constraints.**

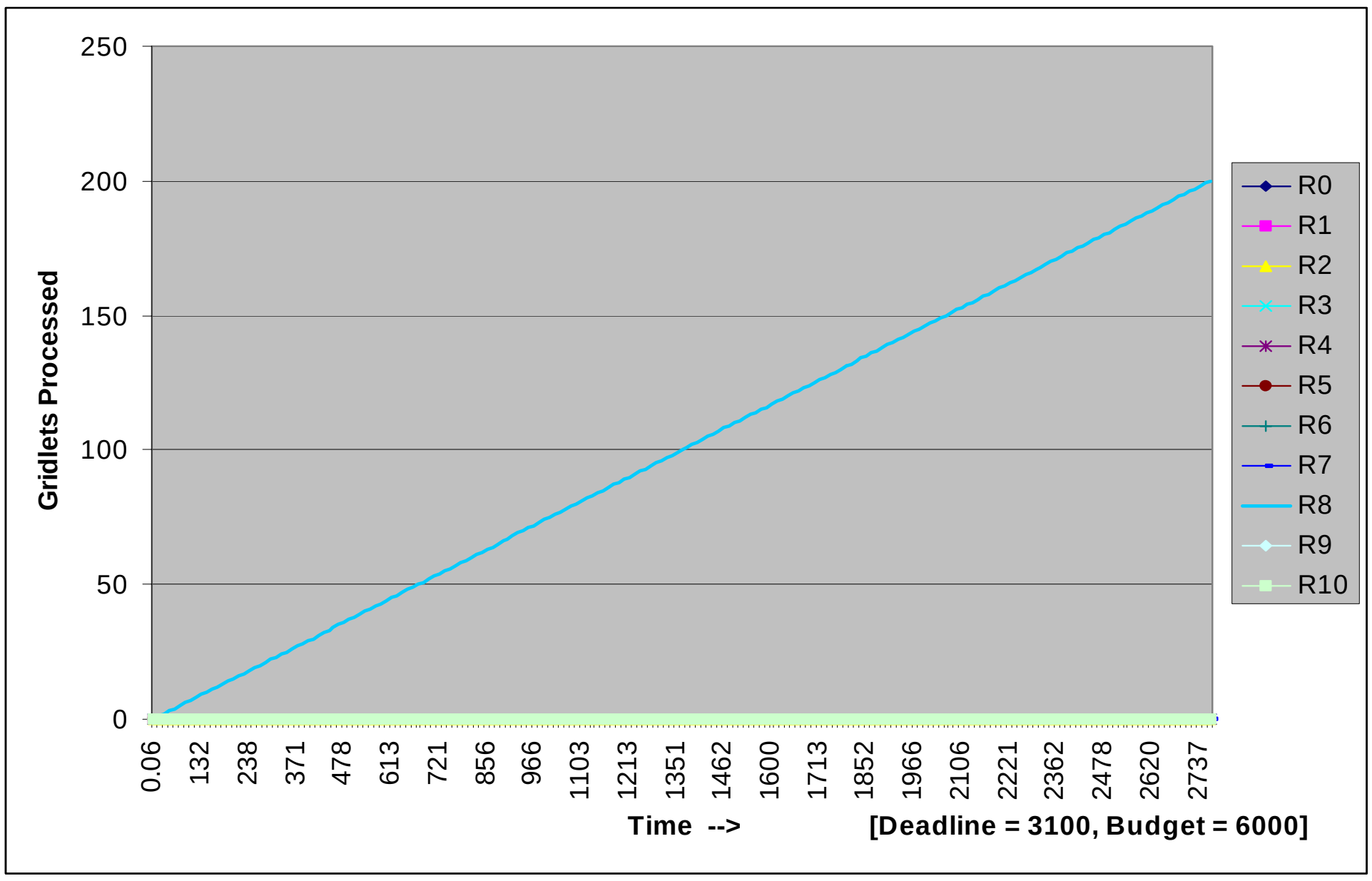

**Figure 6.14: Trace of No. of Gridlets processed for a high deadline and low budget constraints.**

A trace of the number of Gridlets committed to resources at different times depending on their performance, cost, and the user constraints (deadline and budget) and optimization requirements is shown in Figure 6.15 and Figure 6.16 for deadline values of 100 and 1100 time units respectively. In both graphs it can be observed the broker committed Gridlets to expensive resources only when it is required. It committed as many Gridlets as the cheaper resources can process by the deadline. The remaining Gridlets were assigned to expensive resources. The broker used expensive resources in the beginning and continued to use cheaper resources until the end of the experiment. This ability of economic Grid broker to select resources dynamically at runtime demonstrates its adaptive capability driven by the user's quality of service requirements.

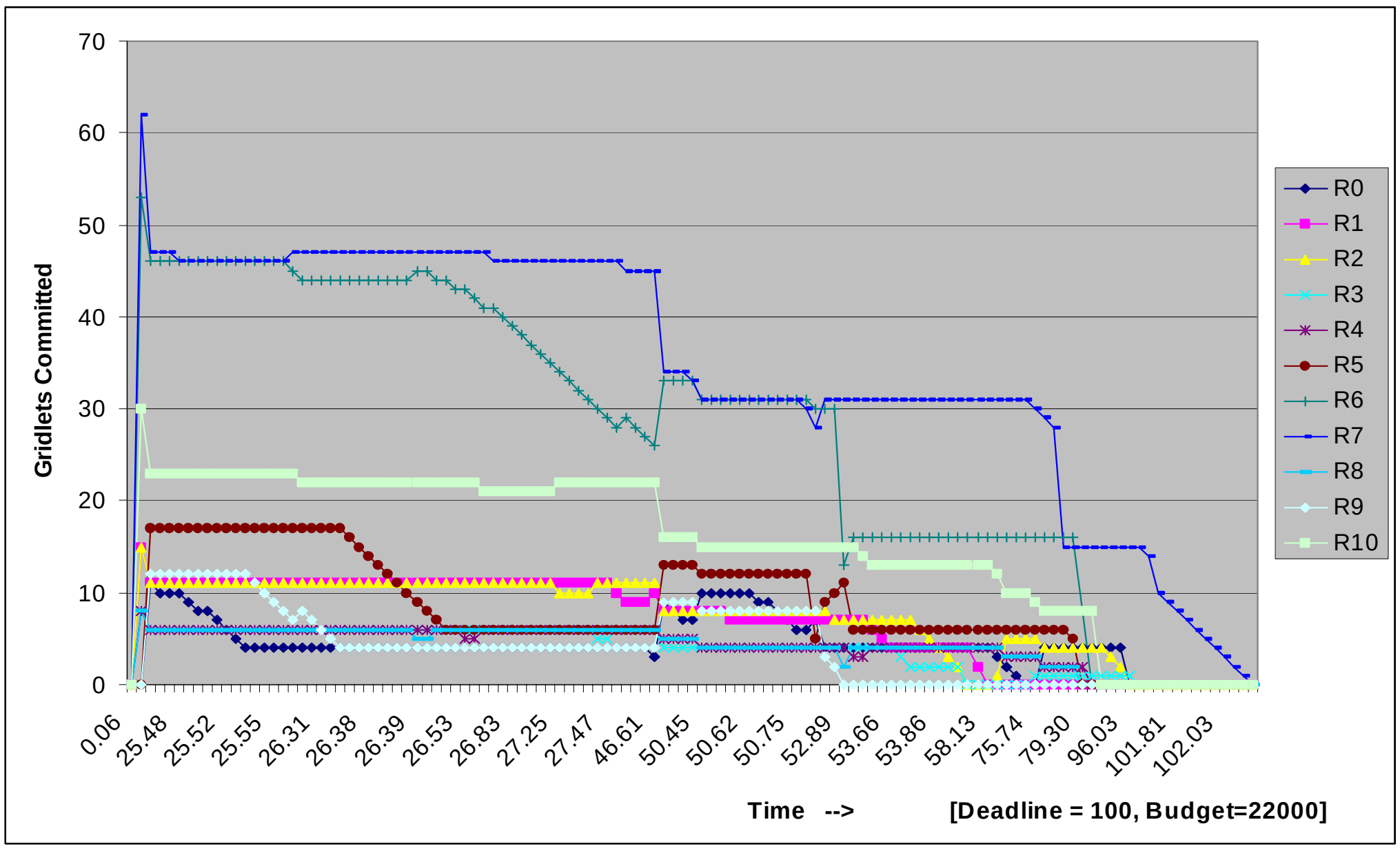

**Figure 6.15: Trace of the number of Gridlets committed to resources for a low deadline and high budget constraints.**

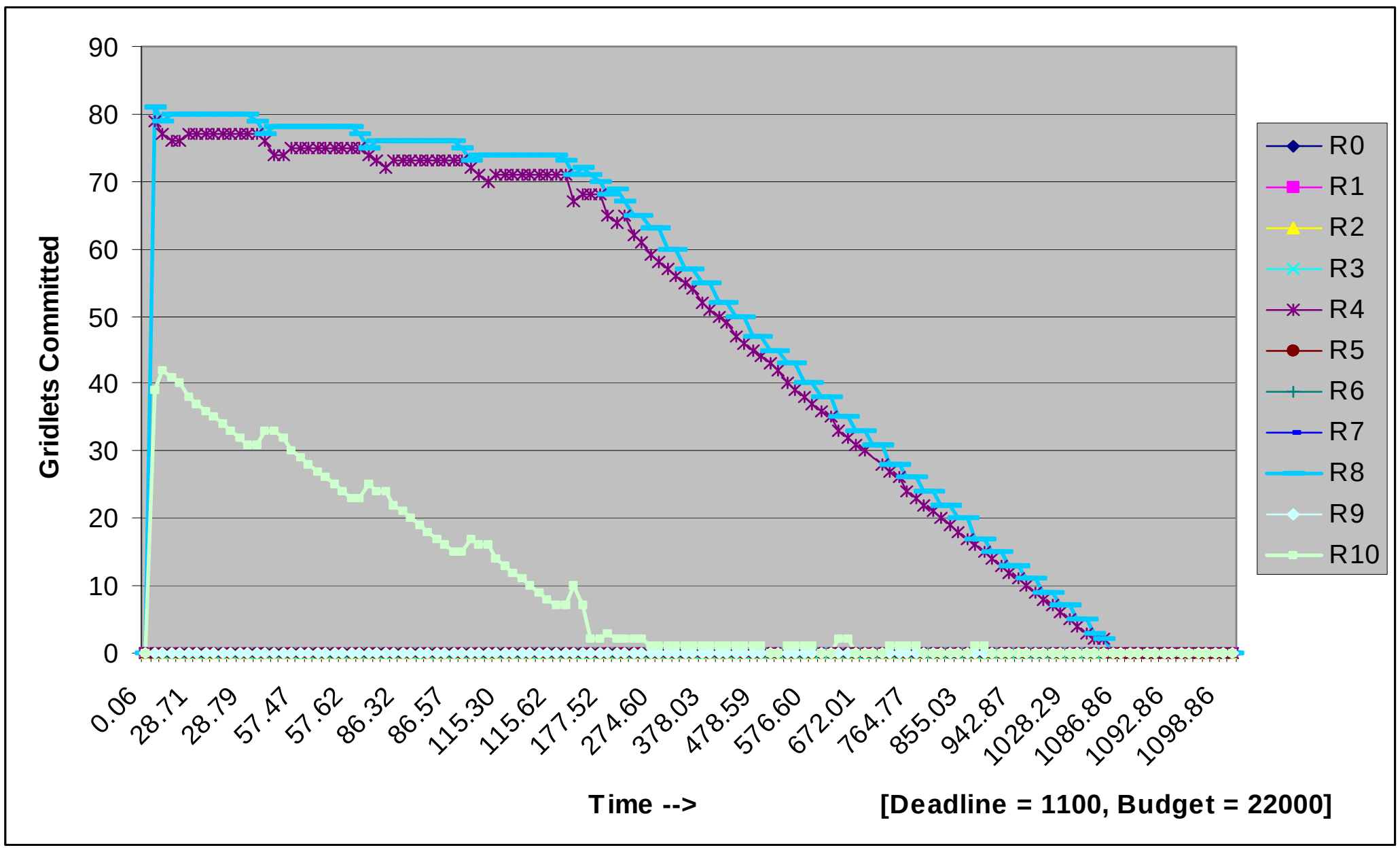

**Figure 6.16: Trace of the number of Gridlets committed to resources for a medium deadline and high budget constraints.**

### 6.3.2 Scheduling Experiments with Multiple Competing Users

In the second experiment, we explore distributed economic scheduling for a varying number of users competing for the same set of resources using the DBC constrained cost-optimisation scheduling algorithm. All users are modeled to have similar requirements to enable comparison among them and understand the overall scenario. Each user application contains 200 Gridlets with small variation as explained in application modeling section. We modeled varying number of users in series from 1, 10, 20, and so on up to 100 and each with their own broker scheduling Gridlets on simulated WWG testbed resources (listed in Table 6.2). We explored scheduling of Gridlets for different budget values varied from 5000 to 22000 in step of 1000. For this scenario, we performed two scheduling experiments with two different values of deadline for DBC constrained *cost minimization* algorithm.

### *User Deadline = 3100 time unit*

The number of Gridlets processed, average time at which simulation is stopped, and budget spent for different scheduling scenario for each user with the deadline constraint of 3100 time units is shown in Figure 6.17, Figure 6.18, and Figure 6.19. From Figure 6.17, it can be observed that as the number of users competing for the same set of resources increase, the number of Gridlets processed for each user is decreasing because they have tight deadline. Whether there are few users (e.g., 1 or 10 users in this case), they are able to process all jobs in most cases when the budget is increased. Figure 6.18 shows the time at which broker terminated processing of Gridlets. When a large number of users are competing (e.g., 100) for resources, it can be observed that the broker exceeded the deadline. Because, the broker initially planned scheduling Gridlets for the period of deadline, but that schedule had to be terminated because competing users had already occupied high resource share well before the recalibration phase (the first establishment of the amount of resource share available to the user, which of course can change). Figure 6.19 shows the average budget spent by each user for processing Gridlets shown in Figure 6.17, which is also clear from the graphic similarity between both diagrams when a large number of users are competing for resources.

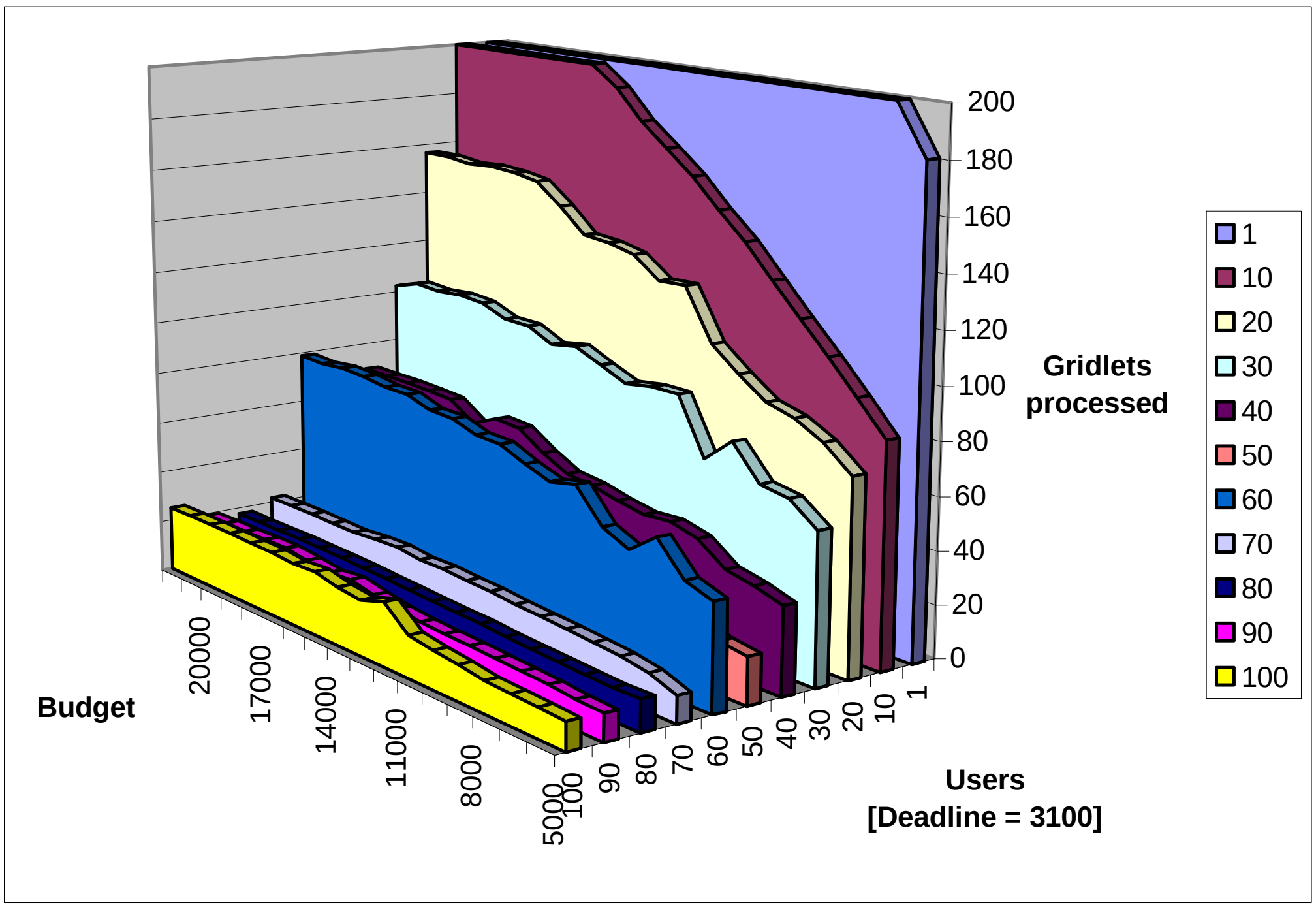

**Figure 6.17: No. of Gridlets processed for each user when a varying number of users competing for resources.**

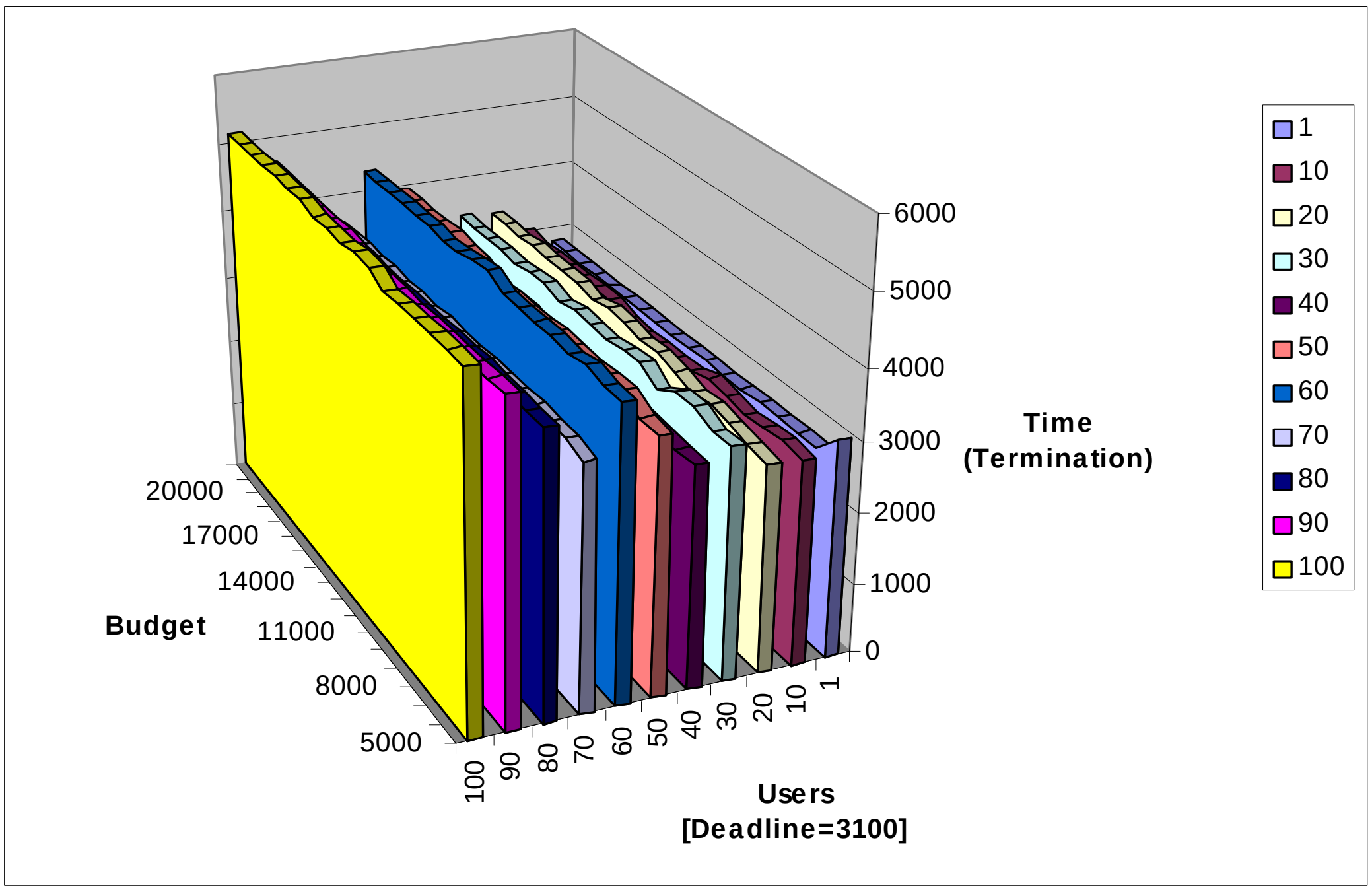

**Figure 6.18: The average time at which the user experiment is terminated with varying number of users competing for resources.** When there are a large number of users arriving at different times, they are likely to impact on the schedule and the execution time of jobs already deployed on resources. The broker waiting for the return of jobs that are deployed on resources leads to the termination time exceeding the soft deadline unless the execution of jobs is cancelled immediately.

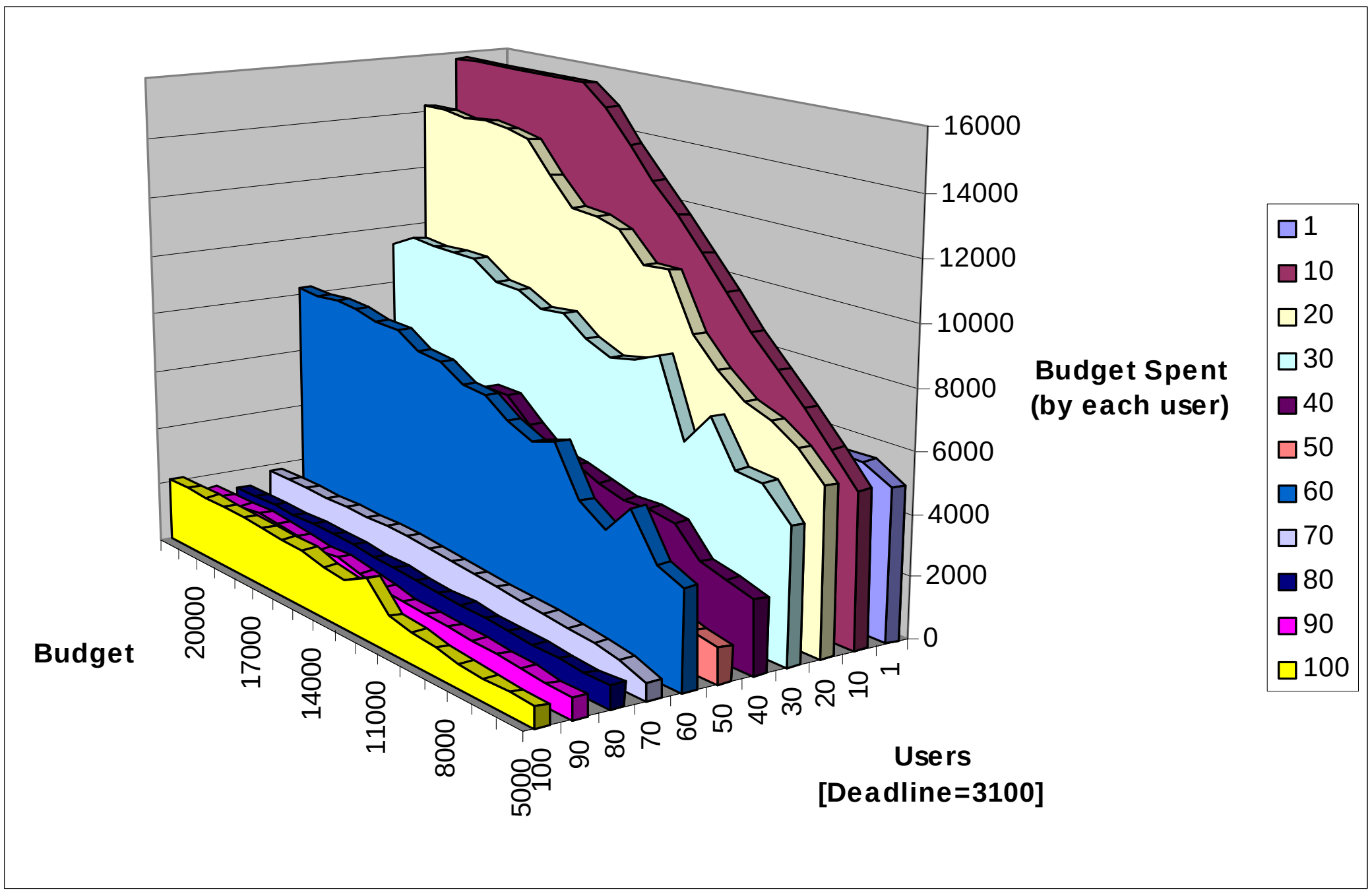

**Figure 6.19: The average budget spent by each user for processing Gridlets.**

***User Deadline = 10000 time unit***

The number of Gridlets processed, average time at which simulation is stopped, and budget spent for different scheduling scenario for each user with the deadline constraint of 10000 time units is shown in Figure 6.20, Figure 6.21, and Figure 6.22. In this experiment, the number of Gridlets processed for each user improved substantially due to the relaxed deadline constraint compared to the previous experiment (see Figure 6.17 and Figure 6.20). As the number of users competing for resources increased, the number of Gridlets processed for each user decreased. But when the budget is increased, the number of Gridlets processed increased as well. Unlike the previous experiment, the broker is able to learn and make better predictions on the availability of resource share and the number of Gridlets that can be finished by deadline. As the deadline was sufficient enough to revisit the past scheduling decisions, the broker is able to ensure that the experiment is terminated within the deadline for most of the time (see Figure 6.21). The average budget spent by each user for processing Gridlets is shown in Figure 6.22, which is also clear from the graphic similarity between Figure 6.20 and Figure 6.22 when a large number of users are competing for resources. However, up to the medium number of users (1-40 users), they were able to get many Gridlets processed, but decreased with the increasing number of users competing for resources, which means the increase in processing cost.

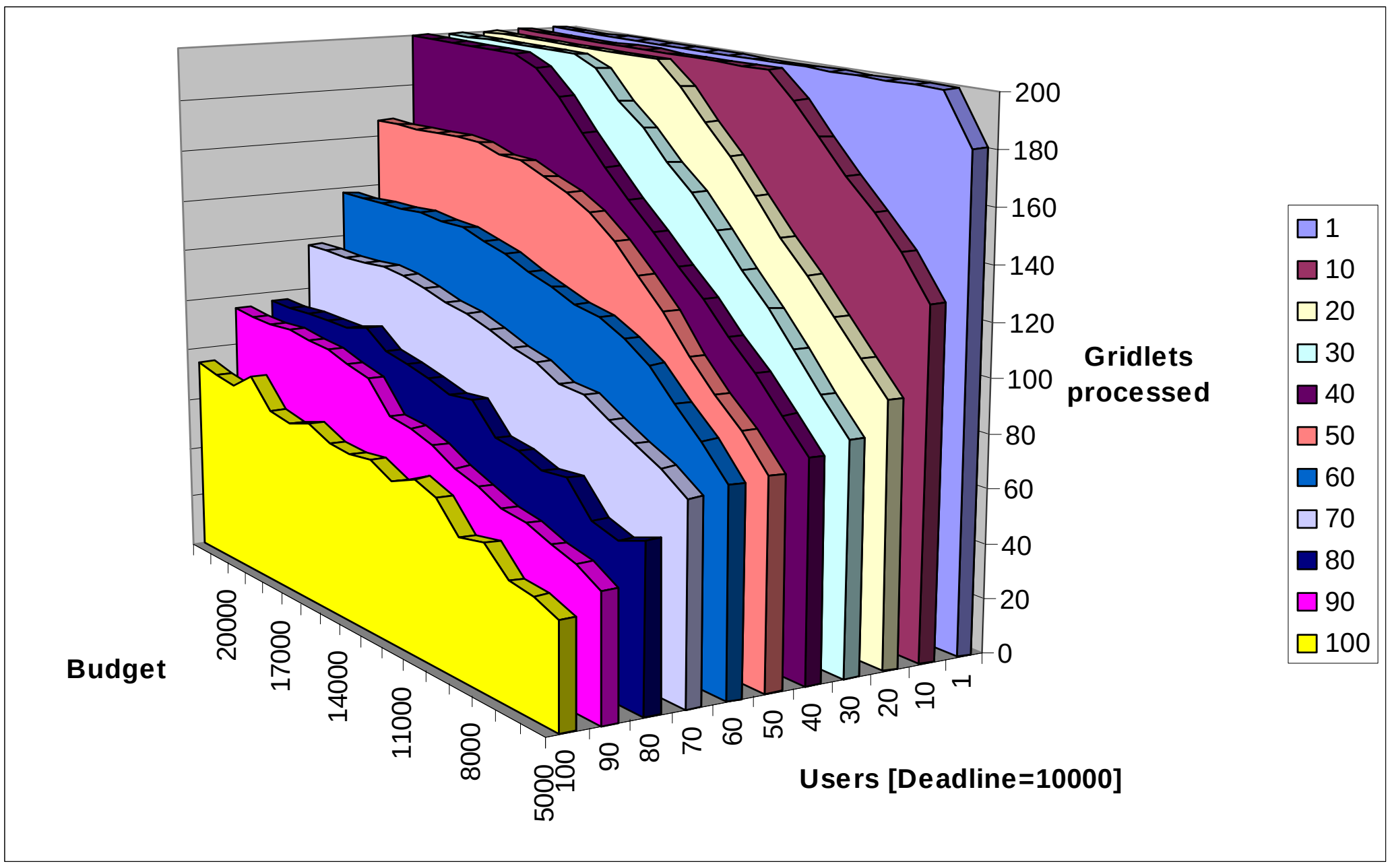

**Figure 6.20: No. of Gridlets processed for each user with varying number of users competing for resources.**

## 6.4 Deadline and Budget Constrained Time Optimisation Scheduling

In this experiment, we perform scheduling experiments with different values of deadline and budget constraints (DBC) for a single user using the DBC constrained time-optimisation scheduling algorithm shown in Figure 6.23. The deadline is varied in simulation time from 100 to 3600 in steps of 500. The budget is varied from G$ 5000 to 22000 in steps of 1000. The number of Gridlets processed, time spent, and budget spent for different scheduling scenario is shown in Figure 6.24, Figure 6.25, and Figure 6.26. The number of Gridlets processed increased with the increase in budget or deadline value (see Figure 6.25 for a tight deadline value say 100). This is because, when a higher budget is available, the broker is able to select expensive resources to process more jobs as quickly as possible. The increase in budget has impact not only on the number of Gridlets completed; it also has impact on the completion time. The application processing *completion time decreases* with the *increase in budget value* (see Figure 6.25). When a higher budget is available, the broker schedules jobs on even expensive resources depending on their capability and availability (e.g., resources R6 and R7) to complete the processing at the earliest possible time (see

Figure 6.27). This also means that as the application processing completion time decreases, the processing cost increases as powerful and expensive resources are used in processing jobs (see Figure 6.28).

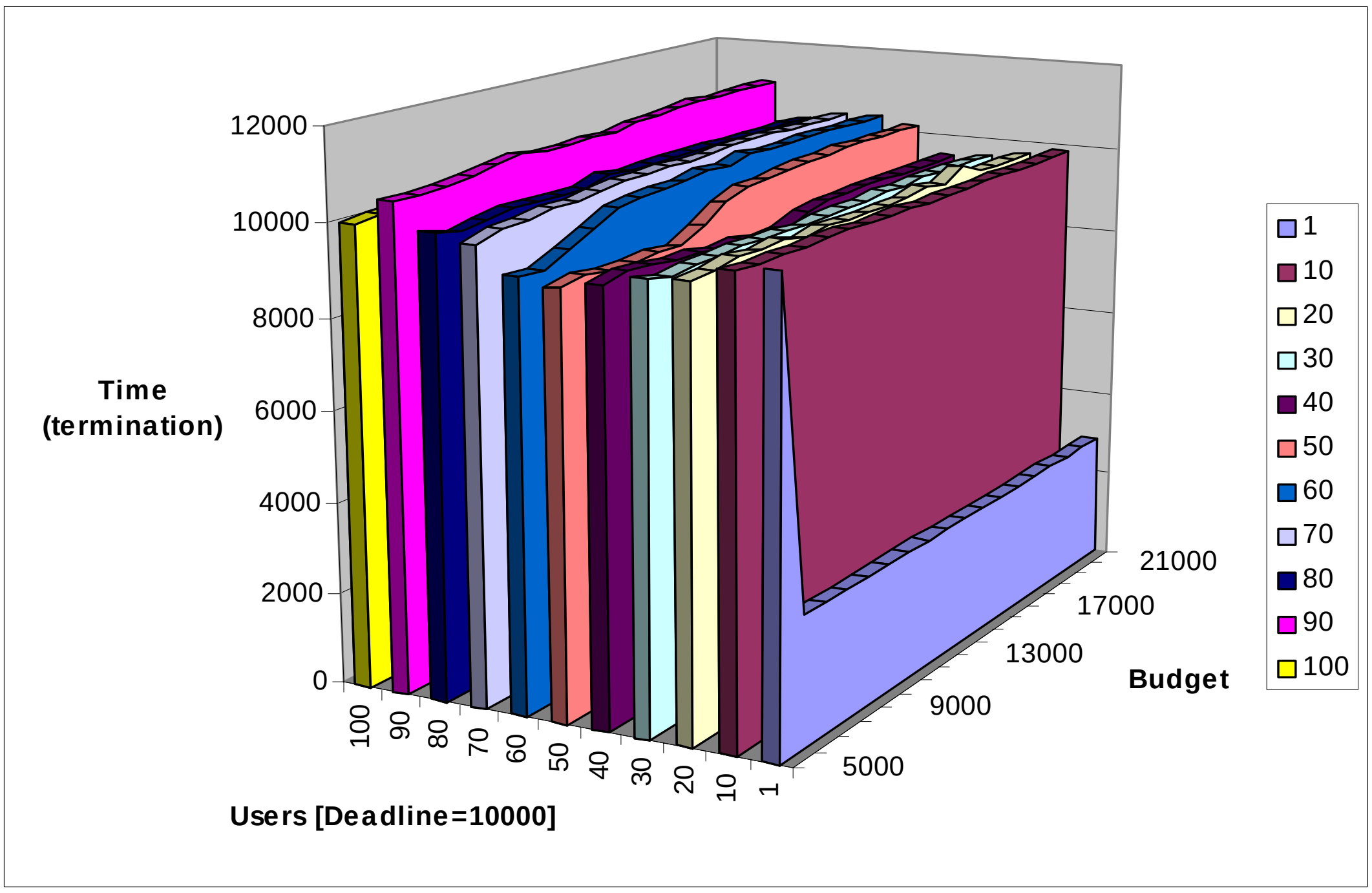

**Figure 6.21: The average time at which the user experiment is terminated with varying number of users competing for resources.**

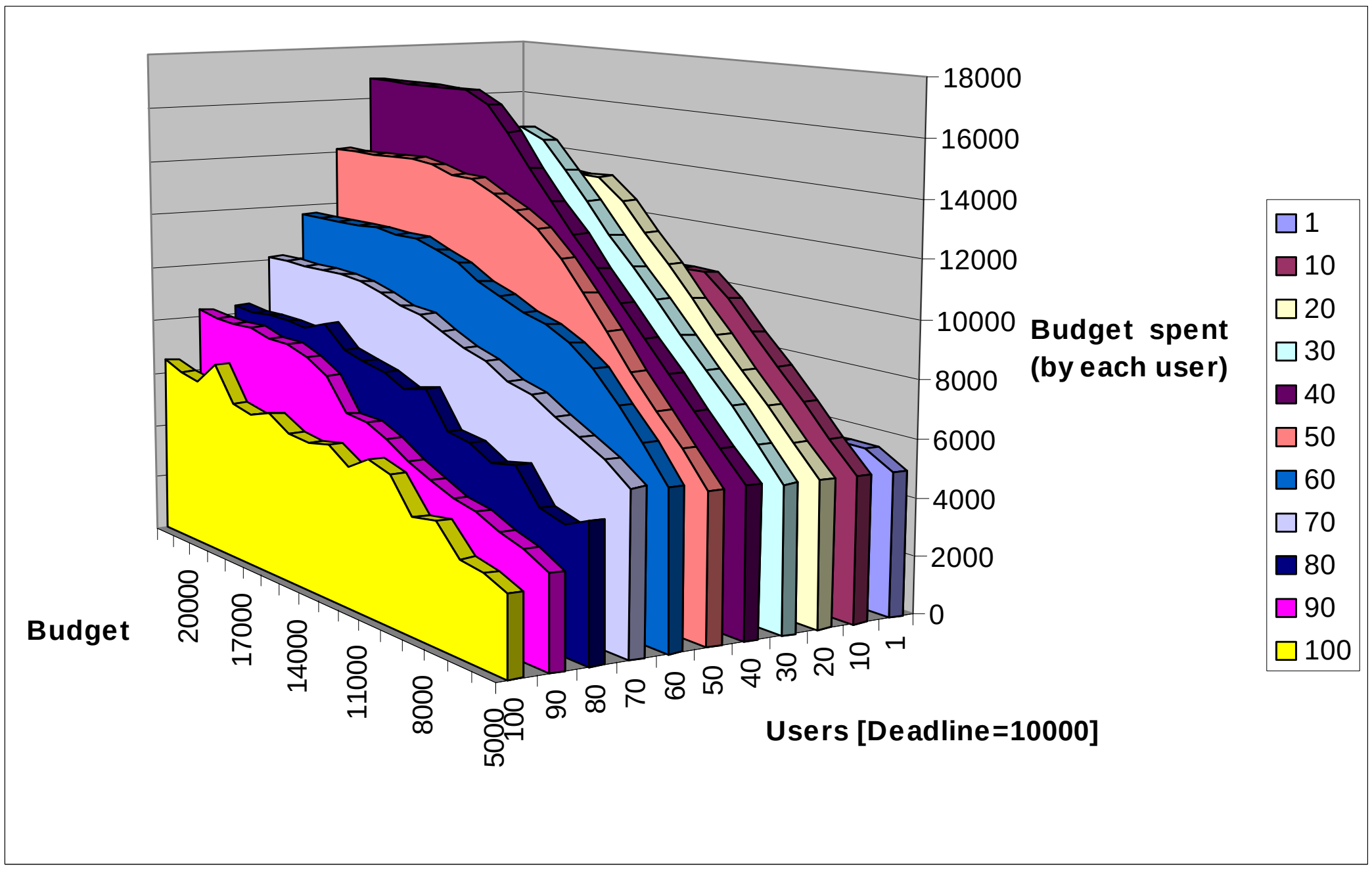

**Figure 6.22: The average budget spent by each user for processing Gridlets.**

*Algorithm: DBC_Scheduling_with_Time_Optimisation()*

1. RESOURCE DISCOVERY: Identify the resources and their capability using the Grid information services.
2. RESOURCE TRADING: Identify *the cost* of all resources and *the capability* to be delivered per cost-unit. The *resource cost* can be expressed in units such as processing cost-per-MI, cost-per-job, CPU cost per time unit, etc. and the scheduler needs to choose suitable unit for comparison.
3. If the user supplies D and B-factors, then determine the absolute deadline and budget based on the capability of resources and their cost, and the application processing requirements (e.g., total MI required).
4. SCHEDULING: Repeat while there exist *unprocessed jobs* and the current time and processing expenses are within the deadline and budget limits. [It is triggered for each scheduling event or whenever a job completes. The event period is a function of deadline, job processing time, rescheduling overhead, resource share variation, etc.]:

[SCHEDULE ADVISOR with Policy]

a.) For each resource, predict and establish the *job consumption rate* or *the available resource share* through the measure and extrapolation strategy taking into account the time taken to process previous jobs.

b.) If any of the resource has jobs assigned to it in the previous scheduling event, but not dispatched to the resource for execution and there is variation in resource availability, then move appropriate number of jobs to the Unassigned-Jobs-List. This helps in updating the whole schedule based on the latest resource availability information.

c.) Repeat the following steps for each job in the Unassigned-Jobs-List:

- Select a job from the Unassigned-Jobs-List.
- Create a *resource group* containing affordable resources (i.e., whose processing price is less than or equal to the remaining budget per job).
- For each resource in the resource group, calculate/predict the job completion time taking into account previously assigned jobs and the job completion rate and resource share availability.
- Sort resources in the resource group by the increasing order of job completion time.
- Assign the job to the first resource in the resource group and remove it from the Unassigned-Jobs-List if the predicted job completion time is less than the deadline.

5. [DISPATCHER with Policy]

*Repeat the following steps for each resource if it has jobs to be dispatched:*

- Identify the number of jobs that can be submitted without overloading the resource. Our default policy is to dispatch jobs as long as the number of user jobs deployed (active or in queue) is less than the number of PEs in the resource.

**Figure 6.23: Deadline and budget constrained (DBC) time optimisation scheduling algorithm.**

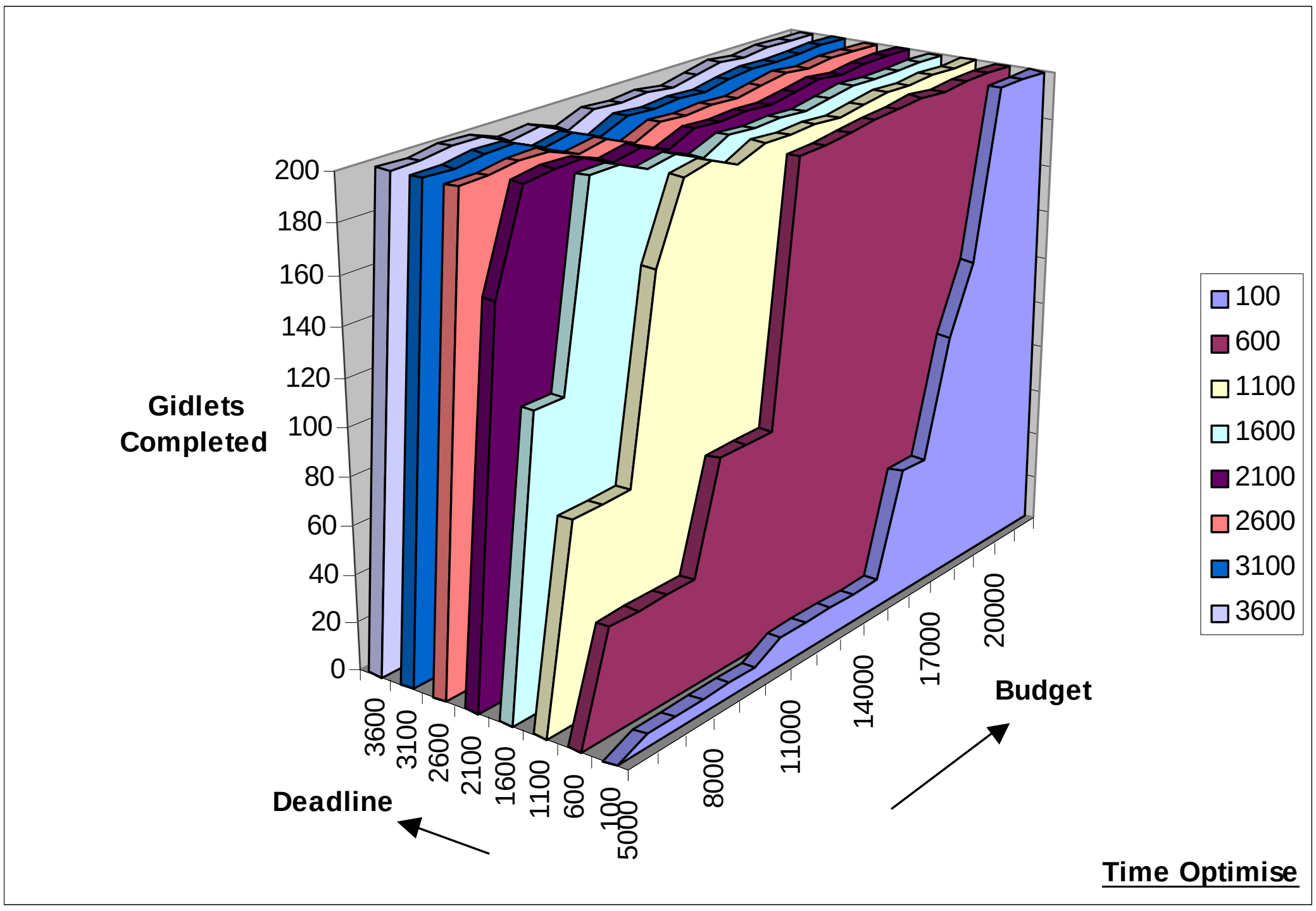

**Figure 6.24: No. of Gridlets processed for different budget and deadline limits.**

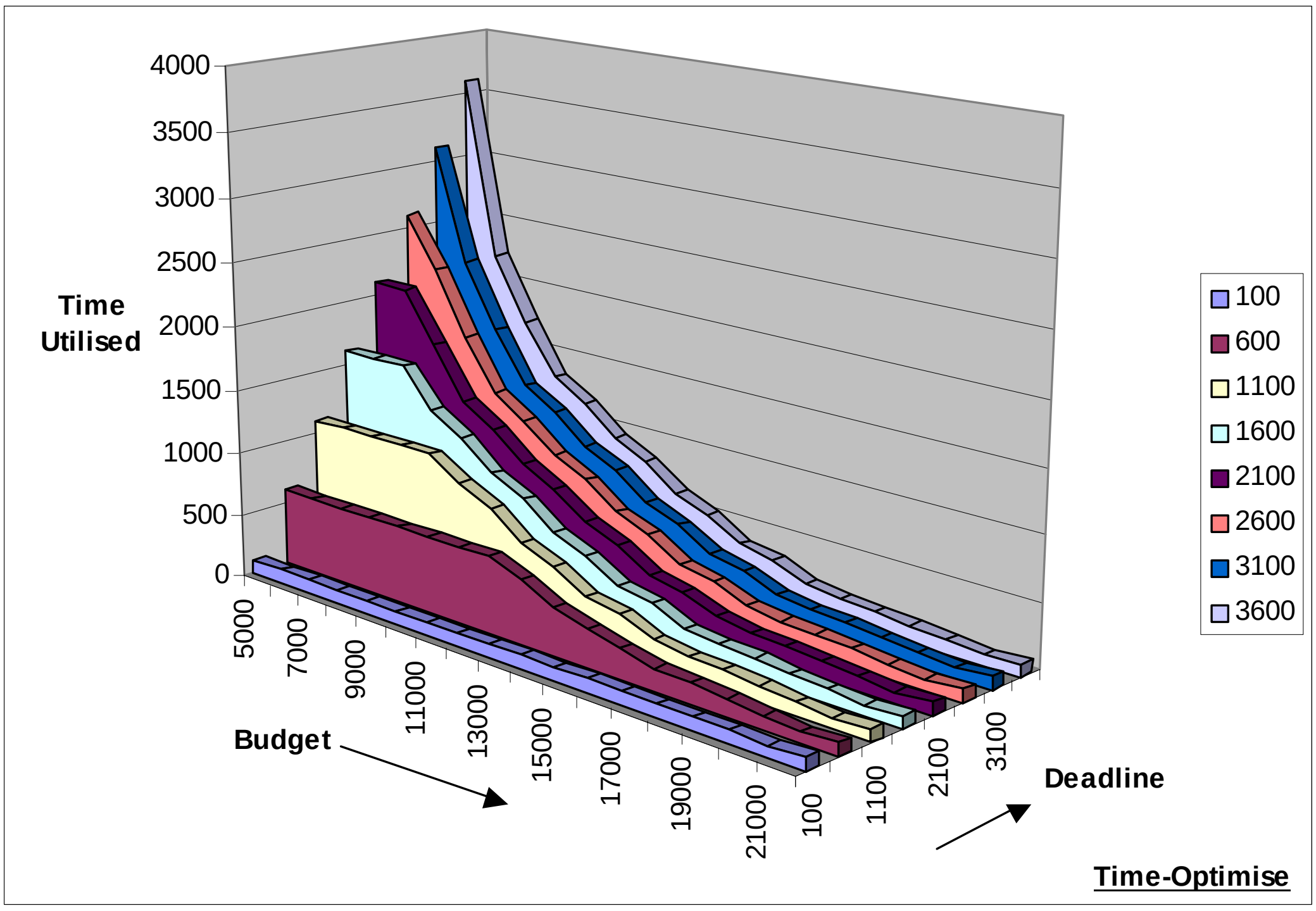

**Figure 6.25: The time spent in processing Gridlets using the DBC time optimisation.**

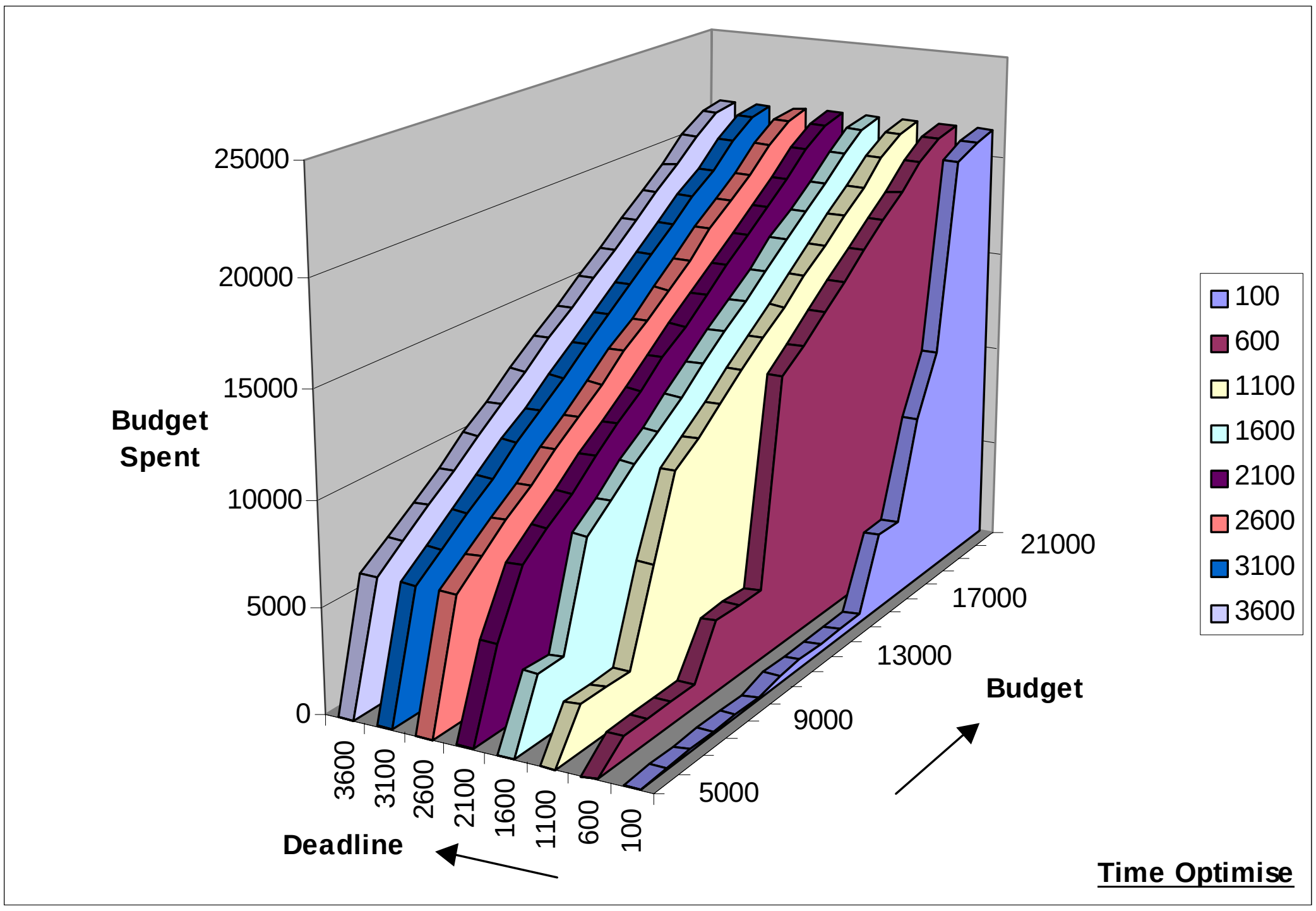

**Figure 6.26: The budget spent in processing Gridlets using the DBC time optimisation.**

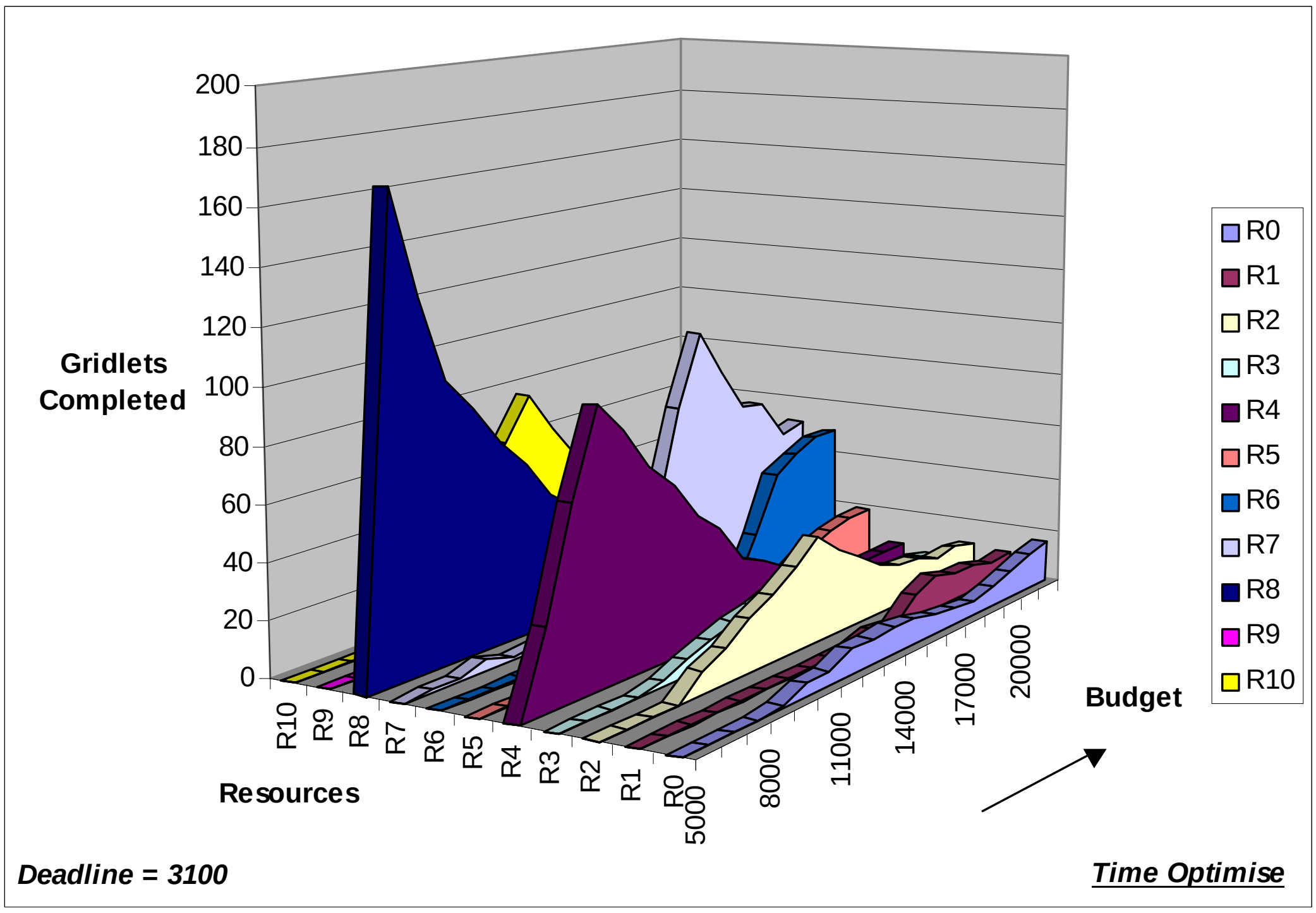

**Figure 6.27: Selection of different resources for processing Gridlets for different budget limits.**

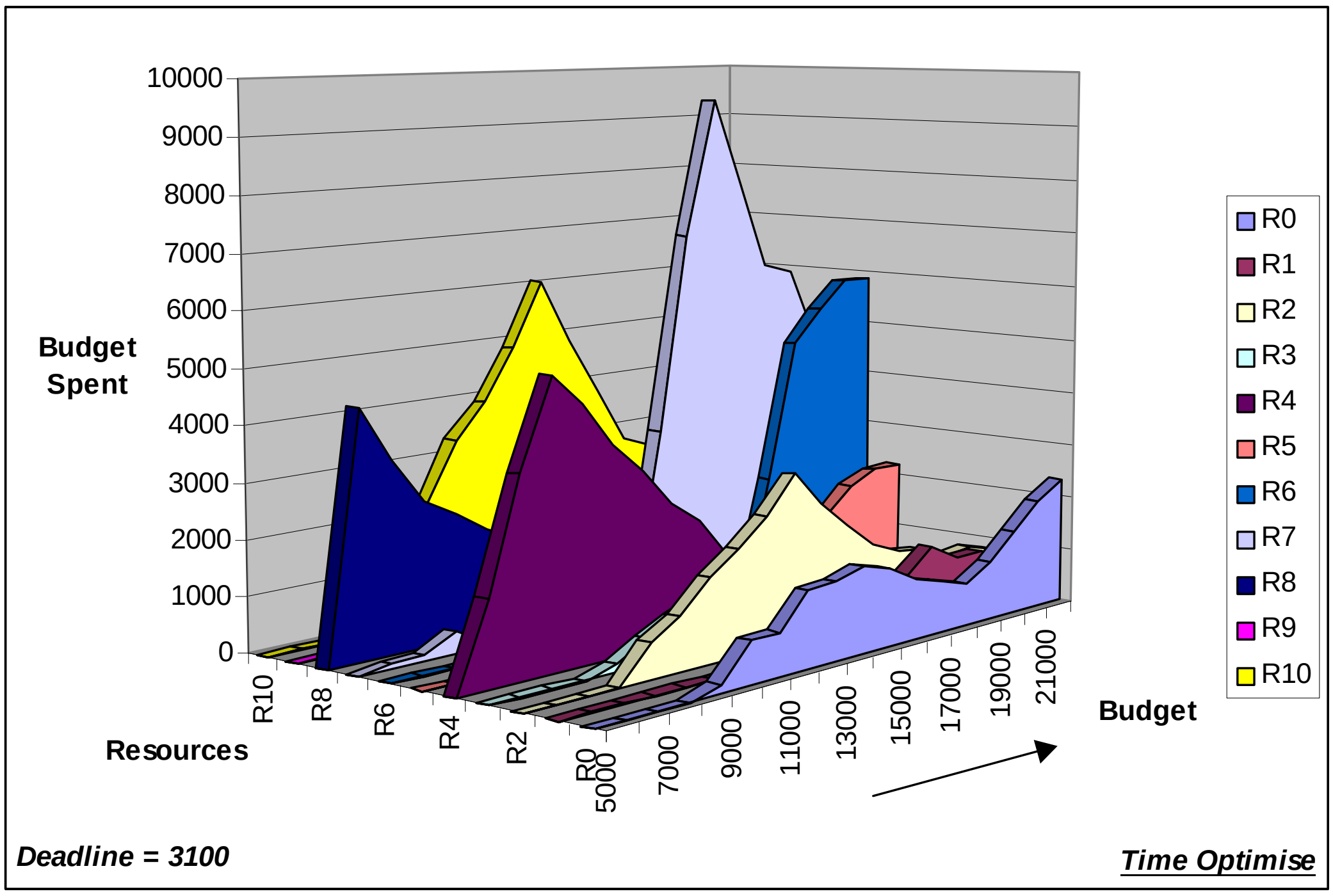

**Figure 6.28: The budget spent in processing Gridlets on different resources for different budgets.**

## 6.5 Comparing the Cost and Time Optimisation Scheduling

The completion time and the budget spent for processing application jobs scheduled using the cost and the time optimisation strategies is shown in Figure 6.29 and Figure 6.30. In both scheduling optimisation scenarios, the deadline value is set to 3100 time units and budget value is varied from 5000 to 22000 in steps of 1000. In general, as the budget value increases, the completion time decreases and the processing cost increases when the time-optimisation scheduling strategy is used; whereas, the completion time remains closer to the deadline and processing cost decreases when the cost-optimisation scheduling is used.

Note that, when the available budget per job is less than the cost of processing a job on any resource, no jobs are scheduled for processing in the case of time-optimisation scheduling. This can be observed from Figure 6.30 when the budget is 5000—the budget per job is less than the cost of processing even on the cheapest resource and no jobs are processed, hence the budget spent in processing is shown as 0. Such a condition can also be strictly enforced within the cost-optimisation strategy.

The time-optimisation scheduling algorithm uses as many resources as it can in parallel as long as the budget is available since minimizing the completion time is a major goal. Whereas, the cost-optimisation scheduling algorithm uses resources, giving the first preference to cheaper resources, as long as the deadline can be met, since minimizing the processing cost is a major goal. That means, the users can choose a scheduling strategy that meets their quality of service requirements. When the work is urgent and the results are needed as quickly as possible, they can choose the time optimisation strategy and place a limit on the processing expenses. If they do not have immediate requirement for results, they can choose the cost-optimisation scheduling strategy and minimize the processing cost.

In the DBC time optimization scheduling, the increase in budget value has much impact on resource selection and the completion time. When a higher budget is available, the increase in deadline to a larger value does not have much impact on a completion time or budget spent (see Figure 6.25 and Figure 6.26). This situation is very much different for the cost optimisation scheduling where deadline parameter drives the selection of resources.

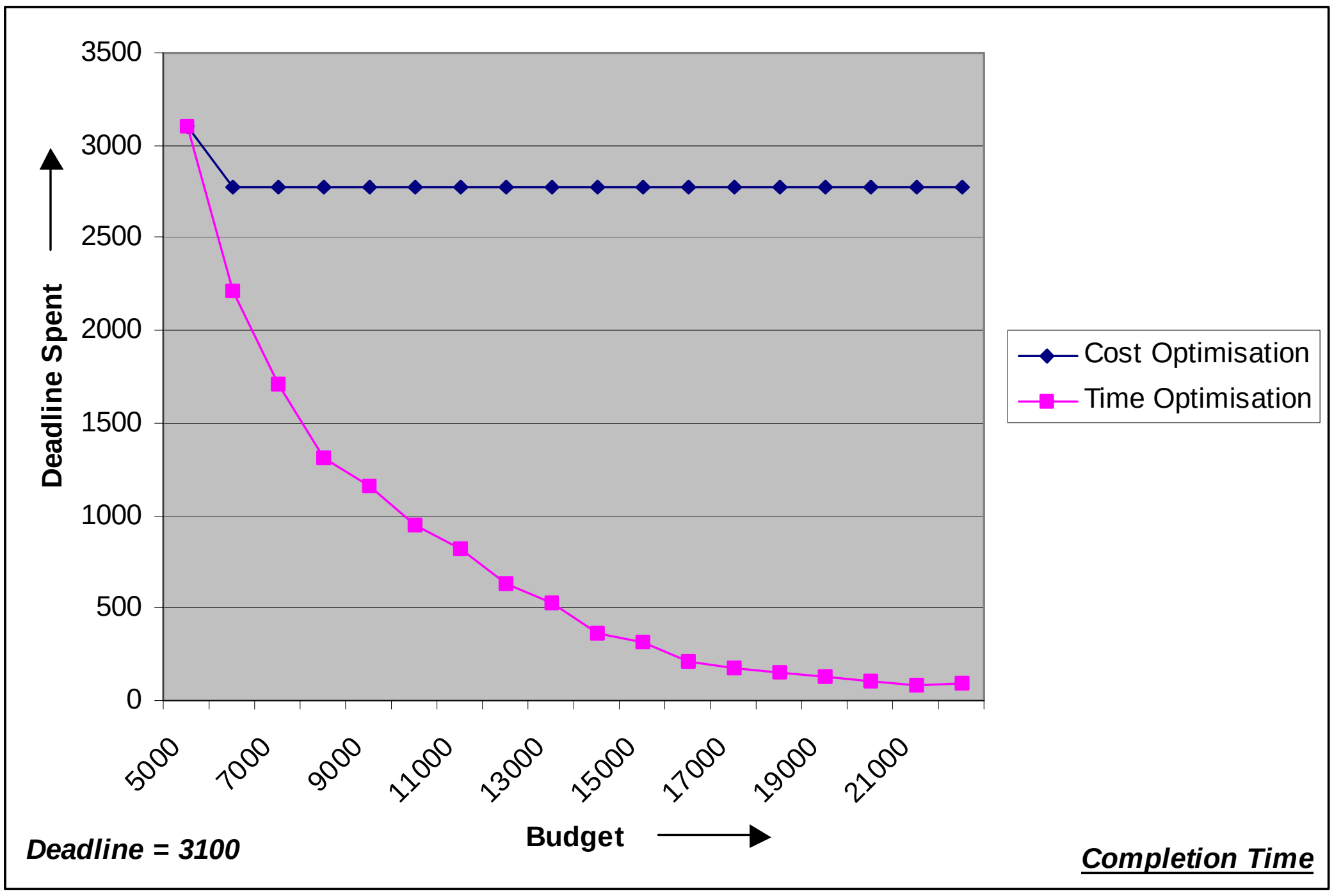

**Figure 6.29: The time spent in processing application jobs using time cost and time optimisation scheduling algorithms given different budget limits.**

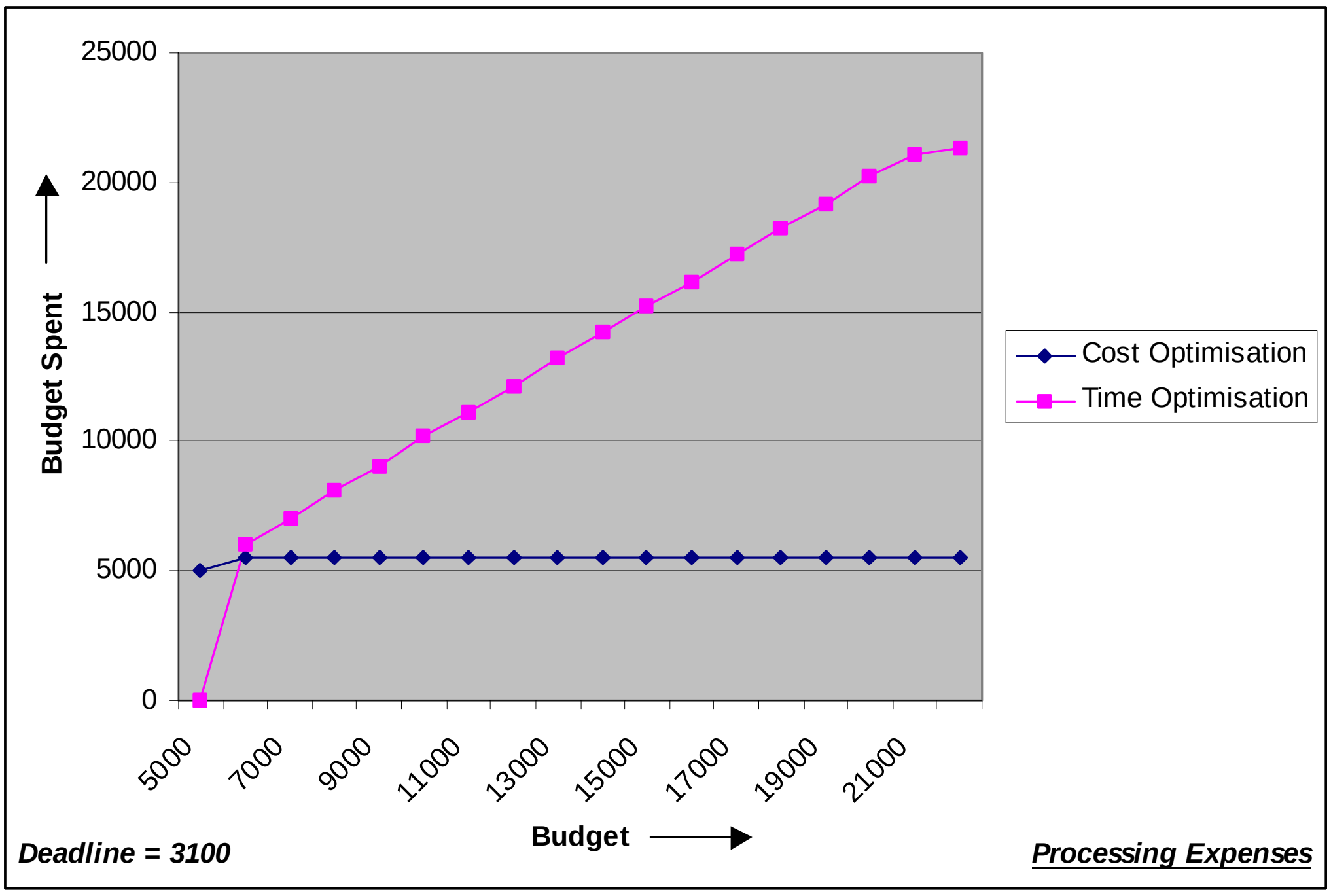

**Figure 6.30: The budget spent in processing application jobs using time cost and time optimisation scheduling algorithms given different budget limits.**

## 6.6 DBC Cost-Time Optimisation Scheduling

The DBC cost-time optimisation scheduling algorithm (shown in Figure 6.31) extends the cost-optimisation algorithm to optimise the time without incurring additional processing expenses. This is accomplished by applying the time-optimisation algorithm to schedule jobs on resources having the same processing cost.

*Algorithm: DBC_Scheduling_with_Cost_Time_Optimisation()*

1. RESOURCE DISCOVERY: Identify the resources and their capability using the Grid information services.
2. RESOURCE TRADING: Identify the cost of all resources and the capability to be delivered per cost-unit. The resource cost can be expressed in units such as processing cost-per-MI, cost-per-job, CPU cost per time unit, etc. and the scheduler needs to choose suitable unit for comparison.
3. If the user supplies D and B-factors, then determine the absolute deadline and budget based on the capability of resources and their cost, and the application processing requirements (e.g., total MI required).
4. SCHEDULING: Repeat while there exist unprocessed jobs and the current time and processing expenses are within the deadline and budget limits. [It is triggered for each scheduling event or whenever a job completes. The event period is a function of deadline, job processing time, rescheduling overhead, resource share variation, etc.]:

   [SCHEDULE ADVISOR with Policy]

   a. For each resource, predict and establish the job consumption rate or the available resource share through the measure and extrapolation strategy taking into account the time taken to process previous jobs.
   b. SORT the resources by increasing order of cost. If two or more resources have the same cost, order them such that powerful ones (e.g., higher job consumption rate or resource share availability, but the first time based on the total theoretical capability, say the total MIPS) are preferred first.
   c. Create resource groups containing resources with the same cost.
   d. SORT the resource groups with the increasing order of cost.
   e. If any of the resource has jobs assigned to it in the previous scheduling event, but not dispatched to the resource for execution and there is variation in resource availability, then move appropriate number of jobs to the Unassigned-Jobs-List. This helps in updating the whole schedule based on the latest resource availability information.
   f. Repeat the following steps for each resource group as long as there exist unassigned jobs:

      *Repeat the following steps for each job in the Unassigned-Jobs-List depending on the processing cost and the budget availability: [It uses the time optimisation strategy.]*

      - Select a job from the Unassigned-Jobs-List.
      - For each resource, calculate/predict the job completion time taking into account previously assigned jobs and the job completion rate and resource share availability.
      - Sort resources by the increasing order of completion time.
      - Assign the job to the first resource and remove it from the Unassigned-Jobs-List if the predicted job completion time is less than the deadline.

5. [DISPATCHER with Policy]

   *Repeat the following steps for each resource if it has jobs to be dispatched:*

   - Identify the number of jobs that can be submitted without overloading the resource. Our default policy is to dispatch jobs as long as the number of user jobs deployed (active or in queue) is less than the number of PEs in the resource.

**Figure 6.31: Deadline and budget constrained (DBC) scheduling with cost-time optimisation.**

### 6.6.1 Experiment Setup

The resources uses in evaluating the performance of cost-time optimisation scheduling are show in Table 6.3. The characteristics of resources is same as those used in previous experiment except that the price of resource R4 is set to the same value as the resource R8 to demonstrate the superior ability of cost-time optimisation scheduling algorithm over the cost optimisation scheduling algorithm. It can be noted some of the resources in Table 6.3 have the same MIPS per G$. For example, both R4 and R8 have the same cost and so resources R2, R3, and R10.

A task farming application containing of 200 jobs used in this scheduling experiment is same as the one used in previous experiments.

**Table 6.3: Resources used in Cost-Time scheduling simulation.**

| Resource Name in Simulation | Simulated Resource Characteristics<br>Vendor, Resource Type, Node OS, No of PEs | Equivalent Resource in Worldwide Grid<br>(Hostname, Location) | A PE SPEC/ MIPS Rating | Resource Manager Type | Price<br>(G$/PE time unit) | MIPS per G$ |
|---|---|---|---|---|---|---|
| R0 | Compaq, AlphaServer, CPU, OSF1, 4 | grendel.vpac.org,<br>VPAC, Melb, Australia | 515 | Time-shared | 8 | 64.37 |
| R1 | Sun, Ultra, Solaris, 4 | hpc420.hpcc.jp,<br>AIST, Tokyo, Japan | 377 | Time-shared | 4 | 94.25 |
| R2 | Sun, Ultra, Solaris, 4 | hpc420-1.hpcc.jp, AIST, Tokyo, Japan | 377 | Time-shared | 3 | 125.66 |
| R3 | Sun, Ultra, Solaris, 2 | hpc420-2.hpcc.jp, AIST, Tokyo, Japan | 377 | Time-shared | 3 | 125.66 |
| R4 | Intel, Pentium/VC820, Linux, 2 | barbera.cnuce.cnr.it, CNR, Pisa, Italy | 380 | Time-shared | 1 | 380.0 |
| R5 | SGI, Origin 3200, IRIX, 6 | onyx1.zib.de, ZIB, Berlin, Germany | 410 | Time-shared | 5 | 82.0 |
| R6 | SGI, Origin 3200, IRIX, 16 | Onyx3.zib.de, ZIB, Berlin, Germany | 410 | Time-shared | 5 | 82.0 |
| R7 | SGI, Origin 3200, IRIX, 16 | mat.ruk.cuni.cz, Charles U., Prague, Czech Republic | 410 | Space-shared | 4 | 102.5 |
| R8 | Intel, Pentium/VC820, Linux, 2 | marge.csm.port.ac.uk, Portsmouth, UK | 380 | Time-shared | 1 | 380.0 |
| R9 | SGI, Origin 3200, IRIX, 4 (accessible) | green.cfs.ac.uk, Manchester, UK | 410 | Time-shared | 6 | 68.33 |
| R10 | Sun, Ultra, Solaris, 8, | pitcairn.mcs.anl.gov, ANL, Chicago, USA | 377 | Time-shared | 3 | 125.66 |

### 6.6.2 Scheduling Experiments with Cost and Cost-Time Optimisation Strategies

We perform both cost and cost-time optimisation scheduling experiments with different values of deadline and budget constraints (DBC) for a single user. The deadline is varied in simulation time from 100 to 3600 in steps of 500. The budget is varied from G$ 5000 to 22000 in steps of 1000. The number of Gridlets processed, deadline utilized, and budget spent for the DBC cost-optimisation scheduling strategy is shown in Figure 6.32a, Figure 6.32c, and Figure 6.32e, and for the cost-time optimisation scheduling strategy is shown in Figure 6.32b, Figure 6.32d, and Figure 6.32f. In both cases, when the deadline is low (e.g., 100 time unit), the number of Gridlets processed increases as the budget value increases. When a higher budget is available, the broker leases expensive resources to process more jobs within the deadline. Alternatively, when scheduling with a low budget value, the number of Gridlets processed increases as the deadline is relaxed.

The impact of budget for different values of deadline is shown in Figure 6.32e and Figure 6.32f for cost and cost-time strategies. For a larger deadline value (see the time utilization for deadline of 3600), the increase in budget value does not have much impact on resource selection. When the deadline is too tight (e.g., 100), it is likely that the complete budget is spent for processing Gridlets within the deadline.

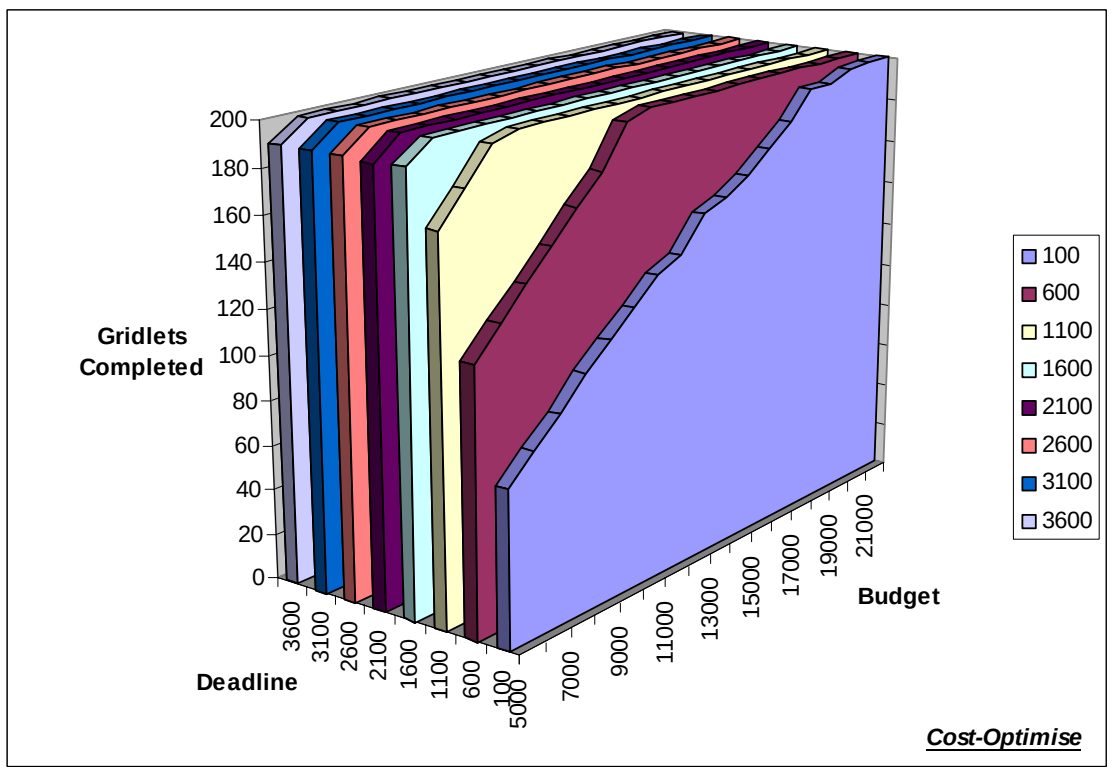

**(a) No. of Gridlets processed.**

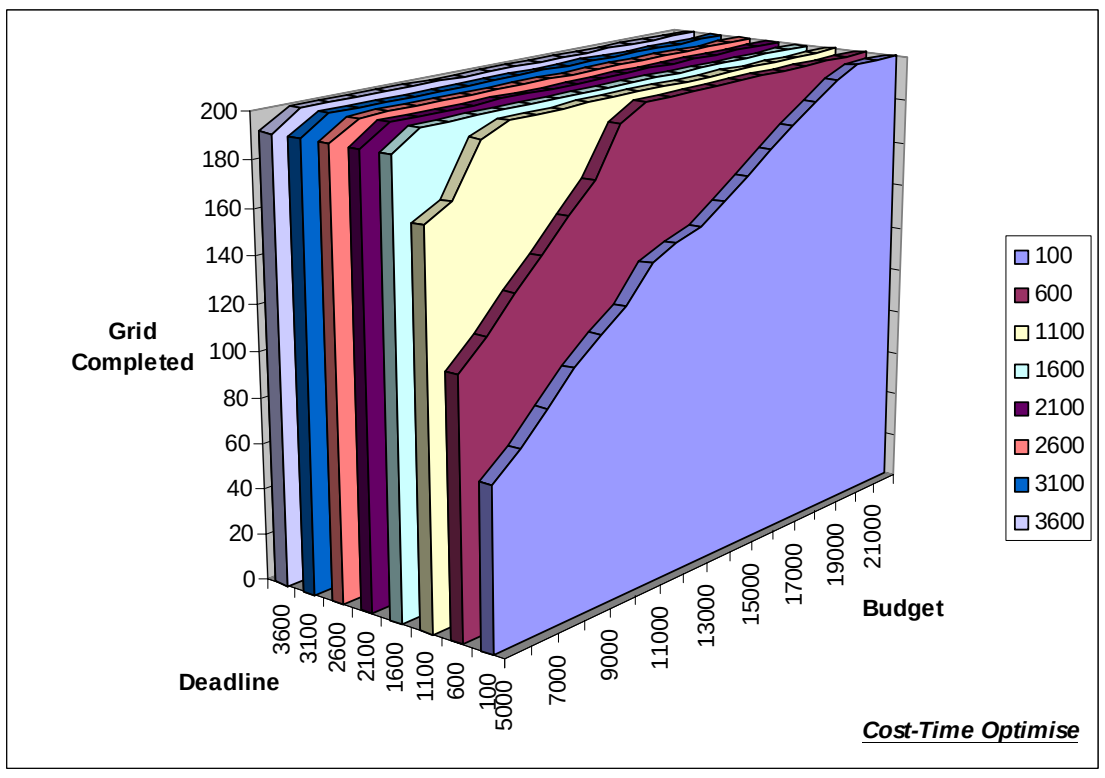

**(b) No. of Gridlets processed**

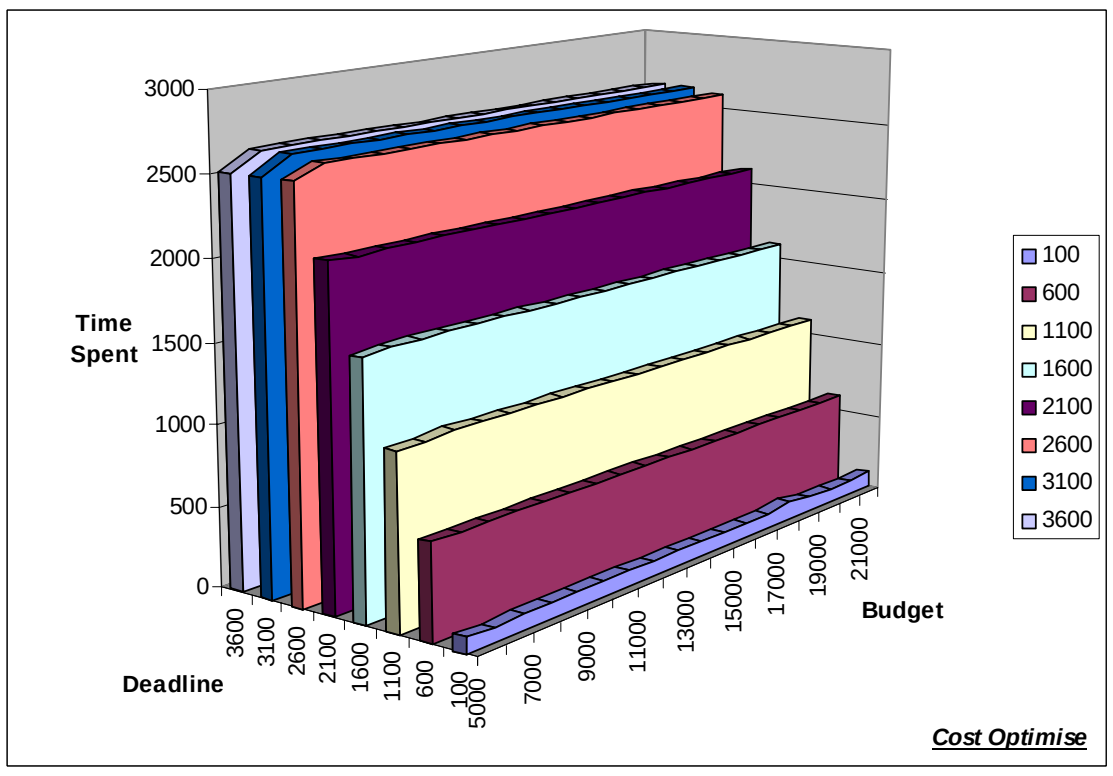

**(c) Time spent for processing Gridlets.**

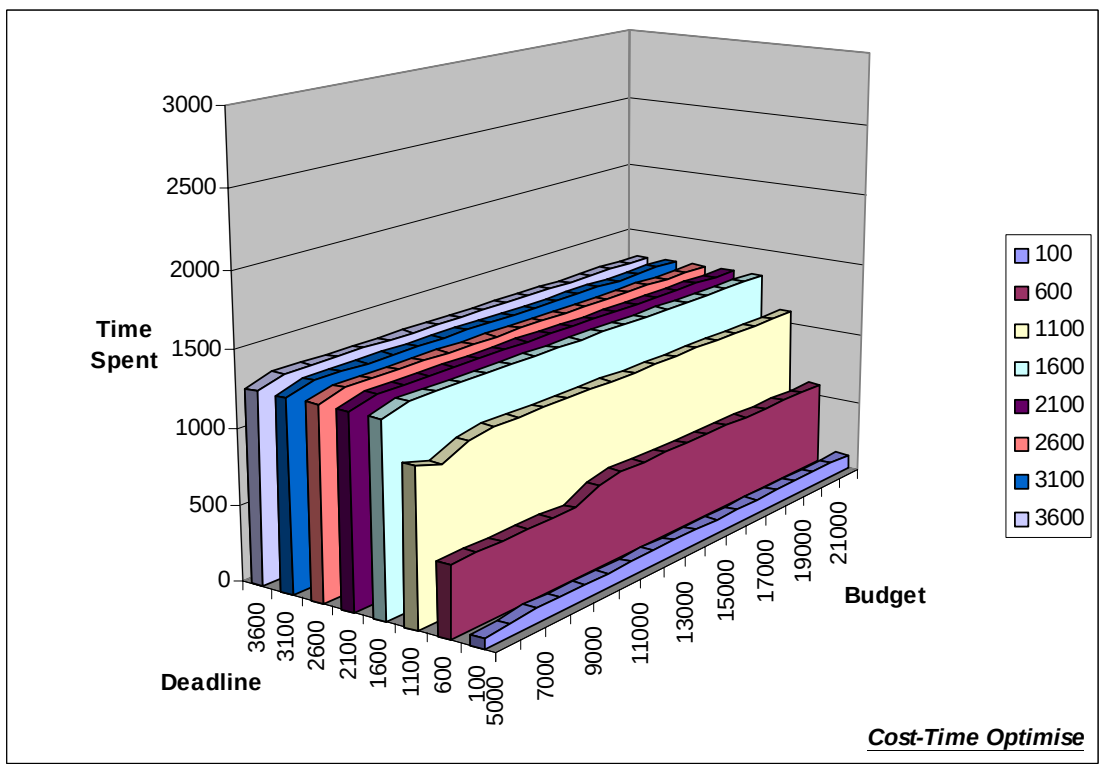

**(d) Time spent for processing Gridlets.**

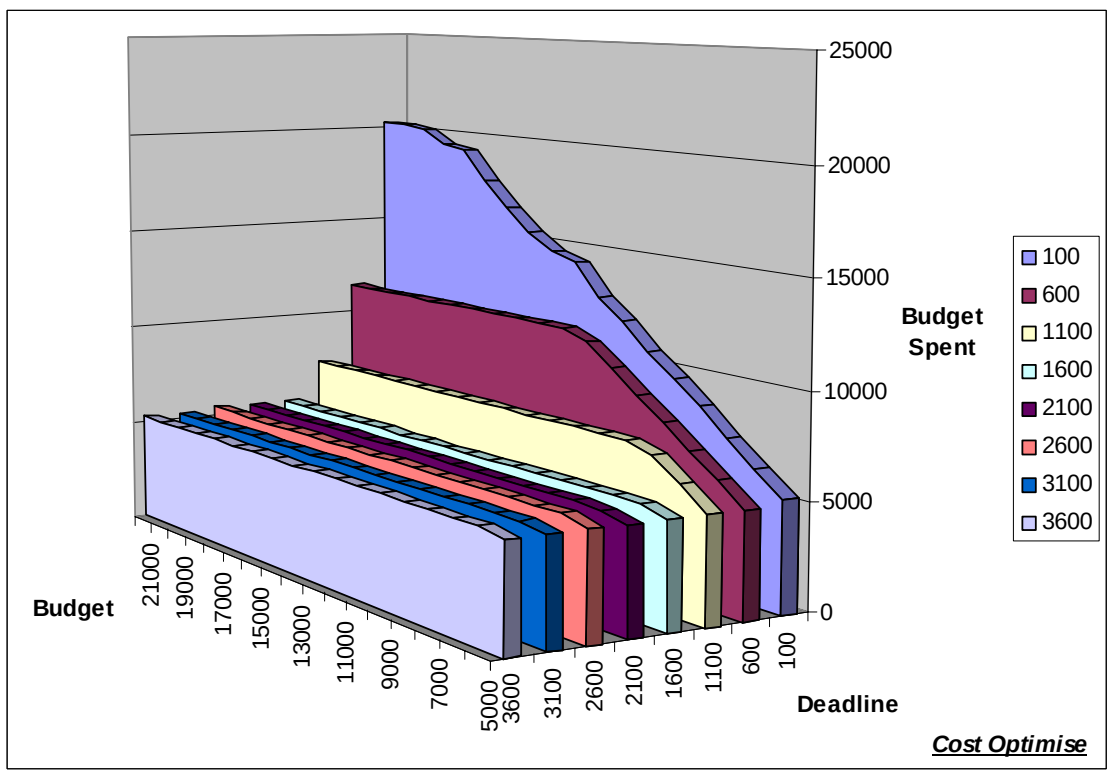

**(e) Budget spent for processing Gridlets.**

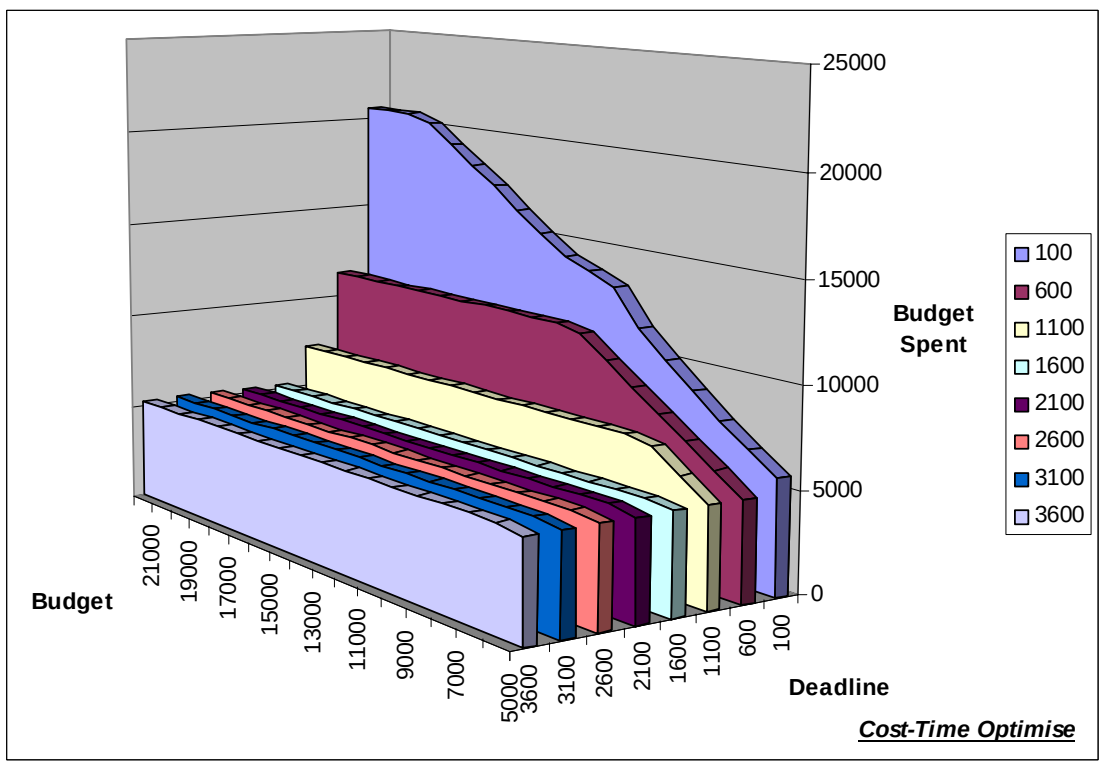

**(f) Budget spent for processing Gridlets.**

**Figure 6.32: The number of Gridlets processed, time, and budget spent for different deadline and time limits when scheduled using the cost and cost-time optimisation algorithms.**

It can be observed that the number of Gridlets processed and the budget-spending pattern is similar for both scheduling strategies. However, the time spent for the completion of all the jobs is significantly different (see Figure 6.32c and Figure 6.32d), as the deadline becomes relaxed. For deadline values from 100 to 1100, the completion time for both cases is similar, but as the deadline increases (e.g., 1600 to 3600), the experiment completion time for cost-time scheduling optimisation strategy is much less than the cost optimisation scheduling strategy. This is because when there are many resources with the same MIPS per G$, the cost-time optimisation scheduling strategy allocates jobs to them using the time-optimisation

strategy for the entire deadline duration since there is no need to spent extra budget for doing so. This does not happen in case of cost-optimisation strategy—it allocates as many jobs that the first cheapest resource can complete by the deadline and then allocates the remaining jobs to the next cheapest resources.

A trace of resource selection and allocation using cost and cost-time optimisation scheduling strategies shown in Figure 6.33 indicates their impact on the application processing completion time. When the deadline is tight (e.g., 100), there is high demand for all the resources in short time, the impact of cost and cost-time scheduling strategies on the completion time is similar as all the resources are used up as long as budget is available to process all jobs within the deadline (see Figure 6.33a and Figure 6.33b). However, when the deadline is relaxed (e.g., 3100), it is likely that all jobs can be completed using the first few cheapest resources. In this experiment there were resources with the same cost and capability (e.g., R4 and R8), the cost optimisation strategy selected resource R4 to process all the jobs (see Figure 6.33c); whereas the cost-time optimisation strategy selected both R4 and R8 (see Figure 6.33d) since both resources cost the same price and completed the experiment earlier than the cost-optimisation scheduling (see Figure 6.32c and Figure 6.32d). This situation can be observed clearly in scheduling experiments with a large budget for different deadline values (see Figure 6.34). Note that the left most solid curve marked with the label “All” in the resources axis in Figure 6.34 represents the aggregation of all resources.

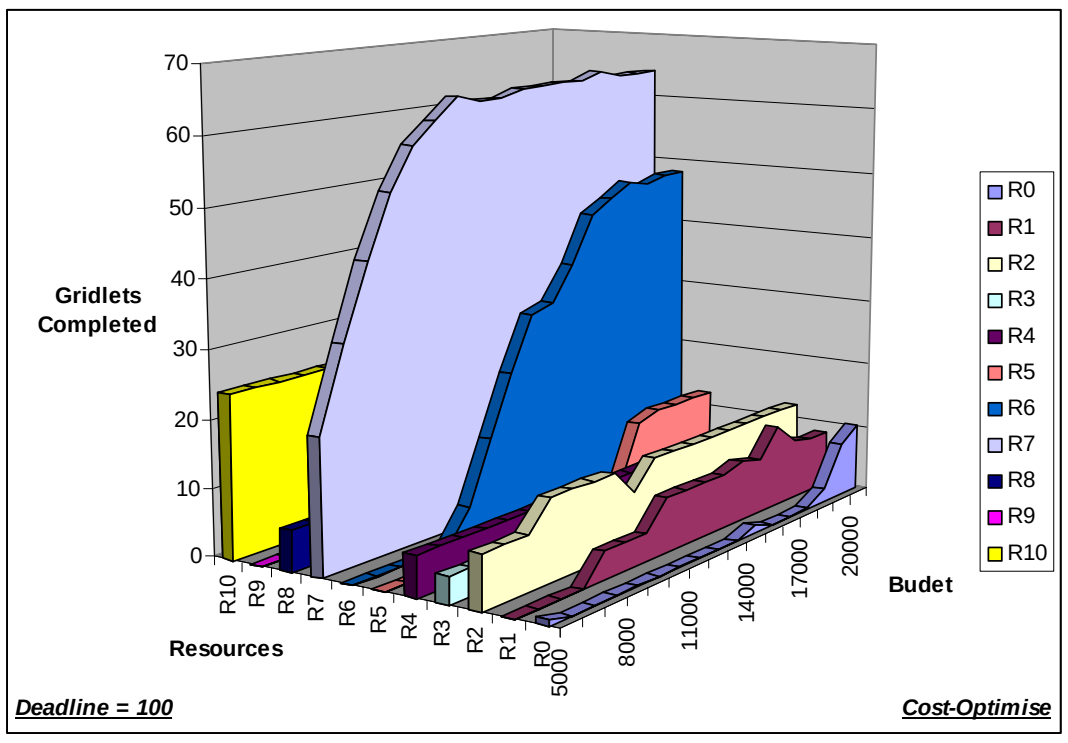

**(a) Cost optimisation with a low deadline.**

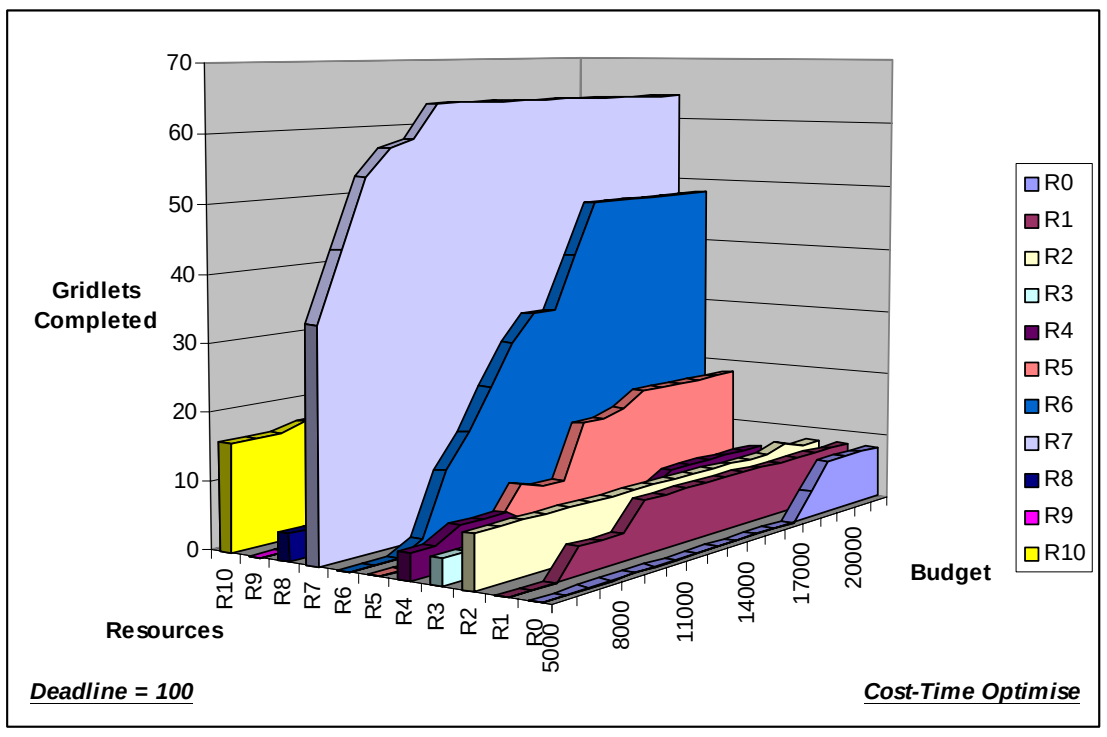

**(b) Cost-time optimisation with a low deadline.**

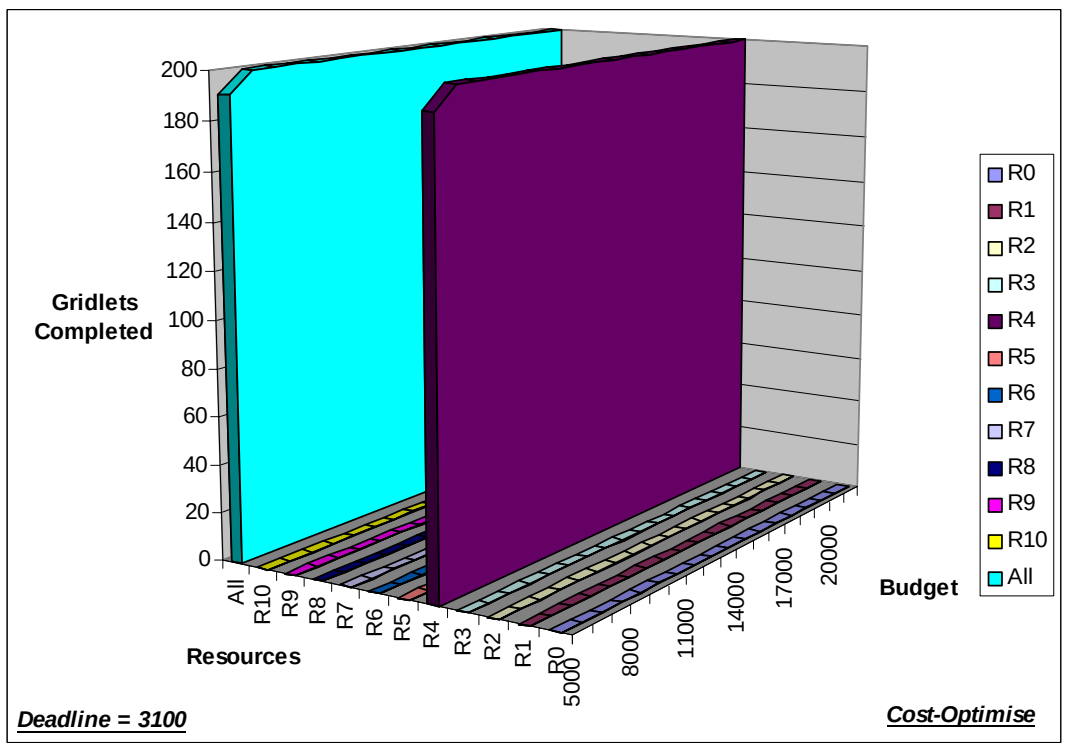

**(c) Cost optimisation with a high deadline.**

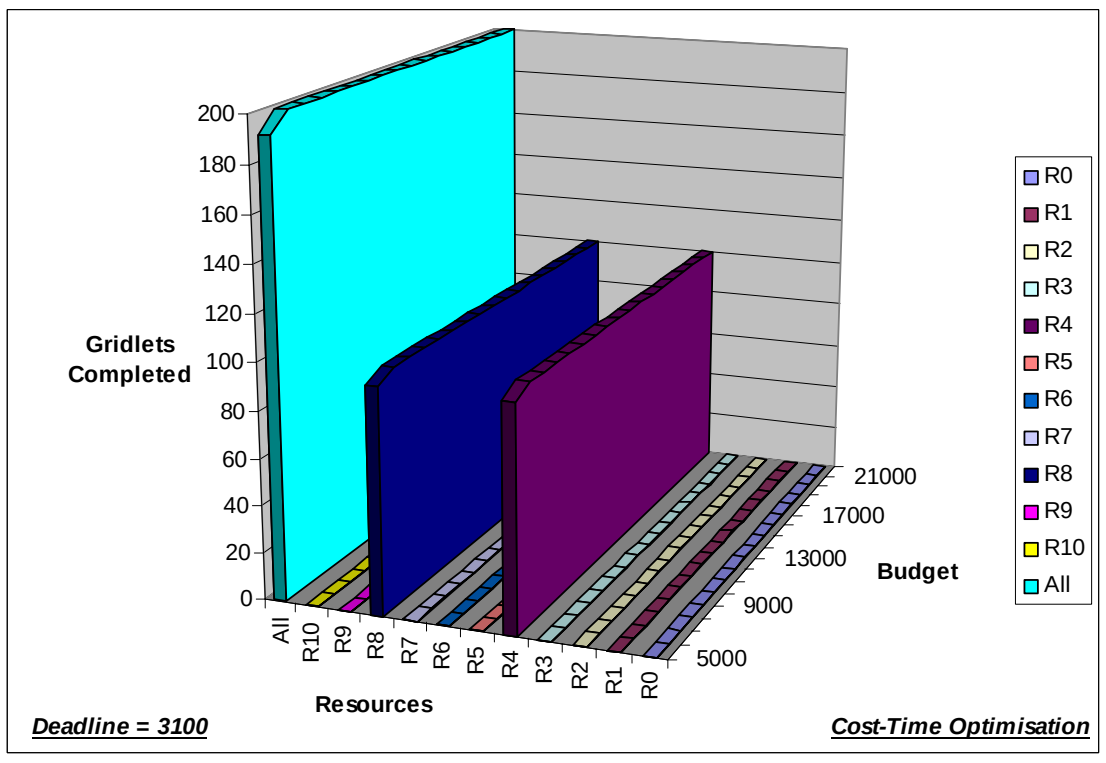

**(d) Cost-time optimisation with a high deadline.**

**Figure 6.33: The number of Gridlets processed and resources selected for different budget values with a high deadline value when scheduled using the cost and cost-time optimisation algorithms.**

As the deadline increases, the cost optimisation algorithm predominantly scheduled jobs on the resource R4 (see Figure 6.34a) whereas, the cost-time optimisation algorithm scheduled jobs on resources R4 and R8 (see Figure 6.34a), the first two cheapest resources with same cost. Therefore, the application scheduling using the cost-time optimisation algorithm is able to finish earlier compared to the one scheduled using the cost optimisation algorithm (see Figure 6.35) and both strategies have spent the same amount of budget for processing its jobs (see Figure 6.36). The completion time for cost optimisation scheduling continued to increase with increase of the deadline as the broker allocated more jobs to the resource R4 and less to the resource R8. However, the completion time for deadline values 3100 and 3660 is the same as the previous one since the broker allocated jobs to only resource R4. This is not the case with the cost-time optimisation scheduling since jobs are allocated proportionally to both resources R4 and R8 and thus minimizing the completion time without spending any extra budget.

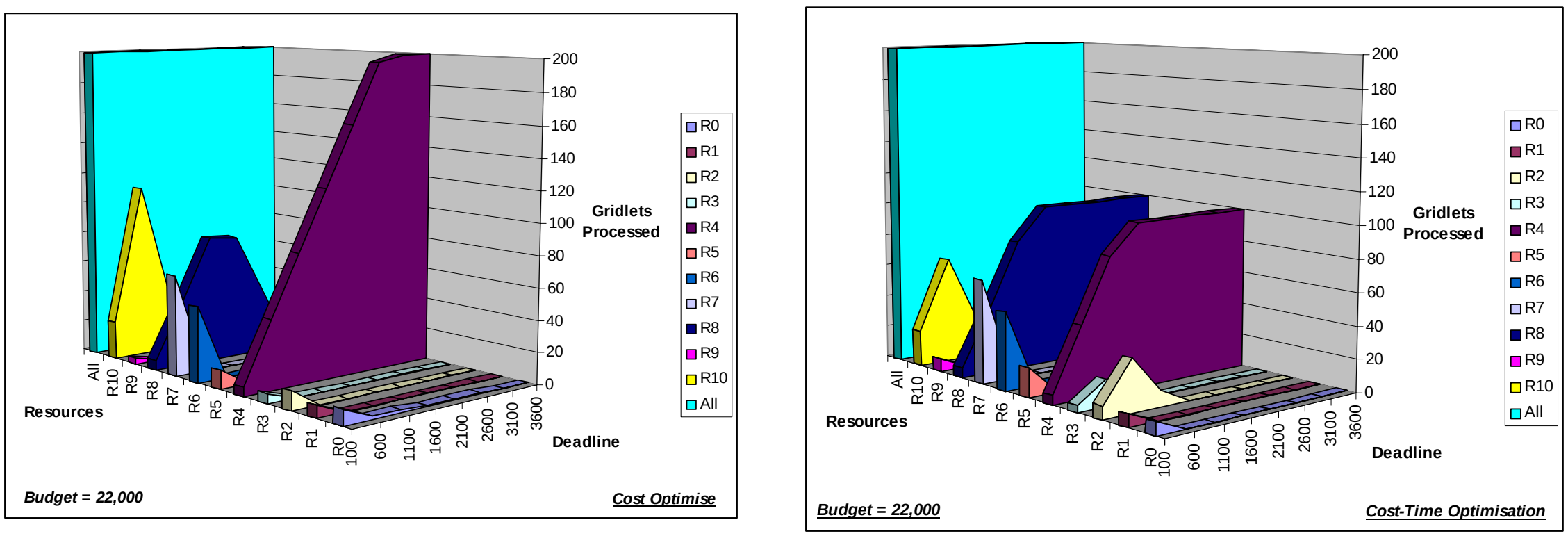

**(a) Resource selection when the budget is high.** **(b) Resource selection when the budget is high.**

**Figure 6.34: The number of Gridlets processed and resources selected for different deadline values with a large budget when scheduled using the cost and cost-time optimisation algorithms.**

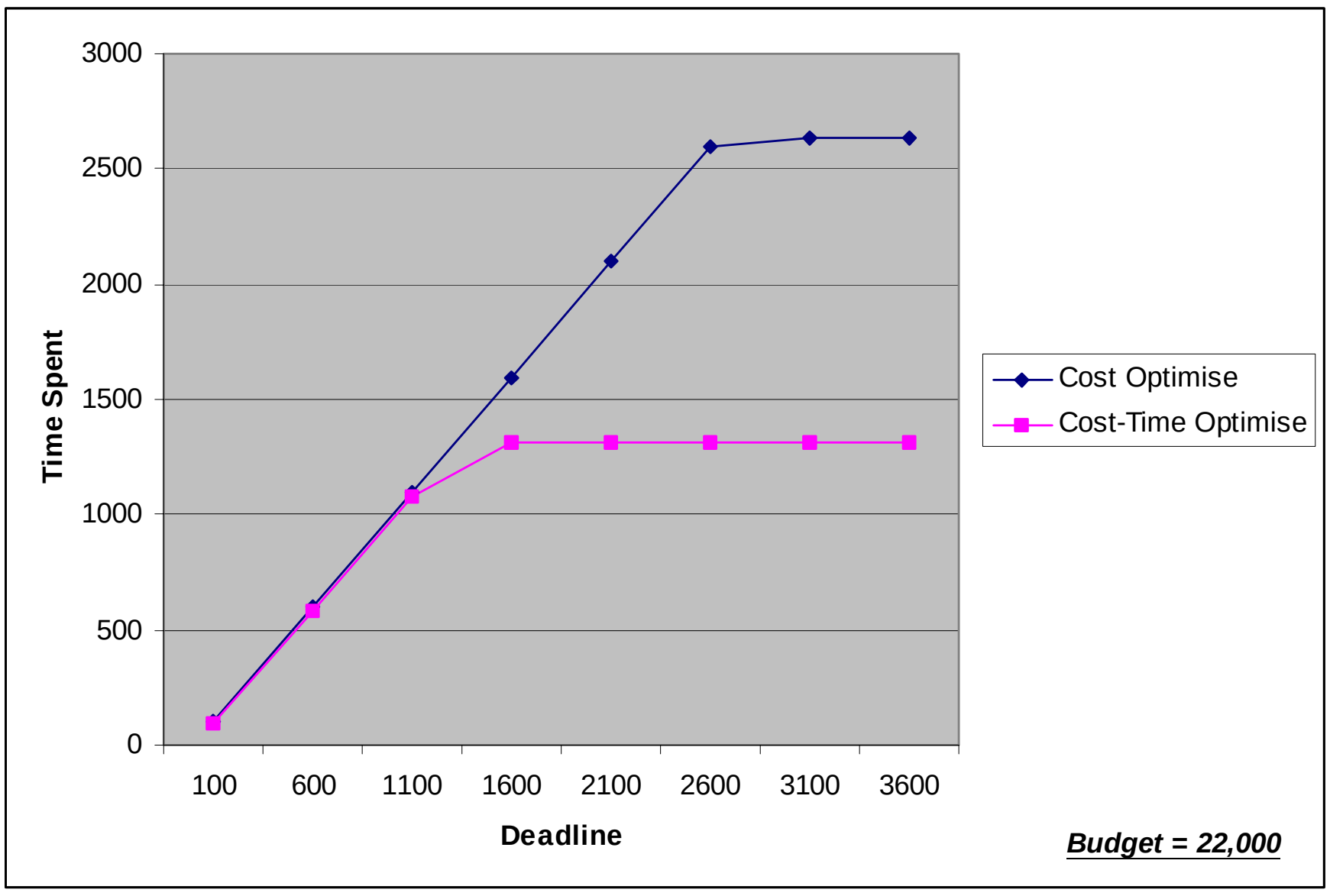

**Figure 6.35: The time spent for processing application jobs for different deadline constraints with a large budget when scheduled using the cost and cost-time optimisation algorithms.**

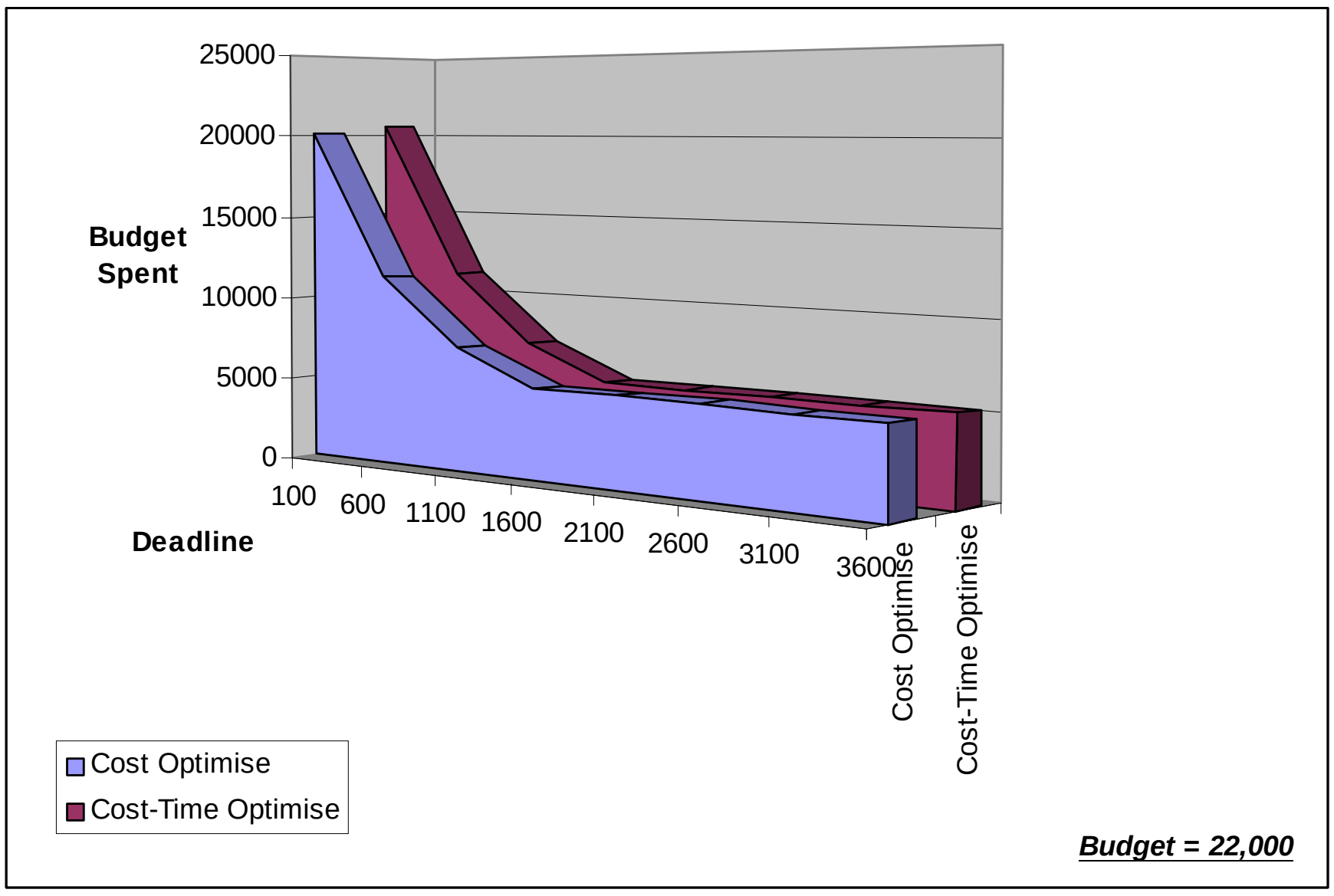

**Figure 6.36: The budget spent for processing application jobs for different deadline constraints with a large budget when scheduled using the cost and cost-time optimisation algorithms.**

Let us now take a microscopic look at the allocation of resources when a moderate deadline and large budget is assigned. A trace of resource allocation and the number of Gridlets processed at different times when scheduled using the cost and cost-time optimisation algorithms is shown in Figure 6.37 and Figure 6.38. It can be observed that for both the strategies, the broker used the first two cheapest resources, R4 and R8 fully. Since the deadline cannot be completed using only these resources, it used the next cheapest resources R2, R3, and R10 to make sure that deadline can be meet. The cost optimisation strategy allocated Gridlets to resource R10 only, whereas the cost-time optimisation allocated Gridlets to resources R2, R3, and R10 as they cost the same price. Based on the availability of resources, the broker predicts the number of Gridlets that each resource can complete by the deadline and allocates to them accordingly (see Figure 6.39 and Figure 6.40). At each scheduling event, the broker evaluates the progress and resource availability and if there is any change, it reschedules some Gridlets to other resources to ensure that the deadline can be meet. This can be observed in Figure 6.39 and Figure 6.40—the broker allocated a few extra Gridlets to resource R10 (cost-optimisation strategy) and resources R2, R3, and R10 (cost-time optimisation strategy) during the first few scheduling events.

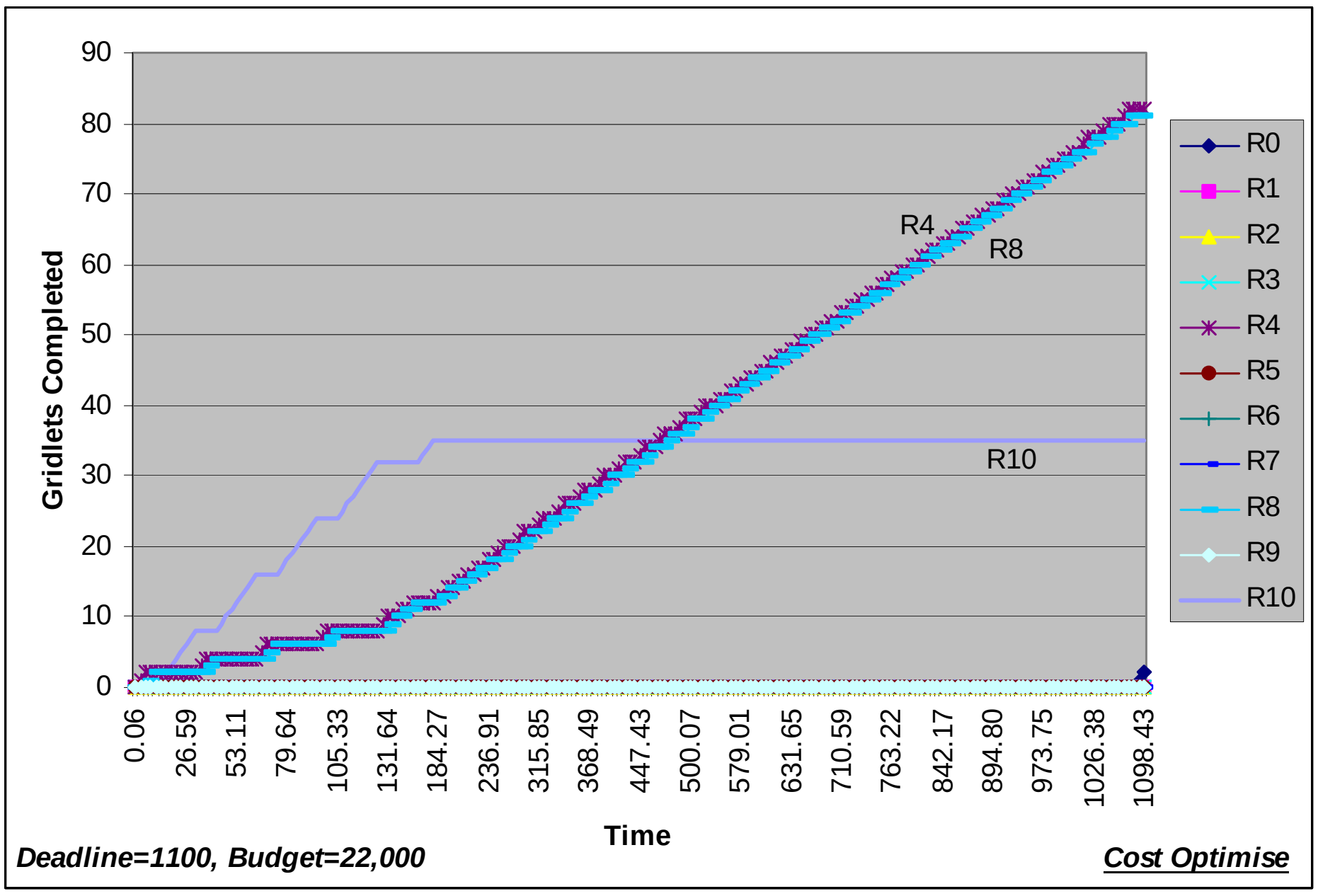

**Figure 6.37: Trace of No. of Gridlets processed on resources for a medium deadline and high budget constraints when scheduled using the cost optimisation strategy.**

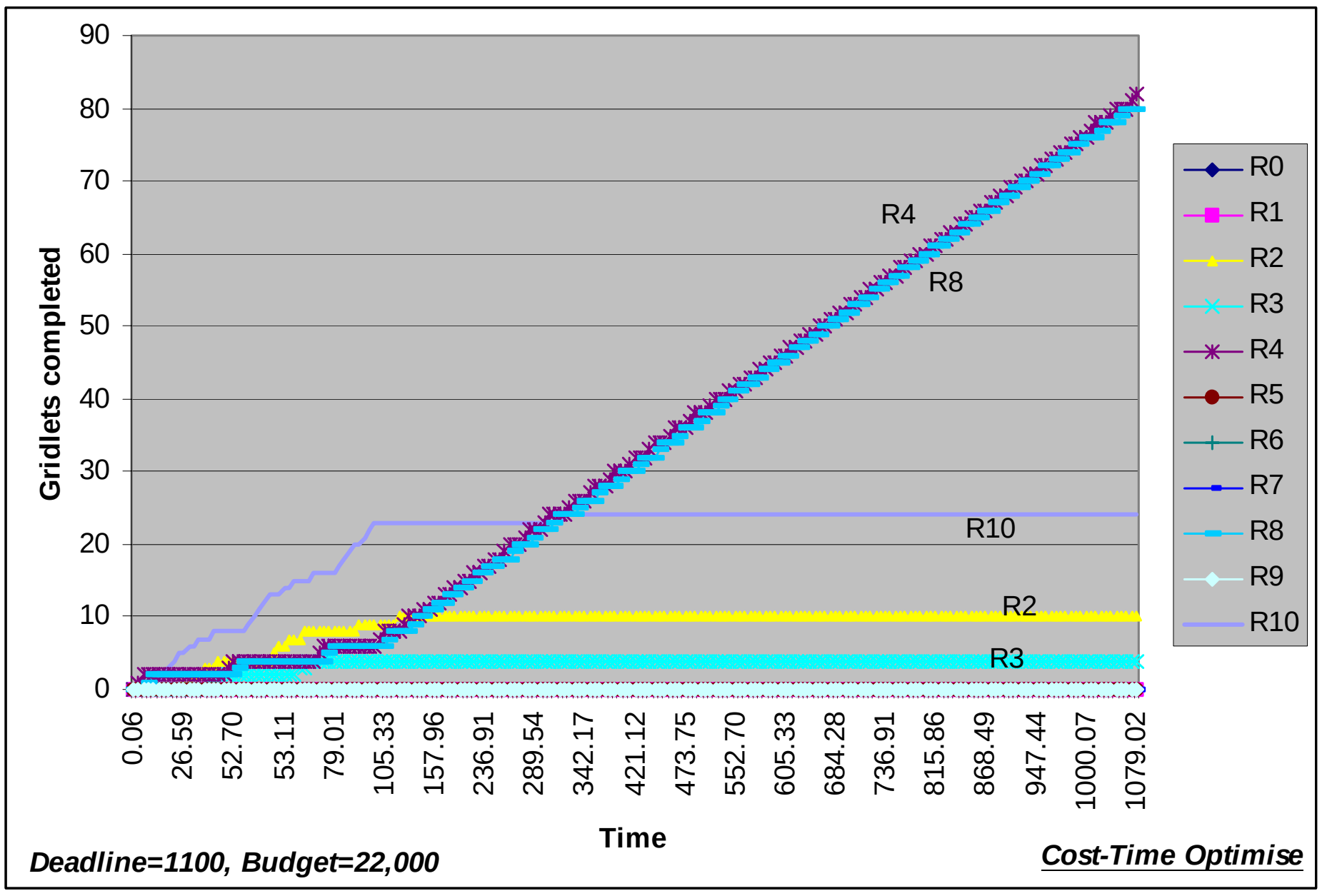

**Figure 6.38: Trace of No. of Gridlets processed on resources for a medium deadline and high budget constraints when scheduling using the cost-time optimisation strategy.**

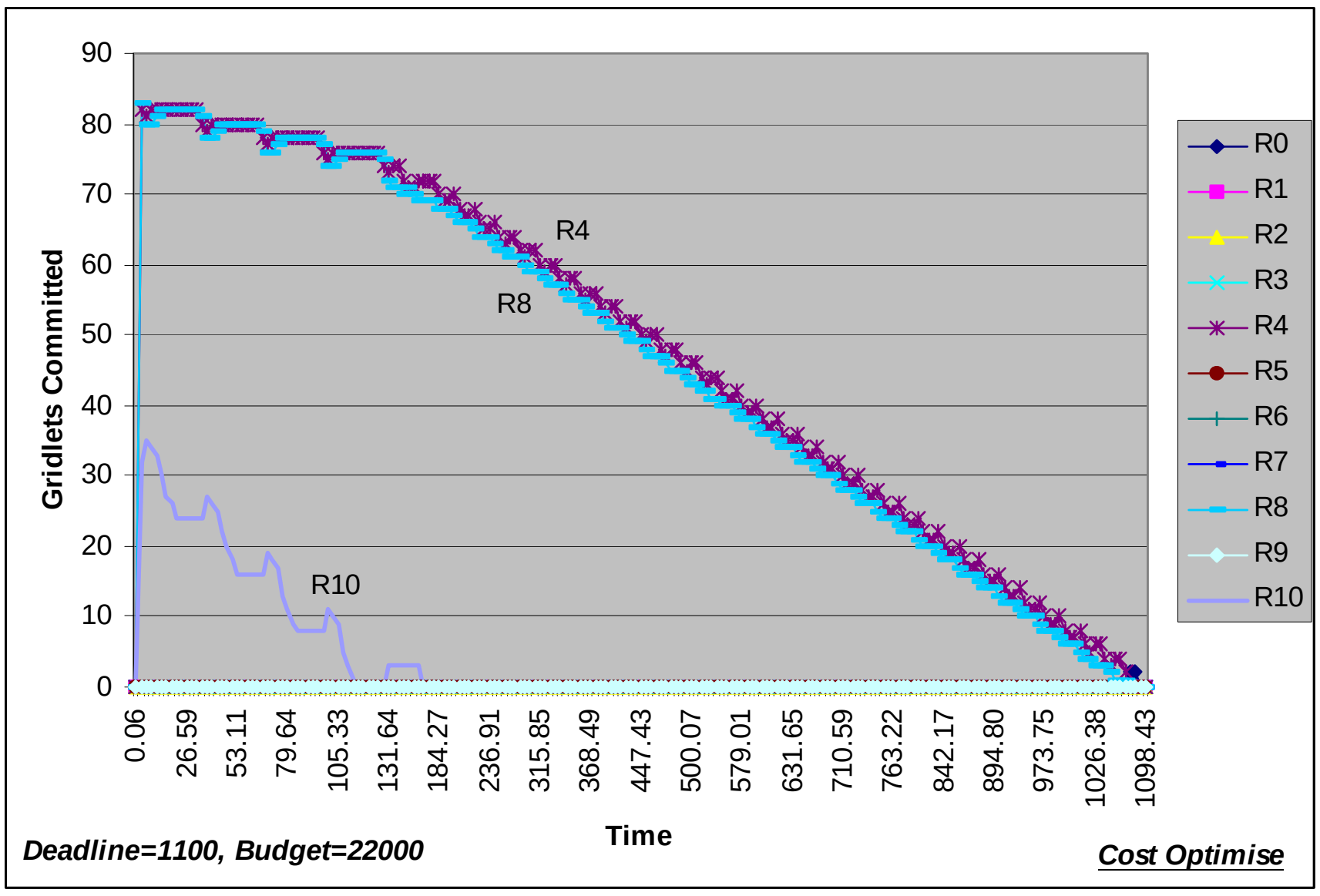

**Figure 6.39: Trace of the number of Gridlets committed to resources for a medium deadline and high budget constraints when scheduled using the cost optimisation strategy.**

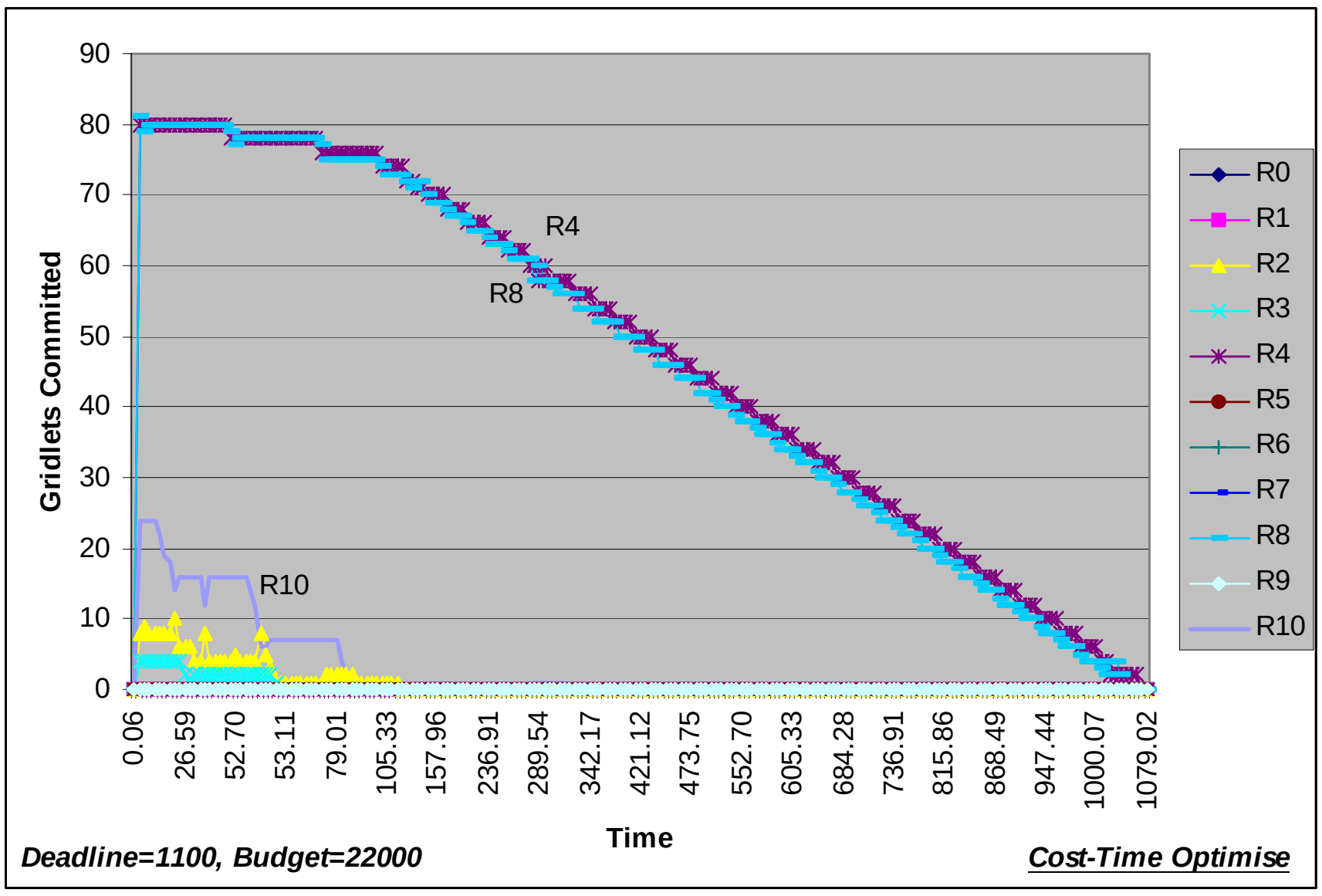

**Figure 6.40: Trace of the number of Gridlets committed to resources for a medium deadline and high budget constraints when scheduled using the cost-time optimisation strategy.**

In summary, when there are multiple resources with the same cost and capability, the cost-time optimisation algorithm schedules jobs on them using the time-optimisation strategy for the deadline period. The results of scheduling experiments for many scenarios with a different combination of the deadline and budget constraints, we observe that applications scheduled using the *cost-time* optimisation are able to complete earlier than those scheduled using the cost optimisation algorithm without incurring any extra expenses. This proves the superiority of the new deadline and budget constrained cost-time optimisation algorithm in scheduling jobs on global Grids.

## 6.7 Summary and Conclusion

We discussed the use of computational economy as a metaphor for devising scheduling strategies for large-scale applications on distributed resources. We used the GridSim toolkit in simulating an economic-based Grid resource broker that supports deadline and budget-based scheduling. We simulated and evaluated performance of scheduling algorithms with cost, time, and cost-time optimisation strategies for a variety of scenarios. The broker is able allocate resources depending on the users' quality of service requirements such as the deadline, budget, and optimisation strategy. The performance evaluation results at microscopic level reveal their impact on the application processing cost and time; and demonstrate the usefulness of allowing uses to trade-off between the timeframe and processing cost depending on their QoS requirements. Also, these extensive simulation studies demonstrate the suitability of GridSim for developing simulators for scheduling in parallel and distributed systems.

## Software Availability

The economic Grid resource broker simulator with source code can be downloaded from the GridSim project website:

`http://www.buyya.com/gridsim/`

# Chapter 7

# The Virtual Laboratory: Enabling Drug Design on the Grid

This chapter presents the design and development of a virtual laboratory environment that enables molecular modelling for drug design on geographically distributed data and computational resources. It leverages existing Grid technologies such as the Nimrod-G parameter specification language to transform the existing molecular docking application into a parameter sweep application and the Nimrod-G broker for scheduling jobs on distributed resources. It provides additional tools for enabling and providing distributed access to ligand records/molecules in the chemical databases (CDB) located remotely. The results of scheduling experiments docking jobs on the World-Wide Grid (WWG) resources is being presented to demonstrate the ease of use and power of the Nimrod-G and virtual laboratory tools for Grid computing.

## 7.1 Introduction

Computational Grids serve as a scalable computing platform for executing large-scale computational and data intensive applications in parallel through the *aggregation* of geographically distributed computational resources. They enable exploration of large problems in science, engineering, and business with huge data sets, which is essential for creating new insights into the problem. Molecular modelling for drug design is one of the scientific applications that can benefit from the availability of such a large computational capability.

### *Drug Discovery Process*

Drug discovery is an extended process that can take as many as 15 years from the first compound synthesis in the laboratory until the therapeutic agent, or drug, is brought to market [25]. Reducing the research timeline in the discovery stage is a key priority for pharmaceutical companies worldwide. Many such companies are trying to achieve this goal through the application and integration of advanced technologies such as computational biology, chemistry, computer graphics, and high performance computing (HPC). Molecular modelling has emerged as a popular methodology for drug design—it can combine computational chemistry and computer graphics. Molecular modelling can be implemented as a master-worker parallel application, which can take advantage of HPC technologies such as clusters [93] and Grids for large-scale data exploration.

Drug design using molecular modelling techniques involve screening a very large number (of the order of a million) of ligand[4] records or molecules of compounds in a chemical database (CDB) to identify those that are potential drugs. This process is called molecular *docking*. It helps scientists in predicting how small molecules, such as drug candidates, bind to an enzyme or a protein receptor of known 3D structure (see Figure 7.1). Docking each molecule in the target chemical database is both a compute and data intensive task. It is our goal to use Grid technologies to provide cheap and efficient solutions for the execution of molecular docking tasks on large-scale, wide-area parallel and distributed systems.

While performing docking, information about the molecule must be extracted from one of the many large chemical databases. As each chemical database requires storage space in the order of hundreds of megabytes to terabytes, it is not feasible to transfer the chemical database to all resources in the Grid. Also, each docking job only needs a ligand or module record, not the whole database. Therefore, access to a chemical database must be provided as a *network service* (see Figure 7.2). The chemical databases need to be selectively replicated on a few nodes to avoid any bottleneck, which might happen due to providing access to the database from a single source. Intelligent mechanisms (e.g., CDB broker) need to be supported for selecting optimal sources for CDB services depending on the location of resources selected for processing docking jobs.

[4] An ion, a molecule, or a molecular group that binds to another chemical entity to form a larger complex.

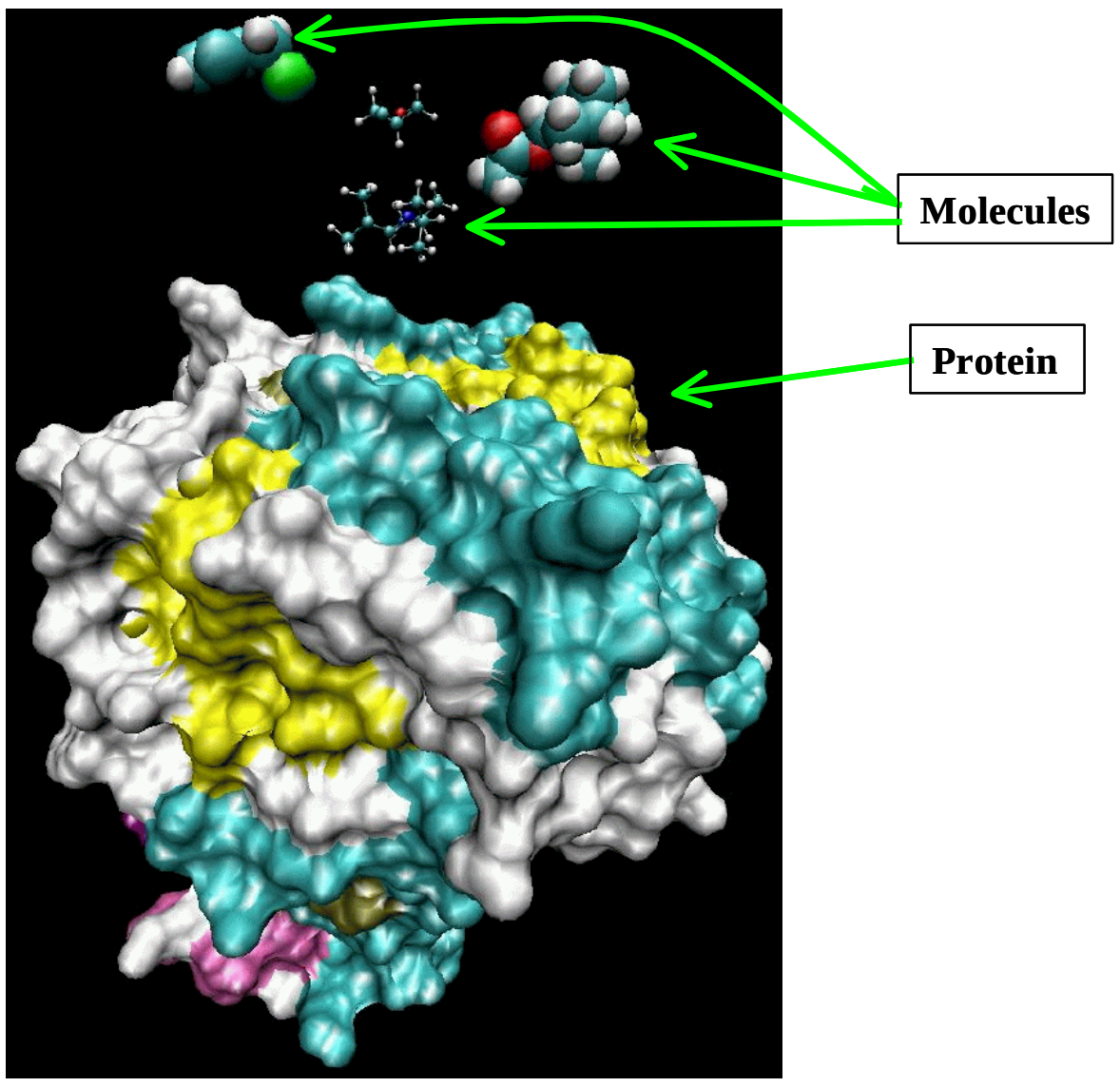

**Figure 7.1: X-ray crystal structure of a target protein receptor and small molecules to be docked.**

Fundamentally, drug design is a *computational and data challenge* problem since it involves screening millions of compounds in chemical databases. Screening each compound, depending on structural complexity, can take from a few minutes to hours on a standard PC, which means screening all compounds in a single database can take years! For example, we are looking into a drug design problem that involves screening 180,000 compounds. Each job screening a compound is expected to take up to 3 hours of execution time on a desktop computer (e.g., Pentium-based Linux/Windows PC). That means, if we aim to screen all these compounds on a single PC, it can take up to 540000 hours, which is roughly equivalent to 61 years! If we use a typical cluster-based supercomputer with 64 nodes, we can solve this problem in one year. The problem can be solved with a large scale Grid of hundreds of supercomputers within a day. If we use a massive network of peer-to-peer style Grid computing infrastructure such as SETI@Home [143], the drug discovery problem could be solved within a few hours.

The rest of this chapter is organized as follows. A high-level operational model for molecular modelling on the Grid is presented in Section 7.2. A layered architecture for building the Virtual Laboratory environment is discussed in Section 7.3. It leverages the existing Grid technologies and supports new tools that are essential for Grid-enabling the chemical database and the docking application on distributed resources. Formulation of molecular docking as a parameter sweep application is presented in Section 7.4. The results of two experiments on scheduling molecular docking jobs for processing on the WWG (World Wide Grid) [113] testbed resources are presented in Section 7.5. The final section summarizes the chapter along with suggestions for future works.

## 7.2 Operational Model

The Virtual Laboratory tools transform the existing molecular modelling application (without the need for making any changes to it) into a parameter sweep application for executing jobs docking molecules in the CDBs in parallel on distributed resources. The parameterized application contains multiple independent jobs, each screening different compounds to identify their drug potential. These jobs are computationally intensive in nature and only a small proportion of the execution time is spent on data communication (e.g., fetching molecular information on demand from remote databases). Applications expressed with this task-farming computational model have high *computation to communication* ratio. Hence, they can tolerate high network latency, which makes them suitable for executing in parallel on Internet-wide distributed resources.

A high-level operation model of docking molecules on the Grid is shown in Figure 7.2. The drug designer formulates the molecular docking problem, submits the application to the Grid resource broker (e.g., Nimrod-G [100]) along with performance and optimisation requirements—"screen 2000 molecules within 30 minutes and the available budget for processing is $10". The broker discovers resources, establishes their cost and capability, and then prepares a schedule to map docking jobs to resources. Let us say, it identified a GSP (Grid Service Provider), say GSP2, and assigned a job of screening a molecule 5 to it. A job has a task specification that specifies a list of operations to be performed. To process a job on GSP2, the broker dispatcher deploys its Agent on resource GSP2. The agent executes a list of commands specified in the job's task specification. A typical task specification contains necessary commands to copy executables and input files from the user machine, substitution of parameters declared in the input file, execution of the program, and finally copying results back to the user. It can also contain special commands for accessing the input data from the remote database. For example, a docking task can contain a special command (e.g., an instruction to fetch molecule record from the CDB) to make a request to the data broker (e.g., CDB broker) for a molecule record. The data broker looks at the replica catalogue for a list of sites providing CDB services, checks the status of those sites, and selects a suitable site (e.g., a node with fast network connectivity) and recommends the same. The molecule fetch command can then request the CDB service provider for a molecule record and write the molecule structure to a file that acts as an input to the docking program. After executing the docking program, the agent executes commands related to copying docking results to the user home node.

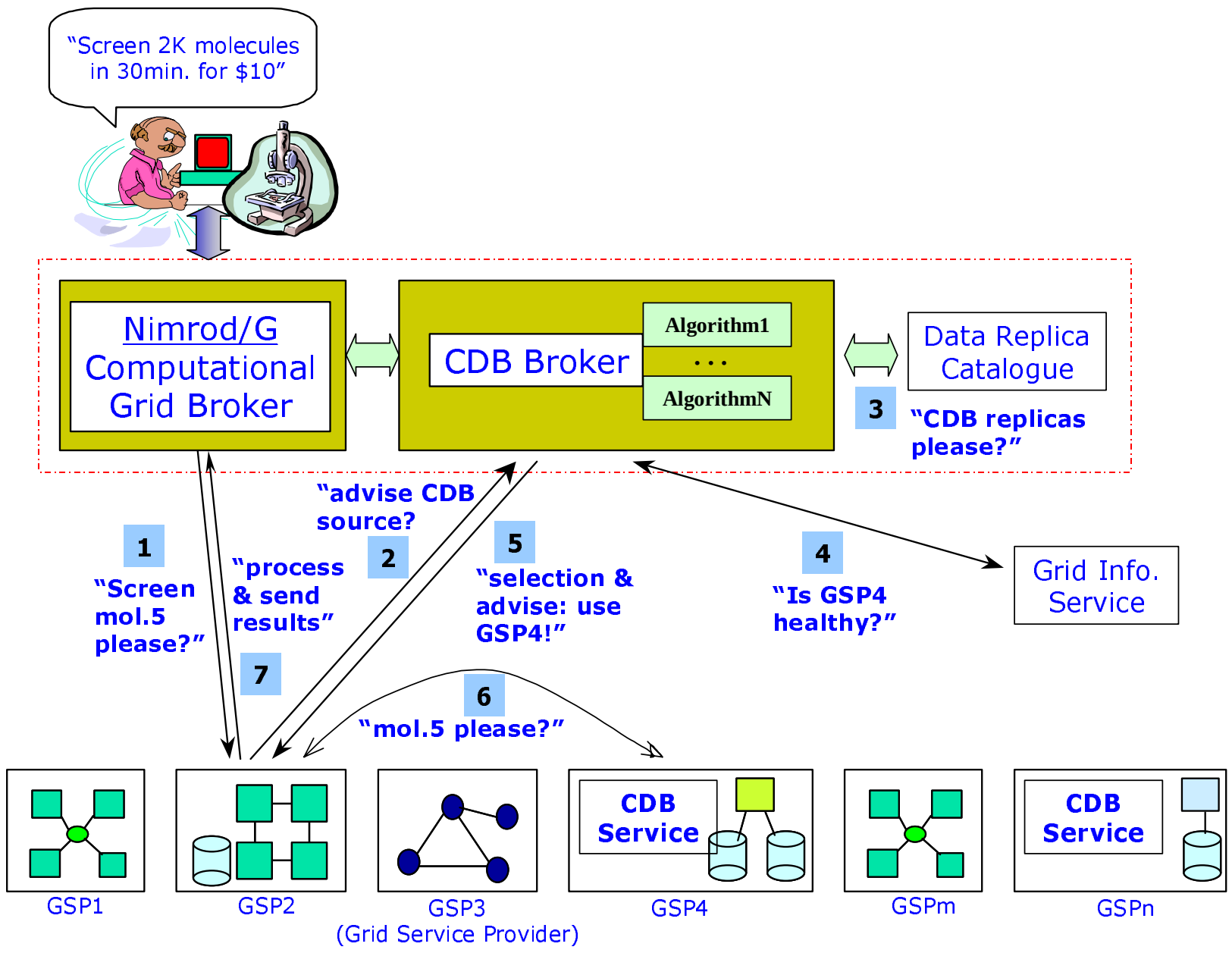

**Figure 7.2: Resource brokering architecture for screening molecules on distributed resources.**

## 7.3 Architecture – The Software Stack

The Virtual Laboratory builds on the existing Grid technologies and tools for performing data intensive computing on distributed resources. It provides new tools for managing and accessing remote chemical databases as a network service. There are many scientific and commercial applications (e.g., molecular modelling, high-energy physics events processing, and financial investment risk-analysis) that explore a range of scenarios. Instead of explicitly developing them as parallel applications using interfaces such as MPI, they can be composed as parameter sweep applications using tools such as Nimrod [21]. Such application jobs can be executed in parallel on distributed resources using the Nimrod-G resource broker (see Figure 7.2). A layered architecture and the software stack essential for performing molecular modelling on distributed resources is depicted in Figure 7.3. The components of the Virtual Laboratory software stack are:

- The DOCK software for Molecular Modelling [11].
- The Nimrod Parameter Modelling Tools [137] for enabling DOCK as a parameter sweep application.
- The Nimrod-G Grid Resource Broker [100] for scheduling DOCK jobs on the Grid.
- Chemical Database (CDB) Management and Intelligent Access Tools:
  - CDB database lookup/Index table generation.
  - CDB and associated index-table replication.
  - CDB replica catalogue for CDB resource discovery.
  - CDB servers for providing CDB services
  - CDB broker for selecting a suitable CDB service (Replica Selection).
  - CDB clients for fetching molecular records (Data Movement).
- The GrACE software for resource trading toolkit [99].
- The Globus middleware for secure and uniform access to distributed resources [49].

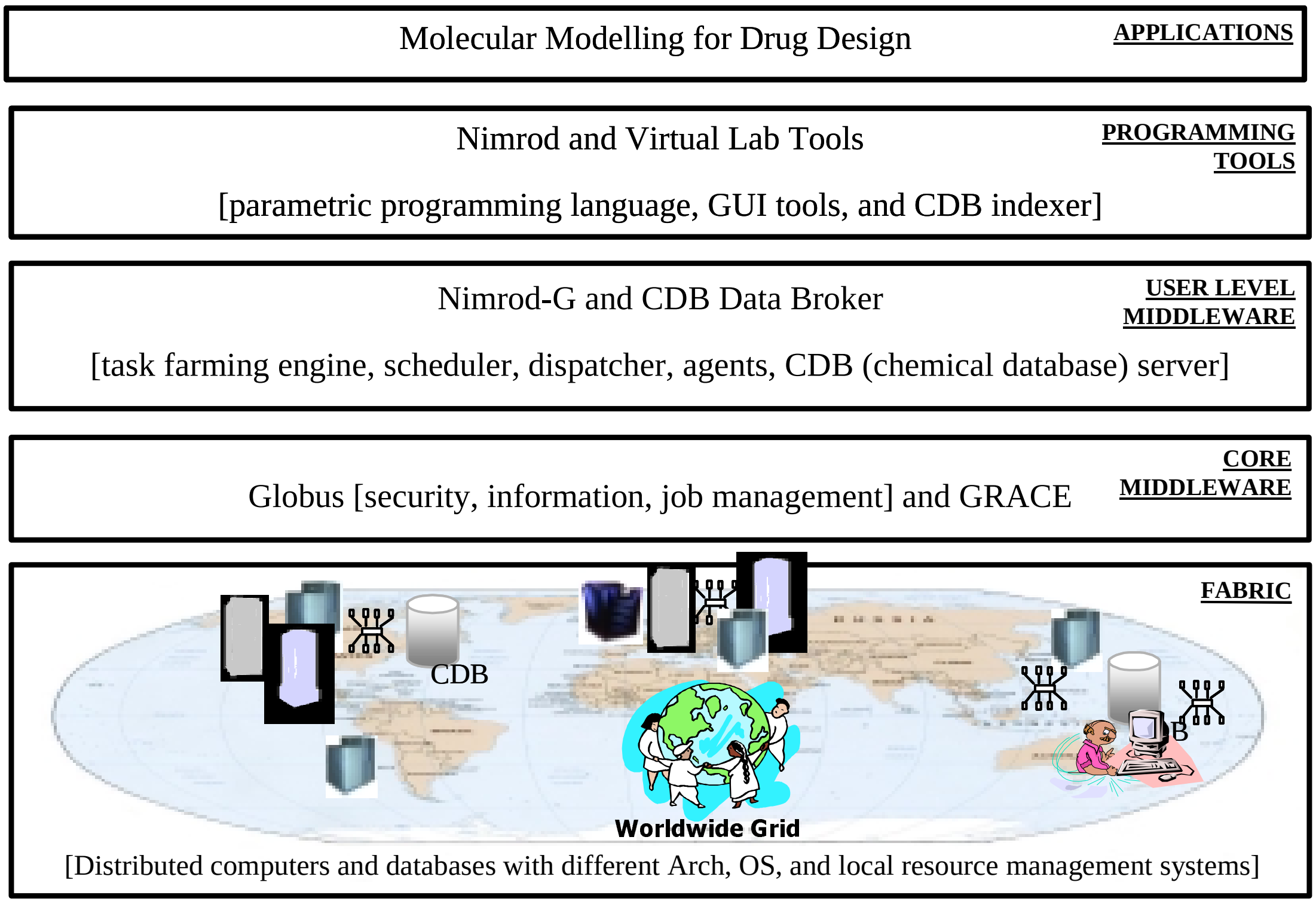

**Figure 7.3: Layered architecture of Virtual Laboratory for drug design.**

The Grid resources (e.g., multiprocessors or clusters) at each location are generally presented as a single entity using resource management systems such as OS-fork, LSF, Condor, and SGE.

In the current implementation, operational steps 2-6 (shown in Figure 7.2) are integrated within the chemical database server. That is, the CDB server deployed at one of the resource sites directly provides the remote access to molecules in the selected databases.

### 7.3.1 Docking Code

The original docking code developed by researchers at the University of California in San Francisco (UCSF) is one of the most popular molecular docking applications [132]. The docking program evaluates the chemical and geometric complementarities between a small molecule and a macromolecular binding site. It explores ways in which two molecules, such as a drug and an enzyme or protein receptor, might fit together. Compounds that might bind tightly to the target receptor must have complementary chemical and spatial natures. Thus docking can be seen as a 3 dimensional puzzle searching for pieces that will fit into the receptor site. It is important to be able to identify small molecules (compounds), which may bind to a target macromolecule. This is because a compound, which binds to a biological macromolecule, may modulate its function, and with further development eventually become a drug candidate. An example of such a drug is the anti influenza drug Relenza which functions by binding to influenza virus attachment proteins thus preventing viral infection.

The relationship between the key programs in the dock suite is depicted in Figure 7.4 (source [132]). The receptor coordinates at the top represent the three-dimensional (3D) structure of a protein. The molecular modeller identifies the active site, and other sites of interest, and uses the program *sphgen* to generate the sphere centers, which fill the site [52]. The program *grid* generates the scoring grids [11]. The program *dock* matches spheres (generated by sphgen) with ligand atoms and uses scoring grids (from grid) to evaluate ligand orientations [11]. It also minimizes energy-based scores [26]. The focus of our work is on docking molecules in CDB with receptors to identify potential compounds that act as a *drug*. Hence, discussion in this chapter is centered on the execution of the program *dock* as a parameter sweep application on world-wide distributed resources.

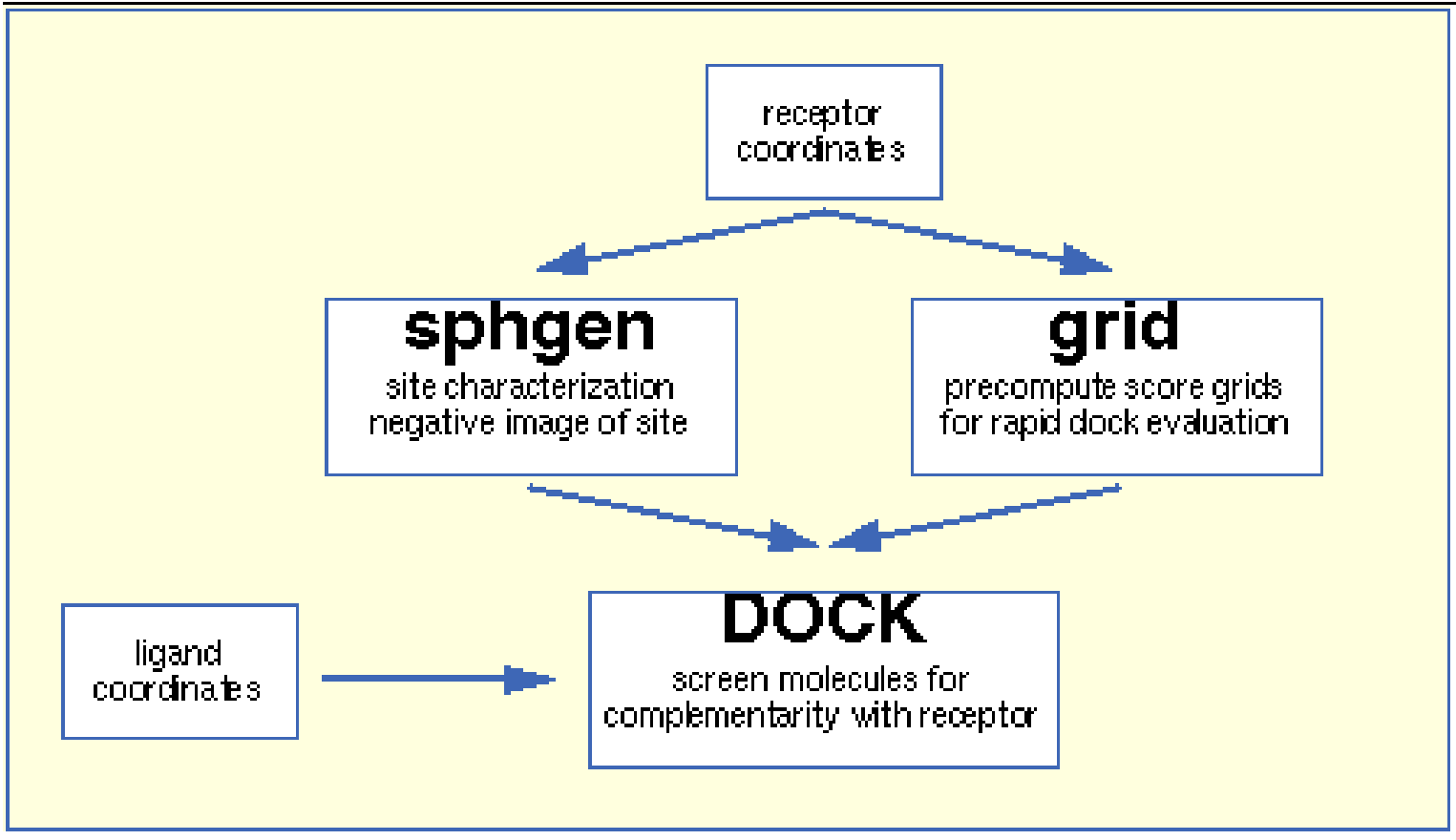

**Figure 7.4: Relation between key programs in the *dock* suite.**

The docking code is highly portable—we have been able to compile and produce executables for Sun-Solaris, PC Linux, SGI IRIX, and Compaq Alpha/OSF1 architectures. For docking on heterogeneous resources, the Nimrod-G broker selects the correct executable automatically based on the resources it discovers at runtime.

### 7.3.2 Nimrod-G Tools

The Nimrod-G toolkit provides a suite of tools and services for creating parameter sweep applications, managing resources and scheduling applications on the world-wide distributed resources. It provides a simple *declarative programming language* or GUI tools for parameterization of application input data files and creation of task-script to be performed by each job; and a programmable *Grid resource broker* for processing jobs on Grid resources.

The steps involved in distributed parametric execution are:

a) parameterise input files,
b) prepare a plan file containing the commands that define parameters and their values,
c) generate a run file, which converts the generic plan file to a detailed list of jobs,
d) schedule jobs for processing on distributed machines, and
e) control and monitor the execution of the jobs.

The application execution environment handles online creation of input files and command line arguments through parameter substitution. The GUI tools supported by enFuzion, a commercial version of Nimrod, can also be used for parameterising applications. enFuzion uses the same syntax as Nimrod for parameter specification [137]. In Section 7.4, we discuss the capabilities of Nimrod-G tools by composing a molecular modelling program as a parameter sweep application for docking compounds in CDB databases. In Section 7.5, we discuss the results of the Nimrod-G broker scheduling a molecular modelling application on the Grid with DBC time and cost optimization scheduling algorithms.

### 7.3.3 Chemical Database Management and Intelligent Access Tools

The chemical databases contain records of a large number of molecules from commercially available organic synthesis libraries, and natural product databases. The molecules in the CDB are represented in MOL2 file (.mol2) format [134], which is a portable representation of a SYBYL [135] molecule. The MOL2 file is an ASCII file that contains all the information needed to reconstruct a SYBYL molecule. Each ligand record in a chemical database represents the three-dimensional structural information of a compound. The number of compounds in each CDB can be in the order of tens of thousands and the database size be anywhere from tens of Megabytes to Gigabytes and even Terabytes. We have developed tools for turning the CDB into a network service and accessing them from remote resources. They include tools for indexing ligand records in the CDB, a multithreaded CDB Server for serving requests for molecule records, and a tool for fetching molecule/ligand record from remote CDB via the network [111].

When a chemical database is available from more than one source (replica site), a suitable strategy such as a source with high network speed or lightly loaded, can be used for selecting one of them. It is likely that multiple users from different locations issue requests for accessing the CDB, the server should be able to process such simultaneous requests concurrently. Therefore, we have developed a multithreaded CDB server that can service requests from multiple users concurrently. An interaction between a Grid node and a node running the CDB server while performing docking is shown in Figure 7.5. We developed and implemented protocols shown in Figure 7.6 for interaction between interaction between the CDB clients and the server. Both figures illustrate the operational model and the flow of control between CDB clients and servers.

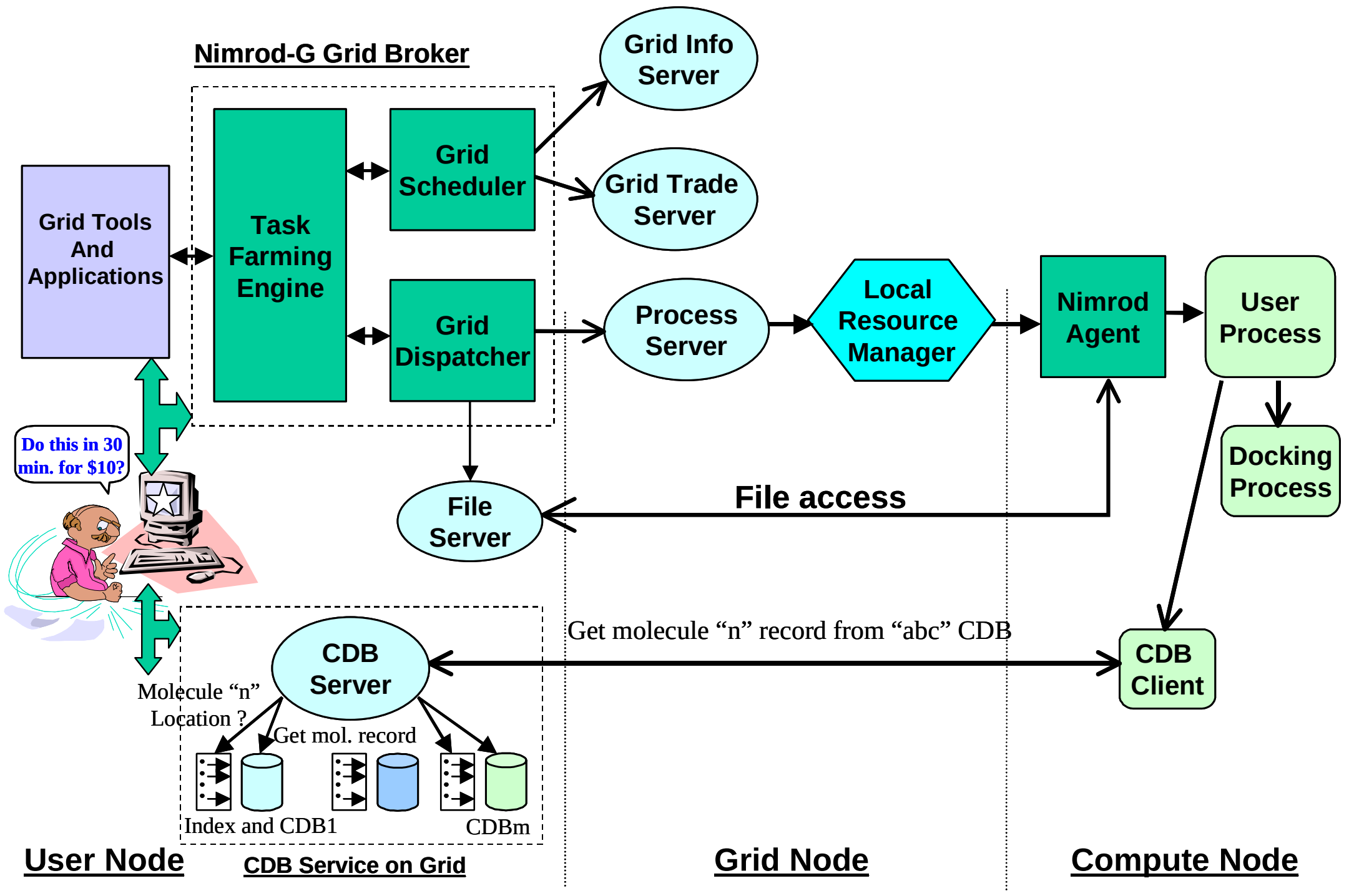

**Figure 7.5: Deployment of Virtual Laboratory components at runtime and their interaction.**

When the Nimrod-G schedules a docking job for processing on one of the Grid resources, it actually submits an agent to Globus GRAM, which acts as a process server. The process server either executes it by forking it as process on time shared system or submits to the site resource manager such as PBS, LSF, and Condor, which allocates a compute node for the agent and starts its execution. The agent contacts the Nimrod-G dispatcher for job task information, which contains instructions for executing a job. It copies input files, performs parameter substitution, executes programs (e.g., CDB client to fetch a molecule record from the remote CDB server and docking program), and ships results back to the Nimrod-G user. When the CDB server receives a request for molecule record, it reads the molecule record from the chemical database and sends back to the client.

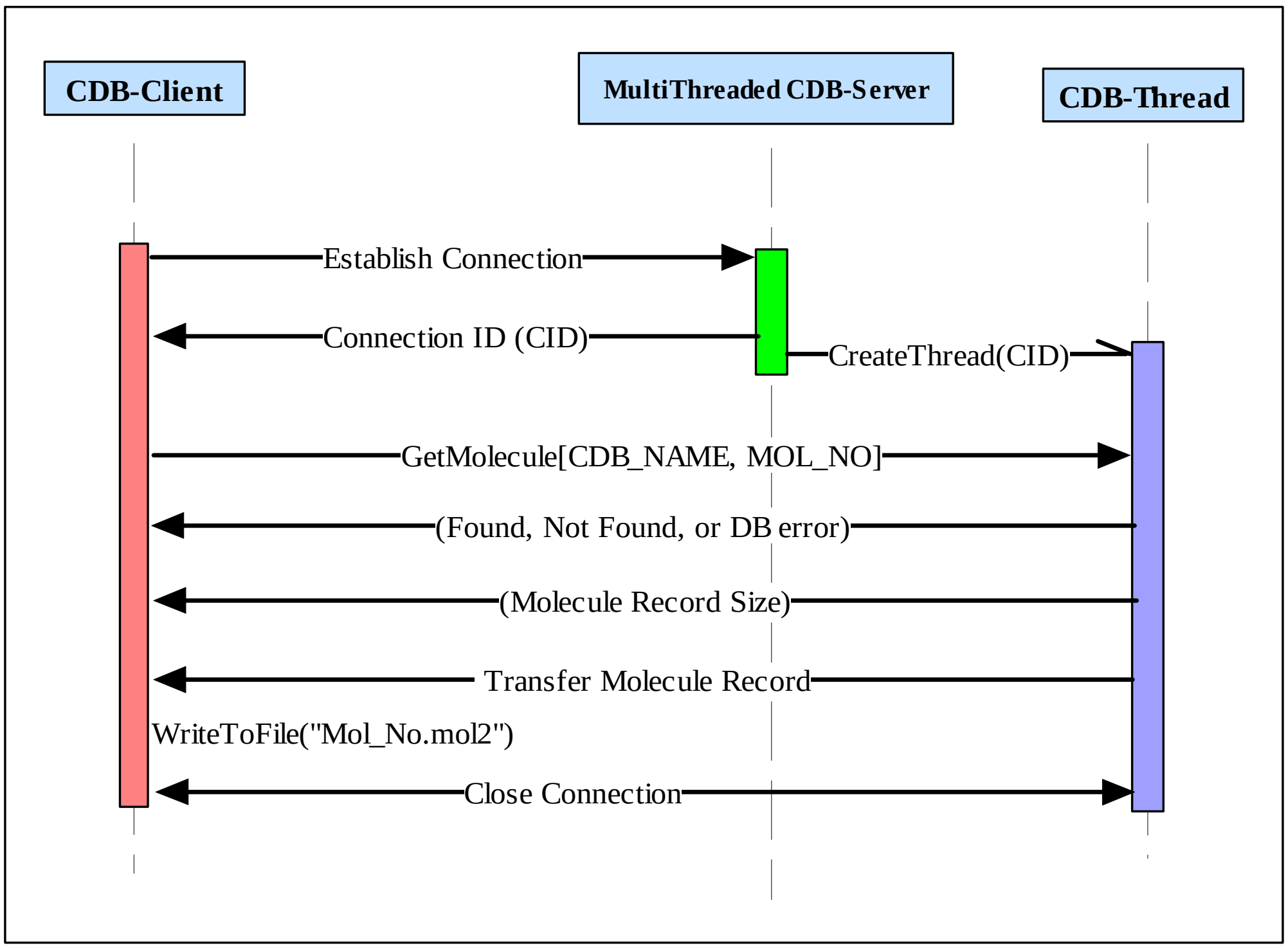

**Figure 7.6: Protocols for Interaction between the CDB clients and the server.**

Instead of searching molecule records sequentially in a database, we have built tools for creating index-tables for each CDB represent using the MOL2 format along with the record size information. The CDB index file, organized in in binary format, contains the starting address (byte location) of a molecule record and record size of all molecules in sequence. When a molecule record is requested, the CDB server first looks at the CDB index file to identify the record location and its size. It then fetches the molecule record from the CDB file with a single read operation; thus improving the access and response speed.

It is possible to screen virtual combinatorial databases in their entirety. This methodology allows only the potential compounds to be subjected to physical (manual) screening and/or synthesis in laboratories, which is extremely time-consuming and resource-intensive.

## 7.4 Application Composition

A docking application having an ability to screen a molecule for each execution can be composed as a task-farming, parameter sweep application for distributed execution. This can be achieved by using Nimrod-G parameter specification language to parameterize docking application input data and files. There is no need to make any changes to the existing (sequential) docking application nor it needs to be developed as parallel application explicitly for distributed execution. The users just need to parameterize the input data and files appropriately and define a Nimrod-G plan file once. Note that the values of parameters can be changed while launching the application execution. The plan file specifies the parametric tasks and the types of the parameters and their values for these tasks. A parametric task consists of a script defined using a sequence of simple commands, providing the ability to copy files to and from the remote node, perform parameter substitutions in input files, execute certain programs, and copy output files back to the user home node. The parametric plan can be submitted to the Nimrod-G runtime machinery, which creates independent docking jobs, and schedules these jobs for concurrent execution on distributed resources. It takes care of replacing the actual value of parameters in the parameterized input files before executing docking jobs.

A sample configuration input file of the docking application is shown in Figure 7.7. It specifies docking configuration parameters and molecule to be docked by indicating a name of the file in which molecule record is stored using the parameter variable "ligand_atom_file". To perform a parameter sweep of different molecules, the value specified by the parameter variable "ligand_atom_file" needs to be parameterized. This is accomplished by replacing the current value, which represents the name of a file containing molecule record, by a substitution place marker. The place marker T consists of a dollar-sign ($) followed by the name of the parameter controlling the substitution, optionally surrounded by braces.

```
score_ligand                yes
minimize_ligand             yes
multiple_ligands            no
random_seed                 7
anchor_search               no
torsion_drive               yes
clash_overlap               0.5
conformation_cutoff_factor  3
torsion_minimize            yes
match_receptor_sites        no
random_search               yes
    . . . . . .
    . . . . . .
maximum_cycles              1
ligand_atom_file            S_1.mol2
receptor_site_file          ece.sph
score_grid_prefix           ece
vdw_definition_file         parameter/vdw.defn
chemical_definition_file    parameter/chem.defn
chemical_score_file         parameter/chem_score.tbl
flex_definition_file        parameter/flex.defn
flex_drive_file             parameter/flex_drive.tbl
ligand_contact_file         dock_cnt.mol2
ligand_chemical_file        dock_chm.mol2
ligand_energy_file          dock_nrg.mol2
```

Molecule to be screened

**Figure 7.7: A configuration input file for docking application.**

A parameterized input file with several attributes replaced by substitution place markers is shown in Figure 7.8. For example, a place marker called for the parameter “ligand_number” has replaced the first part of the “ligand_atom_file” attribute value. The actual value of these parameters is defined in the Nimrod-G plan file that contains parameter definition and task specification. The parameter definition section of the plan file is shown in Figure 7.9. Each parameter is defined by a keyword "parameter", followed by the parameter name, an optional label, and a parameter type. The remaining information on each line defines valid values for the parameter.

The parameter, for example, "database_name" has a label, and is of type *text*. Its valid values are listed, and the user will be able to select one of the values for the duration of the entire experiment. Most of the remaining parameters are single values, either text strings or integers, selected by the user, but with default values provided if the user does not wish to choose a value.

The range parameter, "ligand_number", used to select the molecule, is defined as an integer variable with bounds. For example, to process the first 2000 molecules in the CDB, this range parameter can vary from 1 to 2000 with the step size of 1.

```
score_ligand                    $score_ligand
minimize_ligand                 $minimize_ligand
multiple_ligands                $multiple_ligands
random_seed                     $random_seed
anchor_search                   $anchor_search
torsion_drive                   $torsion_drive
clash_overlap                   $clash_overlap
conformation_cutoff_factor      $conformation_cutoff_factor
torsion_minimize                $torsion_minimize
match_receptor_sites            $match_receptor_sites
random_search                   $random_search
    . . . . . .
    . . . . . .
maximum_cycles                  $maximum_cycles
ligand_atom_file                ${ligand_number}.mol2
receptor_site_file              $HOME/dock_inputs/${receptor_site_file}
score_grid_prefix               $HOME/dock_inputs/${score_grid_prefix}
vdw_definition_file             vdw.defn
chemical_definition_file        chem.defn
chemical_score_file             chem_score.tbl
flex_definition_file            flex.defn
flex_drive_file                 flex_drive.tbl
ligand_contact_file             dock_cnt.mol2
ligand_chemical_file            dock_chm.mol2
ligand_energy_file              dock_nrg.mol2
```

Molecule to be screened

**Figure 7.8: Parameterisation of a configuration input file**.

```
parameter database_name label "database_name" text select oneof "aldrich"
    "maybridge" "maybridge_300" "asinex_egc" "asinex_epc" "asinex_pre"
    "available_chemicals_directory" "inter_bioscreen_s" "inter_bioscreen_n"
    "inter_bioscreen_n_300" "inter_bioscreen_n_500" "biomolecular_research_institute"
    "molecular_science" "molecular_diversity_preservation"
    "national_cancer_institute" "IGF_HITS" "aldrich_300" "molecular_science_500"
    "APP" "ECE" default "aldrich_300";
parameter CDB_SERVER text default "bezek.dstc.monash.edu.au";
parameter CDB_PORT_NO text default "5001";
parameter score_ligand text default "yes";
parameter minimize_ligand text default "yes";
parameter multiple_ligands text default "no";
parameter random_seed integer default 7;
parameter anchor_search text default "no";
parameter torsion_drive text default "yes";
parameter clash_overlap float default 0.5;
parameter conformation_cutoff_factor integer default 5;
parameter torsion_minimize text default "yes";
parameter match_receptor_sites text default "no";
    . . . . . .
    . . . . . .
parameter maximum_cycles integer default 1;
parameter receptor_site_file text default "ece.sph";
parameter score_grid_prefix text default "ece";
parameter ligand_number integer range from 1 to 2000 step 1;
```

Molecules to be screened

**Figure 7.9: A plan file defining parameters type and their values.**

The parameters "receptor_site_file" and "score_grid_prefix" indicate the data input files. Their values indicate that data input files are located in the user home directory on Grid nodes. Instead of pre-staging, these files can be copied at runtime by defining necessary "copy" operations in the job's "nodestart" or "main" task (see Figure 7.10). However, it is advisable to copy or "pre-stage" large input files in the beginning of application execution instead of copying them during execution of every job. This saves transmission time particularly when those files are going to be used for docking with many databases.

The plan file is submitted to a job generation tool, such as the EnFuzion Generator, in order to create a run file that contains specific instances of jobs to be run, which is then submitted to the Nimrod-G runtime machinery for processing on the Grid. The run file contains a job for each combination of parameters. Hence the number of jobs is the product of the number of values chosen for each parameter. Since most of the parameters except "ligand_number" are single-valued, they have no effect on the number of jobs.

```
task nodestart
      copy ./parameter/vdw.defn node:.
      copy ./parameter/chem.defn node:.
      copy ./parameter/chem_score.tbl node:.
      copy ./parameter/flex.defn node:.
      copy ./parameter/flex_drive.tbl node:.
      copy ./dock_inputs/get_molecule node:.
      copy ./dock_inputs/dock_base node:.
endtask
task main
           node:substitute dock_base dock_run
           node:substitute get_molecule get_molecule_fetch
           node:execute sh ./get_molecule_fetch
           node:execute $HOME/bin/dock.$OS -i dock_run -o dock_out
           copy node:dock_out ./results/dock_out.$jobname
           copy node:dock_cnt.mol2 ./results/dock_cnt.mol2.$jobname
           copy node:dock_chm.mol2 ./results/dock_chm.mol2.$jobname
           copy node:dock_nrg.mol2 ./results/dock_nrg.mol2.$jobname
endtask
```

**Figure 7.10: Task definition of docking jobs.**

It is also possible to set concrete values for each of the parameters at runtime when job Generator is invoked. For the parameter "ligand_number", the user may choose not to select all values from 1 to 2000, but may select a subset of these values. By default, this generated 2000 jobs, each docking a single molecule.

The second part of Nimrod-G plan file is task specification that defines a series of operations that each job needs to perform to dock a molecule (see Figure 7.10). The "nodestart" task is performed once for each remote node. Following that, the files copied during that stage are available to each job when it is started. The "main" task controls the actions performed for each job.

The first line of the "main" task performs parameter substitution on the file "dock_base", creating a file "dock_run". This is the action that replaces the substitution place markers in our input file with the actual values for the job.

As each docking operation is performed on a selected molecule in the CDB database, it is not necessary to copy such large databases on all Grid nodes. Hence, not only is the molecule file named in the configuration file, we also go to particular lengths to copy only the data for the molecule being tested. The executable script "get_molecule_fetch" (see Figure 7.11) is also created using parameter substitution, and runs the "vlab-cdb-get-molecule" executable, which fetches the molecule record from the *CDB molecule server* based on the parameter "ligand_number". The molecule record is saved in a file whose name is the same as integer value of the “ligand_number” parameter and “mol2” as its extension. For instance, if the parameter ligand_number value is 5, then molecule record will be saved in a file “5.mol2”.

```
#!/bin/sh
$HOME/bin/vlab-cdb-get-molecule.$OS $CDB_SERVER $CDB_PORT_NO ${database_name}.db $ligand_number
```

**Figure 7.11: Parameterisation of script for extracting molecule from CDB.**

The main code is the "dock" executable. Note that in the “execute” command, there are pseudo-parameters that do not appear in the plan file. These include environment variables, such as "HOME", as well as other useful parameters, such as "OS" indicating the operating system on the node. This allows us to select the correct executable for the node. If the “dock” executable files do not exist on Grid nodes, they need to be copied at runtime as part of the job’s “nodestart” task similar to copying input files.

The dock_run file created in the substitution step previously is now provided as the input configuration file for the docking process. The output files are then copied back to the local host, and renamed with another pseudo-parameter, the unique "jobname" parameter.

## 7.5 Scheduling Experimentations

We have performed scheduling experiments from a Grid resource in Australia along with four resources available in Japan and one in USA. Table 7.1 shows the list of resources and their properties, Grid services, access cost or price in terms of Grid dollar (G$) per CPU-second, and the number of jobs processed on resources with deadline-and-budget constrained (DBC) time optimization (TimeOpt) or cost optimization (CostOpt) strategies. The resource price in terms of G$ is assigned arbitrarily at runtime in these experiments, however, they can be set to match the power of resources and job turn around time as valued in supercomputing centers such as the Manchester computing services. The G$ can be equated to real money or tokens charged to users for accessing resources. In the current scenario, the users get allocation of tokens via funding from the project sponsoring agents or partnerships. There are supercomputing centers that sell tokens to commercial users and the value of tokens correspond to the quantity of resource allocations. It is also possible to price resources based on the real world economic models [102] that are driven by the supply and demand for resources.

**Table 7.1: The WWG testbed resources used in scheduling experiments, job execution and costing.**

| Organization & Location | Vendor, Resource Type, # CPU, OS, hostname | Grid Services and Fabric, Role | Price (G$/CPU sec.) | Number of Jobs Executed | |
|---|---|---|---|---|---|
| | | | | TimeOpt | CostOpt |
| Monash University, Melbourne, Australia | Sun: Ultra-1, 1 node, bezek.dstc.monash.edu.au | Globus, Nimrod-G, CDB Server, Fork (Master node) | -- | -- | -- |
| AIST, Tokyo, Japan | Sun: Ultra-4, 4 nodes, Solaris, hpc420.hpcc.jp | Globus, GTS, Fork (Worker node) | 1 | 44 | 102 |
| AIST, Tokyo, Japan | Sun: Ultra-4, 4 nodes, Solaris, hpc420-1.hpcc.jp | Globus, GTS, Fork (Worker node) | 2 | 41 | 41 |
| AIST, Tokyo, Japan | Sun: Ultra-4, 4 nodes, Solaris, hpc420-2.hpcc.jp | Globus, GTS, Fork (Worker node) | 1 | 42 | 39 |
| AIST, Tokyo, Japan | Sun: Ultra-2, 2 nodes, Solaris, hpc220-2.hpcc.jp | Globus, GTS, Fork (Worker node) | 3 | 11 | 4 |
| Argonne National Lab, Chicago, USA | Sun: Ultra -8, 8 nodes, Solaris, pitcairn.mcs.anl.gov | Globus, GTS, Fork (Worker node) | 1 | 62 | 14 |
| | | Total Experiment Cost (G$) | | 17702 | 14277 |
| | | Time to Finish Experiment (Min.) | | 34.00 | 59.30 |

We have performed a trial screening 200 molecules (from the *aldrich_300* CDB) on a target receptor called endothelin converting enzyme (ECE), which is involved in hypotension. The three dimensional structure of the receptor is derived from homology modelling using related receptor structures. In these experimentations, for faster evaluation purpose, the range parameter "ligand_number" is defined with the bounds 1 and 200 and the step size as 1, which produces 200 jobs for docking molecules. As shown in Figure 12, the *dock* program takes two different types of inputs files: a) *common input files*, the same files are required for all docking jobs and b) *ligand specific input files*, which vary from one job to another. The large common input files (receptor structure and pre-calculated score potentials) are pre-staged on resources instead of copying them at runtime. The files are copied using the *globus-rcp* command and stored in the directory location "$HOME/dock_inputs/" on resources as specified by the parameters "receptor_site_file" and "score_grid_prefix" (see Figure 7.8). The two application-specific executable files, "dock" and "vlab-cdb-get-molecule" invoked in the task scripts (see Figure 7.10 and Figure 7.11) are also pre-staged. The executable files are stored in the "$HOME/bin/" directory on resources.

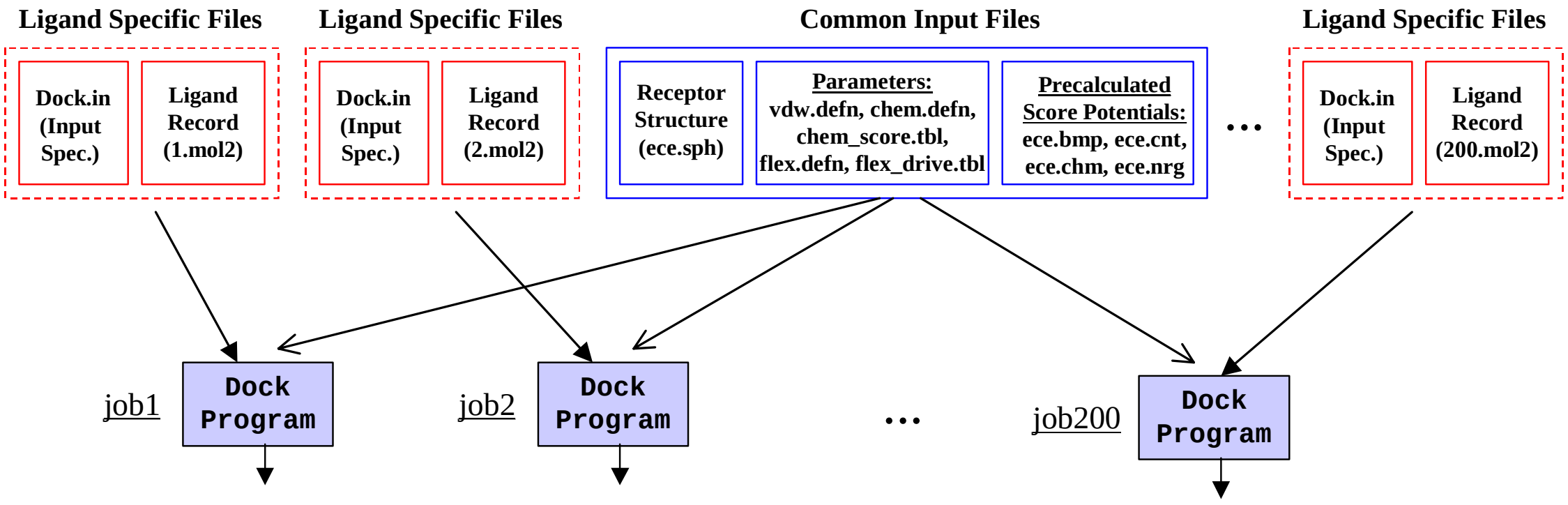

**Figure 7.12: Static and Dynamic Input Files of Docking program.**

We conducted deadline and budget constrained scheduling experiments for two different optimization strategies [107]:

1. Optimize for Time - this strategy aims to produce results at the earliest possible time before a deadline, and within a budget limit. It process as many jobs as possible cheapest resources for the deadline period and uses expensive ones just to meet the deadline.
2. Optimize for Cost - this strategy aims to minimize the cost of execution (spending from the given budget) and complete the experiment on or before the deadline. It uses all resources aggressively as long as it can afford them and tries to process all jobs at the earlier possible time.

In both experiments, we have set 60 minutes as the deadline limit and 50,000 G$ as the budget limit at runtime using the Nimrod-G scheduler steering and control monitor. The value of these constraints can be changed at anytime during the execution, of course not less than the time and budget that is already spent!

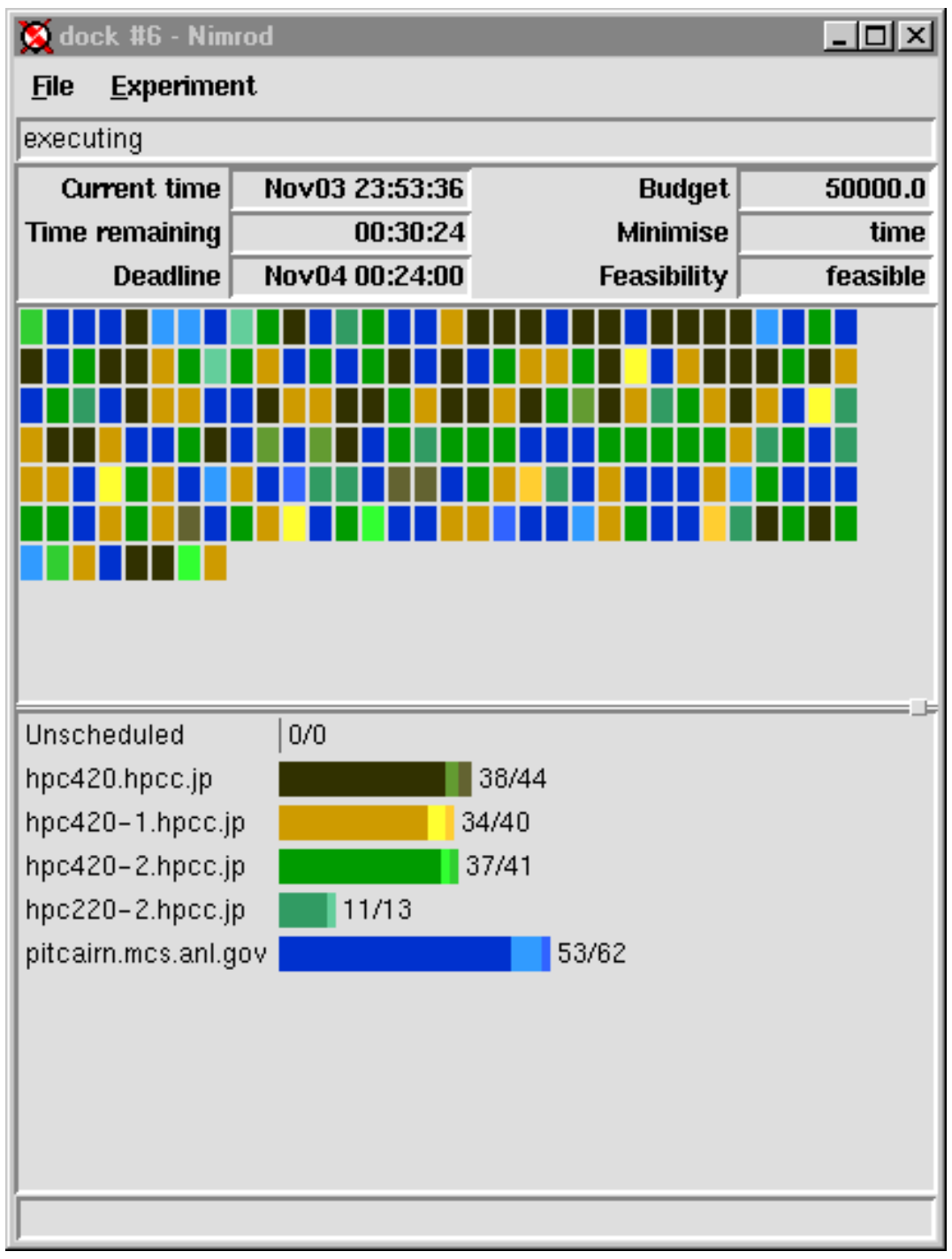

**Figure 7.13: A snapshot of the Nimrod-G monitor during "Optimize for Time" scheduling.**

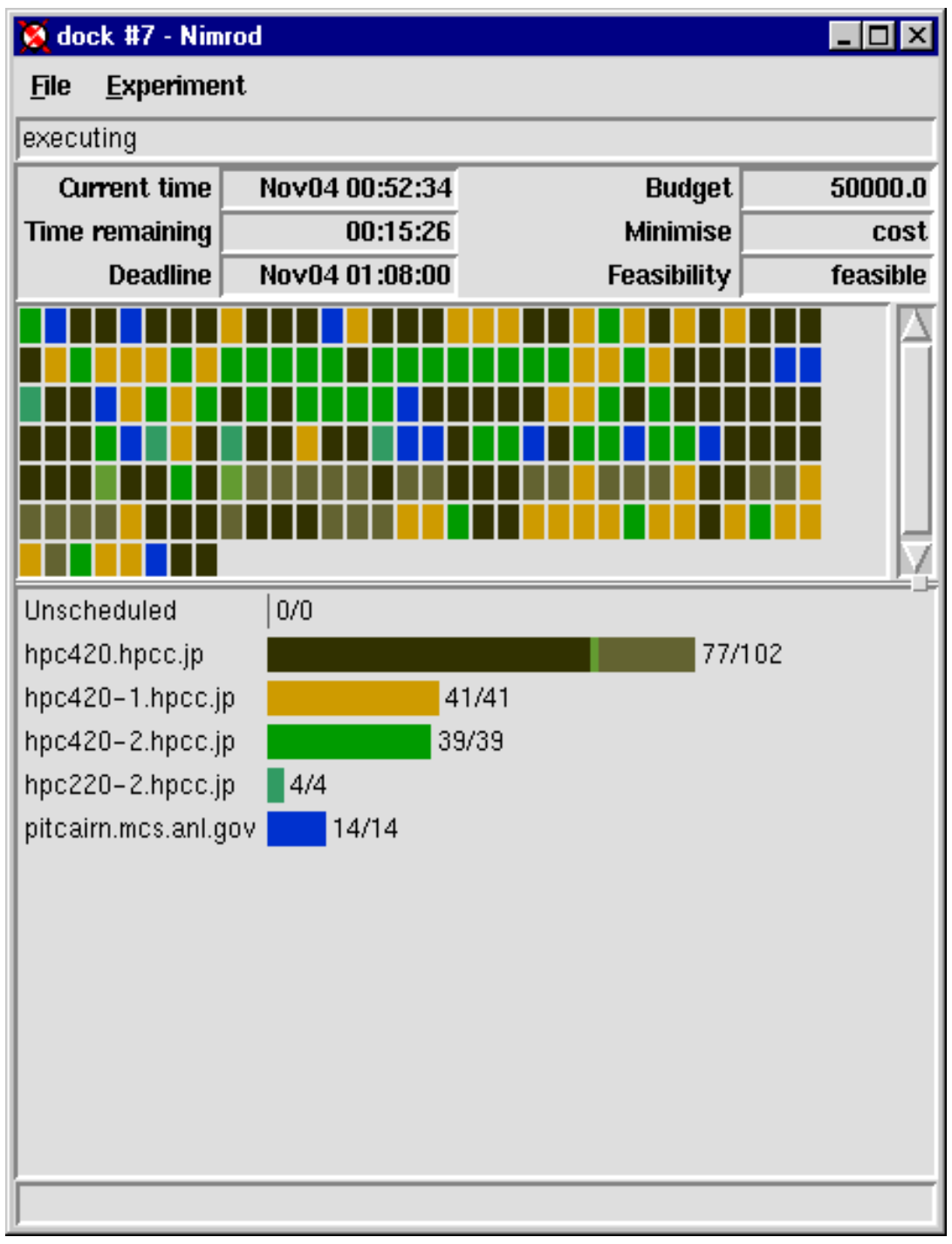

**Figure 7.14: A snapshot of the Nimrod-G monitor during "Optimize for Cost" scheduling.**

The first experiment, *Optimize for Time* scheduling, was performed on November 3, 2001 at 23:23:00,

Australian Eastern Standard Time (AEST), with a 60-minute deadline and finished on November 3, 2001 by 23:57:00. A snapshot of the Nimrod-G monitoring and steering client taken a few minutes (~5min.) before the completion of application processing is shown in Figure 7.13. This experiment took 34 minutes to finish the processing of all jobs using resources available at that time with an expense of 17,702 G$. Figure 7.15 shows the number of jobs processed on different resources selected depending on their cost and availability. Figure 7.16 shows the corresponding expenses of processing on resources. Figure 7.17 shows the number of jobs in execution on resources at different times. From the graphs it can be observed that the broker selected resources to ensure that the experiment was completed at the earliest possible time given the current availability of resources and the budget limitations. After 30 minutes, it discovered that it could still complete early without using the most expensive resource, hpc220-2.hpcc.jp.

It should be noted that for each job scheduled for execution on the Grid, the Nimrod-G runtime machinery (actuator) deploys Nimrod-G agents on remote resources. The Nimrod agents setup runtime environments (generally in scratch file space, "/tmp") on remote resources and execute commands specified in the task definition script (see Figure 7.10). The docking parameter files and ligand specific files are transferred from the home node, bezek.dstc.monash.edu.au in this case. The agent uses *http* protocols to fetch files via the http-based file server running on the home node. All parameter variables in the parameterized input files (see Figure 7.8 and Figure 7.9) are substituted by their concrete values before processing. The ligand record is fetched from the CDB database server running on the home node. The agent then executes the *dock* program and stores output files in the scratch area. The required output files are then transferred to the home node and stored with the job number as their extension. All these steps involved in the execution of the *dock* program on Grid resources were completely hidden from the user.

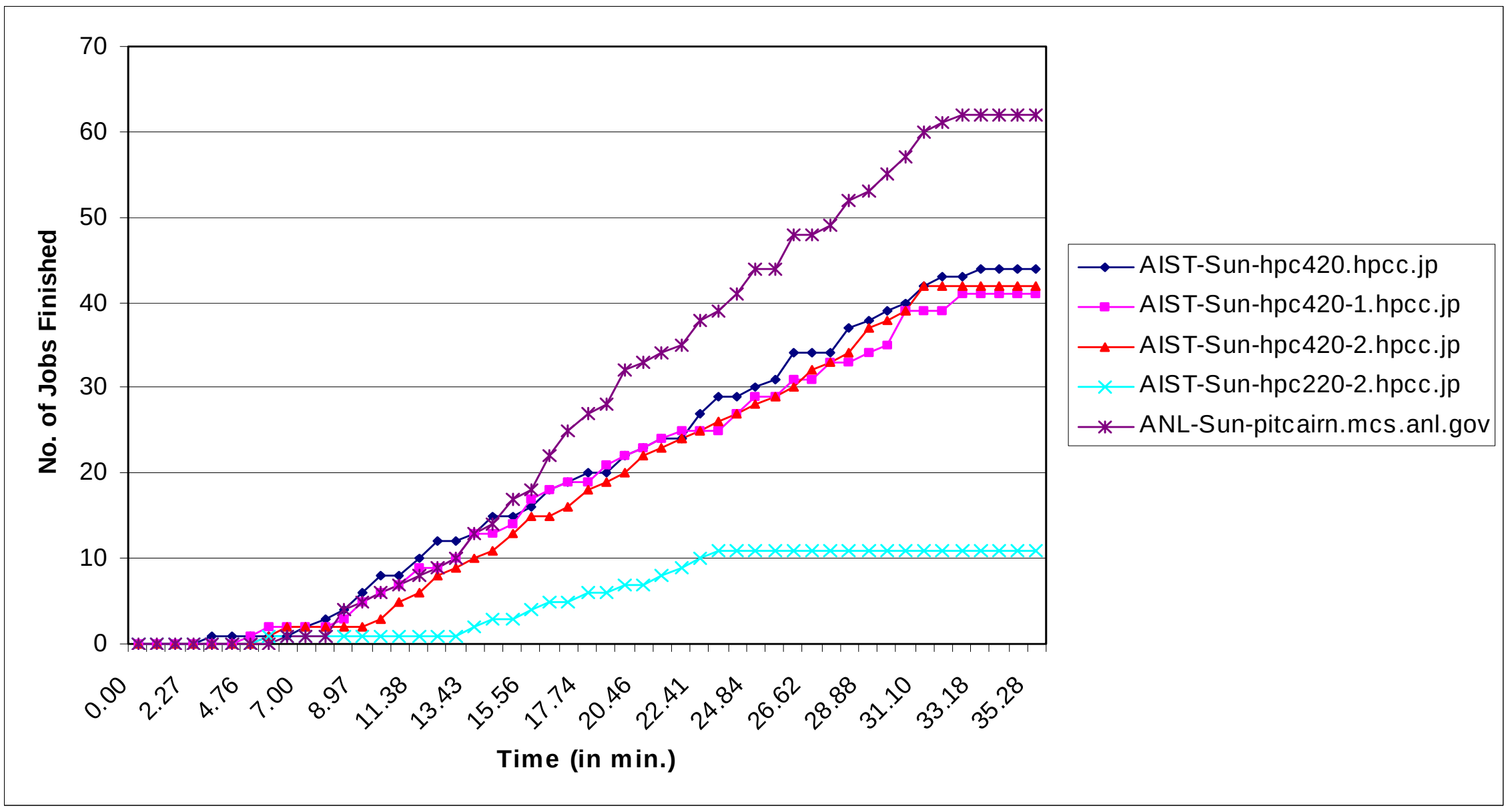

**Figure 7.15: No. of jobs processed on Grid resources during DBC time optimization scheduling.**

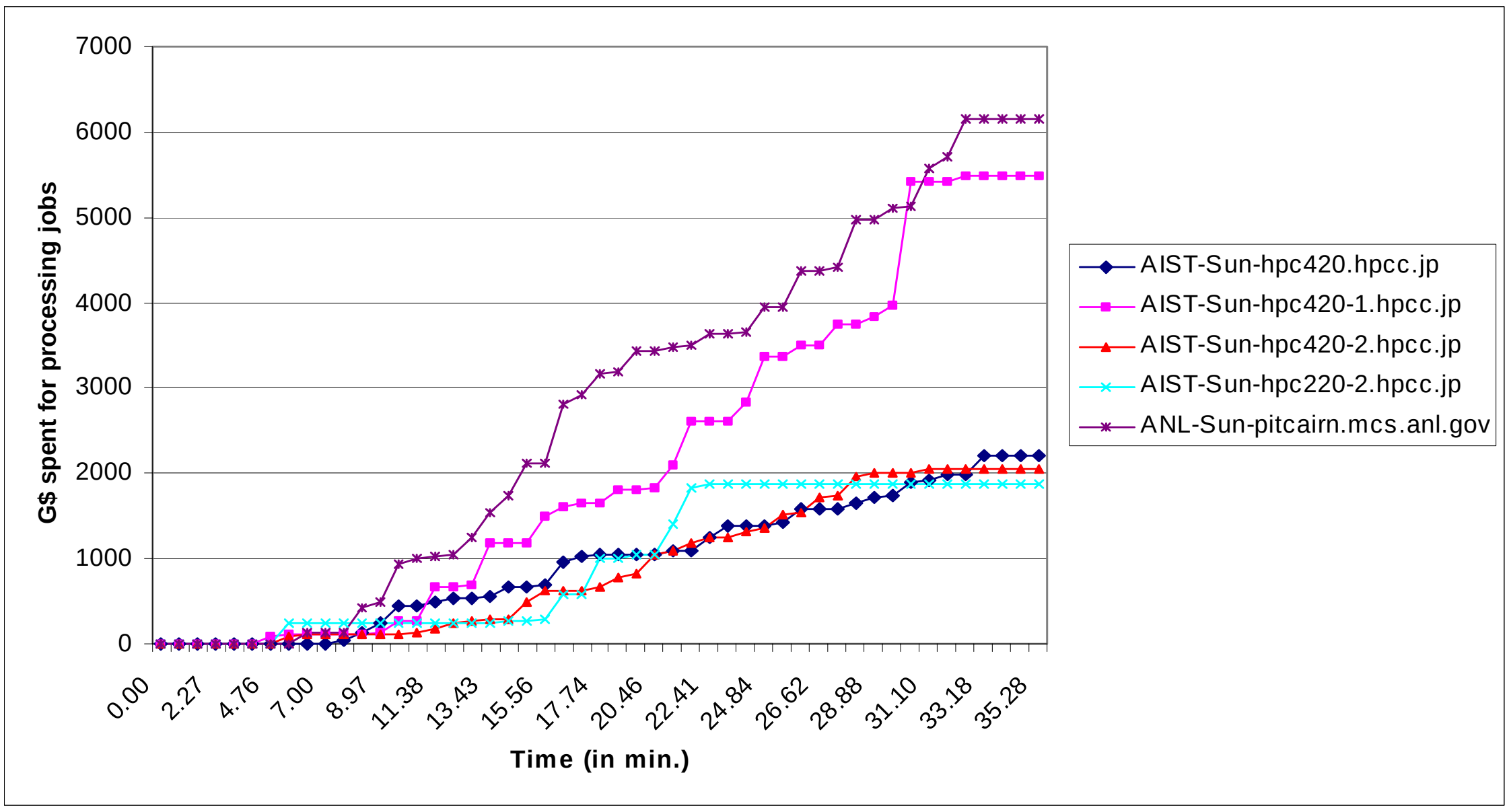

**Figure 7.16: The amount spent on resources during DBC time optimization scheduling.**

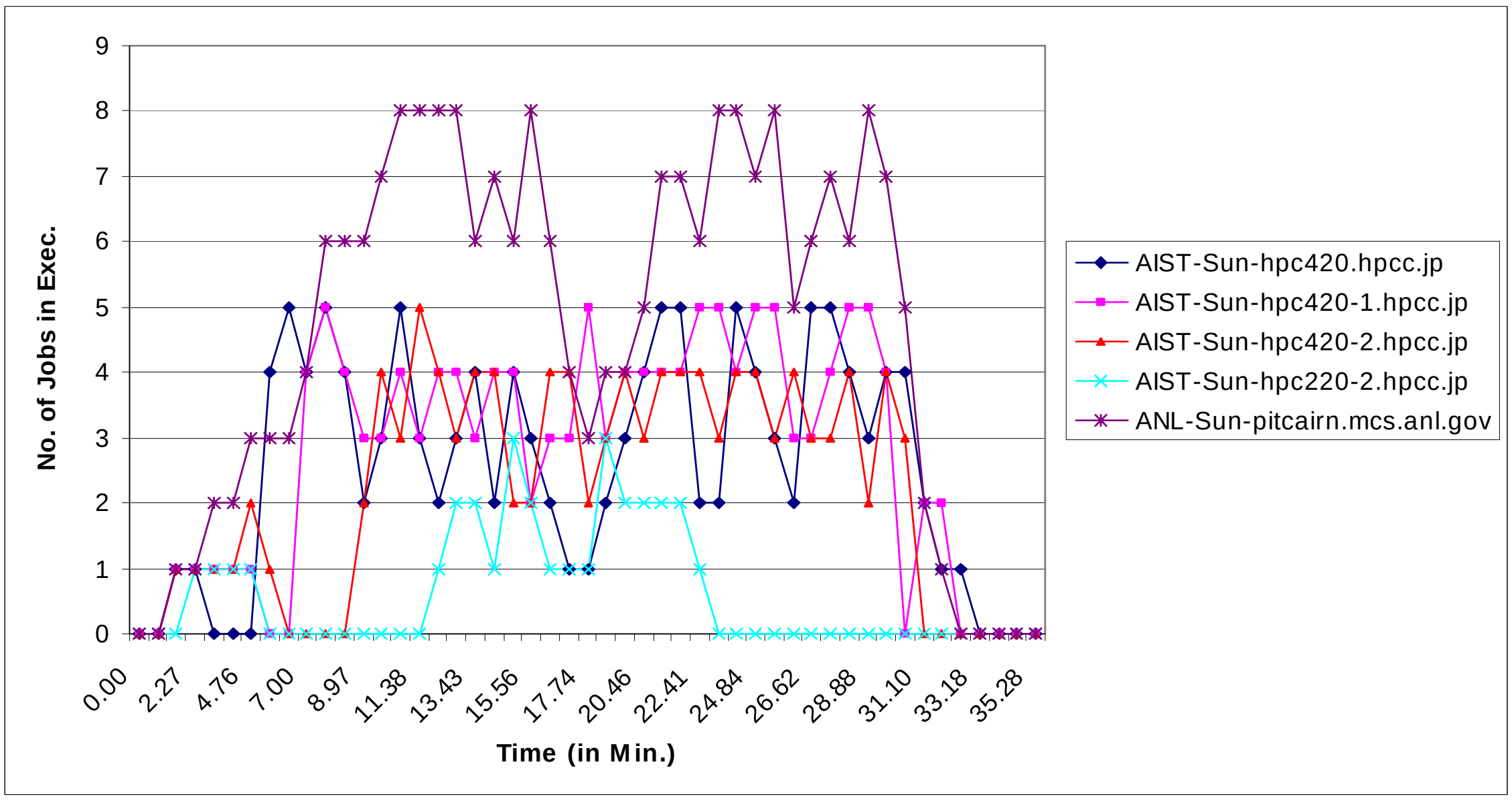

**Figure 7.17: No. of jobs in execution on Grid resources during DBC time optimization scheduling.**

The second experiment, *Optimize for Cost* scheduling, was performed on November 4, 2001 at 00:08:00, AEST, with a 60-minute deadline and finished on November 4, 2001 by 01:07:30. A snapshot of the Nimrod-G monitoring and steering client taken few minutes (~5min.) before the completion of application processing is shown in Figure 7.14. This experiment took almost 59.30 minutes to finish the processing of all jobs using resources available at that time with an expense of 14,277 G$. It is interesting to note that the second experiment took an extra 25.30 minutes and saved 3,425 G$ in the process. Figure 7.18 shows the number of jobs processed on different resources selected depending on their cost and availability. Figure 7.19 shows the corresponding expenses of processing on resources. Figure 7.20 shows the number of jobs in execution on resources at different times. From the graphs it can be observed that the broker selected the cheapest resources to ensure that the experiment was completed with minimum expenses, but before the deadline limit. In the beginning expensive resources are used to ensure that the deadline can be met. If for any reason cheapest resources are unable to deliver expected performance, then the broker seeks the help of expensive resources to meet the deadline.

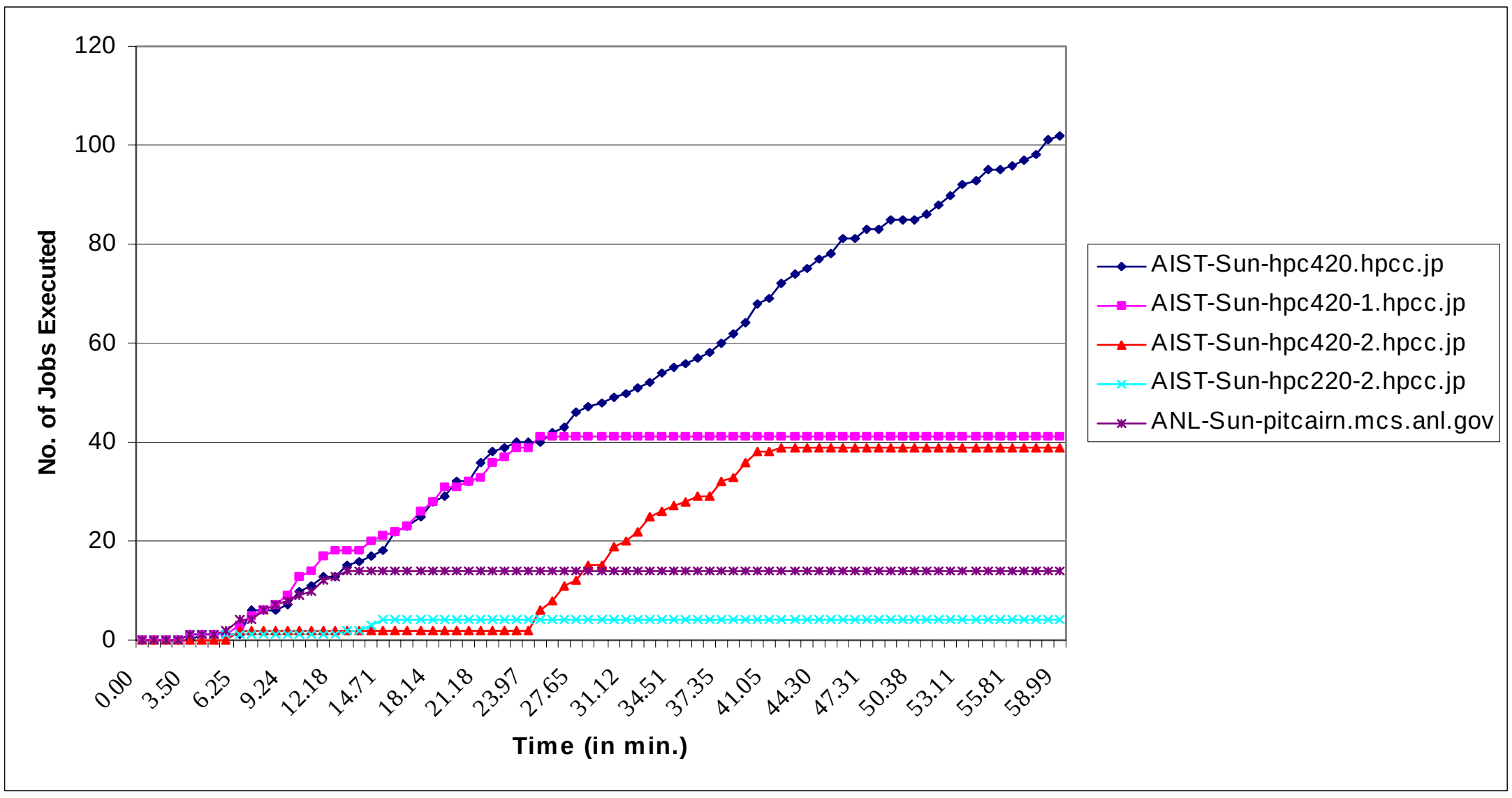

**Figure 7.18: No. of jobs processed on Grid resources during DBC Cost optimization scheduling.**

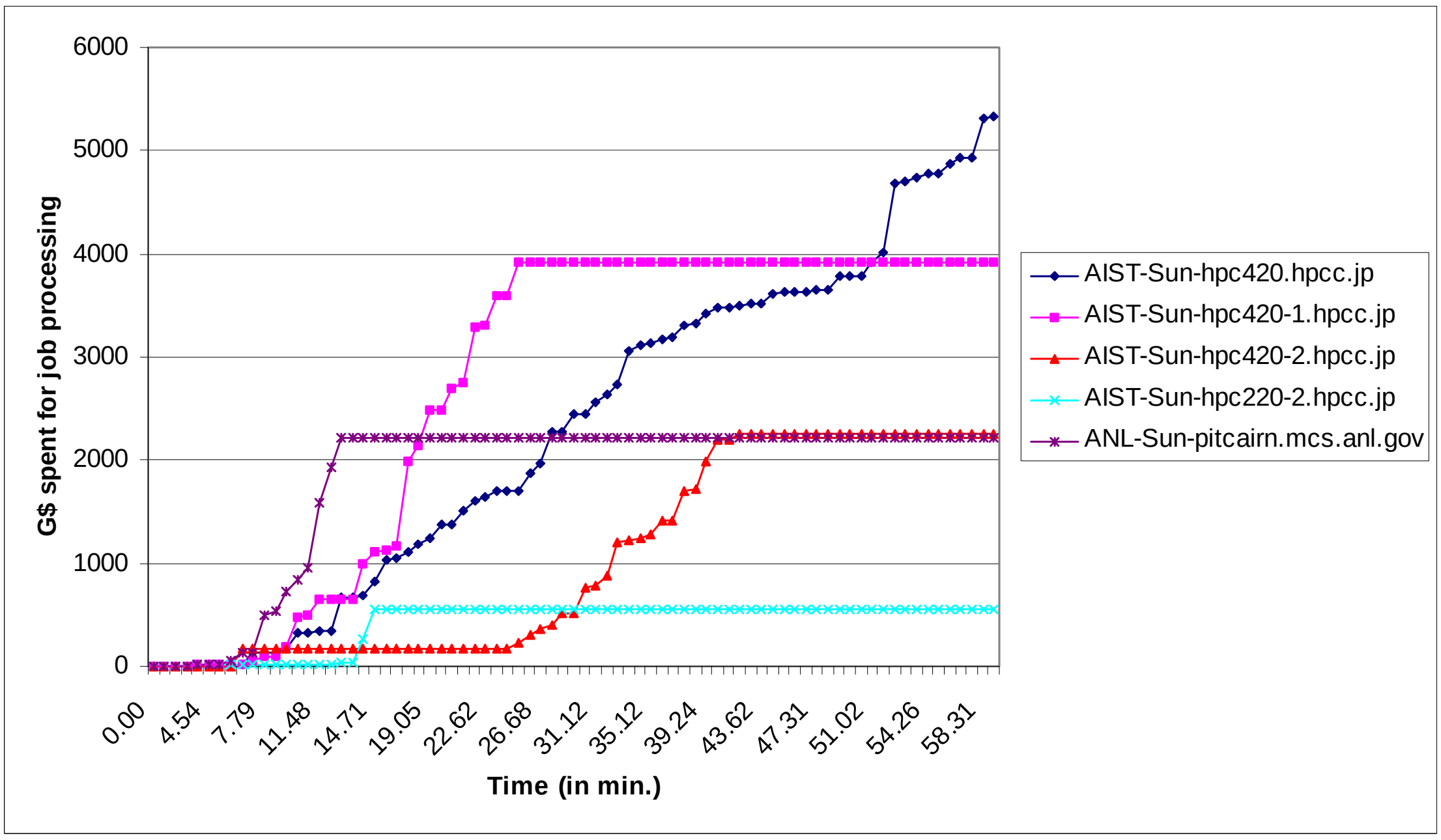

**Figure 7.19: The amount spent on resources during DBC Cost optimization scheduling.**

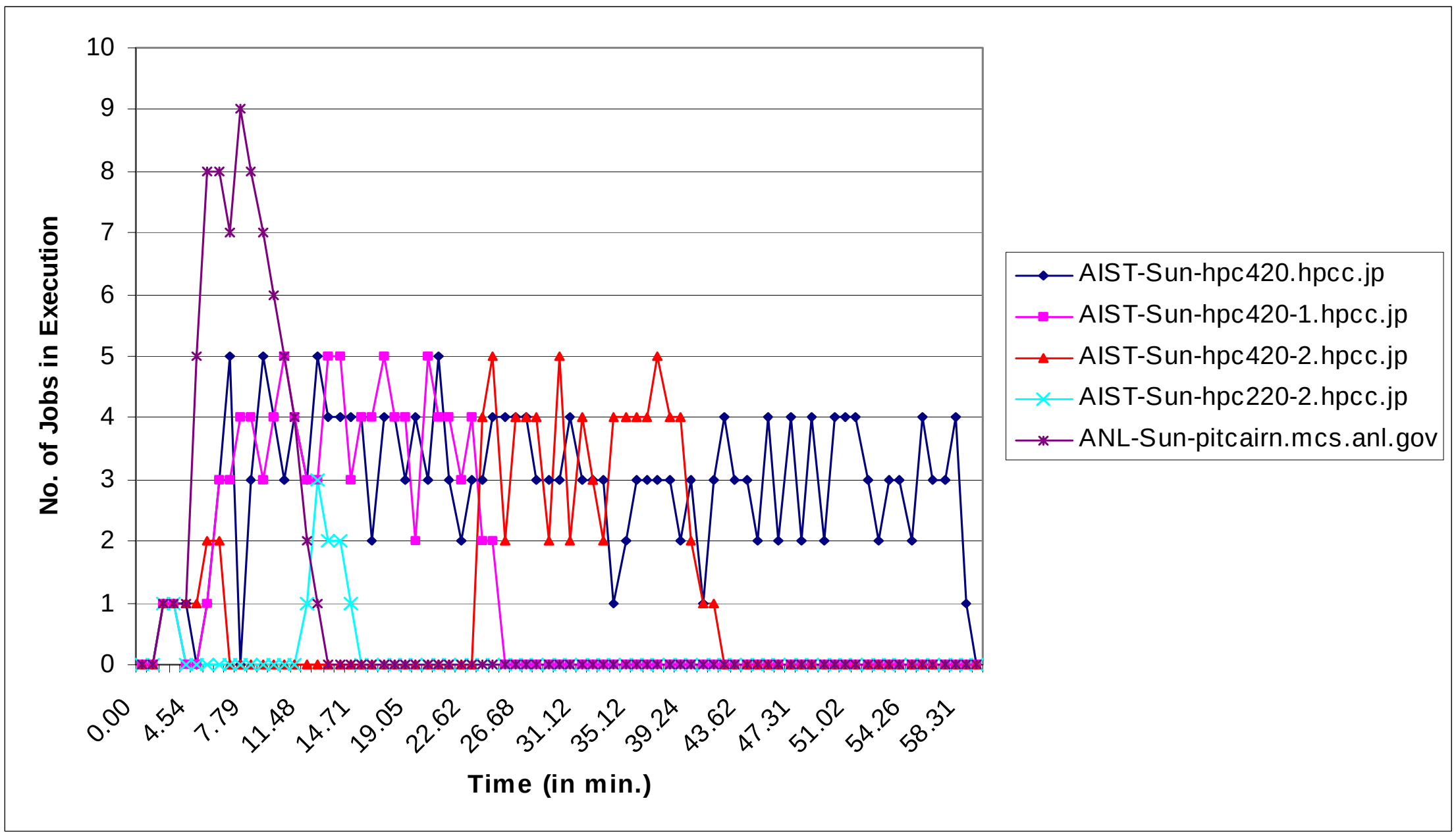

**Figure 7.20: No. of jobs in execution on Grid resources during DBC Cost optimization scheduling.**

## 7.6 Related Work

Although many researchers have explored the use of parallel computing techniques in molecular docking for drug design [35][118], there are only a few efforts that use the world wide distributed computers for processing docking jobs in parallel. One of the most related efforts is the FightAIDS@Home project [27], which is based on the Entropia's distributed computing network and the Scripps Research Institute's docking application. In this case, volunteers need to download Entropia's screen saver program that runs in the background on the volunteer computer. The volunteer PC contacts the Entropia server to download the data to perform docking. When docking on an assigned data is completed, it uploads the results to the server. This execution model is different from our model where the scheduler (Nimrod-G) assigns the work to computers that are available and initiates the execution.

Most of the efforts explicitly develop docking application as a parallel application using a special purpose, legacy or standard, parallel programming languages and interfaces such as PVM and MPI, which requires extra development effort and time. The scalability of such applications and runtime systems is limited to resources available in a single domain and they need powerful computers and networks for faster processing.

Our techniques are novel in many ways. To perform parallel and distributed docking using our tools, there is no need to develop docking application as a parallel application. Our framework supports the composition of the existing molecular docking application as a parameter sweep application without making any changes to it. Our runtime machinery, the Nimrod-G resource broker, creates independent docking jobs automatically and launches their parallel execution on world-wide distributed computers. It hides all the complexity associated with scheduling jobs, shipping appropriated input files, starting and monitoring their execution, and shipping results back to the user. Our scheduler also supports the deadline and budget based scheduling, which prioritizes the processing depending on the user requirements—how quickly they need results, how much they want to spend, and which one to optimise.

## 7.7 Summary and Conclusion

Computational Grids enable the *sharing* and *aggregation* of geographically distributed resources for solving large-scale, resource and data-intensive problems faster and cheaper. However, application development, resource management, and scheduling in these environments is a complex undertaking. We have developed a Virtual Laboratory environment and tools for formulating molecular docking for drug design as a parameter sweep application, chemical database management, and scheduling docking jobs for processing on a wide area distributed resources by leveraging existing Grid technologies. The new tools developed include a chemical database indexer, CDB server for providing access to molecules in chemical databases as a network service, clients for accessing CDB services from a selected CDB service. We have used the Nimrod-G parameter specification language for composing an existing docking application as a parameter sweep application and the Nimrod-G Grid resource broker for processing molecular docking jobs on distributed resources.

We have conducted deadline and budget constrained scheduling experiments for concurrent processing of docking jobs on the WWG testbed under two different optimization scenarios. The results of this molecular docking application scheduling on a large-scale distributed resources demonstrate the potential of the Virtual Laboratory tools for service oriented computing. They demonstrate the suitability of Grids and Grid technologies (like Globus and Nimrod-G) for computational and data intensive computing and at the same time prove the effectiveness of computational economy and quality of services (QoS) driven scheduling as an efficient mechanism for the management of supply-and-demand for resources depending on the value delivered to the user. Also, this framework with economic incentive encourages the users to reveal their true requirements and allows them to trade-off between the deadline and budget. Thus allowing the allocation of resources to users with high priority jobs and requirements.

Chapter 8

# Conclusions and Future Directions

## 8.1 Summary

Grids are emerging as the infrastructure for next generation computing. In Grid environments, the resources are heterogeneous and geographically distributed with varying availability and a variety of usage and cost policies for diverse users at different times and, priorities as well as goals that vary with time. The management of resources and application scheduling in such a large and distributed environment is a complex task. It is envisioned that the use of a distributed computational economy is an effective metaphor for the management of resources and we have developed an architectural framework called the Grid Architecture for Computational Economy (GRACE).

To support the thesis that an economic-based Grid resource management and scheduling system can deliver significant value to users, resource providers and consumers, compared to traditional approaches, we have:

- identified the key requirements that an economic-based Grid system needs to support,
- developed a distributed computational economy framework called the GRACE, which is generic enough to accommodate different economic models and maps well onto the architecture of wide-area distributed systems,
- designed deadline and budget constrained scheduling algorithms with four different strategies: cost, time, conservative-time, and cost-time optimisations,
- developed a Grid resource broker called Nimrod-G that supports deadline and budget constrained algorithms for scheduling parameter sweep applications on the Grid,
- developed a Grid simulation toolkit, called GridSim, that supports discrete-event based simulation of Grid environments to allow repeatable performance evaluation under different scenarios,
- evaluated the performance of deadline and scheduling algorithms through a series of simulations by varying the number of users, deadlines, budgets, and optimisation strategies and simulating geographically distributed Grid resources, and
- demonstrated the effectiveness and application of Grid technologies for solving real-world problems such as molecular modelling for drug design on WWG (World-Wide Grid) testbed.

## 8.2 Conclusions

The resource management and scheduling systems used in Grid environments need to be adaptive so that they can handle dynamic changes in availability of resources and user requirements. At the same time, they need to provide scalable, controllable, measurable, and easily enforceable policies for management of the aforementioned resources. To address these requirements, a resource broker called Nimrod-G has been developed. The Nimrod-G resource broker supports the deadline and budget constrained algorithms for scheduling task-farming applications on large-scale distributed systems. The use of a component-based architecture has enabled the Nimrod-G broker implementations for different middleware technologies such as Globus, Legion, and Condor with a minimal development effort.

The Nimrod tools for modeling parametric experiments are maturing and in production use for cluster computing. A prototype version Nimrod-G resource broker is publicly available [101] and efforts are currently underway to deploy it on production Grids. The Nimrod-G task farming engine (TFE) services have been used for developing customized clients and applications. In addition, the TFE job management protocols and services can be used for developing new scheduling policies. We have developed a number of market-driven deadline and budget constrained scheduling algorithms, namely, cost, time, conservative-time, and cost-time optimizations. The results of scheduling experiments with different QoS requirements show promising insights into the effectiveness of distributed computational economy for management of resources and its usefulness in application scheduling with optimizations.

A number of deadline and budget constrained scheduling experiments have been performed with a variety of requirements at different times by selecting different sets of resources available in the WWG testbed during each experiment. They can be categorised into the following areas:

1. Cost optimisation scheduling during Australian peak and off-peak times,
2. Cost and time optimisation scheduling using cheap local and expensive remote resources,
3. Large scale scheduling using cost and time optimisation algorithms, and
4. Molecular docking application scheduling using cost and time optimisation algorithms.

In a competitive commodity-market economy, the resources are priced differently at different times based on the supply and demand. For example, they are priced higher during peak hours and lower during off-peak hours. In the first experiment, for the same deadline, the results of application scheduling show that the broker is able to process jobs with low cost during Australian off-peak hours and more during peak hours. This means that pricing resources higher during peak hours and lower during off-peak hours motivates users to process their low priority jobs during off-peak. In the second experiment, for a given deadline and budget, we explored application scheduling with the cost and time optimisation strategies. The results show that the broker is able to process earlier with a time-optimisation strategy than the cost-optimisation, but at the expense of a higher financial cost. Similar results have been observed in the third experiment with a large number of geographically distributed heterogenous resources.

The fourth scheduling experiment demonstrates the effectiveness and application of Grid technologies for solving real-world problems by creating a Virtual Laboratory environment. The molecular modelling for drug design application has been formulated as a parameter-sweep application using the Nimrod-G parameter specification language. The experiment then used the Nimrod-G broker to process molecular docking jobs on the Grid. The scheduling experiments with cost and time optimisations using Nimrod-G, demonstrates that the users can indeed express their valuations naturally by defining deadline, budget limits, and optimisation preference. It also proves that Grids indeed enable the sharing and aggregation of geographically distributed resources for solving real-world data-intensive computing problems faster and cheaper.

The GridSim toolkit has been used to evaluate the performance of deadline and scheduling algorithms through a series of simulations by varying the number of users, deadlines, budgets, and optimisation strategies and then simulating geographically distributed Grid resources. The scheduling simulations with varying deadlines and budgets for cost-optimisation strategy show that as the deadline is increased the cost of computation decreases until it reaches the optimal level—i.e., processing all jobs on the cheapest resources. The time-optimisation scheduling showed that as the budget is increased, the completion time decreases and the cost increases. Also, when the number of users competing for the same set of resources increases, there will be proportional impact on others depending on each user’s strategies and constraints. Apart from complementing and strengthening the results of Nimrod-G scheduling studies, the simulations demonstrate the capability of GridSim and the ease with which it can be used to develop and evaluate the performance of new scheduling algorithms.

The results of scheduling applications with different QoS requirements demonstrates that market-based systems, such as the Nimrod-G broker, allow users to trade-off QoS parameters, deadlines and computational costs, and offer an incentive for relaxing their requirements—reduced computational cost for relaxed deadline when timeframe for earliest results delivery is not too critical. This approach of offering an economic incentive for resource owners to share their resources and resource users to trade-off between the deadlines and budgets, promotes the Grid as a platform for mainstream computing. This could in turn help lead to the emergence of a new service oriented computing industry.

The results of these experiments also demonstrate that the computational economy framework in the Grid environment helps in regulating the supply-and-demand for resources and offers an incentive to resource owners to share their resources and resource users to think about trade-off between the deadline and budget. It provides a decentralized resource management capability and is adaptable to changes in the Grid environment and user requirements. The economic-based Grid system is scalable, controllable, measurable, and uses easily understandable policy for management of resources.

The realization of the GRACE framework by utilising existing technologies such as Globus and providing new services that are essential for resource trading and the aggregation of resources, demonstrates that an economic-based Grid resource management systems can be developed and deployed. It also demonstrates that future network computing applications will have the capability to select resources/services dynamically at runtime depending on their availability, capability, cost, and the users QoS requirements.

## 8.3 Future Directions

This thesis formulated a comprehensive distributed computational economy architectural framework and strategies for service-oriented Grid computing. It demonstrated the benefits of developing economic-based Grid systems for distributed resource management and scheduling. A number of deadline and budget constrained scheduling algorithms for different optimisation strategies have been developed to meet user's QoS requirements.

This work has laid a foundation for the Grid economy and it opens up several avenues for future work in economic-based Grid resource management and scheduling.

### 8.3.1 Supporting Different Application Models

While a parameter-based parallel application model is a dominant model for many applications (e.g., molecular modelling, protein folding, high-energy physics, data mining, design explorations, and structural engineering) that are being used to explore the use of the Grid, there are applications (e.g., computational fluid dynamics) that need a different application model. These applications have tasks that need to communicate frequently with each other and may have interdependencies. To execute such tasks, resources need to be co-allocated to enable communications between tasks at runtime. This introduces various complexities into resource management and scheduling. To overcome these complexities an advance reservation capability is needed, but it is hard to get resources at multiple sites for co-allocation since each resource has a different allocation policy. We believe that this is where computational economy can play a greater role. It could possibly help in prioritizing allocations by supporting the cancellation of existing allocations, if penalties are less than the benefits, and encourage resource providers to create alliances to support co-reservation and allocation.

### 8.3.2 Supporting Different Economic Models

While a commodity economic model is expected to be the dominant model for pricing resources, similar to that of the Internet where access pricing is based on a flat-price model with some variance and large-slot size, a computational economy approach for Grid resource management requires extensive exploration. For example, currently the Nimrod-G scheduler does not support changes in the price of resources dynamically within a small period, less than the job execution time, once initial scheduling decisions are made. This is because, in scheduling the remaining jobs over the resources within the remaining budget, the scheduler assumes that the price of resources does not change. In addition, the scheduler uses the current price to calculate the cost of jobs that have completed in the past. Hence, using the current scheduler in a system where prices vary and cannot be guaranteed, cost estimations become meaningless and the budget cannot be guaranteed. In order to overcome this limitation, new scheduling strategies and algorithms that learn from historical and market dynamics are needed. These strategies and algorithms should not only be able to adapt dynamically to the changes in resource conditions at runtime, but also to changes in access prices, even during the execution of jobs.

There exist a few other economic models such as auctions, contract net, and bid-based proportional resource allocation for resource trading. Previous research in market-based systems have explored such frameworks, but they expect the user to build applications explicitly using a market-oriented programming framework, where the programmer has to develop a budget allocation strategy for each task and create bids. This makes application development harder and time consuming. Tools similar to Nimrod-G need to be developed that automatically take budget allocations into account and work with different economic models.

It is expected that Nimrod-G will be enhanced to support scheduling with advance resource reservation. Plans are also underway to explore other economic models such as tenders/contract-net and auctions for resource brokering. These new models require new scheduling algorithms.

### 8.3.3 Accounting

Within the Grid community, there is a great interest in building an accounting model and infrastructure. In the GRACE framework, we have proposed the concept of Grid Market Directory, Grid Bank (GB), and automated payment mechanisms. This is similar to a debit/credit card company mediating payments, i.e., the buyer presents the card to purchase items, the seller requests the credit/debit card company to make payment, and the credit/debit card company then claims aggregated amount periodically from the customer. We propose a similar model for implementing the Grid Bank and automating payments.

In a Grid environment, both resource owners and consumers need to have an account in the GB, which can record activities. Under such a framework, the Nimrod-G user, having a unique identity in the Grid, submits their application to the broker along with the deadline and budget. When the broker schedules a job on the resource, it can inform the resource owner about its GB account details to which expenses should be charged. The GB admission control model having the record of resource usage details can then charge the user account. The Nimrod-G broker agent also maintains the record of resource usage details for each job, to which it can refer for verifying charges if there is a fraud.

To support anonymous online payments, we need digital currency. Although we have not developed digital currency, we note that electronic currency technology is rapidly progressing with emerging e-commerce infrastructures [98] such as NetCash, NetCheque, and Paypal. When electronic currency is available, Nimrod-G can be enhanced to automate the payments with minimal effort.

### 8.3.4 Enhancing GridSim to Support QoS based Resource Entities

The GridSim toolkit is rapidly evolving. The network model needs to be enhanced by supporting various types of networks with different static and dynamic configurations and cost-based QoS services. To enhance resource model entity with file I/O operation, off-the-shelf storage I/O simulators need to be incorporated. GridSim currently supports a framework for a resource model with advance reservation. To enable the simulation of Grid resource management with economic models such as tenders and auctions, the FIPA (Foundation for Intelligent Physical Agents) standards [30] based interaction protocol infrastructure can be integrated, along with necessary enhancements to the resource model to support admission control.

### 8.3.5 Wide-Area Data-Intensive Programming and Scheduling Framework

An experience in developing a prototype Virtual Laboratory environment for distributed drug design demonstrates the ease of use and applicability of the Nimrod-G tools for data intensive computing on the Grid. The current system can be extended to support adaptive mechanisms for the selection of the best CDB service depending on the access speed and cost. A new project, called the HEPGrid (High Energy Physics and the Grid Network), has been initiated to develop a virtual laboratory environment for enabling high-energy physics events processing on distributed resources on a larger scale [96].

We expect that the economy driven approach to resource management and scheduling will make a great impact on the eventual success and widespread adoption of the Grid in day-to-day computational activities.

# Appendix A

## Tools automate computer sharing♦

How many economists does it take to inaccurately forecast a recession?

To answer that impertinent puzzler, you might try tapping into a Grid of computing power made up of spare cycles from sources as disparate as a university supercomputer across town, a cluster of servers in another state, and a scattering of workstations around the world. You'll also need some stray disk storage and spare networking resources to tie it all together.

Grid computing started as a response to scientific users' need to pull together large amounts of computing power to tackle complex applications. These ad hoc assemblages of distributed resources are coordinated by software that mediates different computer operating systems and manages things like scheduling and security to create sophisticated, virtual computers.

Grid computing, still generally confined to the research community, is one manifestation of utility-style data processing services made possible by the Internet. Peer-to-peer computing, which allows disparate users to dedicate portions of their computers to cooperative processing via the Internet, is a related phenomenon used mostly by consumers and businesses.

Both models harness a potentially vast amount of computing power in the form of excess, spare or dedicated system resources from the entire range of computers spread out across the Internet. The University of California at Berkeley, for example, coordinates one popular scientific example of Grid computing -- an Internet community application that uses background or downtime resources from thousands of systems, many of them home PCs, to analyze telescope data for the search for extraterrestrial intelligence (SETI) project.

A group of researchers at Monash University in Australia and the European Council For Nuclear Research (CERN) in Switzerland has proposed a scheme that has the potential to increase the reach of Grid computing by applying traditional economic models - from barter to monopoly - to manage Grid resource supply and demand.

The researchers have built a software architecture and mapped out policies for managing grid computing resources; these could also work with peer-to-peer applications, according to Rajkumar Buyya, a graduate student in the computer science department at Monash University.

♦ This article by Ted Smalley Bowen appeared in the Technology Review News, September 12, 2001. Its content is driven by our work on economic models for grid computing. It has interesting remarks from an American Law & Diplomacy professor.

The methods could facilitate a broad range of computing services applications, said Buyya.

“They can be used in executing science, engineering, industrial, and commercial applications such as drug design, automobile design, crash simulation, aerospace modeling, high energy physics, astrophysics, earth modeling, electronic CAD, ray tracing, data mining, financial modeling, and so on,” he said.

Although peer-to-peer and grid computing are not new, there hasn't been an overarching scheme for handling the massive amount of bargaining and staging required to carry out such on-demand jobs with reliable levels of quality, and pricing to match, Buyya said.

The researchers' scheme is aiming to fill that gap, he said. “We are focusing on the use of economics as a metaphor for management of resources and scheduling in peer-to-peer and grid computing, as... a mechanism for regulating supply-and-demand for resources depending on users'... requirements.”

The researchers scheme allows consumers and computing service providers to connect and hammer out pricing and service levels. It would allow the parties to agree on one price for quick delivery of services during times of peak demand, and another for less urgent delivery, for example.

Resource brokering/sharing tools analogous to Napster will eventually handle the trade in access to computers, content, scientific and technical instruments, databases, and software, Buyya said.

“With new technologies, the users need not own expensive [computer] resources. Resource brokers [can] lease services that are necessary to meet... requirements such as deadline, spending limit, and importance of the work. Our technologies help both resource consumers and providers to manage the whole scenario automatically,” he said.

In a grid computing scheme, consumers usually enlist brokers to procure computing resources for a given project. Grid service providers make their systems available by running specialized applications and resource trading services. A grid market directory links brokers and providers.

The researchers' grid architecture goes a step further, using standard economic pricing models, such as commodity market, posted price, bargaining, tendering and auctions, to hash out the terms of broker-provider deals.

The researchers' tools, Nimrod-G Computational Resource Broker, DataGrid broker, Grid Trading Services, Grid Market Directory, and Grid Bank, work with existing grid middleware like the Globus toolkit.

The researchers have tested the tools on the World Wide Grid (WWG), a global network testbed of different types of computers including PCs, workstations and servers.

Two types of tests simulated brokering, scheduling and execution computing jobs, and emphasized speed and cost, respectively. The tests used a commodity market pricing, or fixed-price model. One application scheduled computations needed for a drug design application that screened molecules, he said.

The researchers used Nimrod-G to aggregate the systems resources as they were needed. “The resource broker automatically leases necessary resources competitively, depending on the [users'] requirements, such as deadline and budget constraints," Buyya said.

Using a more common systems-centric approach would make it more difficult to provide service levels that can vary from user to user and application to application, depending on the importance of the problem at the time of execution, he said.

As the tools get established, they could be deployed for use in production systems such as Australian Partnership for Advanced Computing (APAC) and Victorian Partnership for Advance Computing (VPAC) resources for routine use, said Buyya. “Depending on market forces, we believe that it will take two or three years for widespread use of economic models for Grid and [peer-to-peer] computing,” said Buyya.

The researchers plan next to test the methods’ scalability, improve scheduling algorithms, and update the Nimrod-G broker software to handle more sophisticated task allocation and management, Buyya said.

The study makes a good start at hashing out ways in which disparate computing resources can be made available and consumed, according to Lee McKnight, a professor at Tufts University’s Fletcher School of Law &Diplomacy.

The researchers' contribution is "imagining and testing a standards or protocol-based framework through which computing resources may be accessed or shared on the basis of one of a variety of different models for brokering or trading resources,” he said.

But the way the researchers used the models is artificially limited to narrowly defined grid computing resources and doesn't address networked computing services like application hosting and bandwidth brokering, and quality controls like service level agreements, said McKnight.

The work "is but one element of a yet-to-be defined economic model of pervasive computing and communications environments,” he said. "The 'data economy' as the authors call it will ultimately include both [peer-to-peer and] a variety of other interaction and resource access modes.”

Buyya’s research colleagues were Jonathan Giddy and David Abramson of Monash University and Heinz Stockinger of CERN.

The work was funded by the Australian Government, Monash University, Cooperative Research Center (CRC) for the Enterprise Distributed Systems Technology (DSTC), and the Institute of Electrical and Electronics Engineers (IEEE) Computer Society. Heinz Stokinger's work was funded by CERN and the European Union.

The researchers are scheduled to present their work at the International Society for Optical Engineering (SPIE) International Symposium on The Convergence of Information Technologies and Communications (ITCom 2001) in Denver, August 20-24, 2001.

# Appendix B

## Toolset teams computers to design drugs♦

Computational grids provide the raw material for assembling temporary, virtual computers from sometimes far-flung resources connected to the Internet or private networks. They came about because researchers often require processing power, storage, and bandwidth far beyond the scope of their own systems.

This type of distributed computing, which can also include scientific instruments, makes the means to tackle complex applications available on an ad hoc basis, and allows researchers to draw on widely-dispersed stores of information.

The molecular modeling programs used to design drugs are especially data-hungry and computationally intensive applications. Designing a drug involves screening massive databases of molecules to identify pairs that can be combined, and figuring out the best way to combine them to achieve a certain affect. The molecules could be enzymes, protein receptors, DNA, or the drugs designed to act on them.

During this molecular docking process, researchers try to match the generally small molecules of prospective drugs with the larger biological molecules they are designed to affect, such as proteins or DNA. These searches can entail sifting through millions of files that contain three-dimensional representations of the molecules.

A group of researchers in Australia has put together a set of software tools to perform molecular docking over a computational grid. The tools tap into remote databases of chemical structures in order to carry out the molecular matching process.

Grid computing software finds and accesses resources from networked computers that can be physically located almost anywhere. It coordinates scheduling and security among systems that may be running different operating systems, to combine, for example, the processing capabilities of half a dozen Unix servers and a supercomputer with databases stored in a collection of disk drives connected to yet another computer.

The researchers adapted a molecular docking program to work on a grid configuration by having it run several copies of a molecular matching program on different systems or portions of systems. The software performed many computations at once on different subsets of the data, then combined the results. This type of parallel processing, also known as a parameter sweep, enabled the grid application to work through the

♦ This article by Ted Smalley Bowen appeared in the Technology Review News, September 12, 2001. Its content is driven by our work on virtual laboratory for drug design on Grid. It has interesting remarks on our work from American researchers.

matching process more quickly.

The complexity of each molecule record and the scale of the database searches involved in molecular docking put such applications beyond the reach of most labs' conventional computing resources, according to Rajkumar Buyya, a research scientist at Monash University in Australia. "Screening each compound, depending on structural complexity, can take hours on a standard PC, which means screening all compounds in a single database can take years."

Even on a supercomputer, "large-scale exploration is still limited by the availability of processing power," he said.

Using a computational grid, however, researchers could feed extensive computing jobs to a coordinated mix of PCs, workstations, multiprocessor systems and supercomputers, in order to crunch the numbers simultaneously.

A drug design problem that requires screening 180,000 compounds at three hours each would take a single PC about 61 years to process, and would tie-up a typical 64-node supercomputer for about a year, according to Buyya. "The problem can be solved with a large scale grid of hundreds of supercomputers in a day," he said.

To run the docking application on a computational grid, the researchers developed a program to index chemical databases, and software for accessing the chemical databases.

To speed the scheme, the researchers replicated the chemical database so that more requests for database information could be processed at once. To further speed the process, the researchers wrote a database server program that allowed computers to field more than one database query at a time.

The researcher's scheme compensates for the uneven bandwidth, processing speeds, and available resources among grid-linked systems by mapping the location of files and selecting the optimal computer to query, according to Buyya. "The data broker assists in the discovery and selection of a suitable source... depending on... availability, network proximity, load, and the access price," he said.

Because the performance of database applications suffers over network connections, the researchers generated indices for each chemical database, including references to each record's size.

This allowed each computer to respond to queries by first checking the index file for the record's size and location and then accessing the record directly from the database file, rather than sequentially sifting through the database, said Buyya.

The application requirements and the tools used to meet them are specific to molecular docking, but similar software would speed compute-intensive tasks like high-energy physics calculations and risk analysis, according to Buyya.

The researchers tested the scheduling portion of their scheme on the World Wide Grid test-bed of systems in Australia, Japan and the US, and successfully estimated the time and cost required to run the applications in configurations optimized for speed and for budget, Buyya said.

Using the test bed, they screened files of 200 candidate molecules for docking with the target enzyme endothelin-converting enzyme (ECE), which is associated with low blood pressure.

The researchers' use of grid computing tools to automate molecular docking is "an excellent application of grid computing," said Julie Mitchell, an assistant principal research scientist at the San Diego Supercomputer Center. Features like "deadline- and budget-constrained scheduling should make the software very attractive to pharmaceutical companies" and to companies interested in such computationally demanding applications as risk analysis, scientific visualization and complex modeling said Mitchell. "There's nothing specific to molecular biology in their tools, and I imagine they could be applied quite readily in other areas."

The researchers also handled the process management aspects of adapting the applications to grids well, she added.

"The [researchers'] approach is obviously the way to go for those type of applications on the Computational Grid," said Henri Casanova, a research scientist in the computer science and engineering department of the University of California at San Diego. "The notion of providing remote access to small portions of domain-specific databases is clearly a good idea and fits the molecular docking applications," he said.

The economic concepts underlying the scheduling and costing of grid applications application are still immature, Casanova added. "The results concerning application execution are based on a Grid economy model and policies that are not yet in place. There are only vague notions of "Grid credit unit" in the community and the authors of the paper assume some arbitrary charging scheme for their experiments. This is an interesting avenue of research, but...there is very little in terms of Grid economy that is in place at the moment," he said.

The data access and computation techniques are technically ready to be used in practical applications today, according to Buyya.

Buyya's research colleagues were Jon Giddy, and David Abramson of Monash University in Australia and Kim Branson of the Walter and Eliza Hall Institute, in Australia. The research was funded by the Australian Cooperative Research Center for Enterprise Distributed Systems Technology (EDST), Monash University, the Walter and Eliza Hall Institute of Medical Research, the IEEE Computer Society, and Advanced Micro Devices Corp.